\documentclass[12pt,a4paper,notitlepage]{article}

\usepackage{graphicx}
\usepackage{amssymb}
\usepackage{amsmath}

\usepackage{amsthm}
\usepackage{amsfonts}
\usepackage{times}

\usepackage[format=hang]{subcaption}
\usepackage{color}
\usepackage[T1]{fontenc}
\usepackage{hyperref}
\usepackage[square]{natbib}
\usepackage{upgreek}

\usepackage{pgfplots}
\usepackage{bm}

\usepackage{float}

\usepackage{todonotes}

\usepackage{subcaption} 
\usepackage{siunitx}
\usepackage{tikz}
\usepackage{tikz-3dplot}
\usetikzlibrary{calc,backgrounds,spy}

\defcitealias{KlockeZeisEtAl2014}{Klocke, Zeis et~al.}
\defcitealias{KlockeKlinkEtAl2014}{Klocke, Klink et~al.}
\defcitealias{HolthusenBrepolsEtAl2022a}{Holthusen et~al.}
\defcitealias{HolthusenBrepolsEtAl2022b}{Holthusen et~al.}
\defcitealias{BarfuszBrepolsEtAl2021}{Barfusz et~al.}
\defcitealias{BarfuszvanderVeldenEtAl2021}{Barfusz et~al.}
\defcitealias{BarfuszvanderVeldenEtAl2021}{Barfusz et~al.}
\defcitealias{vanderVeldenRommesEtAl2021}{van der Velden et~al.}
\defcitealias{vanderVeldenRitzertEtAl2023}{van der Velden et~al.}

\pgfrealjobname{paper} 
\pgfplotsset{compat=1.14}

\definecolor{rwth1}{RGB}{0,84,159}      
\definecolor{rwth2}{RGB}{142,186,229}   
\definecolor{rwth3}{RGB}{0,97,101}      
\definecolor{rwth4}{RGB}{0,152,161}     
\definecolor{rwth5}{RGB}{87,171,39}     
\definecolor{rwth6}{RGB}{189,205,0}     
\definecolor{rwth7}{RGB}{255,237,0}     
\definecolor{rwth8}{RGB}{246,168,0}     
\definecolor{rwth9}{RGB}{227,0,102}     
\definecolor{rwth10}{RGB}{204,7,30}     
\definecolor{rwth11}{RGB}{161,16,53}    
\definecolor{rwth12}{RGB}{97,33,88}     
\definecolor{rwth13}{RGB}{122,111,172}  

\definecolor{modelC}{RGB}{0,84,159}      
\definecolor{modelA}{RGB}{87,171,39}     
\definecolor{modelB}{RGB}{246,168,0}     

\definecolor{sfb1}{RGB}{0,84,165}      
\definecolor{sfb2}{RGB}{201,0,35}      
\definecolor{sfb3}{RGB}{231,95,1}      
\definecolor{sfb4}{RGB}{127,127,127}   
\definecolor{sfb5}{RGB}{217,217,217}   

\tikzstyle{dashpattern0} = [dash pattern = ]
\tikzstyle{dashpattern1} = [dash pattern = on 4.25pt off 0.75pt]
\tikzstyle{dashpattern2} = [dash pattern = on 1.5pt off 0.5pt]
\tikzstyle{dashpattern3} = [dash pattern = on 0.75pt off 0.4pt]
\tikzstyle{dashpattern4} = [dash pattern = on 3pt off 1pt on 1pt off 1pt]
\tikzstyle{dashpattern5} = [dash pattern = on 3.75pt off 0.5pt on 0.75pt off 0.5pt on 0.75pt off 0.5pt]
\tikzstyle{dashpattern6} = [dash pattern = on 3.25pt off 0.5pt on 0.75pt off 0.5pt on 0.75pt off 0.5pt on 0.75pt off 0.5pt]
\tikzstyle{dashpattern7} = [dash pattern = on 3.25pt off 0.5pt on 0.75pt off 0.5pt on 0.75pt off 0.5pt on 0.75pt off 0.5pt on 0.75pt off 0.5pt]
\tikzstyle{dashpattern8} = [line cap=round, dash pattern = on 3.25pt off 2.75pt]
\tikzstyle{dashpattern9} = [line cap=round, dash pattern = on 0.01pt off 2pt]
\tikzstyle{dashpattern10}= [line cap=round, dash pattern = on 3.25pt off 2pt on 0.01pt off 2pt]
\tikzstyle{dashpattern11}= [line cap=round, dash pattern = on 3.5pt off 1.75pt on 0.01pt off 1.75pt on 0.01pt off 1.75pt]
\tikzstyle{dashpattern12}= [line cap=round, dash pattern = on 3.5pt off 1.75pt on 0.01pt off 1.75pt on 0.01pt off 1.75pt on 0.01pt off 1.75pt]
\tikzstyle{dashpattern13}= [line cap=round, dash pattern = on 3.5pt off 1.75pt on 0.01pt off 1.75pt on 0.01pt off 1.75pt on 0.01pt off 1.75pt on 0.01pt off 1.75pt]

\newcommand{\dV}{\mathrm{d}V}

\newcommand{\pd}[2]{\displaystyle\frac{\partial #1}{\partial #2}}

\newcommand{\D}{\bm{D}}

\newcommand{\I}{\mathbf{I}}

\renewcommand{\d}{d}

\newcommand{\Div}[1]{{\rm Div} \hspace{-0.5mm} \left( #1 \right)}
\newcommand{\dev}[1]{{\rm dev} \hspace{-0.5mm} \left( #1 \right)}
\renewcommand{\exp}[1]{{\rm exp} \hspace{-0.5mm} \left( #1 \right)}

\newcommand{\Grad}[1]{{\rm Grad} \hspace{-0.5mm} \left( #1 \right)}
\renewcommand{\ln}[1]{{\rm ln} \hspace{-0.5mm} \left( #1 \right)}
\newcommand{\sym}[1]{{\rm sym} \hspace{-0.5mm} \left( #1 \right)}
\newcommand{\tr}[1]{{\rm tr} \hspace{-0.5mm} \left( #1 \right)}
\newcommand{\bigtr}[1]{{\rm tr} \hspace{-0.0mm} \big( #1 \big)}
\newcommand{\Bigtr}[1]{{\rm tr} \hspace{-0.0mm} \Big( #1 \Big)}

\newcommand{\Tzd}{^{\stackrel{23}{\mathrm{T}}}}

\newcommand{\dA}{\mathrm{d}A}
\renewcommand{\dV}{\mathrm{d}V}

\newcommand{\intGtn}{\int_{\Gamma_{t0}}}

\newcommand{\intOn}{\int_{\Omega_0}}

\renewcommand{\pd}[2]{\displaystyle\frac{\partial #1}{\partial #2}}

\newcommand{\ah}{a_h}

\newcommand{\cd}{{c_d}}
\renewcommand{\d}{\mathbf{d}}
\newcommand{\dbar}{\bar{\mathbf{d}}}
\newcommand{\dbardot}{\dot{\bar{\mathbf{d}}}}
\newcommand{\dbari}{\bar{d}_i}

\newcommand{\e}{\bm{e}}
\newcommand{\ed}{{e_d}}

\newcommand{\fn}{\bm{f}_0}

\newcommand{\gddot}{\dot{\gamma}_d}
\newcommand{\gdbar}{g_{\bar{d}}}
\newcommand{\gu}{g_u}
\newcommand{\Hd}{H_d}

\newcommand{\Kh}{K_h}
\newcommand{\ndbar}{{n_{\bar{d}}}}

\newcommand{\nh}{n_h}
\newcommand{\niY}{\bm{n}_i^{\Y}}
\newcommand{\Phid}{\Phi_d}
\newcommand{\psidot}{\dot{\psi}}
\newcommand{\psie}{\psi_e}
\newcommand{\psid}{\psi_d}
\newcommand{\psih}{\psi_h}
\newcommand{\psidbar}{\psi_{\bar{d}}}

\newcommand{\rd}{r_d}
\newcommand{\Rd}{R_d}

\newcommand{\sd}{s_d}
\newcommand{\tn}{\bm{t}_0}
\renewcommand{\u}{\bm{u}}
\newcommand{\vardbar}{\delta\dbar}
\newcommand{\varE}{\delta\bm{E}}
\newcommand{\varu}{\delta\bm{u}}
\newcommand{\xid}{\xi_d}
\newcommand{\xiddot}{\dot{\xi}_d}
\newcommand{\xinc}{\bm{\xi}_{0_c}}
\newcommand{\xine}{\bm{\xi}_{0_e}}
\newcommand{\xini}{\bm{\xi}_{0_i}}
\newcommand{\Xine}{\bm{\Xi}_{0_e}}
\newcommand{\Xini}{\bm{\Xi}_{0_i}}

\newcommand{\Yn}{Y_0}

\newcommand{\A}{\mathbb{A}}

\newcommand{\C}{\bm{C}}

\renewcommand{\D}{\bm{D}}
\renewcommand{\Ddot}{\dot{\bm{D}}}

\newcommand{\F}{\bm{F}}

\renewcommand{\I}{\bm{I}}

\renewcommand{\S}{\bm{S}}

\newcommand{\Y}{\bm{Y}}
\newcommand{\Ydbar}{\bm{Y}_{\bar{d}}}
\newcommand{\Ye}{\bm{Y}_e}
\newcommand{\Yh}{\bm{Y}_h}
\newcommand{\Ypos}{\bm{Y}_+}

\AtBeginDocument{}

\date{}

\graphicspath{{./02_Figures/}}

\begin{document}

\author{\large {\parbox{\linewidth}{\centering Tim van der Velden\footnote{Corresponding author: \\ phone: +49 (0) 241 80 25016, fax: +49 (0) 241 80 22001, email: tim.van.der.velden@ifam.rwth-aachen.de}\hspace{1.0mm},
                                               Tim Brepols,
                                               Stefanie Reese,
                                               Hagen Holthusen}}\\[0.5cm]
  \hspace*{-0.1cm}
  \normalsize{\em \parbox{\linewidth}{\centering
    \vspace{2mm}
    Institute of Applied Mechanics, RWTH Aachen University,\\Mies-van-der-Rohe-Str. 1, D-52074 Aachen, Germany
%
    }
  }
}

\title{\LARGE A comparative study of micromorphic gradient-extensions for anisotropic damage\\at finite strains}

\maketitle

\vspace{-3mm}

\small
{\bf Abstract.} {
Modern inelastic material model formulations rely on the use of tensor-valued internal variables. When inelastic phenomena include softening, simulations of the former are prone to localization. Thus, an accurate regularization of the tensor-valued internal variables is essential to obtain physically correct results. Here, we focus on the regularization of anisotropic damage at finite strains. Thus, a flexible anisotropic damage model with isotropic, kinematic, and distortional hardening is equipped with three gradient-extensions using a full and two reduced regularizations of the damage tensor. Theoretical and numerical comparisons of the three gradient-extensions yield excellent agreement between the full and the reduced regularization based on a volumetric-deviatoric regularization using only two nonlocal degrees of freedom.

}


\normalsize

\section{Introduction}
\label{sec:intro}

\textbf{Motivation.\quad}
The prediction of complex material phenomena is, nowadays, based on inelastic material models with tensor-valued internal variables for the description of e.g.~plasticity, viscoelasticity, anisotropic damage, or growth. Yet, finite element simulations of inelastic phenomena without a regularization method suffer from the occurrence of localization when modeling softening in e.g.~plasticity and damage (\cite{deBorstPaminEtAl1999}), or viscoelasticity (\cite{SteifSpaepenEtAl1982}). Analogously to \cite{PohPeerlingsEtAl2011} for small strain plasticity, this work is concerned with the open research question of choosing a regularization for tensor-valued internal variables and focuses on the specific inelastic phenomenon of anisotropic damage.

\textbf{Anisotropic damage modeling.\quad}
Various modeling methodologies have evolved to describe the induced anisotropy due to material degradation.
Formulations based on a split of the volumetric (isotropic) and deviatoric (anisotropic) material response that are separately degraded by two scalar damage variables may be found in e.g.~\cite{CarolRizziEtAl2002}, \cite{LeukartRamm2003}.
Microplane models, see  e.g.~\cite{CarolBazantEtAl1991}, \cite{KuhlRammEtAl2000}, project the macroscopic strain state onto different material planes, where unidirectional constitutive laws are evaluated, and afterwards obtain the macroscopic material response by a homogenization process (cf.~\cite{LeukartRamm2003}).
A multiplicative split of the deformation gradient into elastic and damage related components is used by e.g.~\cite{VoyiadjisPark1999}, \cite{SchuetteBruhns2002}, and \cite{DornWulfinghoff2021}, where the latter consider the inelastic part to consist out of a normal crack and a shear crack contribution.
An effective or fictitious undamaged configuration is introduced by e.g.~\cite{MenzelEkhEtAl2002}, \cite{LangenfeldMosler2020}, \cite{SpraveMenzel2023} to formulate the modeling equations.
Finally, anisotropic damage can be interpreted as an evolving structural tensor, see e.g.~\cite{DesmoratGatuingtEtAl2007} (whose localization properties were investigated in \cite{JirasekSuarez2016}), \cite{BadreddineSaanouniEtAl2015}, \cite{Desmorat2016}, \cite{FassinEggersmannEtAl2019a,FassinEggersmannEtAl2019b}, \cite{ReeseBrepolsEtAl2021}, \cite{HegdeMulay2022}, \citetalias{HolthusenBrepolsEtAl2022a}[\citeyear{HolthusenBrepolsEtAl2022a}], which is also the approach followed in this work.

\textbf{Regularization techniques.\quad}
To remedy the mesh dependence and localization, different approaches can be pursued to account for a nonlocal behavior.
Spatial averaging techniques for a specific quantity are employed in nonlocal integral-type formulations, see e.g.~\cite{PijaudierCabotBazant1987} for a spatial average of the damage driving variable, \cite{BazantPijaudierCabot1988} for a spatial average of the damage variable, and \cite{BazantJirasek2002} for an overview of nonlocal integral-type formulations. Viscous regularization approaches may be found in e.g.~\cite{Needleman1988}, \cite{GeersBrekelmansEtAl1994}, \cite{NiaziWisselinkEtAl2013}, \cite{LangenfeldJunkerEtAl2018} and peridynamics based formulations that are inherently nonlocal in e.g.~\cite{Silling2000}, \cite{JaviliMcBrideEtAl2021}, \cite{LaurienJaviliEtAl2023}.

Gradient-extended models provide another effective regularization method that incorporate the gradient of a (local) quantity into the formulation. In e.g.~\cite{PeerlingsdeBorstEtAl1996}, the gradient of the equivalent strain and, in \cite{deBorstBenallalEtAl1996}, the gradient of an internal variable are considered. Moreover, the gradient-extension of an anisotropic microplane damage model is presented in \cite{KuhlRammEtAl2000}.
A decisive advancement for gradient-extended material models with respect to their straightforward model incorporation is associated with the works of \cite{DimitrijevicHackl2008,DimitrijevicHackl2011} and \cite{Forest2009,Forest2016}, who introduce a nonlocal counterpart for the local variable, which is to be regularized. The gradient effects act on the nonlocal field and the coupling between the local variable and its nonlocal counterpart is achieved by a penalty approach. Thereby, the local material model formulation is equipped with an additional driving force, but remains otherwise unaltered, which is from the authors' point of view an elegant incorporation of the nonlocal character.

\textbf{Current and future works.\quad}
The search for efficient gradient-extension of tensor-valued internal variables is an active field of research and not restricted to anisotropic damage, but also for e.g.~plasticity still an open question. After the works of e.g.~\cite{WulfinghoffBoehlke2012}, \cite{WulfinghoffForestEtAl2015} for strain gradient plasticity, novel scalar-based gradient plasticity models are presented in \cite{JebahiForest2021}, \cite{AbatourForestEtAl2023}, and \cite{AbatourForest2023}. Further, \cite{FriedleinMergheimEtAl2023} compare different gradient-extensions in the logarithmic strain space for plasticity coupled to damage.
Moreover, gradient-extensions for fiber-reinforced materials are presented by e.g.~\cite{HolthusenBrepolsEtAl2020}, \cite{PoggenpohlBrepolsEtAl2021} and the search for gradient-extended scale-transitions at severe material softening by \cite{PoggenpohlHolthusenEtAl2022}.
In \cite{LangenfeldKurzejaEtAl2022}, three different regularization concepts for brittle damage are compared and, in \cite{SpraveMenzel2023}, gradient-extensions for anisotropic damage and plasticity at finite strains are investigated.

Following the search for a reduced and effective gradient-extension, the model should then be incorporated into structural elements (e.g.~\cite{AldakheelHudobivnikEtAl2019}, \citetalias{BarfuszvanderVeldenEtAl2021}[\citeyear{BarfuszvanderVeldenEtAl2021},\citeyear{BarfuszvanderVeldenEtAl2022}], \cite{KikisAmbatiEtAl2021}) to avoid locking and be combined with a multiphysical framework  (e.g. \cite{DittmannAldakheelEtAl2020} ,\cite{SarkarSinghEtAl2020}, \citetalias{vanderVeldenRommesEtAl2021}[\citeyear{vanderVeldenRommesEtAl2021},\citeyear{vanderVeldenRitzertEtAl2023}]) for holistic production and process simulations. Furthermore, an incorporation of the reduced gradient-extension into the novel iCANN-framework of \cite{HolthusenLammEtAl2023} is aspired.

\textbf{Outline of the work.\quad}
In Section~\ref{sec:modeling}, the constitutive modeling framework of the anisotropic damage model is elaborated for a general gradient-extension. Then, in Section~\ref{sec:gradientextensions}, three specific gradient-extensions are motivated and compared theoretically. Thereafter, in Section~\ref{sec:examples}, the three gradient-extended models are applied to four structural examples and compared with respect to the structural responses and the resulting damage patterns. Finally, a conclusion is presented in Section~\ref{sec:conclusion}. 

\textbf{Notational conventions.\quad}
In this work, italic characters $a$, $A$ denote scalars and zeroth-order tensors and bold-face italic characters $\bm{b}$, $\bm{B}$ refer to first- and second-order tensors. The operators $\Div{\bullet}$ and $\Grad{\bullet}$ denote the divergence and gradient operation of a quantity with respect to the reference configuration. A $\cdot$ defines the single contraction and a $:$ the double contraction of two tensors. The time derivative of a quantity is given by $\dot{(\bullet)}$.

\section{Constitutive modeling}
\label{sec:modeling}

In this Section~\ref{sec:modeling}, we briefly present the brittle model version of \citetalias{HolthusenBrepolsEtAl2022a}[\citeyear{HolthusenBrepolsEtAl2022a}] without specification of its gradient-extension. The core and novelty part of this paper, i.e.~the choice and comparison of different gradient-extensions, is discussed in detail in Section~\ref{sec:gradientextensions}.

\subsection{Strong and weak forms}
\label{sec:strongweak}

The gradient-extension in this work is incorporated following the micromorphic approach of \cite{Forest2009,Forest2016} using a nonlocal micromorphic tuple (see \citetalias{HolthusenBrepolsEtAl2022a}[\citeyear{HolthusenBrepolsEtAl2022a}]). Thus, the balance of linear momentum, stated in the reference configuration, reads
\begin{align}
  \Div{\F\S} + \bm{f}_0 &= \bm{0}    \hphantom{ \bm{t}_0 \bm{u}'  } \text{in}~\Omega_0    \label{eq:blm}\\
  \F\S \cdot \bm{n}_0   &= \bm{t}_0  \hphantom{ \bm{0}   \bm{u}'  } \text{on}~\Gamma_{t0} \label{eq:blmNBC}\\
  \u                    &= \bm{u}'   \hphantom{ \bm{0}   \bm{t}_0 } \text{on}~\Gamma_{u0} \label{eq:blmDBC}
\end{align}
and, furthermore, the balance of the micromorphic field reads
\begin{align}
  \Div{\Xini - \Xine} - \xini + \xine           &= \bm{0}  \hphantom{ \xinc \dbar'  } \text{in}~\Omega_0          \label{eq:bmm}\\
  \left( \Xini - \Xine \right) \cdot \bm{n}_0   &= \xinc   \hphantom{ \bm{0} \dbar' } \text{on}~\Gamma_{c0}       \label{eq:bmmNBC}\\
  \dbar                                         &= \dbar'  \hphantom{ \bm{0} \xinc  } \text{on}~\Gamma_{\bar{d}0} \label{eq:bmmDBC}
\end{align}
with the primary variables being the displacment $\u$ and the nonlocal micromorphic tuple $\dbar$. Moreover, $\F$ denotes the deformation gradient, $\S$ the Second Piola-Kirchhoff stress, $\bm{f}_0$ the mechanical volume forces, $\bm{n}_0$ the outward normal vector, $\bm{t}_0$ the applied mechanical surface tractions, $\xini$ and $\Xini$ the internal forces related to the micromorphic tuple and its gradient, $\xine$ and $\Xine$ the micromorphic volume forces, and $\xinc$ the micromorphic tractions. Boundary conditions for the primary variables are generally denoted by $(\bullet)'$. However, since $\Gamma_{\bar{d}0} = \emptyset$ is employed, for the micromorphic boundary conditions $\Gamma = \Gamma_{\bar{c}0}$ holds.

Using the test functions $\varu$ and $\vardbar$, the strong forms, Eqs.~\eqref{eq:blm}-\eqref{eq:bmmDBC}, are transferred to their corresponding weak forms under the assumption of a simplified micromorphic balance equation, i.e.~neglecting external and contact forces as well as Dirichlet boundary conditions for the micromorphic tuple, resulting in
\begin{align}
  \gu \left( \u,\dbar,\varu \right)       &:= \intOn \S : \varE \, \dV - \intOn \fn \cdot \varu \, \dV - \intGtn \tn \cdot \varu \, \dA = 0, \\
  \gdbar \left( \u,\dbar,\vardbar \right) &:=  \intOn \xini \cdot \vardbar \, \dV + \intOn \Xini : \Grad{\vardbar} \, \dV = 0.
\end{align}
Finally, for the linearization and finite element discretization the reader is kindly referred to \citetalias{HolthusenBrepolsEtAl2022a}[\citeyear{HolthusenBrepolsEtAl2022a}].

\subsection{Kinematics}
\label{sec:kinematics}

The constitutive framework is based on logarithmic strains and, thereby, facilitates a physically motivated formulation of the elastic energy contribution that distinguishes between isochoric and volumetric damage mechanisms in the finite strain regime considering the damage growth criterion of \cite{WulfinghoffFassinEtAl2017}. Analogously to e.g.~\cite{MieheApelEtAl2002}, the logarithmic strain is defined as
\begin{equation}
  \bm{\eta} := \frac{1}{2} \, \ln{ \C }
\end{equation}
where $\C$ denotes the right Cauchy-Green deformation tensor.

In contrast to the additive split used in finite strain plasticity, see e.g.~\citetalias{HolthusenBrepolsEtAl2022a}[\citeyear{HolthusenBrepolsEtAl2022a}] for ductile damage with logarithmic strains, which is only exactly valid for coaxial loadings, the consideration of solely brittle damage does not rely on kinematic approximations and, hence, the framework in this work is geometrically exact. 

\subsection{Thermodynamically consistent derivation}
\label{sec:thermodynamics}

The model's total Helmholtz free energy $\psi$ is additively split into four contributions 
\begin{equation}
  \psi \left( \bm{\eta}, \D, \xid, \d, \dbar, \Grad{\dbar} \right)
  =
  \psie \left( \bm{\eta}, \D \right) + \psid \left( \xid \right) + \psih \left( \D \right) + \psidbar \left( \d, \dbar, \Grad{\dbar} \right)
  \label{eq:psi}
\end{equation}
where $\psie$ represents the elastic energy depending on the strain $\bm{\eta}$ and the second-order damage tensor $\D$. Next, $\psid$ represents the isotropic damage hardening energy depending on the accumulated damage variable $\xid$. The additional kinematic damage hardening energy $\psih$ (cf.~\cite{HansenSchreyer1994}) ensures that the eigenvalues of the damage tensor are limited to a value of one and that complete failure is described by $\D = \I$ (see \cite{FassinEggersmannEtAl2019a,FassinEggersmannEtAl2019b},\citetalias{HolthusenBrepolsEtAl2022a}[\citeyear{HolthusenBrepolsEtAl2022a}]). Finally, $\psidbar$ represents the micromorphic energy contribution depending on a general local tuple $\d := (d_1, ..., d_{\ndbar})$, a set of $\ndbar$ local invariants formulated in terms of the damage tensor $\D$, and as a corresponding counterpart the nonlocal micromorphic tuple $\dbar := (\bar{d}_1, ..., \bar{d}_{\ndbar})$ and its gradient $\Grad{\dbar}$.

Following a general derivation in this section, the specific forms of the energies are presented in Section~\ref{sec:specificforms}.

The isothermal Clausius-Duhem inequality including the micromorphic extension reads (cf.~\cite{Forest2009,Forest2016})
\begin{equation}
  - \psidot + \bm{\alpha} : \dot{\bm{\eta}} + \xini \cdot \dbardot + \Xini : \Grad{ \dbardot } \geq 0
  \label{eq:CD}
\end{equation}
where the stress power is given in terms of the logarithmic strain rate $\dot{\bm{\eta}}$ and its thermodynamically conjugate force $\bm{\alpha}$.

The rate of the Helmholtz free energy, Eq.~\eqref{eq:psi}, is computed with respect to the rates of its arguments as 
\begin{equation}
  \psidot
  =
  \pd{\psi}{\bm{\eta}} : \dot{\bm{\eta}}
  +
  \pd{\psi}{\D} : \Ddot
  +
  \pd{\psi}{\xid} \, \xiddot
  +
  \pd{\psi}{\dbar} \cdot \dbardot
  +
  \pd{\psi}{\Grad{\dbar}} : \Grad{ \dbardot }.
  \label{eq:rates}
\end{equation}
Please note, that the partial derivative of the energy $\psi$ with respect to the damage tensor $\D$ yields the elastic, the additional damage
hardening and the nonlocal damage driving forces $\Ye$, $\Yh$, and $\Ydbar$ that are defined as
\begin{equation}
  \pd{\psi}{\D}
  = \underbrace{\pd{\psie}{\D}}_{=: - \Ye}
  + \underbrace{\pd{\psih}{\D}}_{=: \Yh}
  + \underbrace{\pd{\psidbar}{\D}}_{=: \Ydbar}
  =:-\Y.
\end{equation}
In Section~\ref{sec:gradientextensions}, we will present and compare the explicit forms of the nonlocal damage driving force $\Ydbar$, since these differ for distinct choices of the micromorphic tuple, i.e.~the gradient-extension, whilst the other damage driving forces $\Ye$ and $\Yh$ remain unchanged.

Thereafter, the rates of Eq.~\eqref{eq:rates} are inserted into the balance equation, Eq.~\eqref{eq:CD}, and yield by repositioning
\begin{align}
  & \hspace{10mm}
      \left( \bm{\alpha} - \pd{\psi}{\bm{\eta}} \right) : \dot{\bm{\eta}}
    + (\underbrace{\Ye - \Yh - \Ydbar}_{=: \Y}) : \Ddot
    + \Rd \, \xiddot \notag \\
  & + \left( \xini - \pd{\psi}{\dbar} \right) \cdot \dbardot
    + \left( \Xini - \pd{\psi}{\Grad{\dbar}} \right) : \Grad{ \dbardot }
  \geq 0.
\end{align}
\textbf{State laws.}~The state laws are obtained by the \cite{ColemanNoll1961} procedure and the argumentation of \cite{Forest2016} as
\begin{equation}
  \bm{\alpha} = \pd{\psi}{\bm{\eta}},
\end{equation}
\begin{equation}
  \xini = \pd{\psi}{\dbar},
\end{equation}
\begin{equation}
  \Xini = \pd{\psi}{\Grad{\dbar}}
\end{equation}
and the reduced dissipation inequality with $\Rd := -\partial\psi/\partial\xid$ as
\begin{equation}
  \Y : \Ddot + \Rd \, \xiddot \geq 0.
\end{equation}
\textbf{Evolution equations.}~For the evolution of the internal variables $\D$ and $\xid$, we define two general convex, zero-valued, and non-negative inelastic potentials $g_{d_1}$ and $g_{d_2}$ in terms of the driving forces $\Y$ and $\Rd$ that yield the evolution equations
\begin{equation}
  \Ddot = \gddot \, \pd{g_{d_1}}{\Y},
\end{equation}
\begin{equation}
  \xiddot = \gddot \, \pd{g_{d_2}}{\Rd}
\end{equation}
where $\gddot$ is the damage multiplier which is obtained by satisfying the damage onset criterion $\Phid(\Y,\Rd) \leq 0$ in accordance with the Karush-Kuhn-Tucker conditions
\begin{equation}
  \gddot \geq 0, \quad \Phid \leq 0, \quad \gddot \Phid = 0.
\end{equation}

\subsection{Specific forms of Helmholtz free energy, damage onset criterion and inelastic potentials}
\label{sec:specificforms}

\textbf{Helmholtz free energy.}~Motivated by e.g.~\cite{Desmorat2016}, \cite{BadreddineSaanouniEtAl2015}, \cite{LeukartRamm2003}, \cite{LemaitreDesmoratEtAl2000} and similar to \cite{Simo1987}, the elastic energy features a physically motivated split into isochoric and volumetric parts to account for the evolution of micro cracks and microvoids separately. Moreover, it fulfills the damage growth criterion (\cite{WulfinghoffFassinEtAl2017}) and reads
\begin{equation}
  \psie =        \mu \, \tr{\dev{\bm{\eta}}^2 \left(\I - \D \right) } \vartheta
        + f_d \, \mu \, \tr{\dev{\bm{\eta}}^2                       } ( 1 - \vartheta )
        + f_d \, \frac{K}{2} \, \tr{\bm{\eta}}^2
\end{equation}
with the isotropic degradation function
\begin{equation}
  f_d = \left( 1 - \frac{\tr{\D}}{3} \right)^{e_d}
\end{equation}
where $\mu$ denotes the elastic shear modulus, $\kappa$ the elastic bulk modulus, $\vartheta$ the degree of damage anisotropy, and $e_d$ the exponent of the isotropic degradation function. Nonlinear and linear isotropic damage hardening are incorporated by
\begin{equation}
  \psid = \rd \left( \xid + \frac{\exp{-\sd \, \xid}-1}{\sd} \right) + \frac{1}{2} \Hd \, \xid^2
\end{equation}
with the damage hardening parameters $\rd$, $\sd$ and $\Hd$. The additional kinematic damage hardening energy is formulated in terms of the eigenvalues $D_i$ of the damage tensor 
\begin{equation}
  \psih = \Kh \sum_{i=1}^3
  \left( - \frac{ \left( 1 - D_i \right)^{1-\frac{1}{\nh}} }{ 1-\frac{1}{\nh} } - D_i + \frac{1}{1-\frac{1}{\nh}} \right)
\end{equation}
where $\Kh$ and $\nh$ are numerical parameters. The micromorphic energy contribution penalizes the difference between the components of the local and the nonlocal tuple by the numerical penalty parameters $H_i$ and incorporates an internal length scale via the gradient of the nonlocal quantity and the materials parameters $A_i$ for each component of the micromorphic tuple up to the total number of nonlocal degrees of freedom $\ndbar$
\begin{equation}
  \psidbar
  =
  \frac{1}{2} \sum_{i=1}^{\ndbar} H_i \left( d_i - \dbari \right)^2
  +
  \frac{1}{2} \sum_{i=1}^{\ndbar} A_i \, \Grad{ \dbari } \cdot \Grad{ \dbari }.
\end{equation}
\textbf{Damage onset criterion.}~The chosen damage onset criterion with damage threshold $\Yn$
\begin{equation}
  \Phid := \sqrt{3} \sqrt{\Ypos : \A : \Ypos} - ( \Yn - \Rd ) \leq 0
  \label{eq:Phid}
\end{equation}
features the option to include distortional damage hardening with the fourth order interaction tensor $\A$ and material parameter $\cd$
\begin{equation}
  \A = \left( \left( \I - \D \right)^\cd \otimes \left( \I - \D \right)^\cd \right)\Tzd
\end{equation}
with the positive semi-definite part of the damage driving force being
\begin{equation}
  \Ypos = \sum_{i=1}^3 \left< Y_i \right> \niY \otimes \niY
\end{equation}
where $\left< \bullet \right> = \mathrm{max}( \bullet, 0 )$.

\textbf{Inelastic potentials.}~The inelastic potential $g_{d_1}$ for the evolution of the damage tensor is chosen in a pseudo-non-associative structure as
\begin{equation}
  g_{d_1} = \frac{ 3 }{ 2 \left( \Yn - \Rd \right) } \, \Ypos : \A : \Ypos
\end{equation}
where the relation $\sqrt{3}\sqrt{\Ypos : \A : \Ypos} = \Yn - \Rd$ obtained from Eq.~\eqref{eq:Phid} for a converged state is utilized to avoid a division by zero in the local iteration (cf.~\cite{ChallamelLanosEtAl2005},\citetalias{HolthusenBrepolsEtAl2022a}[\citeyear{HolthusenBrepolsEtAl2022a}]), when algorithmic differentiation (e.g.~\cite{Korelc2002},\cite{KorelcWriggers2016}) is employed. However, the absolute value and direction of the evolution are identical to choosing an associative evolution equation, i.e.~$\Ddot=\partial \Phid / \partial \Y$. Furthermore, the inelastic potential $g_{d_2}$ for the evolution of the accumulated damage is chosen linearly as
\begin{equation}
  g_{d_2} = \Rd.
\end{equation}

\section{Micromorphic gradient-extensions}
\label{sec:gradientextensions}

\subsection{Motivation}

The novelty of this work lies in the comparison of different gradient-extensions for anisotropic damage with respect to their efficiency and accuracy. To ensure the comparability of the results, the same local anisotropic damage formulation is utilized throughout this work and only the choice of the local micromorphic tuple, i.e.~the selection of local quantities whose localization is prevented by the gradient-extensions, is adapted. Here, we restrict ourselves to invariant-based micromorphic tuples of the damage tensor and are, thus, able to study the effect of different nonlocal damage driving forces.

Other authors, e.g.~\cite{FassinEggersmannEtAl2019a,FassinEggersmannEtAl2019b},\cite{SpraveMenzel2023}, investigated the regularization of a scalar damage hardening variable. However, as pointed out by \cite{FassinEggersmannEtAl2019a}, this procedure can violate the differentiability of the damage onset function when employing associative damage evolution by maximizing the dissipation and is, thus, not considered in this context.

In the following, we present three model formulations (models A, B, and C) with full, using six nonlocal degrees of freedom, and reduced regularization of the damage tensor, using three and two nonlocal degrees of freedom.

Initially, we strive for a rigorous regularization of the damage tensor and, therefore, in model~A, all six independent components of the symmetric second order damage tensor are regularized individually. Thereby, no localization is expected to occur and, furthermore, an accurate reference solution for the reduced regularizations is obtained. A similar procedure can be found in \cite{LangenfeldMosler2020}, where the six independent components of the integrity tensor are regularized. However, a full regularization requires six additional nonlocal micromorphic degrees of freedom and, thus, triples the number of global degrees of freedom compared to the local, purely mechanical problem. Due to this significant increase in degrees of freedom, we aim to reduce the former and to simultaneously maintain the regularization's accuracy.

The idea for the first reduced regularization is based on the uniqueness of the eigenvalues of the damage tensor. A regularization of the former should, thus, also lead to a proper regularization of the entire tensor. For the ease of numerical implementation and since the principal traces of the damage tensor can unambiguously determined from its eigenvalues, model~B utilizes the reduced micromorphic tuple of \citetalias{HolthusenBrepolsEtAl2022a}[\citeyear{HolthusenBrepolsEtAl2022a}]. In this formulation, the micromorphic tuple contains the three principal traces of the damage tensor to each of which a corresponding nonlocal counterpart is introduced. Compared to model~A, model~B requires three nonlocal degrees of freedom less, but still doubles the total number of degrees of freedom compared to the local model.

We, therefore, aim to achieve a further reduction in the required number of nonlocal degrees of freedom and motivate a regularization of the volumetric and deviatoric part of the damage tensor based on two nonlocal degrees of freedom. Since isotropic damage yields by its nature and the sole consideration of microvoids a volumetric damage tensor $D \I$ and requires only a single nonlocal degree of freedom, we aim to capture the damage anisotropy due to the micro cracks by a regularization of the deviatoric part of the damage tensor as is has been suggested for investigation in \citetalias{HolthusenBrepolsEtAl2022b}[\citeyear{HolthusenBrepolsEtAl2022b}]. A further advantage of model~C becomes apparent when considering isotropic damage, since only one nonlocal degree of freedom is non-zero whereas for model~A and B still three nonlocal degrees of freedom are non-zero.

\subsection{Specific micromorphic tuples}

To ensure all models' objectivity, the micromorphic tuples are formulated based on invariants of the damage tensor. For the micromorphic tuple of model~A, we introduce six general structural tensors $\bm{M}_1$, $\bm{M}_2$, $\bm{M}_3$, $\bm{M}_4$, $\bm{M}_5$, and $\bm{M}_6$ that yield
\begin{equation}
  \d^\text{A} = \left(
  \bigtr{\D\bm{M}_1},
  \bigtr{\D\bm{M}_2},
  \bigtr{\D\bm{M}_3},
  \bigtr{\D\bm{M}_4},
  \bigtr{\D\bm{M}_5},
  \bigtr{\D\bm{M}_6}
  \right).
\end{equation}
In order to control the normal and shear components of the damage tensor, we specify the structural tensors according to the Cartesian basis vectors $\e_1$, $\e_2$, and $\e_3$ as
\begin{align}
 \bm{M}_1 &= \e_1 \otimes \e_1, \bm{M}_2 = \e_2 \otimes \e_2, \bm{M}_3 = \e_3 \otimes \e_3, \notag \\
 \bm{M}_4 &= \e_1 \otimes \e_2, \bm{M}_5 = \e_1 \otimes \e_3, \bm{M}_6 = \e_2 \otimes \e_3.
\end{align}
The micromorphic tuple based on the principal traces of the damage tensor of model~B stems from \citetalias{HolthusenBrepolsEtAl2022a}[\citeyear{HolthusenBrepolsEtAl2022a}] and reads
\begin{equation}
  \d^\text{B} = \left(
  \bigtr{\D},
  \bigtr{\D^2},
  \bigtr{\D^3} 
  \right).
\end{equation}
Finally, the micromorphic tuple of model~C with a split of the damage tensor into volumetric and deviatoric part reads
\begin{equation}
  \d^\text{C} = \left(
  \frac{\bigtr{\D}}{3},
  \Bigtr{\dev{\D}^2}
  \right).
\end{equation}

\subsection{Explicit nonlocal damage driving forces}

Next, we compare the explicit forms of the nonlocal damage driving forces that are derived from $\Ydbar = \partial \psidbar / \partial \D$. Their general form depends on the number of elements per micromorphic tuple $\ndbar$ and reads
\begin{equation}
  \Ydbar = \sum_{i=1}^{\ndbar} H_i \left( d_i - \dbari \right) \pd{d_i}{\D}.
\end{equation}
The explicit form of the nonlocal damage driving force of model~A reads under the consideration of the symmetry of $\D$
\begin{align}
  \Ydbar^\text{A} = & H_1 \left( \bigtr{\D\bm{M}_1} - \bar{d}_1 \right) \sym{ \bm{M}_1 } \notag \\
                  + & H_2 \left( \bigtr{\D\bm{M}_2} - \bar{d}_2 \right) \sym{ \bm{M}_2 } \notag \\
                  + & H_3 \left( \bigtr{\D\bm{M}_3} - \bar{d}_3 \right) \sym{ \bm{M}_3 } \notag \\
                  + & H_4 \left( \bigtr{\D\bm{M}_4} - \bar{d}_4 \right) \sym{ \bm{M}_4 } \notag \\
                  + & H_5 \left( \bigtr{\D\bm{M}_5} - \bar{d}_5 \right) \sym{ \bm{M}_5 } \notag \\
                  + & H_6 \left( \bigtr{\D\bm{M}_6} - \bar{d}_6 \right) \sym{ \bm{M}_6 }.
  \label{eq:YdbarA}                                    
\end{align}
With $\partial \, \tr{\bm{D}^i} / \partial \bm{D} = i \, \bm{D}^{i-1}$, the explicit form for model~B reads
\begin{align}
  \Ydbar^\text{B} = & H_1 \left( \bigtr{\D}   - \bar{d}_1 \right) \I   \notag \\
                  + & H_2 \left( \bigtr{\D^2} - \bar{d}_2 \right) 2 \D \notag \\
                  + & H_3 \left( \bigtr{\D^3} - \bar{d}_3 \right) 3 \D^2.
  \label{eq:YdbarB}                                    
\end{align}
And using $\dev{\D} = \D - \tr{\D} / 3 \, \I$, the explicit form for model~C reads
\begin{align}
  \Ydbar^\text{C} = & \frac{H_1}{3} \left( \frac{\bigtr{\D}}{3} - \bar{d}_1 \right) \I   \notag \\
                  +  & H_2 \left( \Bigtr{\dev{\D}^2}   - \bar{d}_2 \right) \left( 2 \D -  \frac{2}{3} \tr{\D} \I \right)
  \label{eq:YdbarC}                  
\end{align}
When comparing the damage driving forces of model~A, B, and C, Eqs.\eqref{eq:YdbarA}-\eqref{eq:YdbarC}, their different structures are evident and, thus, also for identical choices of the parameters \text{$H_1$, ..., $H_{\ndbar}$} and \text{$A_1$, ..., $A_{\ndbar}$} different model responses are to be expected.

\section{Numerical examples}
\label{sec:examples}

The aim of this section is to study the interesting research question whether an accurate regularization of anisotropic damage models can efficiently be obtained by a reduced regularization of the damage tensor. Therefore, we investigate four representative structural examples by utilizing models~A, B, and C and are, thus, able to identify the effect of the gradient-extension with the simulation of the very same boundary value problem with different models. Further, we can directly compare the accuracy of the reduced regularizations (models~B and C) to the reference solution with full regularization (model~A).

The material point behavior of the anisotropic damage model was examined in detail in \citetalias{HolthusenBrepolsEtAl2022a}[\citeyear{HolthusenBrepolsEtAl2022a}] to which we kindly refer the interested reader for further information. The generic material parameters are, unless stated otherwise, adopted from \cite{BrepolsWulfinghoffEtAl2020} and \citetalias{HolthusenBrepolsEtAl2022a}[\citeyear{HolthusenBrepolsEtAl2022a}] and listed in Table~\ref{tab:matpar}. For each example, the internal length scales $A_i$ of models~A and C were identified such that the maximum force of the structural response coincided with the one obtained by model~B. The Taylor series sampling point $\ah$ listed in Table~\ref{tab:matpar} is required for the implementation of the kinematic damage driving force (cf.~\citetalias{HolthusenBrepolsEtAl2022a}[\citeyear{HolthusenBrepolsEtAl2022a}]), but was omitted in the model presentation in Section~\ref{sec:modeling}. In order to avoid snap-backs during the simulation, an artificial viscosity $\eta_v$ is utilized. Comprehensive studies in Sections~\ref{sec:Ex_pwh}~and ~\ref{sec:Ex_an} confirm that the results are unaffected by the artificial viscosity for a choice of $\eta_v = 1~[\si{\MPa\s}]$. The two-dimensional examples in Sections~\ref{sec:Ex_pwh}, \ref{sec:Ex_an} and \ref{sec:Ex_ss} utilize four-node quadrilateral plane-strain elements and the three-dimensional example in Section~\ref{sec:Ex_ts} utilizes eight-node hexahedral elements.

The finite element simulations were conducted using the software \textit{FEAP} (\cite{TaylorGovindjee2020}), new finite element meshes for the example of Section~\ref{sec:Ex_ss} were created with the software \textit{HyperMesh} (\cite{HyperWorks2022}), and post-processing of the simulations' results was carried out with \textit{ParaView} (\cite{AhrensGeveciEtAl2005}).

\begin{table}[htbp]
  \centering
  \caption{Material and numerical parameters}
  \label{tab:matpar}
  \begin{tabular}{l l r r}
    \hline
    Symbol      & Material parameter                           & Value                  & Unit        \\ 
    \hline \hline
    $\mu$       & Elastic shear modulus                        & 55000                  & MPa         \\
    $K$         & Elastic bulk modulus                         & 61666.$\bar{\text{6}}$ & MPa         \\
    $\vartheta$ & Damage anisotropy                            & 0 / 1                  & -           \\
    $\ed$       & Isotropic degradation function exponent        & 1                      & -           \\
    $\Yn$       & Initial damage threshold                     & 2.5                    & MPa         \\
    $\cd$       & Distortional hardening exponent              & 1                      & -           \\
    $\Hd$       & Linear isotropic hardening prefactor         & 1                      & MPa         \\  
    $\rd$       & Nonlinear isotropic hardening prefactor      & 5                      & MPa         \\
    $\sd$       & Nonlinear isotropic hardening scaling factor & 100                    & -           \\
    $\Kh$       & Kinematic hardening prefactor                & 0.1                    & MPa         \\
    $\nh$       & Kinematic hardening exponent                 & 2                      & -           \\  
    $A_i$       & Internal length scales                       & 75 - 1300              & MPa mm$^2$  \\
    \hline
    \vspace{0mm}\\
    \hline
    Symbol      & Numerical parameter                          & Value                  & Unit        \\ 
    \hline \hline
    $\ah$       & Taylor series sampling point                 & 0.999999               & -           \\  
    $H_i$       & Micromorphic penalty parameters              & 10$^\text{4}$          & MPa         \\  
    $\eta_v$    & Artificial viscosity                         & 1                      & MPa s       \\
    \hline
  \end{tabular}
\end{table}

\subsection{Plate with hole specimen}
\label{sec:Ex_pwh}

\begin{figure}[htbp] 
  \centering 
  \begin{subfigure}{.45\textwidth} 
      \centering 
      \begin{tikzpicture}
        \node (pic) at (0,0) {\includegraphics[width=\textwidth]{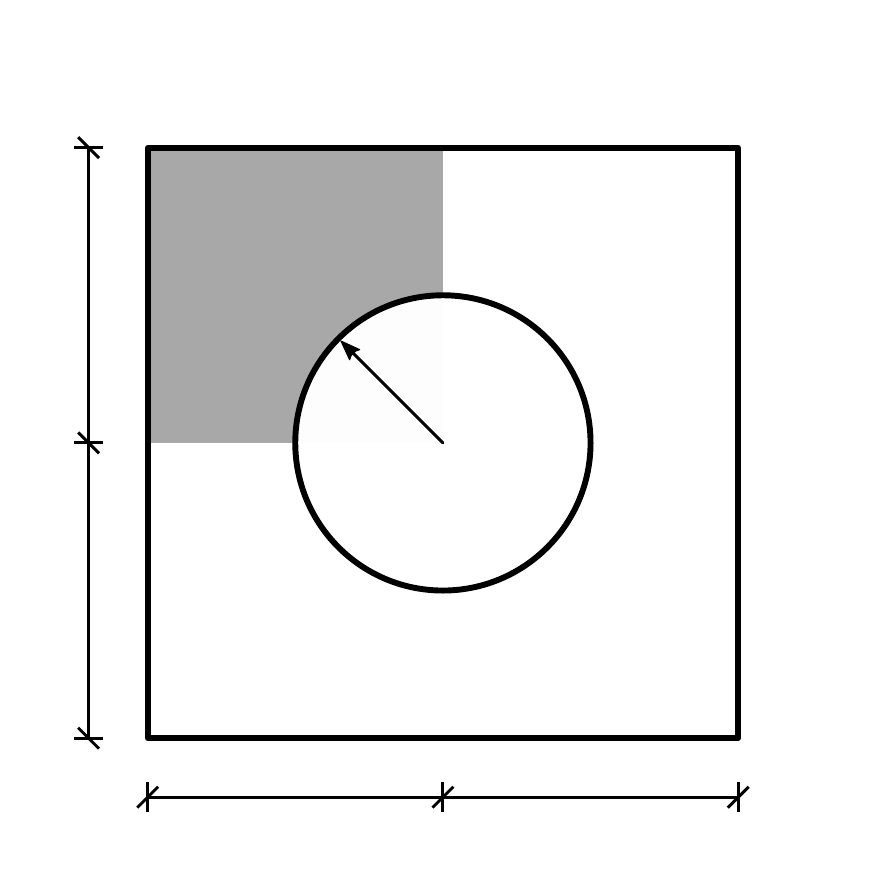}};
        \node (l1)  at ($(pic.west) +( 0.60, 1.20)$)  {$l$};
        \node (l2)  at ($(pic.west) +( 0.60,-1.20)$)  {$l$};
        \node (l3)  at ($(pic.south)+(-1.20, 0.50)$)  {$l$};
        \node (l4)  at ($(pic.south)+( 1.20, 0.50)$)  {$l$};
        \node (r)   at ($(pic.south)+(-0.60, 3.90)$)  {$r$};
      \end{tikzpicture} 
      \caption{Geometry}
      \label{fig:Ex_pwh_geom}
  \end{subfigure}
  \qquad
  \begin{subfigure}{.45\textwidth} 
    \centering 
    \begin{tikzpicture}
      \node[inner sep=0pt] (pic) at (0,0) {\includegraphics[width=\textwidth]
      {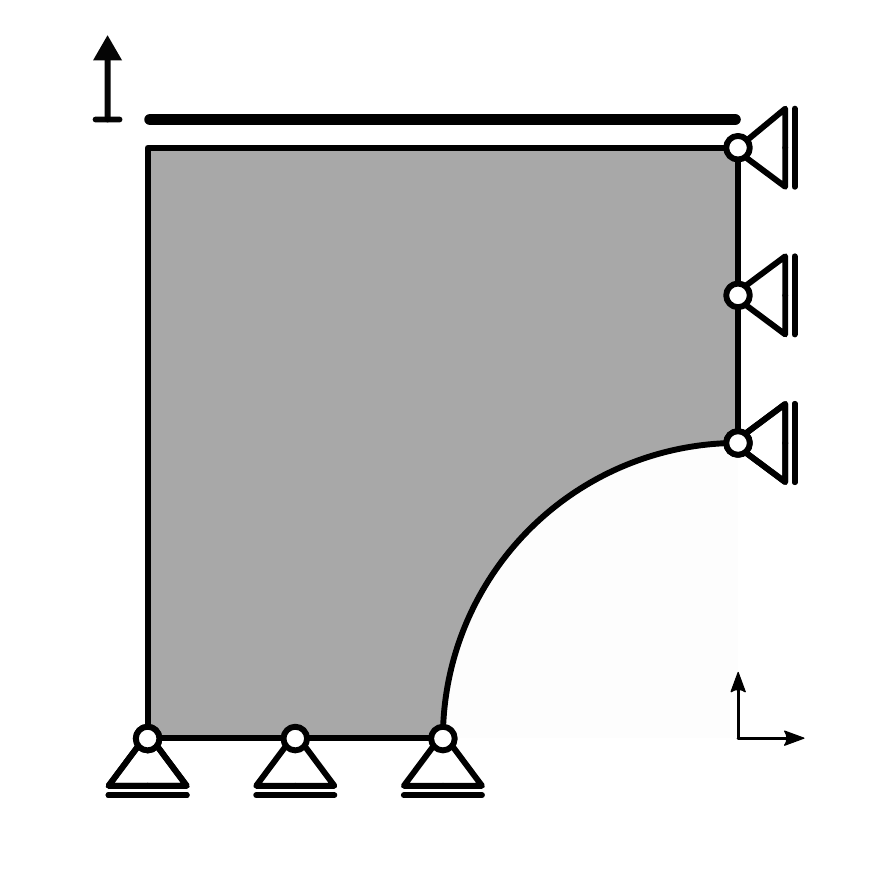}};
      \node[inner sep=0pt] (Fu) at ($(pic.north) +(-3.35,-0.70)$)  {$F,u$};
      \node[inner sep=0pt] (x)  at ($(pic.south) +( 3.20, 1.20)$)  {$x$};
      \node[inner sep=0pt] (y)  at ($(pic.south) +( 2.40, 2.00)$)  {$y$};
    \end{tikzpicture} 
    \caption{Boundary value problem}
    \label{fig:Ex_pwh_bvp}
  \end{subfigure}
  \caption{Geometry and boundary value problem of the plate with hole specimen.} 
  \label{fig:Ex_pwh}
\end{figure}
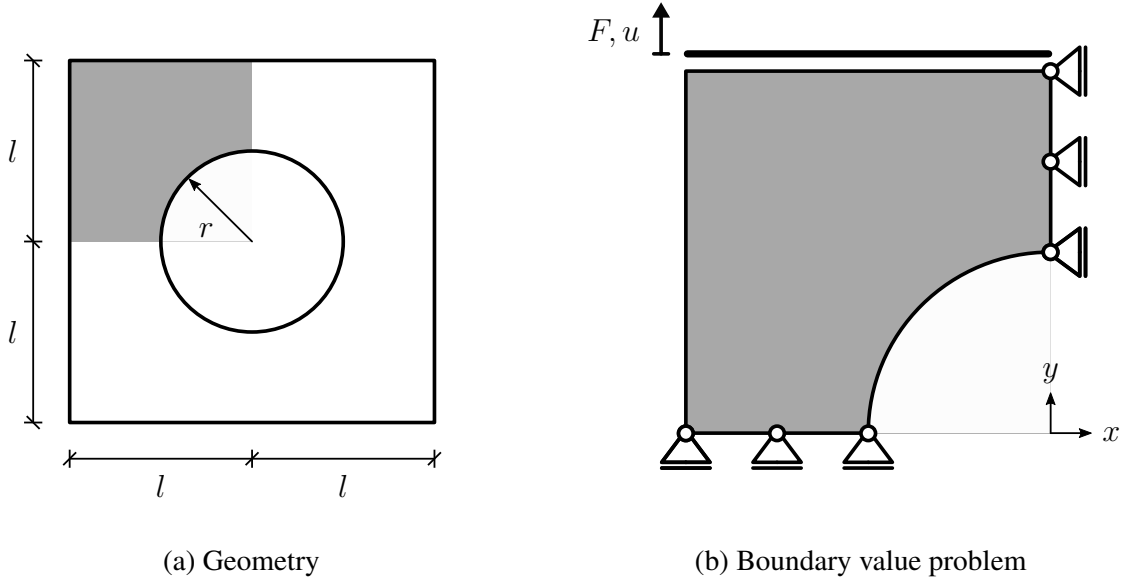

The first example is characterized by a tension dominated loading situation and considers a plate with hole specimen. This example was, in the context of isotropic damage, previously investigated by e.g.~\cite{FriedleinMergheimEtAl2021}, \cite{SpraveMenzel2020}, \cite{KieferWaffenschmidtEtAl2018}, \cite{BrepolsWulfinghoffEtAl2017}, \cite{DimitrijevicHackl2008} and, for anisotropic damage, by e.g.~\cite{SpraveMenzel2023}, \cite{LangenfeldMosler2020}, \cite{FassinEggersmannEtAl2019a}.

Fig.~\ref{fig:Ex_pwh} shows the geometry and the considered boundary value problem. The dimensions read $l = 100~[\si{\mm}]$ and $r = 50~[\si{\mm}]$ with a thickness of $1~[\si{\mm}]$. Due to symmetry, only one quarter of the specimen is modeled in the simulation and the top edge is moved in vertical direction by a prescribed displacement. The finite element meshes stem from \cite{FassinEggersmannEtAl2019a}. The internal length scales of model B are chosen as $A_i^\text{B} = 75~[\si{\MPa\mm\squared}]$ and the parameters of model~A and C are identified as $A_i^\text{A} = 420~[\si{\MPa\mm\squared}]$ and $A_i^\text{C} = 1300~[\si{\MPa\mm\squared}]$.

\begin{figure}[htbp]
  \centering
  \begin{subfigure}{.48\textwidth}
    \centering
    \includegraphics{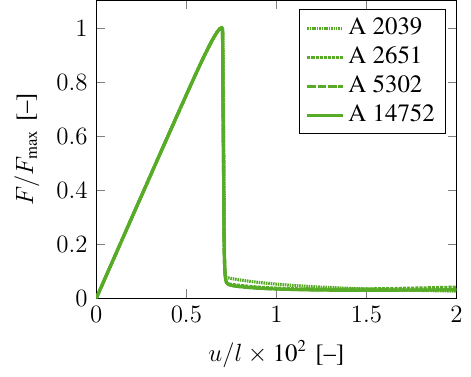}
    \vspace{-7mm}
    \caption{Model~A}
    \label{fig:ExpwhFuA}
  \end{subfigure}
  \quad
  \begin{subfigure}{.48\textwidth}
    \centering
    \includegraphics{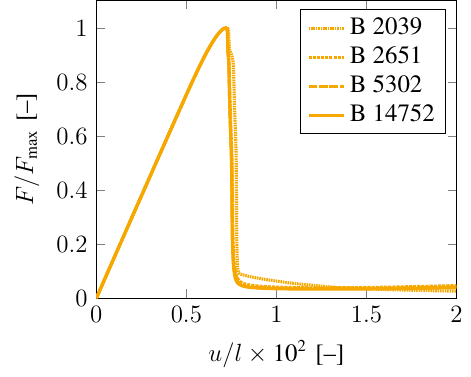}
    \vspace{-7mm}
    \caption{Model~B}
    \label{fig:ExpwhFuB}
  \end{subfigure}%
  \vspace{5mm} 
  \begin{subfigure}{.48\textwidth}
    \centering
    \includegraphics{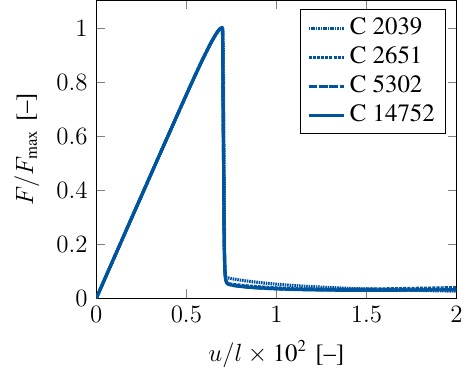}
    \vspace{-7mm}
    \caption{Model~C}
    \label{fig:ExpwhFuC}
  \end{subfigure}
  \quad
  \begin{subfigure}{.48\textwidth}
    \centering
    \includegraphics{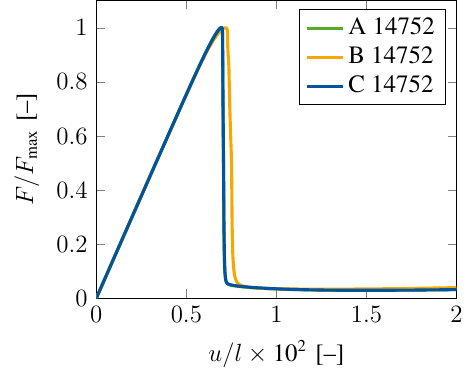}
    \vspace{-7mm}
    \caption{Model comparison}
    \label{fig:ExpwhFuComp}
  \end{subfigure}
  \caption{Mesh convergence studies for the plate with hole specimen and model comparison. The forces are normalized with respect to the maximum force of the finest mesh (14752 elements) of model~B with $F_\text{max} = 5.0767 \times 10^4~[\si{\newton}]$.}
  \label{fig:ExpwhFu}
\end{figure}

In Fig.~\ref{fig:ExpwhFu}, the normalized force-displacement curves prove mesh convergence for all models already upon the first refinement with 2651 elements (see Figs.~\ref{fig:ExpwhFuA}~-~\ref{fig:ExpwhFuC}). Furthermore, Fig.~\ref{fig:ExpwhFuComp} provides the direct comparison of all models using the finest mesh with 14752 elements. Model~A and C yield an identical structural response, while the vertical force drop of model~B is shifted to the right with $u_{0.5 \, F_\text{max}}^\text{B} = 0.751~[\si{\mm}]$ compared to $u_{0.5 \, F_\text{max}}^\text{A,C} = 0.706~[\si{\mm}]$.

\begin{figure}
  \centering 
  \begin{subfigure}{.3\textwidth} 
    \centering 
    \includegraphics[width=\textwidth]{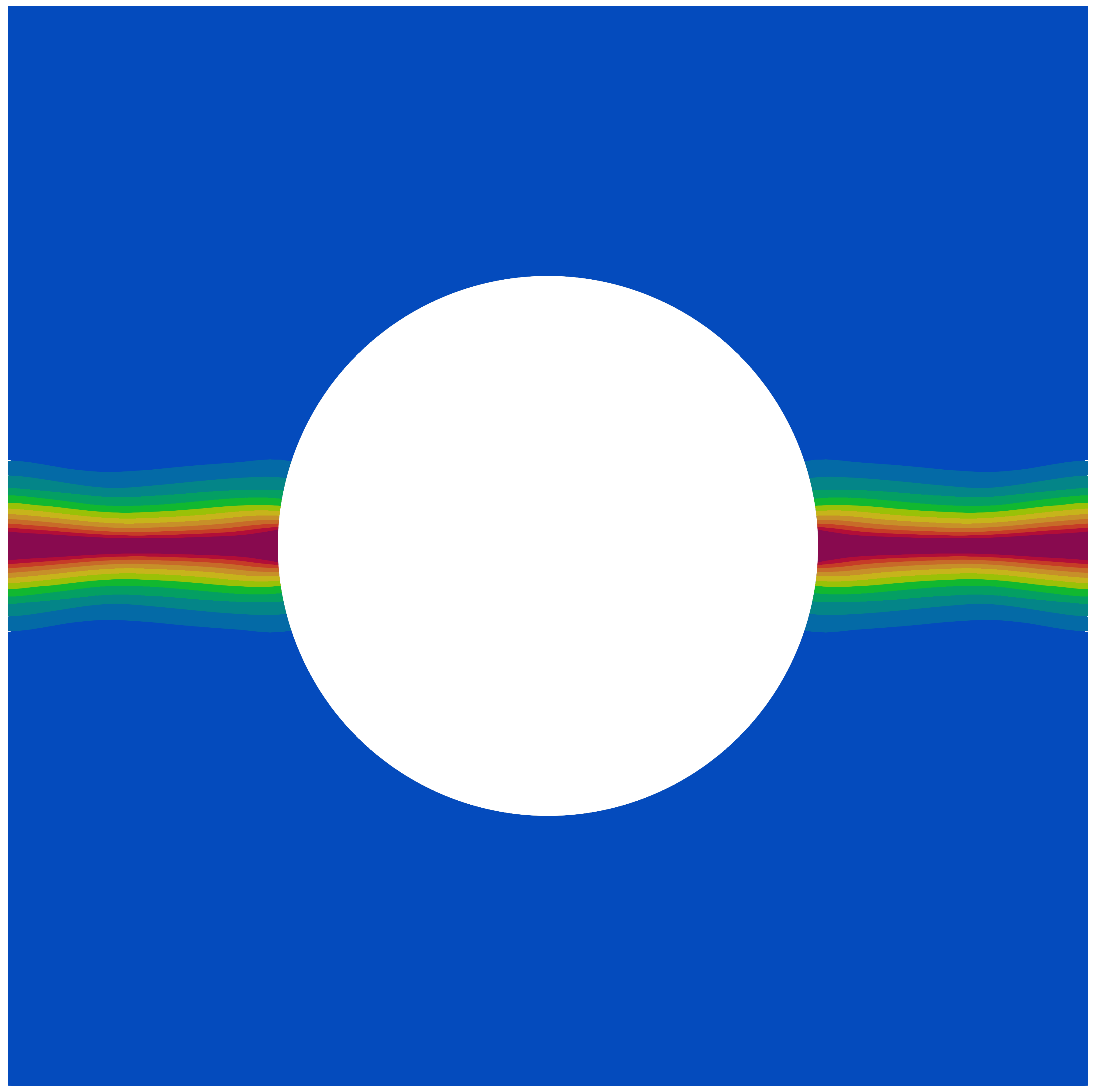}
  \end{subfigure}
  \begin{subfigure}{.3\textwidth} 
    \centering 
    \includegraphics[width=\textwidth]{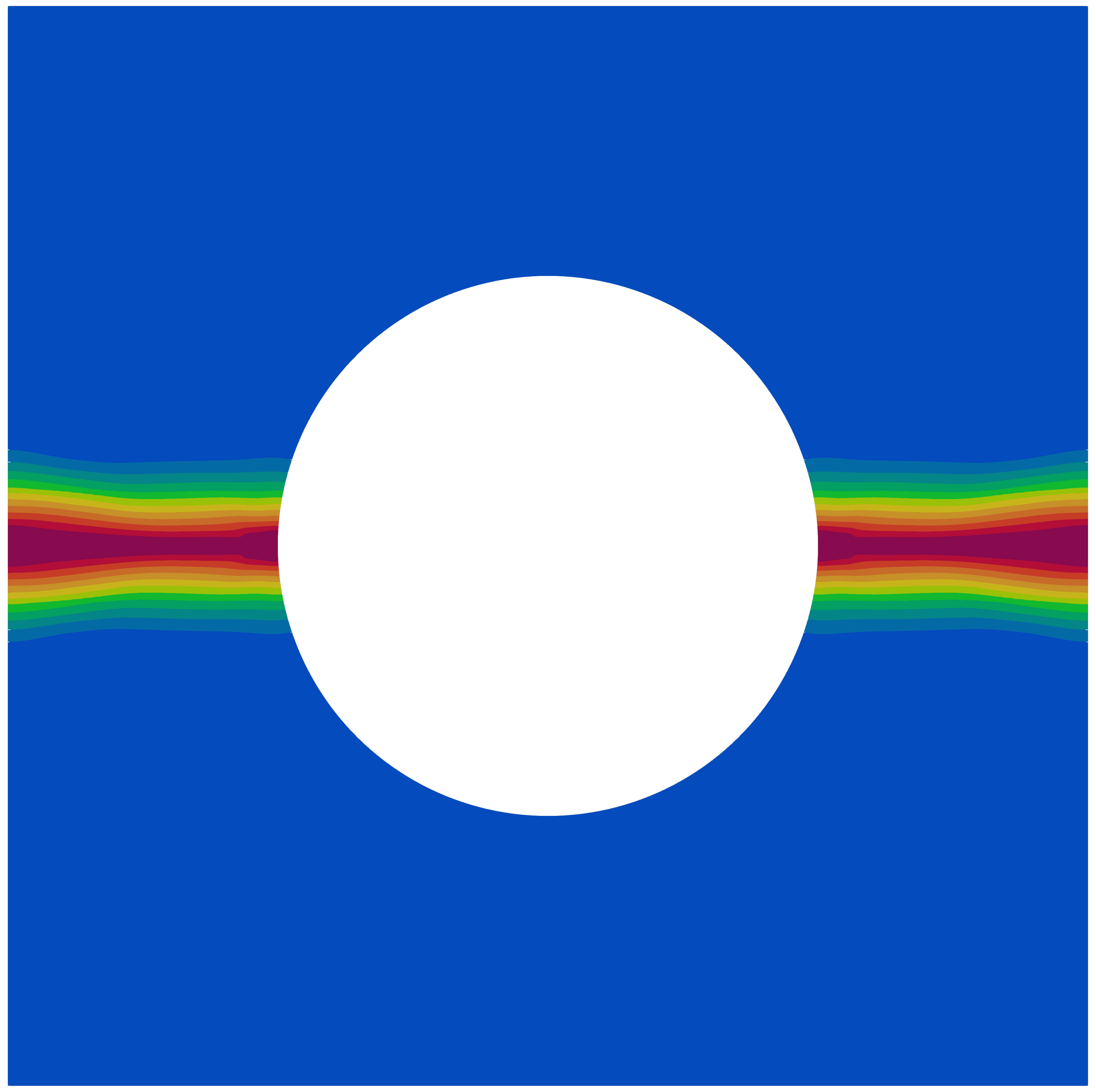}
  \end{subfigure}
  \begin{subfigure}{.3\textwidth} 
    \centering 
    \includegraphics[width=\textwidth]{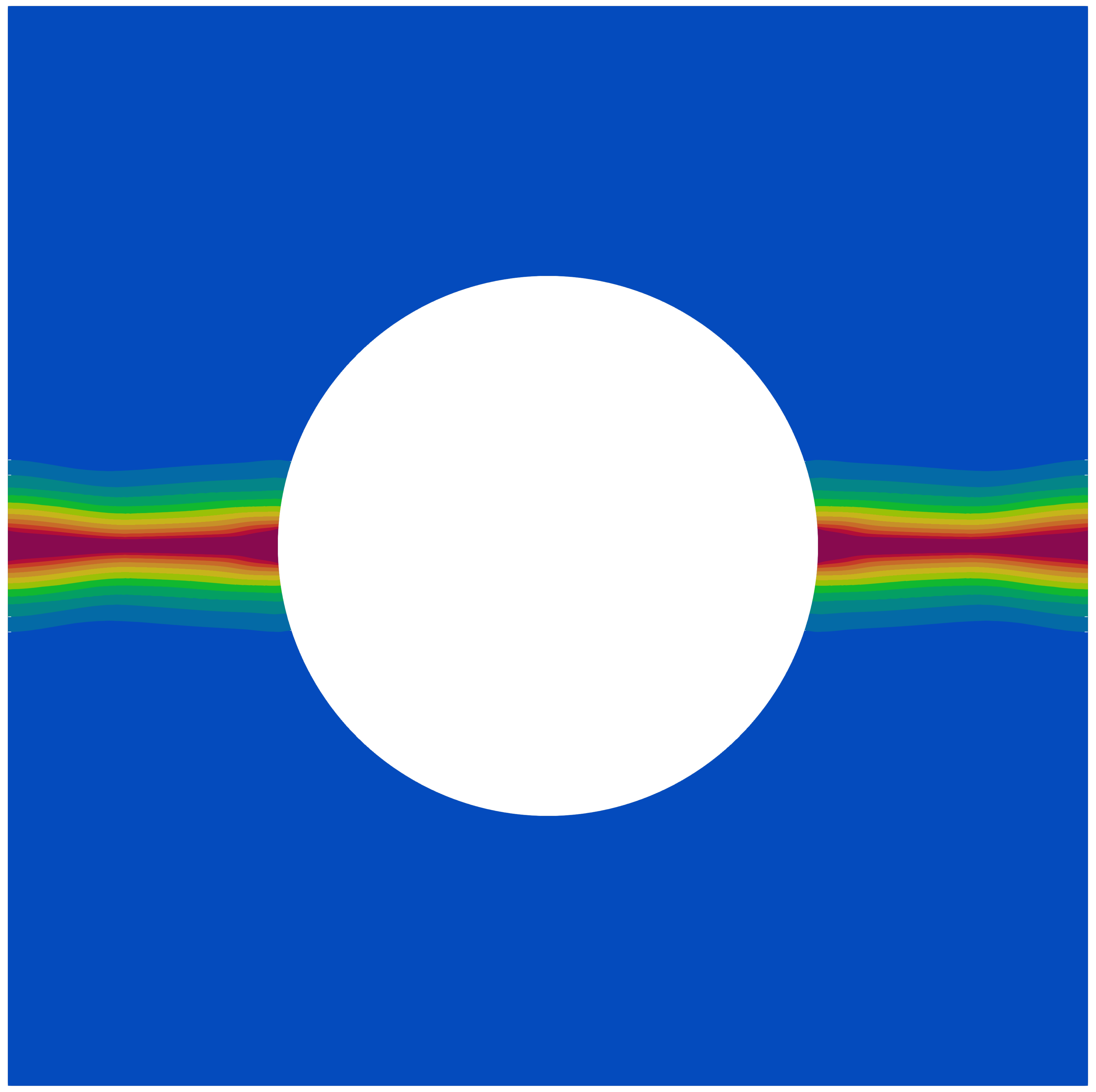}
  \end{subfigure}
  \begin{subfigure}{.08\textwidth} 
    \centering 
    \begin{tikzpicture}
      \node[inner sep=0pt] (pic) at (0,0) {\includegraphics[height=40mm, width=5mm]
      {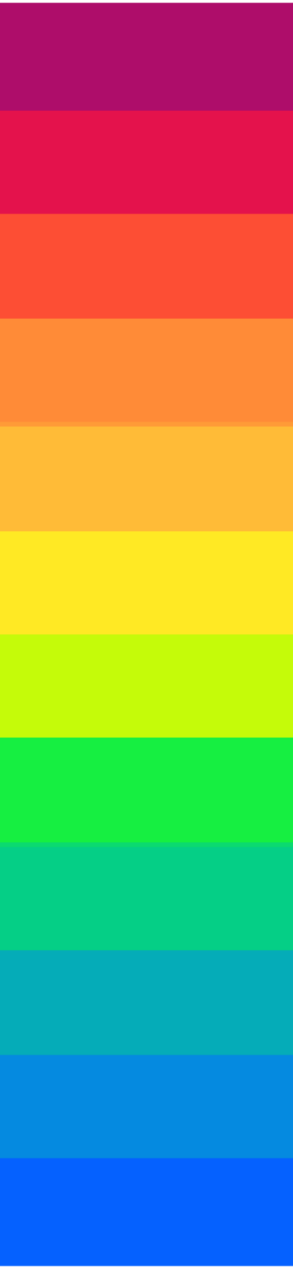}};
      \node[inner sep=0pt] (0)   at ($(pic.south)+( 0.50, 0.15)$)  {$0$};
      \node[inner sep=0pt] (1)   at ($(pic.south)+( 0.50, 3.80)$)  {$1$};
      \node[inner sep=0pt] (d)   at ($(pic.south)+( 0.00, 4.35)$)  {$D_{xx}~\si{[-]}$};
    \end{tikzpicture} 
  \end{subfigure}

  \vspace{1mm}

  \begin{subfigure}{.3\textwidth} 
    \centering 
    \includegraphics[width=\textwidth]{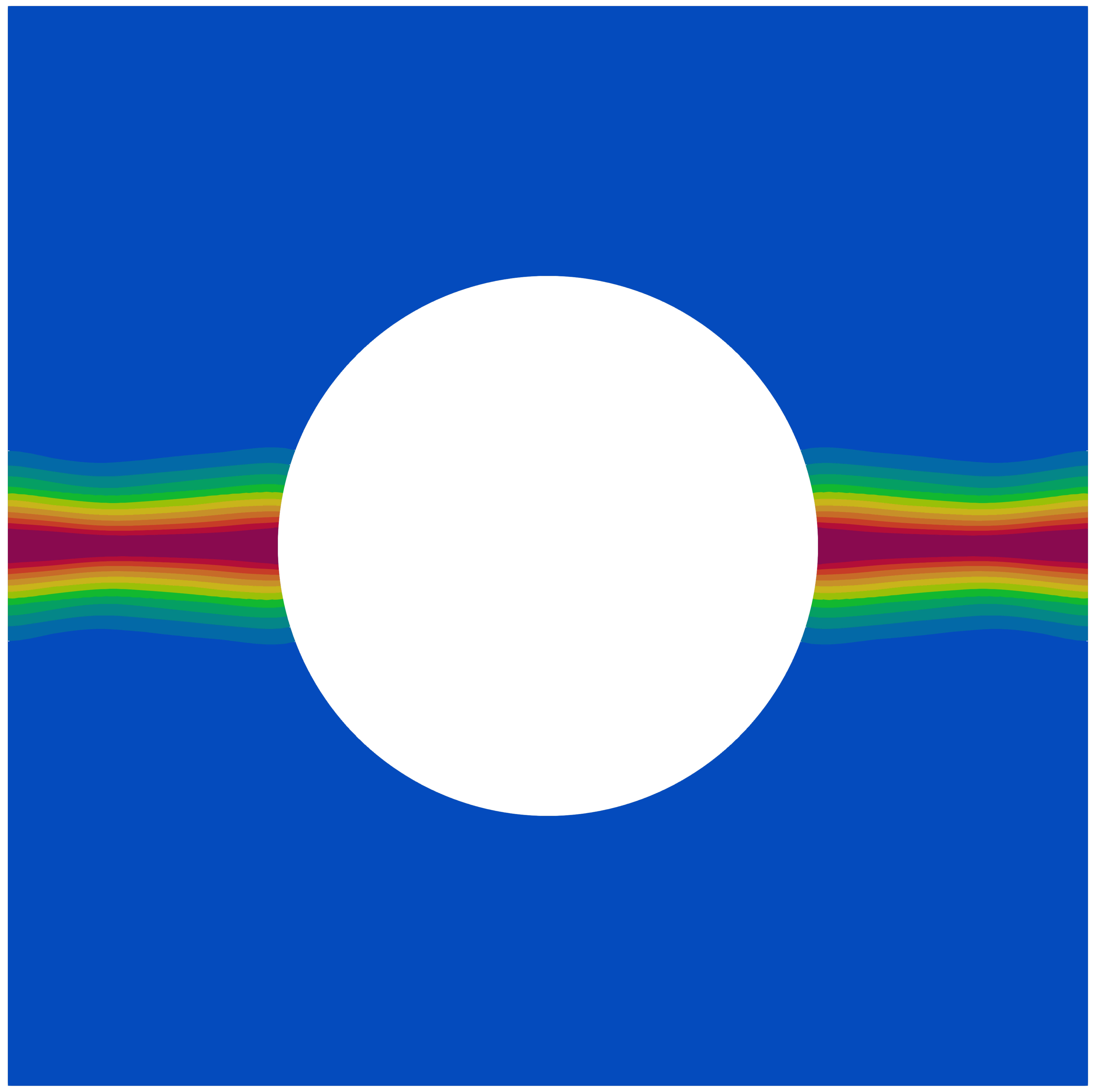}
  \end{subfigure}
  \begin{subfigure}{.3\textwidth} 
    \centering 
    \includegraphics[width=\textwidth]{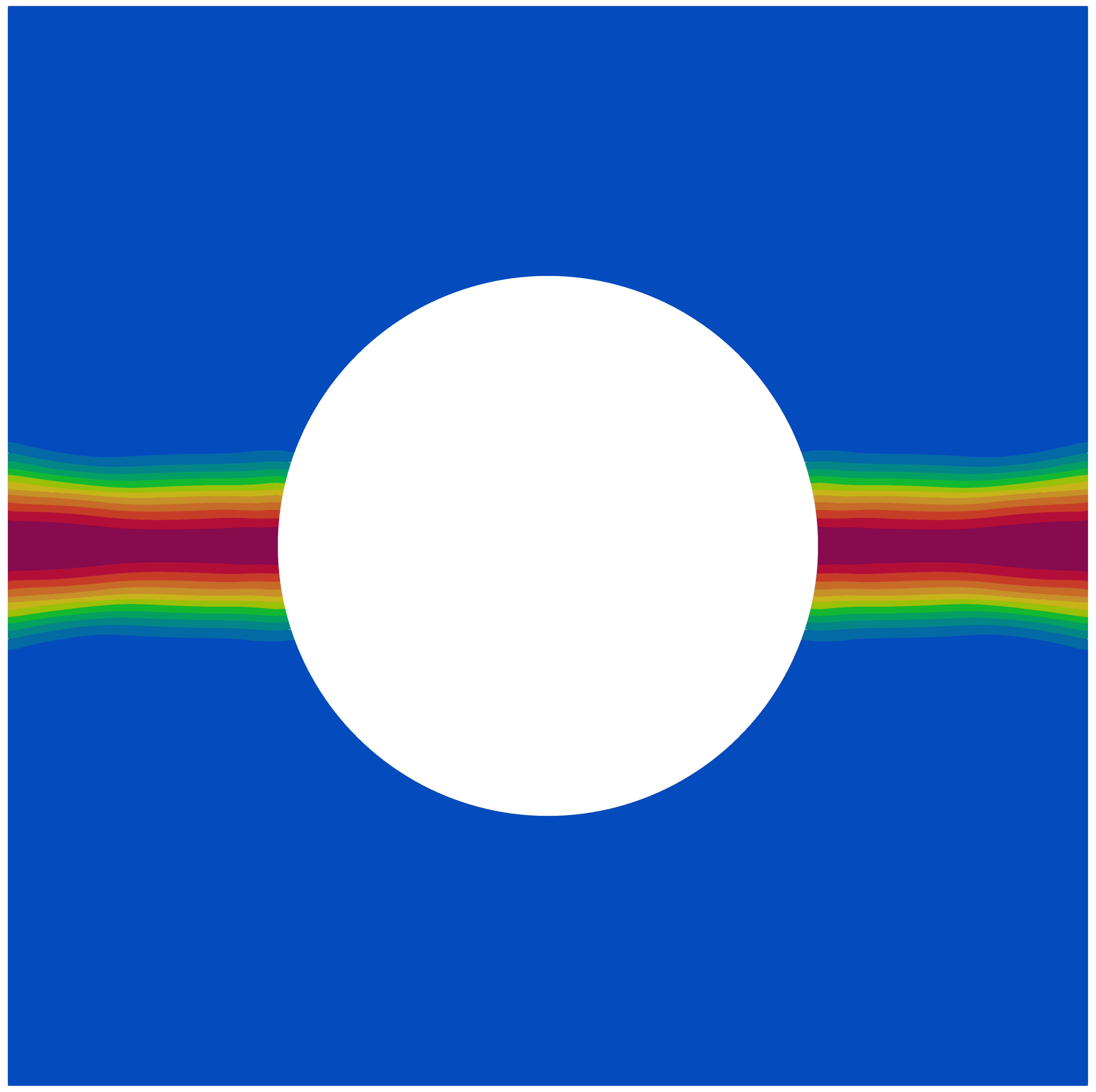}
  \end{subfigure}
  \begin{subfigure}{.3\textwidth} 
    \centering 
    \includegraphics[width=\textwidth]{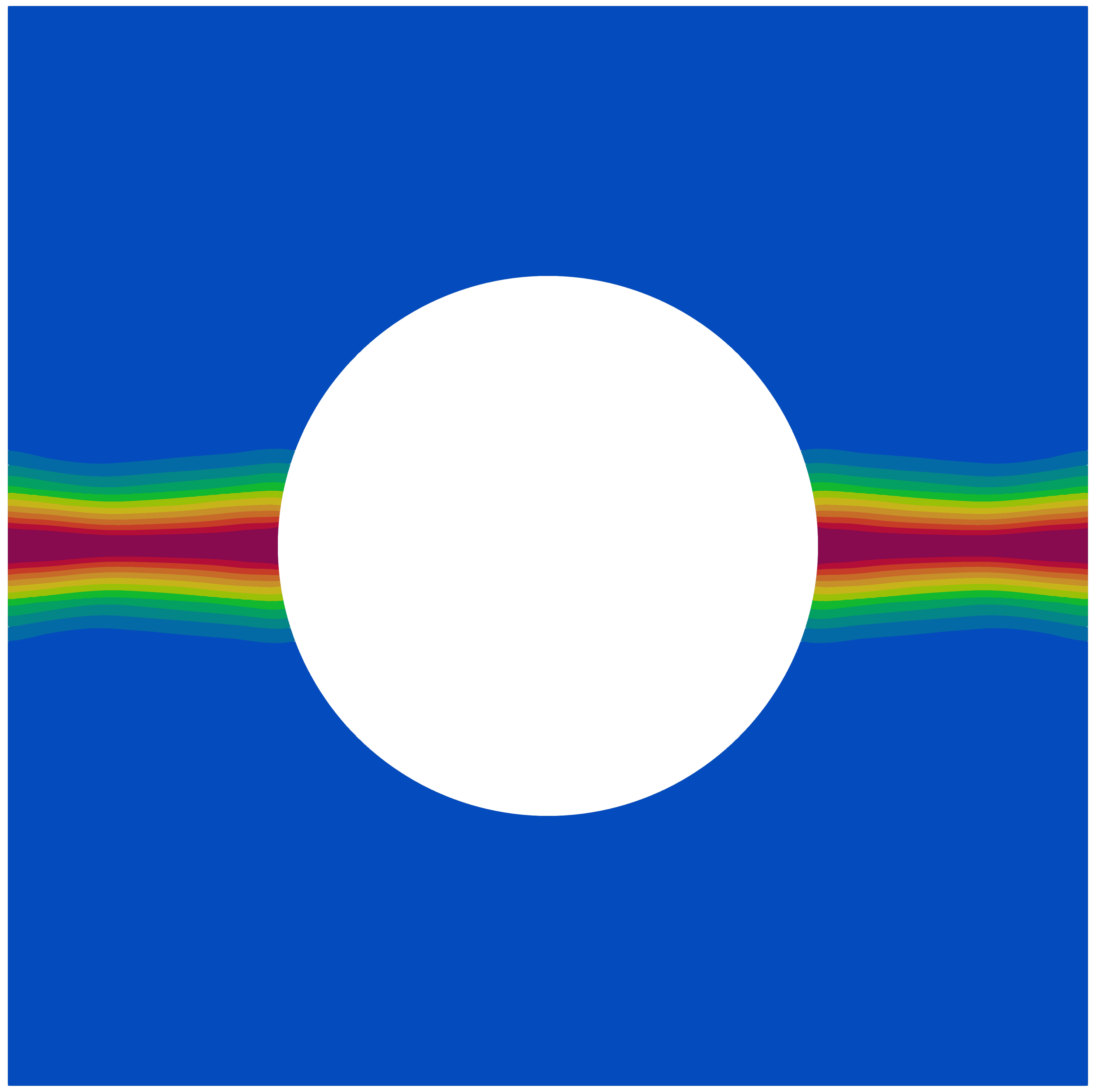}
  \end{subfigure}
  \begin{subfigure}{.08\textwidth} 
    \centering 
    \begin{tikzpicture}
      \node[inner sep=0pt] (pic) at (0,0) {\includegraphics[height=40mm, width=5mm]
      {02_Figures/03_Contour/00_Color_Maps/Damage_Step_Vertical.pdf}};
      \node[inner sep=0pt] (0)   at ($(pic.south)+( 0.50, 0.15)$)  {$0$};
      \node[inner sep=0pt] (1)   at ($(pic.south)+( 0.50, 3.80)$)  {$1$};
      \node[inner sep=0pt] (d)   at ($(pic.south)+( 0.00, 4.35)$)  {$D_{yy}~\si{[-]}$};
    \end{tikzpicture} 
  \end{subfigure}

  \vspace{1mm}

  \begin{subfigure}{.3\textwidth} 
    \centering 
    \includegraphics[width=\textwidth]{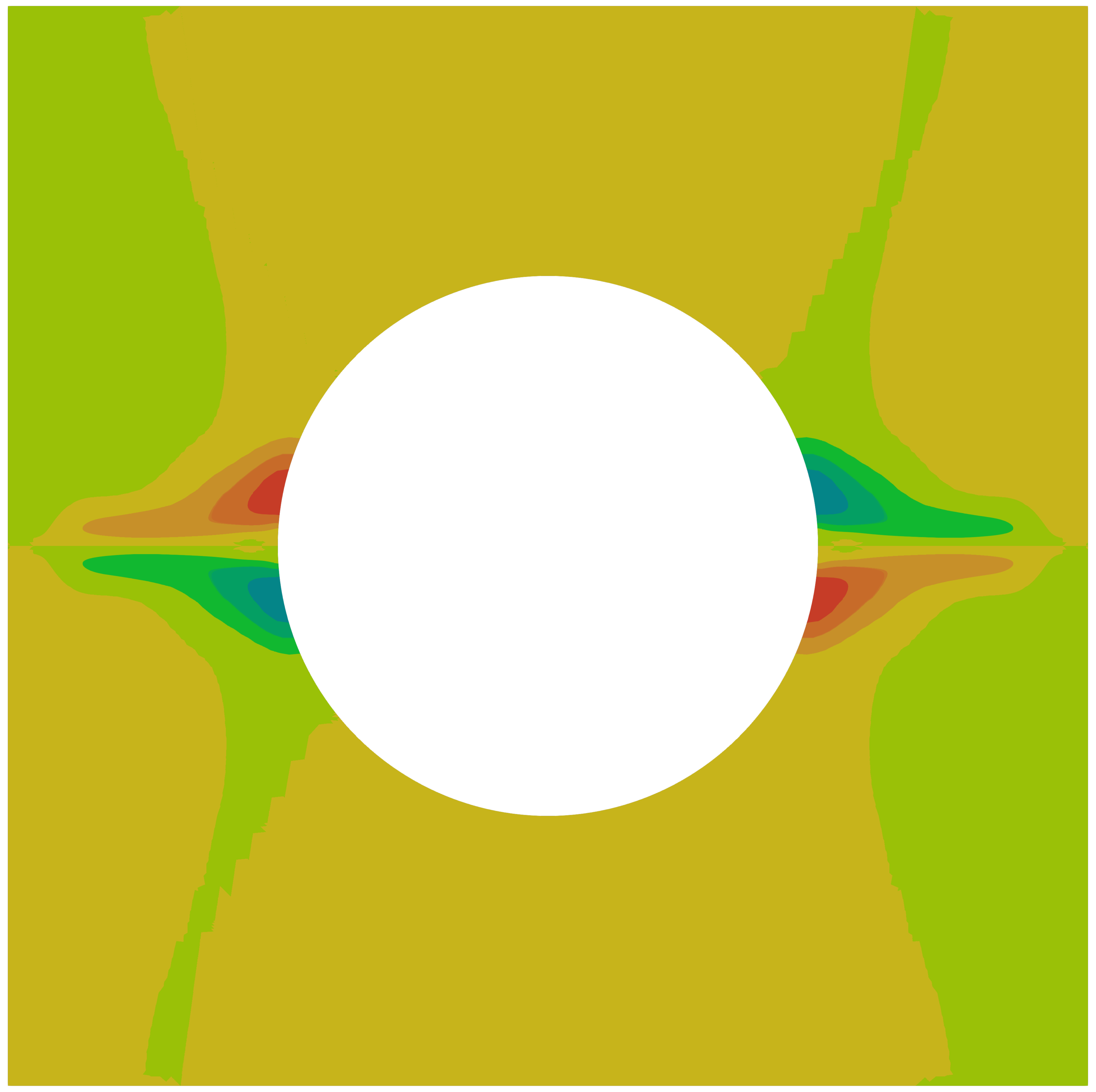}
    \caption{Model~A}
    \label{fig:ExpwhDA}
  \end{subfigure}
  \begin{subfigure}{.3\textwidth} 
    \centering 
    \includegraphics[width=\textwidth]{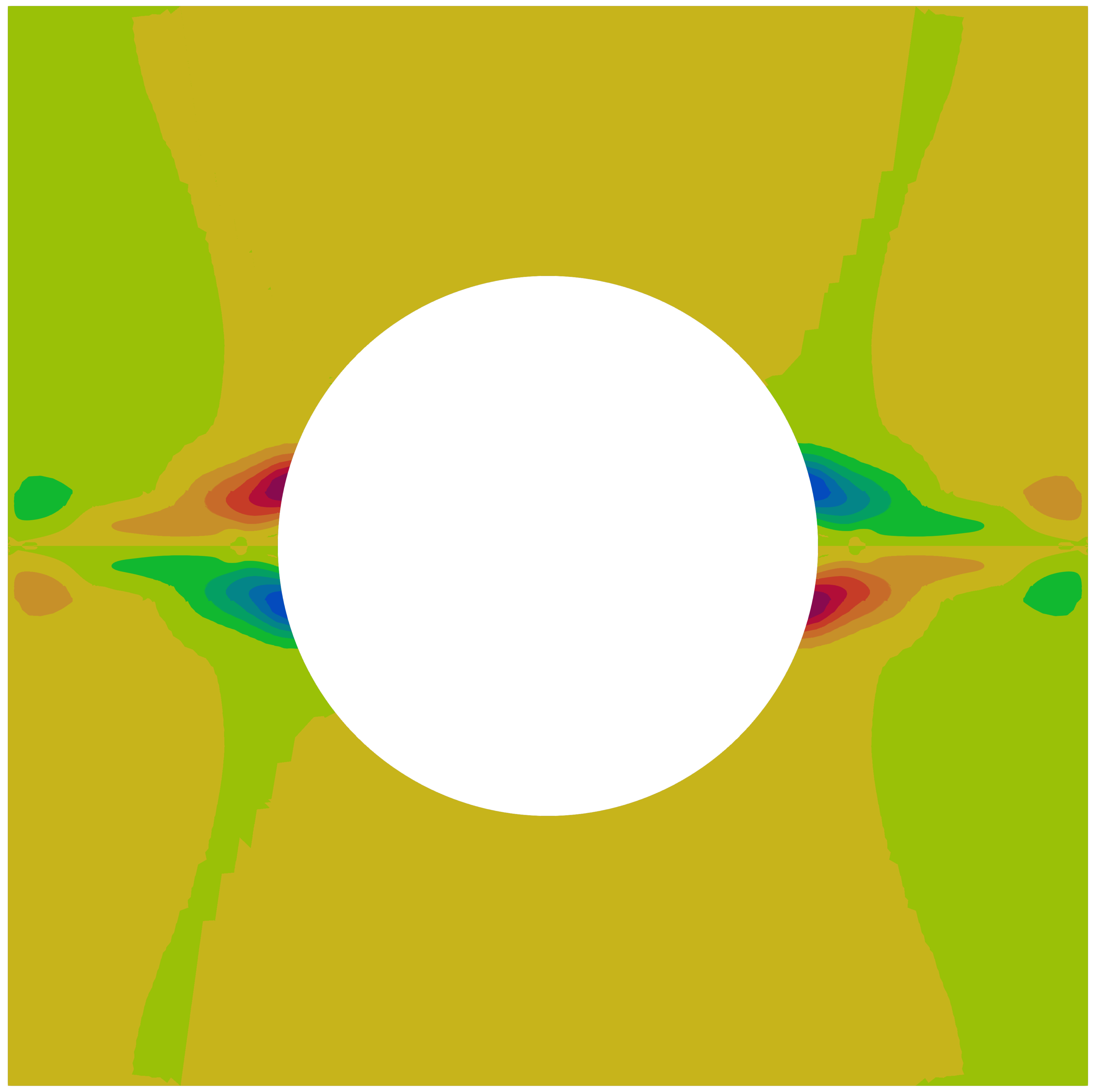}
    \caption{Model~B}
    \label{fig:ExpwhDB}
  \end{subfigure}
  \begin{subfigure}{.3\textwidth} 
    \centering 
    \includegraphics[width=\textwidth]{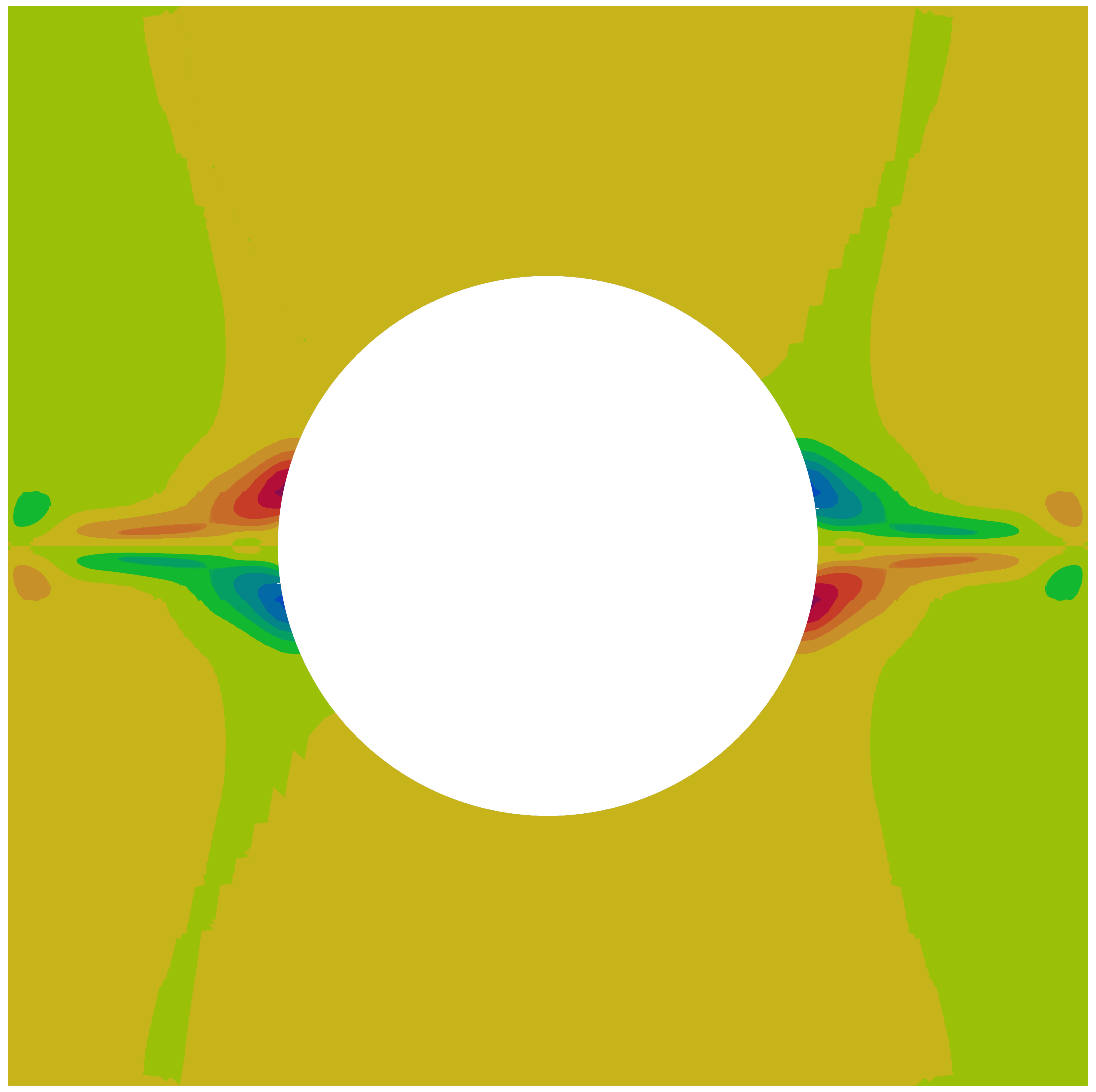}
    \caption{Model~C}
    \label{fig:ExpwhDC}
  \end{subfigure}
  \begin{subfigure}{.08\textwidth} 
    \centering 
    \begin{tikzpicture}
      \node[inner sep=0pt] (pic) at (0,0) {\includegraphics[height=40mm, width=5mm]
      {02_Figures/03_Contour/00_Color_Maps/Damage_Step_Vertical.pdf}};
      \node[inner sep=0pt] (0)   at ($(pic.south)+( 1.00, 0.15)$)  {$-0.039$};
      \node[inner sep=0pt] (1)   at ($(pic.south)+( 1.00, 3.80)$)  {$+0.039$};
      \node[inner sep=0pt] (d)   at ($(pic.south)+( 0.00, 4.35)$)  {$D_{xy}~\si{[-]}$};
    \end{tikzpicture} 
    \hphantom{Model~C}
  \end{subfigure}
  
  \caption{Contour plots of the normal and shear components of the damage tensor for the plate with hole specimen at the end of the simulation.}
  \label{fig:ExpwhD}     
\end{figure}

Fig.~\ref{fig:ExpwhD} shows the damage contour plots at the end of the simulation. For all models, the width of the damage zone of component $D_{yy}$ is thicker than that of component $D_{xx}$, since the specimen is loaded in $y$-direction. Models~A and C yield coinciding results, whilst for model~B, the damage zone for both normal components of the damage tensor are more pronounced. This observation is consistent with the results of Fig.~\ref{fig:ExpwhFuComp}, where, loosely speaking, the area under the force-displacement curve is larger for model~B and, hence, a larger amount of energy is dissipated in this case, which implies that the corresponding damage zones have to be larger as well.

\begin{figure}[htbp]
  \centering
    \includegraphics{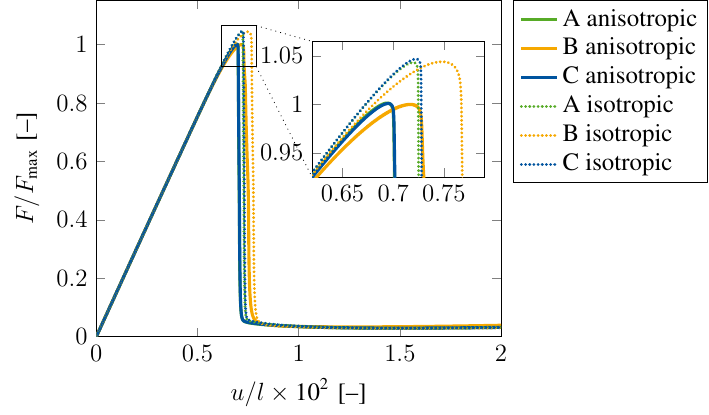}
  \caption{Comparison of the force-displacement curves of the anisotropic and isotropic computation for the plate with hole specimen (14752 elements). The forces are normalized with respect to the maximum force of the anisotropic computation of model~B with $F_\text{max} = 5.0767 \times 10^4~[\si{\newton}]$.}
  \label{fig:ExpwhFuIsotropic}
\end{figure}

\begin{figure}
  \centering 

  \begin{subfigure}{.3\textwidth} 
    \centering 
    \includegraphics[width=\textwidth]{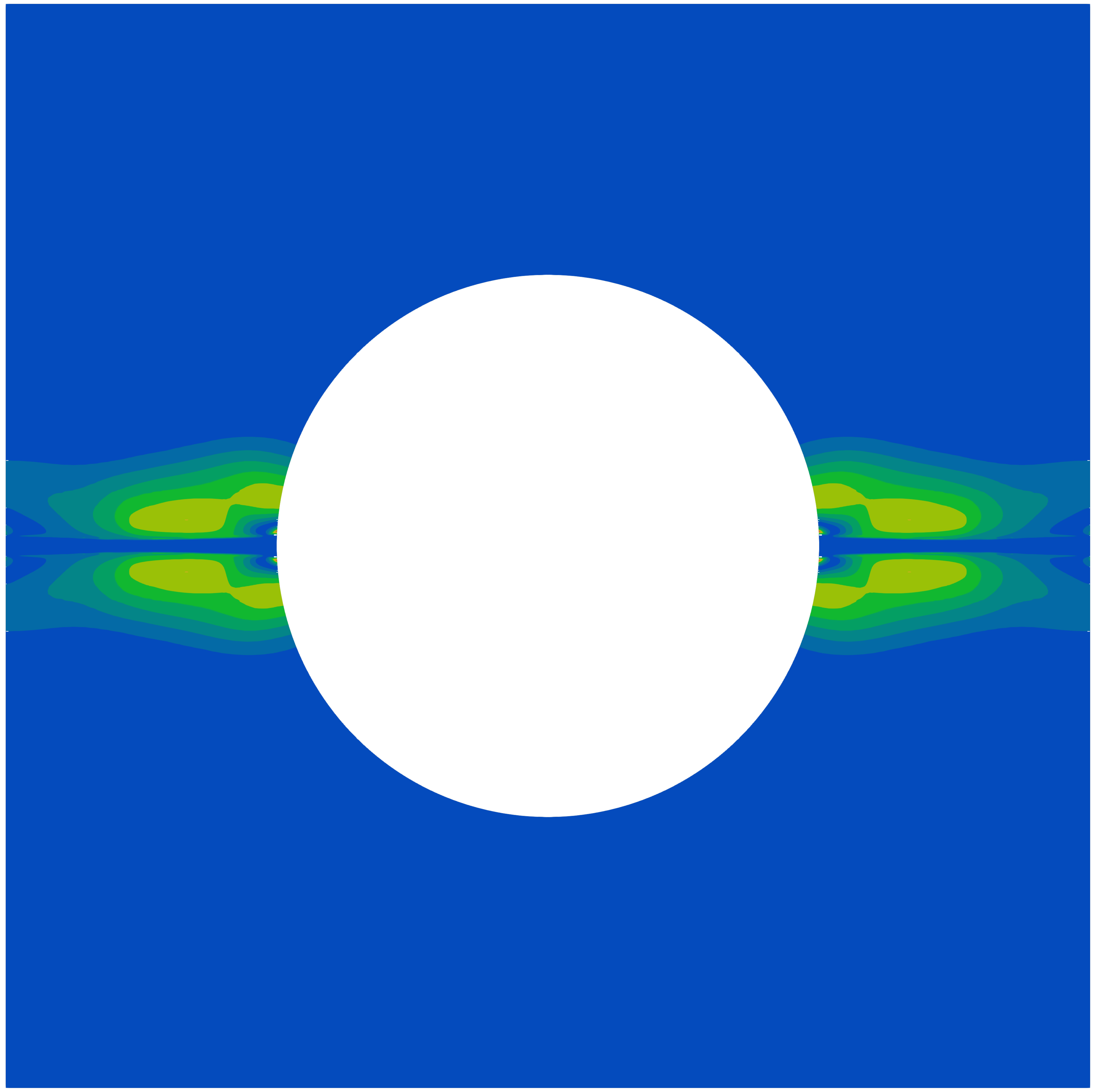}
  \end{subfigure}
  \begin{subfigure}{.3\textwidth} 
    \centering 
    \includegraphics[width=\textwidth]{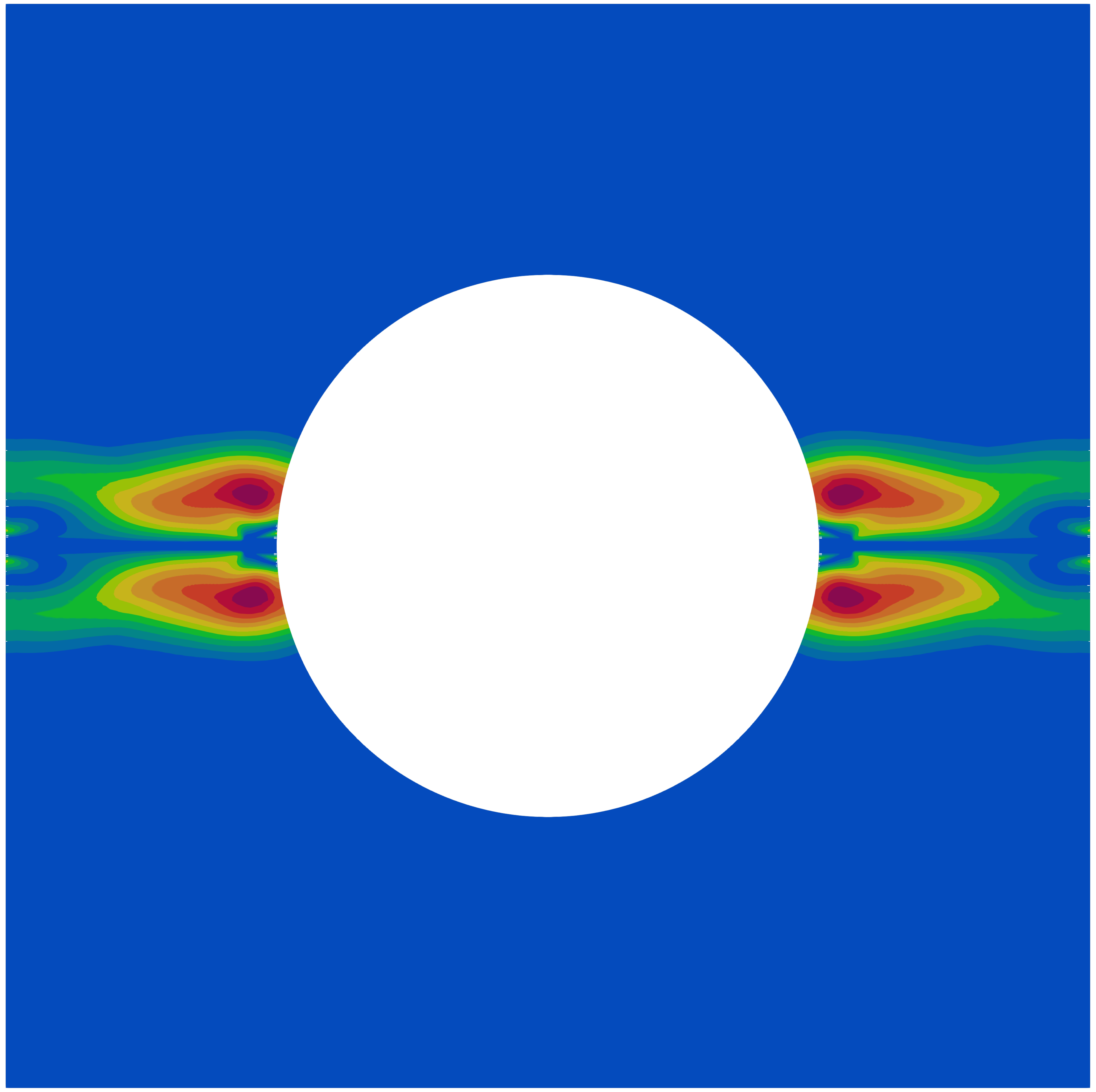}
  \end{subfigure}
  \begin{subfigure}{.3\textwidth} 
    \centering 
    \includegraphics[width=\textwidth]{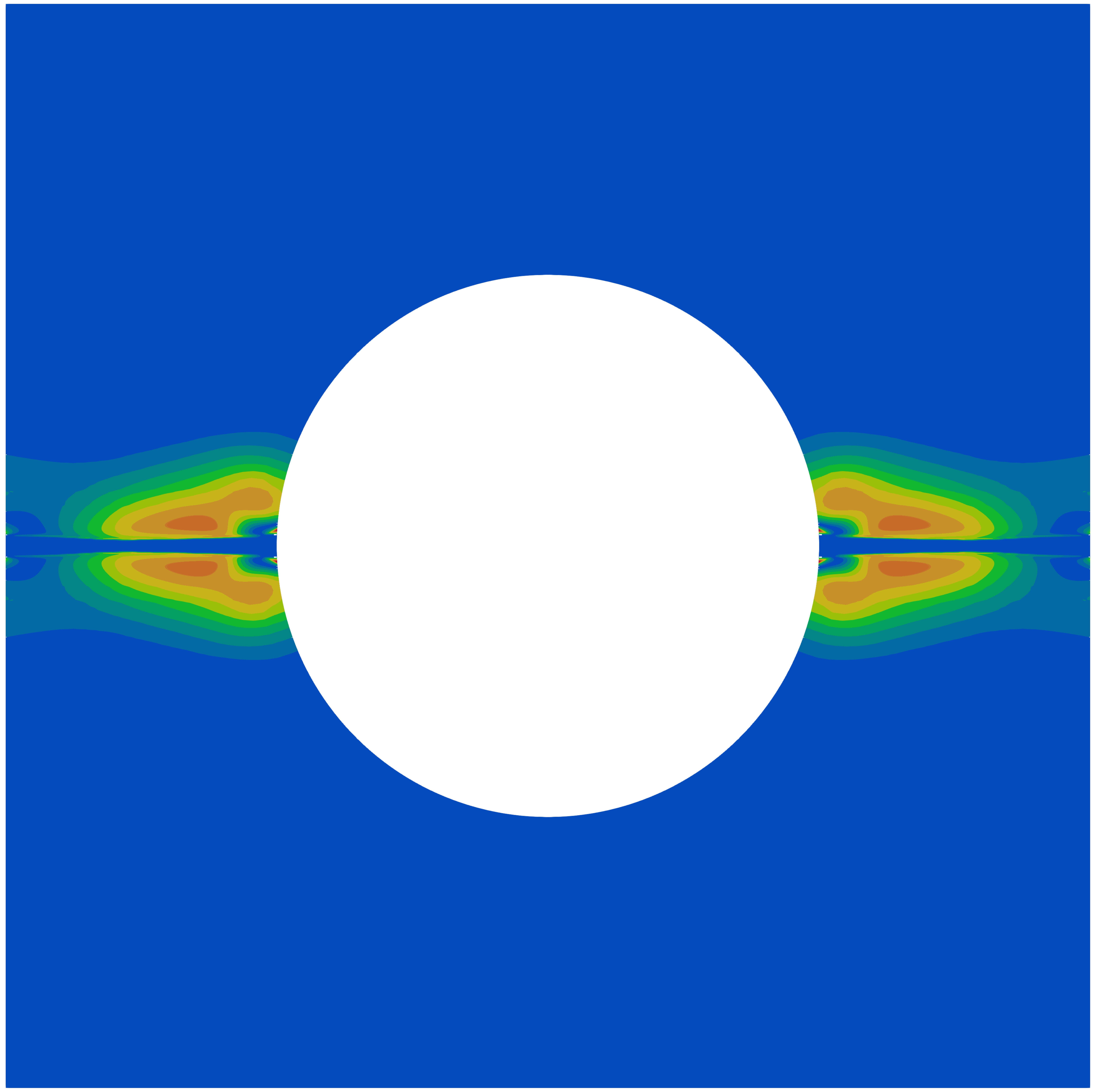}
  \end{subfigure}
  \begin{subfigure}{.08\textwidth} 
    \centering 
    \begin{tikzpicture}
      \node[inner sep=0pt] (pic) at (0,0) {\includegraphics[height=40mm, width=5mm]
      {02_Figures/03_Contour/00_Color_Maps/Damage_Step_Vertical.pdf}};
      \node[inner sep=0pt] (0)   at ($(pic.south)+( 0.50, 0.15)$)  {$0$};
      \node[inner sep=0pt] (1)   at ($(pic.south)+( 0.85, 3.80)$)  {$0.132$};
      \node[inner sep=0pt] (d)   at ($(pic.south)+( 0.60, 4.35)$)  {$|D-D_{xx}|~\si{[-]}$};
    \end{tikzpicture} 
  \end{subfigure}

  \vspace{1mm}

  \begin{subfigure}{.3\textwidth} 
    \centering 
    \includegraphics[width=\textwidth]{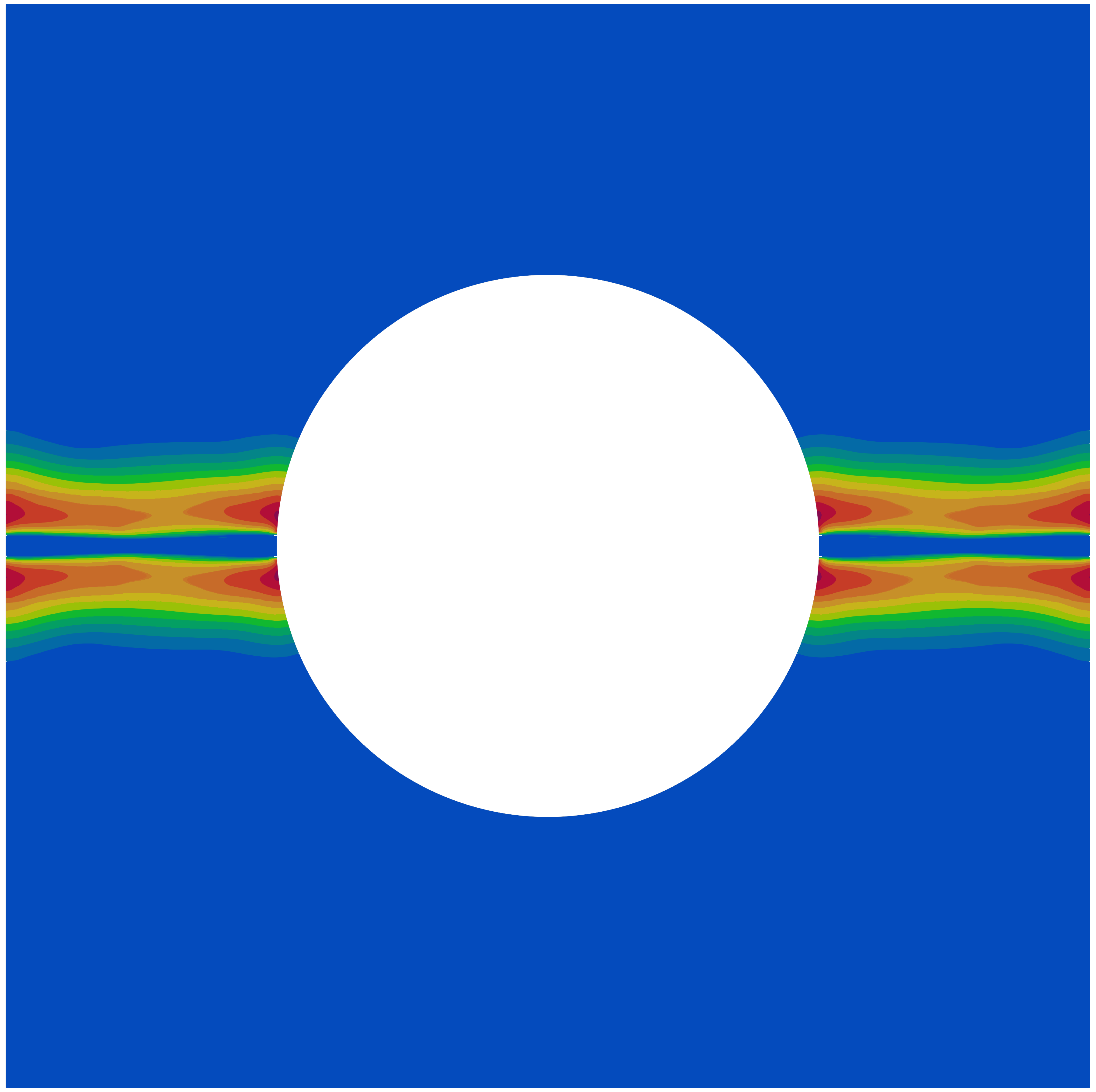}
    \caption{Model~A}
  \end{subfigure}
  \begin{subfigure}{.3\textwidth} 
    \centering 
    \includegraphics[width=\textwidth]{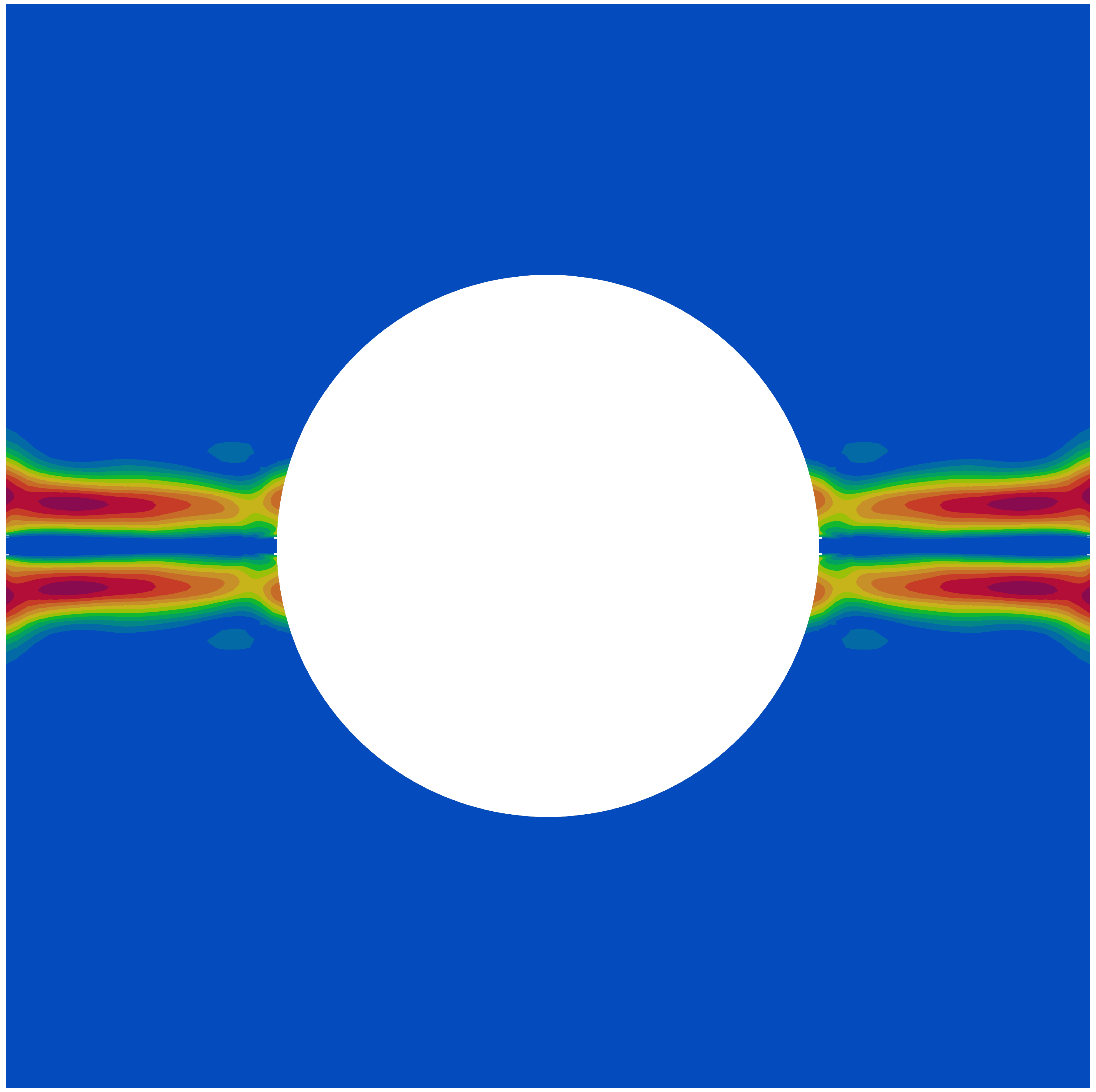}
    \caption{Model~B}
  \end{subfigure}
  \begin{subfigure}{.3\textwidth} 
    \centering 
    \includegraphics[width=\textwidth]{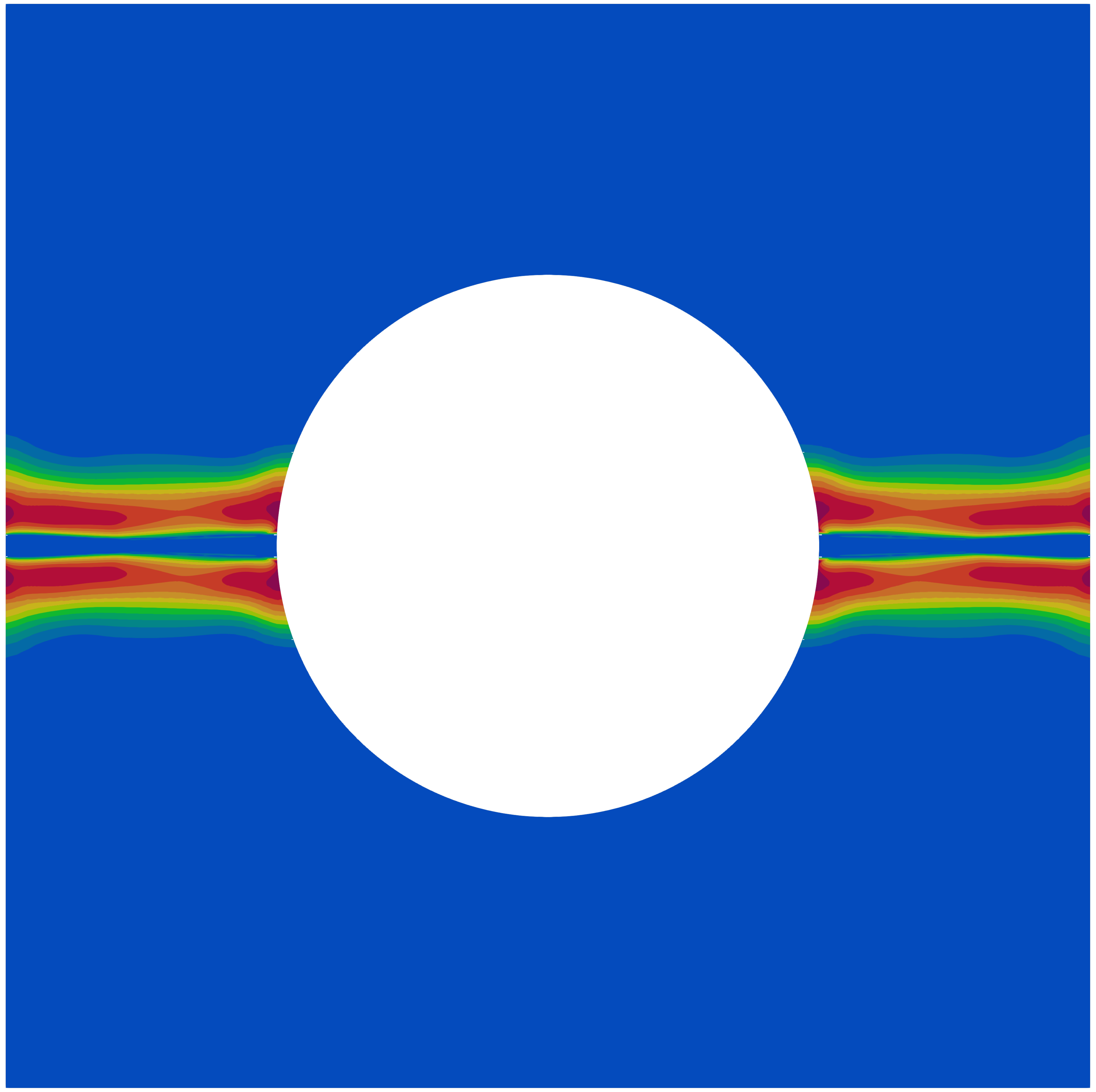}
    \caption{Model~C}
  \end{subfigure}
  \begin{subfigure}{.08\textwidth} 
    \centering 
    \begin{tikzpicture}
      \node[inner sep=0pt] (pic) at (0,0) {\includegraphics[height=40mm, width=5mm]
      {02_Figures/03_Contour/00_Color_Maps/Damage_Step_Vertical.pdf}};
      \node[inner sep=0pt] (0)   at ($(pic.south)+( 0.50, 0.15)$)  {$0$};
      \node[inner sep=0pt] (1)   at ($(pic.south)+( 0.85, 3.80)$)  {$0.117$};
      \node[inner sep=0pt] (d)   at ($(pic.south)+( 0.60, 4.35)$)  {$|D-D_{yy}|~\si{[-]}$};
    \end{tikzpicture} 
    \hphantom{Model~C}
  \end{subfigure}
  
  \caption{Comparison of the damage contour plots of the anisotropic and isotropic computation for the plate with hole specimen by a difference plot of the isotropic damage value $D$ compared to the normal components of the anisotropic damage tensor $D_{xx}$ and $D_{yy}$.} 
  \label{fig:ExpwhDiffIsoDam}
\end{figure}

Next, we examine the necessity of using an anisotropic damage formulation, here i.e.~$\vartheta = 1~[\si{-}]$, compared to an isotropic one, i.e.~$\vartheta = 0~[\si{-}]$. Fig.~\ref{fig:ExpwhFuIsotropic} shows the force-displacement curves of the plate with hole simulation with the finest mesh (14752 elements) for all models using the anisotropic and isotropic model formulation. Evidently, the isotropic damage formulations continuously overestimate the structure's maximum load bearing capacity (A: $+4.18~[\si{\percent}]$, B: $+4.26~[\si{\percent}]$, C: $+4.58~[\si{\percent}]$). Deviations in the resulting damage contour plots for anisotropic and isotropic damage can also be observed in Fig.~\ref{fig:ExpwhDiffIsoDam}, where the shape and intensity are clearly nonconforming at the edges of the damage zone.

\begin{figure}[htbp]
  \centering
    \includegraphics{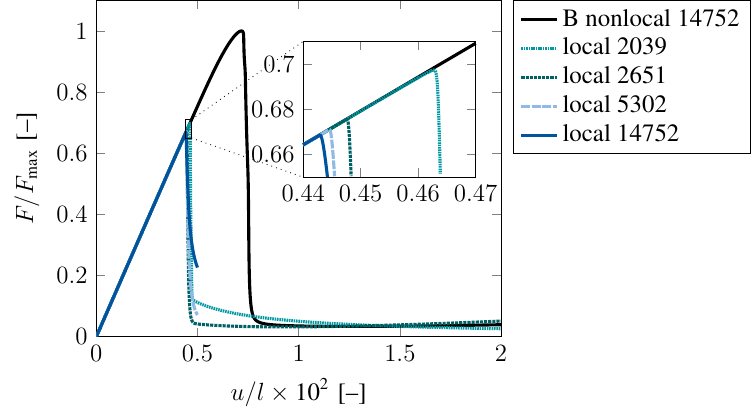}
  \caption{Force-displacement curves for the local damage model without gradient-extension for the plate with hole specimen for increasing degrees of mesh refinement. The forces are normalized with respect to the maximum force of the nonlocal computation of model~B with $F_\text{max} = 5.0767 \times 10^4~[\si{\newton}]$.}
  \label{fig:ExpwhFuLocal}
\end{figure}

\begin{figure}
  \centering 

  \begin{subfigure}{.24\textwidth} 
    \centering 
    \begin{tikzpicture}
      \node[inner sep=0pt] (pic) at (0,0) {\includegraphics[width=\textwidth]{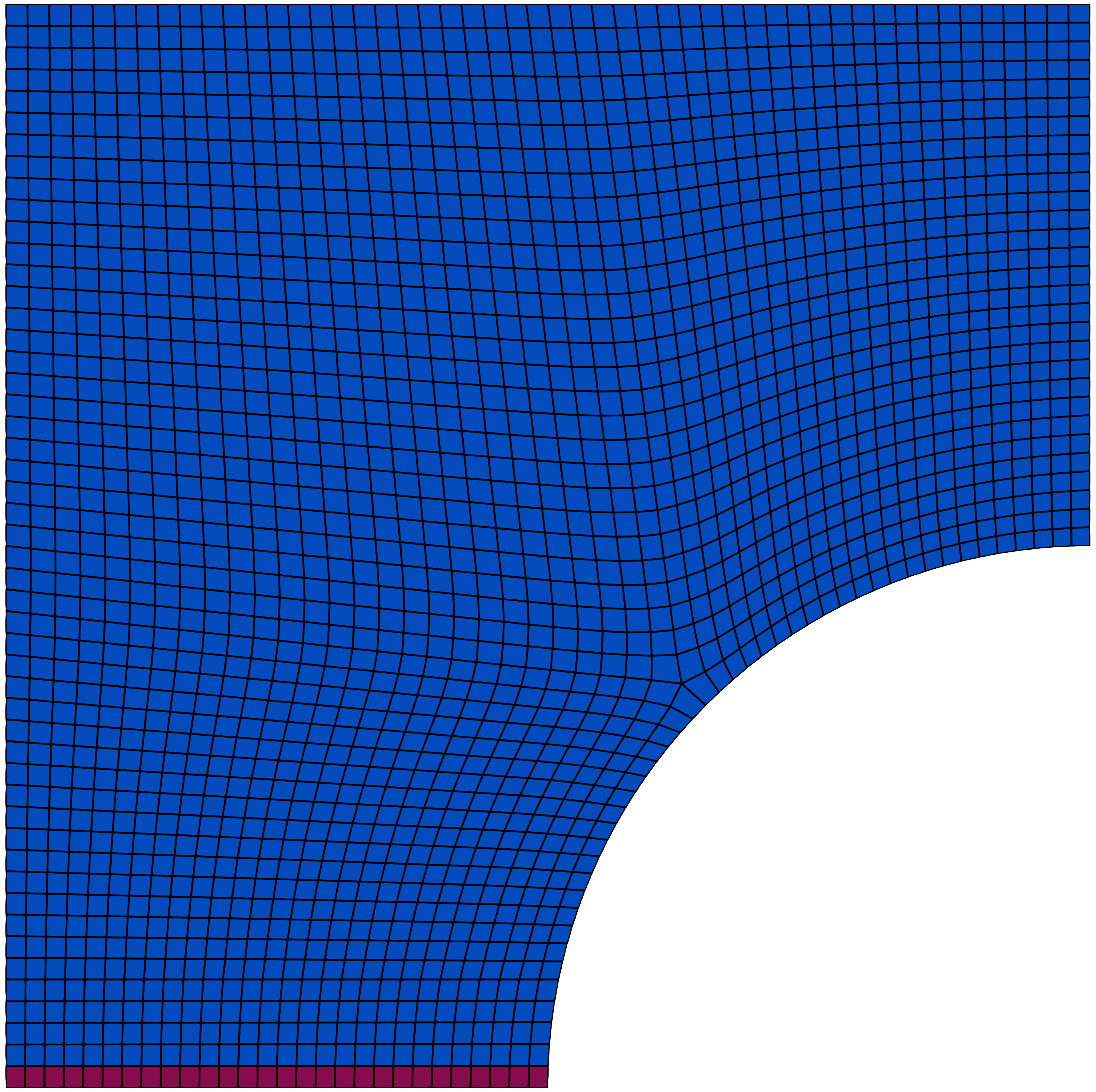}};
      \node[inner sep=0pt] (pic) at (0,-3) {\includegraphics[width=\textwidth]{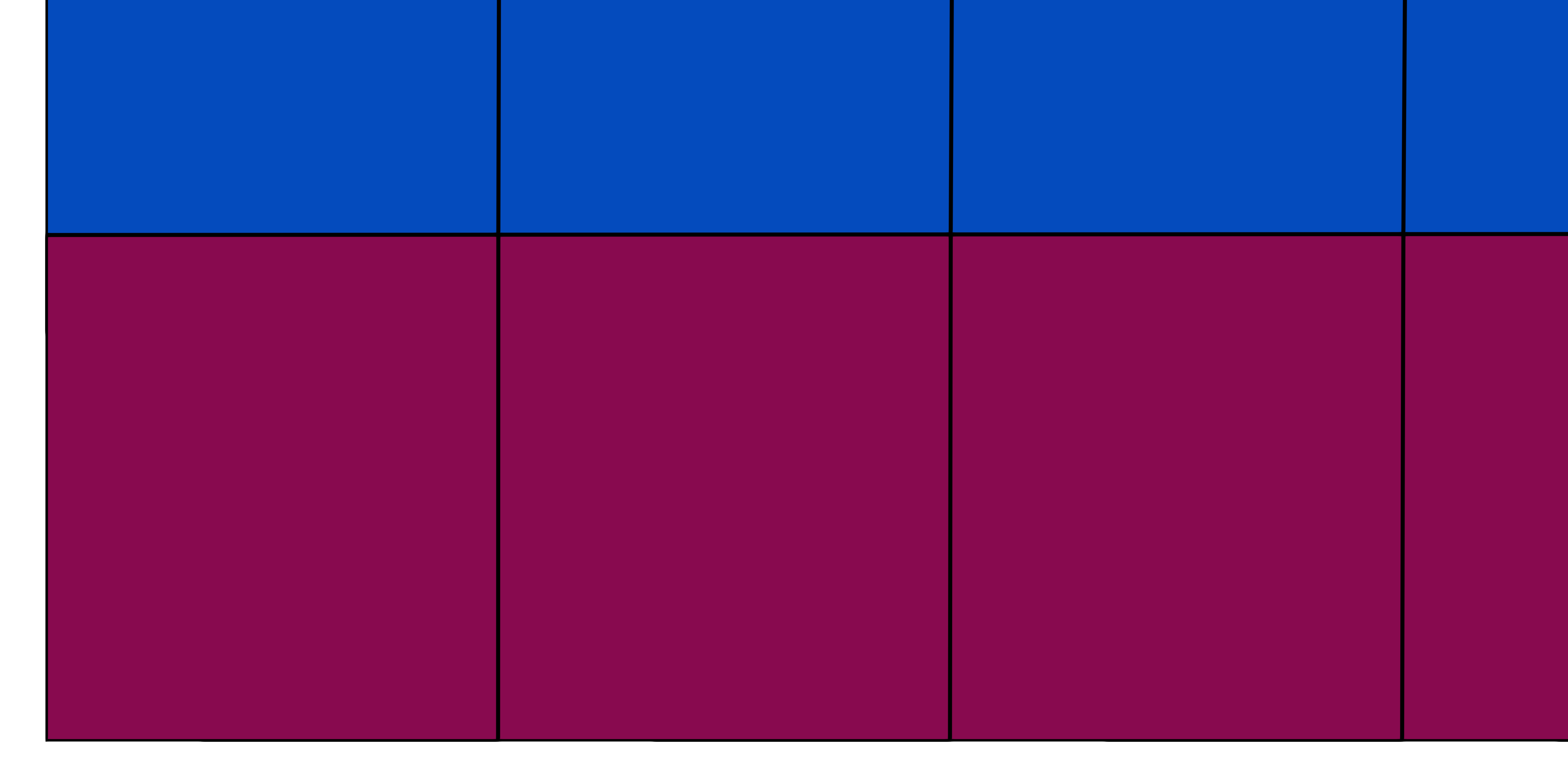}};
      \draw[draw=rwth8, line width=1.5pt] (-1.90,-1.90) rectangle (-1.68,-1.80);
      \draw[draw=rwth8, line width=1.5pt] (-1.82,-3.88) rectangle ( 1.92,-2.05);
      \draw[draw=rwth8, line width=1.5pt] (-1.79,-1.90) -- ( 0.05,-2.05);
    \end{tikzpicture} 
    \caption{2039}
  \end{subfigure}
  \begin{subfigure}{.24\textwidth} 
    \centering 
    \begin{tikzpicture}
      \node[inner sep=0pt] (pic) at (0,0) {\includegraphics[width=\textwidth]{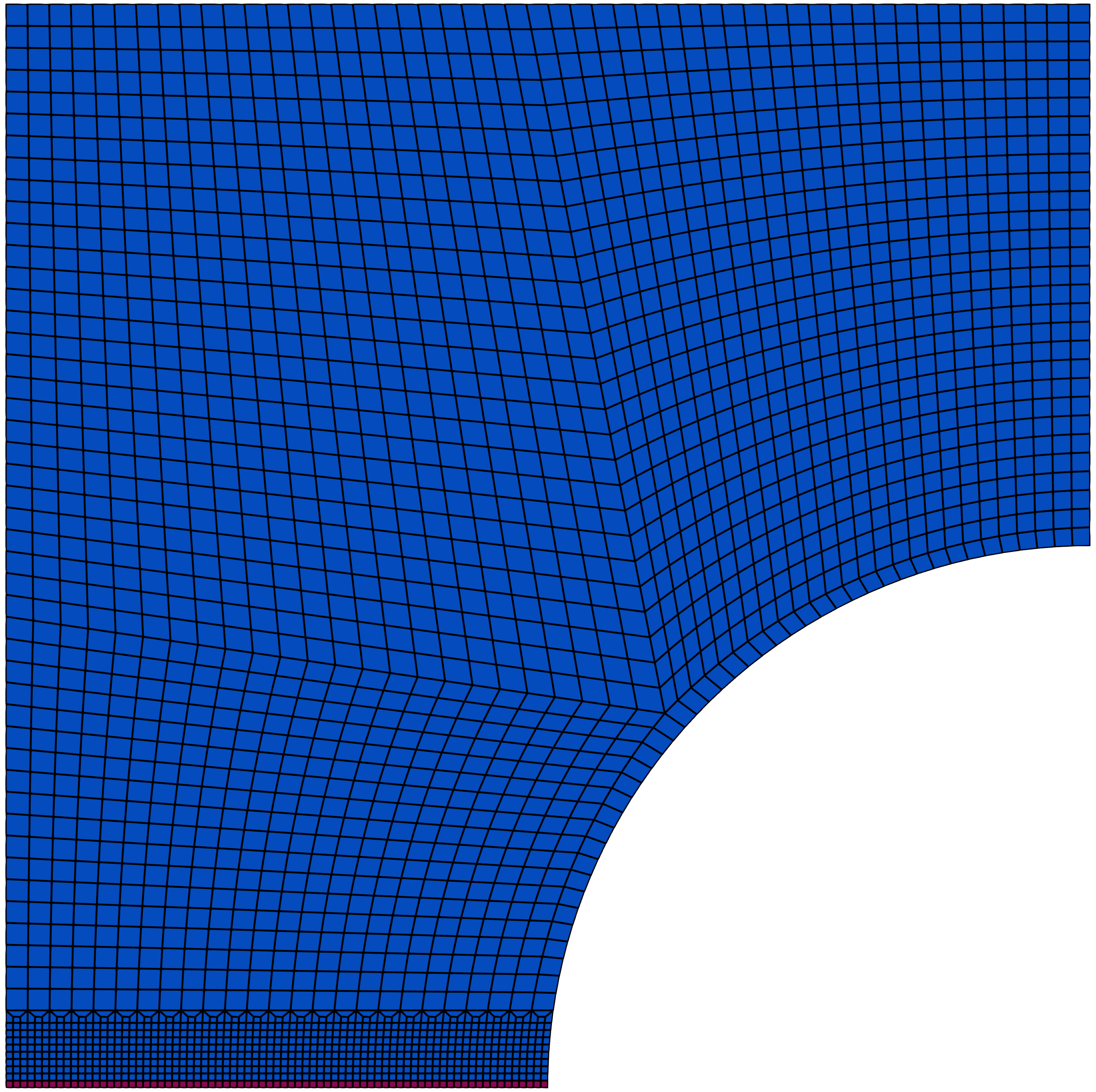}};
      \node[inner sep=0pt] (pic) at (0,-3) {\includegraphics[width=\textwidth]{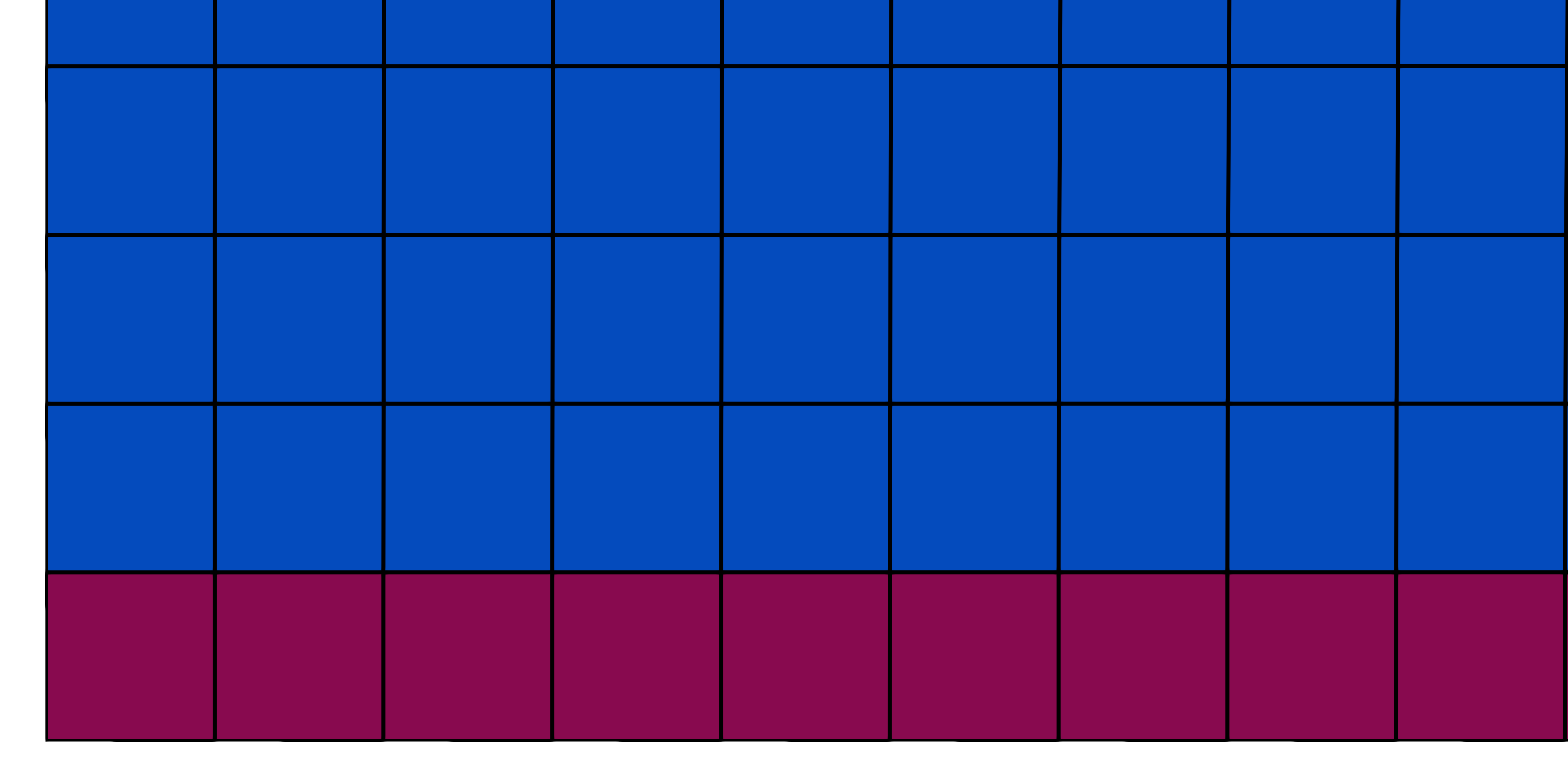}};
      \draw[draw=rwth8, line width=1.5pt] (-1.90,-1.90) rectangle (-1.68,-1.80);
      \draw[draw=rwth8, line width=1.5pt] (-1.82,-3.88) rectangle ( 1.92,-2.05);
      \draw[draw=rwth8, line width=1.5pt] (-1.79,-1.90) -- ( 0.05,-2.05);
    \end{tikzpicture} 
    \caption{2651}
  \end{subfigure}
  \begin{subfigure}{.24\textwidth} 
    \centering 
    \begin{tikzpicture}
      \node[inner sep=0pt] (pic) at (0,0) {\includegraphics[width=\textwidth]{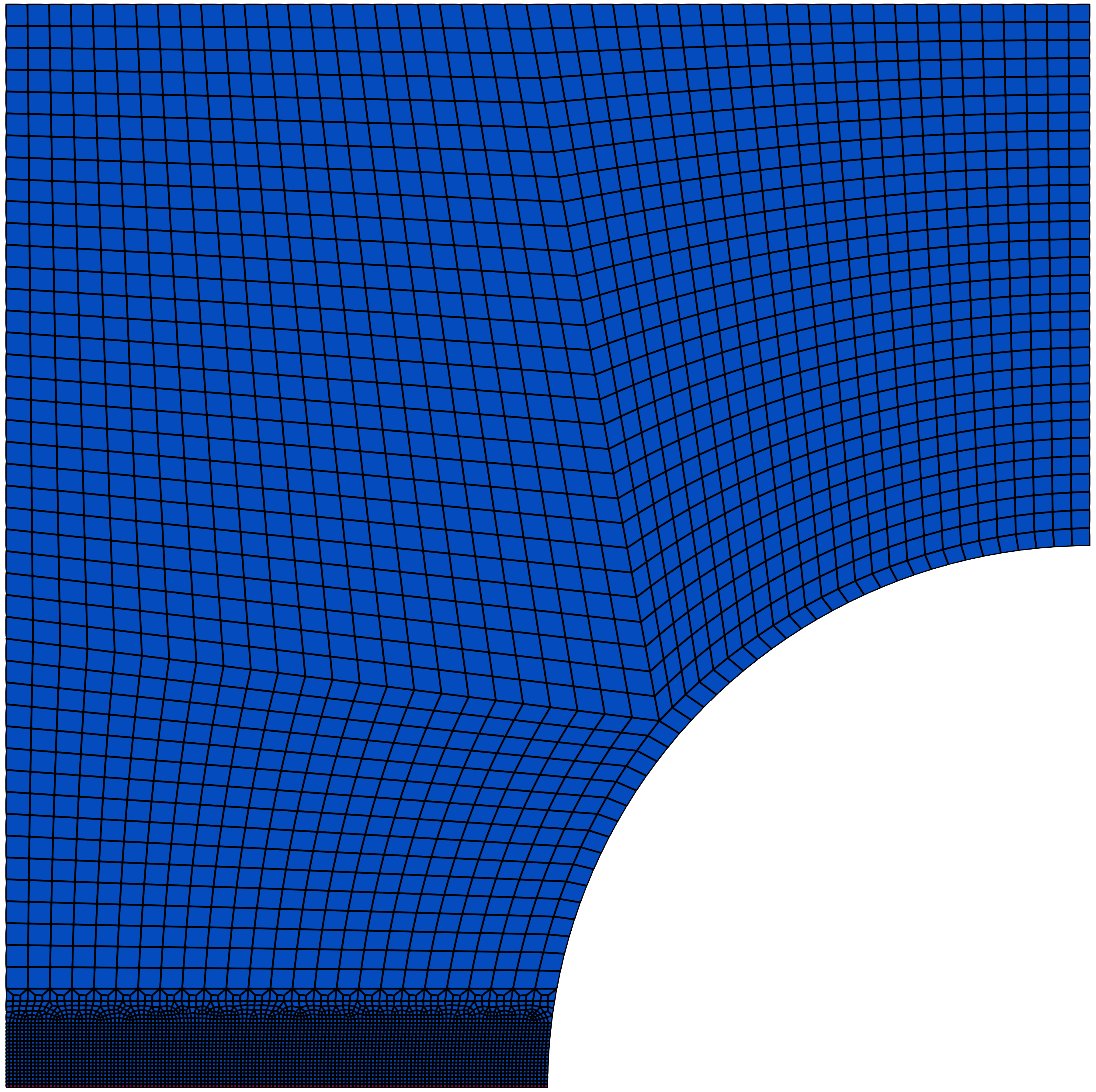}};
      \node[inner sep=0pt] (pic) at (0,-3) {\includegraphics[width=\textwidth]{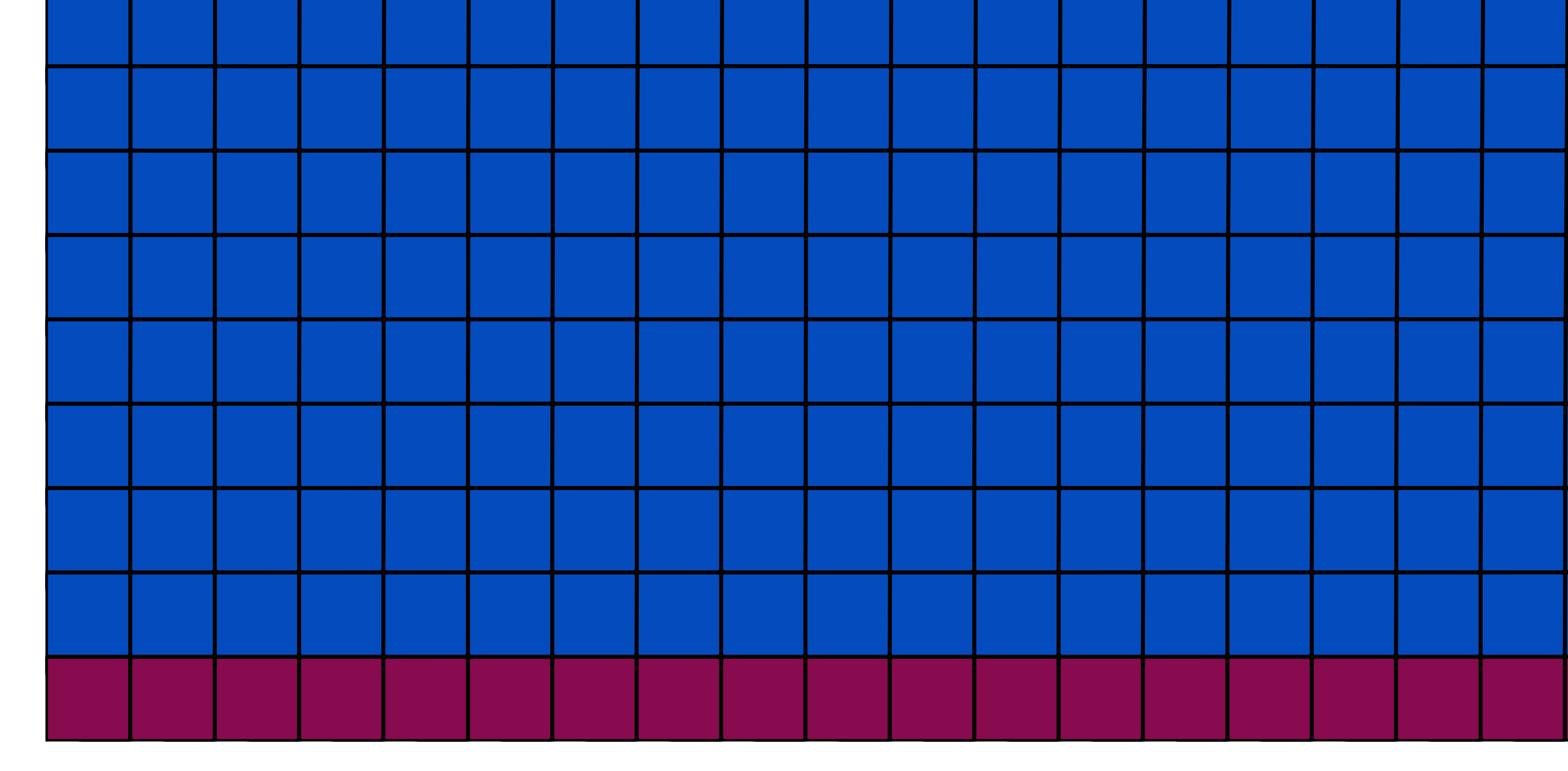}};
      \draw[draw=rwth8, line width=1.5pt] (-1.90,-1.90) rectangle (-1.68,-1.80);
      \draw[draw=rwth8, line width=1.5pt] (-1.82,-3.88) rectangle ( 1.92,-2.05);
      \draw[draw=rwth8, line width=1.5pt] (-1.79,-1.90) -- ( 0.05,-2.05);
      \end{tikzpicture} 
    \caption{5302}
  \end{subfigure}
  \begin{subfigure}{.24\textwidth} 
    \centering 
    \begin{tikzpicture}
      \node[inner sep=0pt] (pic) at (0,0) {\includegraphics[width=\textwidth]{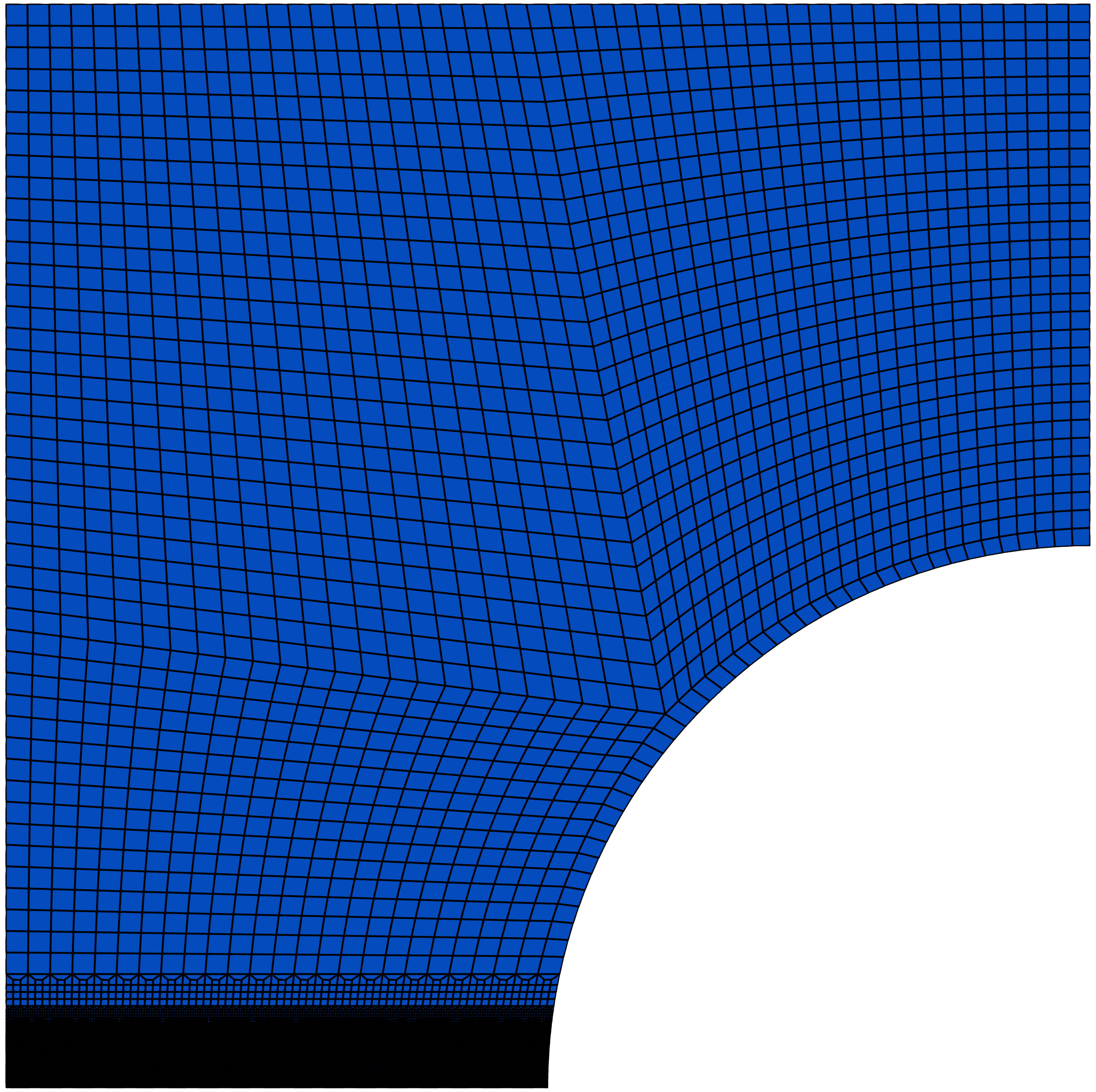}};
      \node[inner sep=0pt] (pic) at (0,-3) {\includegraphics[width=\textwidth]{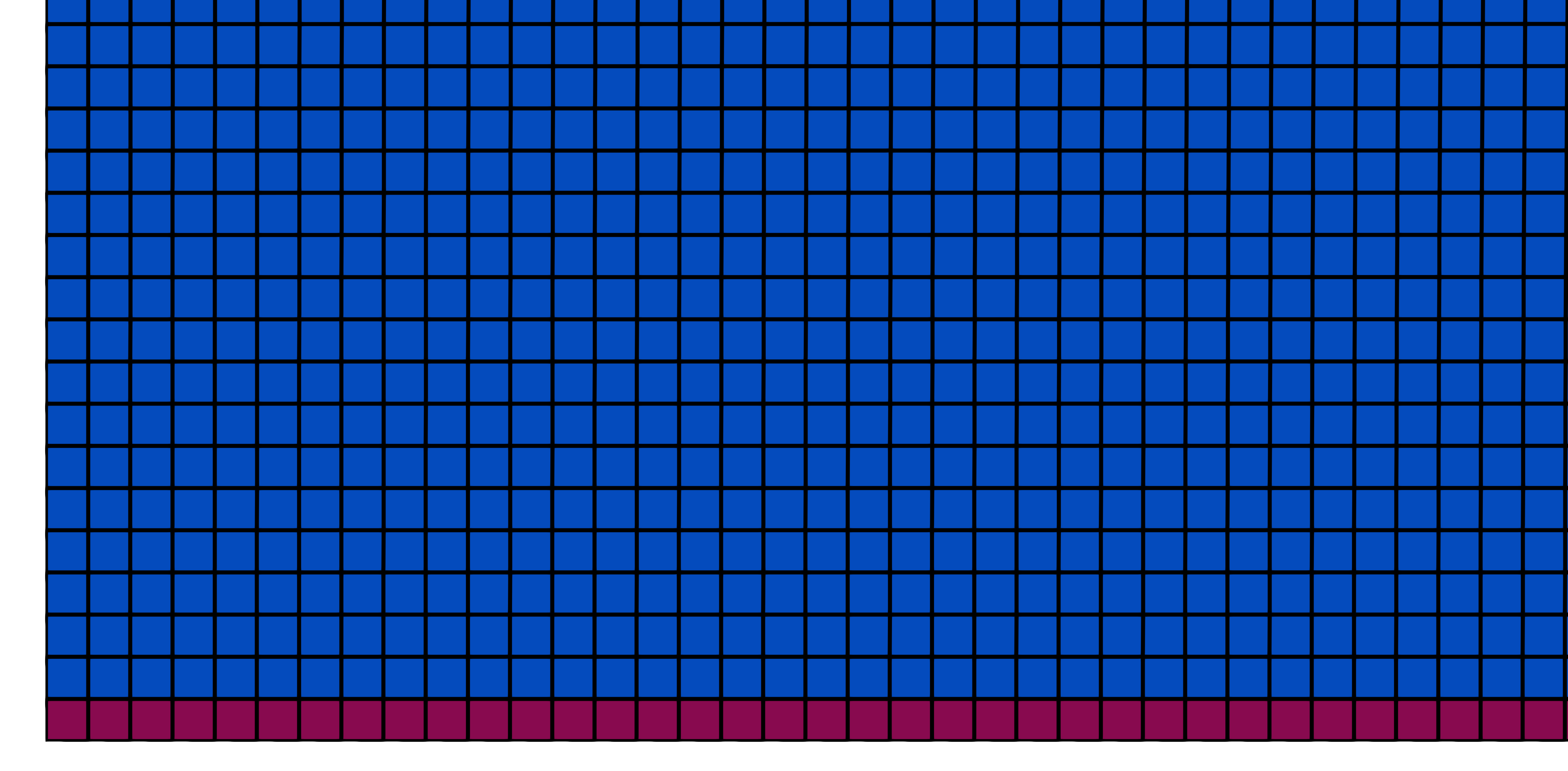}};
      \draw[draw=rwth8, line width=1.5pt] (-1.90,-1.90) rectangle (-1.68,-1.80);
      \draw[draw=rwth8, line width=1.5pt] (-1.82,-3.88) rectangle ( 1.92,-2.05);
      \draw[draw=rwth8, line width=1.5pt] (-1.79,-1.90) -- ( 0.05,-2.05);
    \end{tikzpicture} 
    \caption{14752}
  \end{subfigure}
  
  \vspace{2mm}
  
  \begin{subfigure}{\textwidth} 
    \centering 
    \begin{tikzpicture}
      \node[inner sep=0pt] (pic) at (0,0) {\includegraphics[height=5mm, width=40mm]{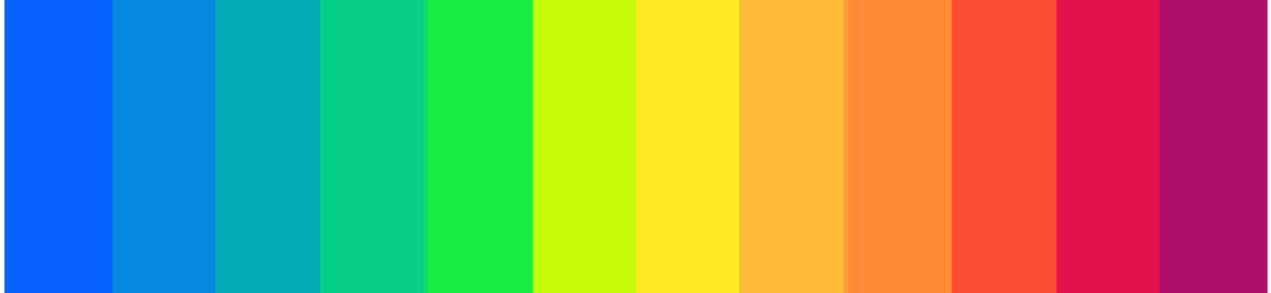}};
      \node[inner sep=0pt] (0)   at ($(pic.south)+(-2.22, 0.26)$)  {$0$};
      \node[inner sep=0pt] (1)   at ($(pic.south)+( 2.22, 0.26)$)  {$1$};
      \node[inner sep=0pt] (d)   at ($(pic.south)+(-3.30, 0.26)$)  {$D_{yy}~\si{[-]}$};
      \node[inner sep=0pt] (d)   at ($(pic.south)+( 3.30, 0.26)$)  {\hphantom{$D_{yy}~\si{[-]}$}};
    \end{tikzpicture} 
  \end{subfigure}
  
  \caption{Damage contour plots for the local anisotropic damage model for the plate with hole specimen captured at position $u/l \times \text{10}^\text{2}=0.5~\text{[--]}$ from Fig.~\ref{fig:ExpwhFuLocal} for different mesh discretizations. The damage values are averaged over all Gauss-points per element.} 
  \label{fig:ExpwhDLocal}     
\end{figure}

Then, we investigate the behavior of the local model formulation without utilizing a gradient-extension, analogously to \cite{FassinEggersmannEtAl2019a}, in order to ensure that no regularizing effects result from the use of an artificial viscosity. Fig.~\ref{fig:ExpwhFuLocal} shows the force-displacement curves for different mesh-discretizations and, as clearly indicated by the enlarged image section, no convergence with respect to the maximum force can be observed upon mesh refinement. This observation suggests the occurrence of localization in the simulation, which is confirmed by the damage contour plots in Fig.~\ref{fig:ExpwhDLocal}, where the crack localizes into a single row of elements for each mesh. From the results of Figs.~\ref{fig:ExpwhFuLocal} and \ref{fig:ExpwhDLocal} we can infer that the consideration of a sufficiently small artificial viscosity, here $\eta_v = 1~[\si{\N\s\per\square\mm}]$, does not cure the mesh dependence of the local damage model and, thus, does not interfere with the investigated regularizations. 

\begin{figure}[htbp]
  \centering
    \includegraphics{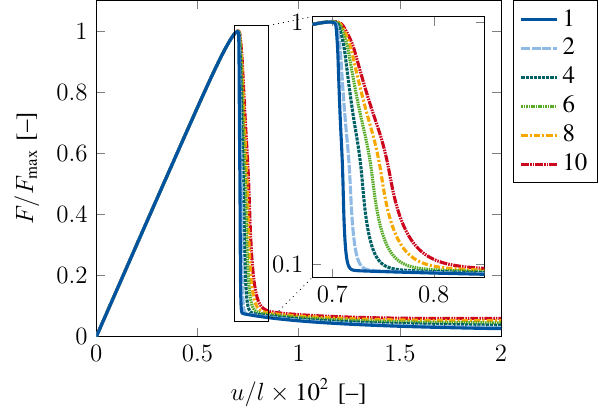}
  \caption{Force-displacement curves for the plate with hole specimen using model~C for a variation of the artificial viscosity $\eta_v$ (mesh 2039). The forces are normalized with respect to the maximum force of $\eta_v=1~[\si{\N\s\per\square\mm}]$ (2039 elements) with $F_\text{max} = 5.0920 \times 10^4~[\si{\newton}]$.}
  \label{fig:ExpwhFuEtav}
\end{figure}

\begin{figure}
  \centering 

  \begin{subfigure}{.3\textwidth} 
    \centering 
    \includegraphics[width=\textwidth]{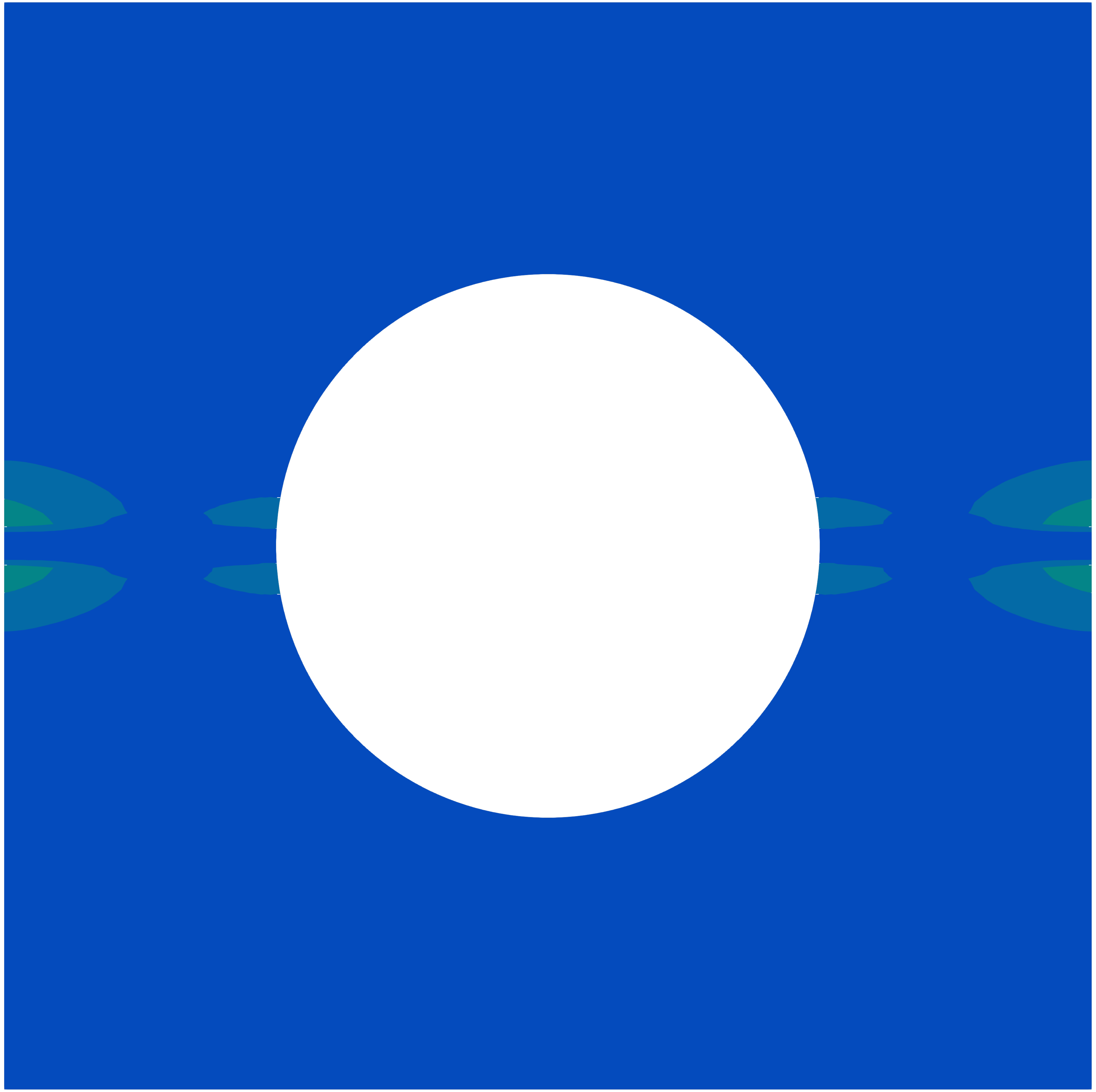}
  \end{subfigure}
  \begin{subfigure}{.3\textwidth} 
    \centering 
    \includegraphics[width=\textwidth]{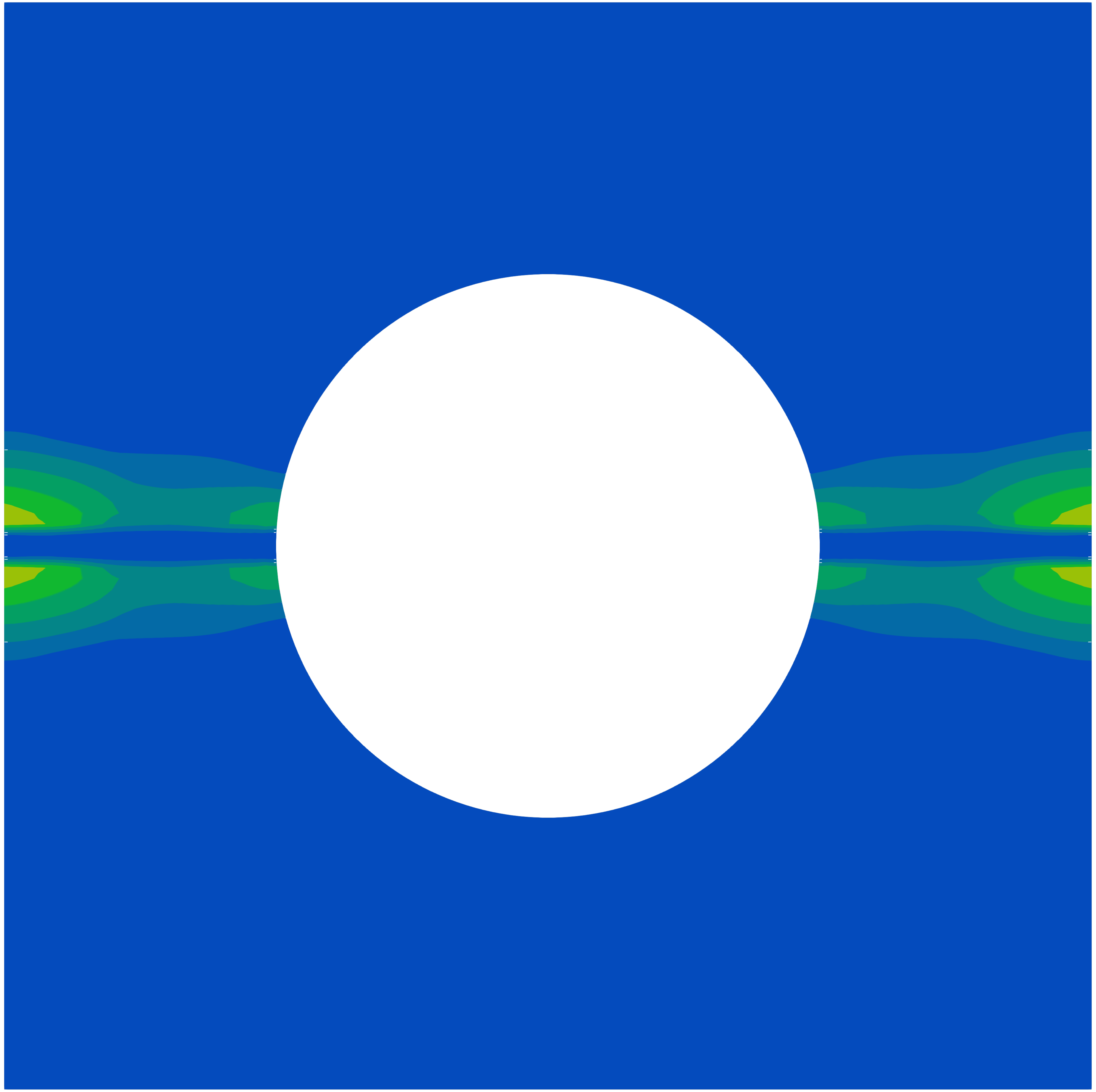}
  \end{subfigure}
  \begin{subfigure}{.3\textwidth} 
    \centering 
    \includegraphics[width=\textwidth]{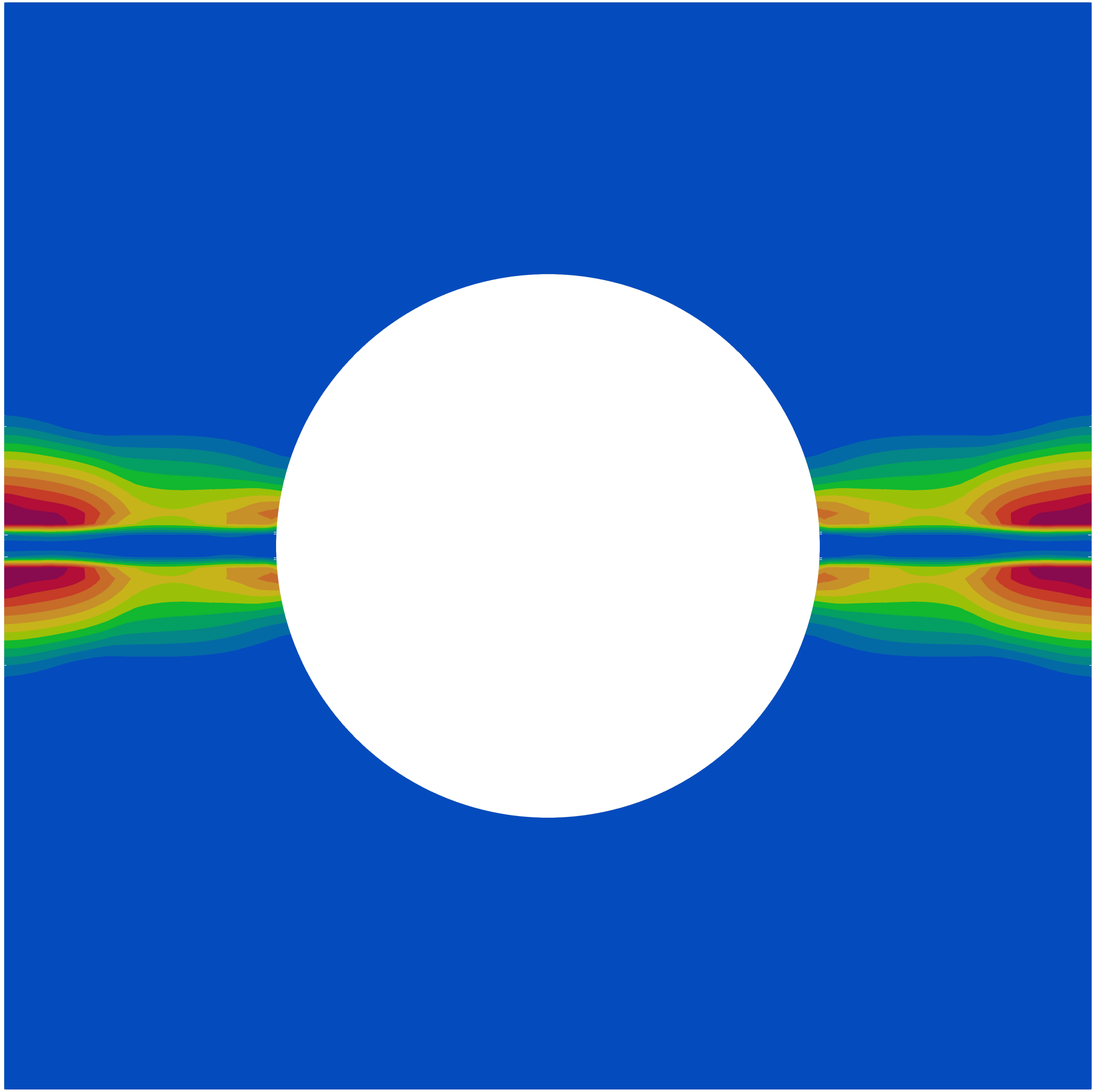}
  \end{subfigure}
  \begin{subfigure}{.08\textwidth} 
    \centering 
    \begin{tikzpicture}
      \node[inner sep=0pt] (pic) at (0,0) {\includegraphics[height=40mm, width=5mm]
      {02_Figures/03_Contour/00_Color_Maps/Damage_Step_Vertical.pdf}};
      \node[inner sep=0pt] (0)   at ($(pic.south)+( 0.50, 0.15)$)  {$0$};
      \node[inner sep=0pt] (1)   at ($(pic.south)+( 1.00, 3.80)$)  {$0.0386$};
      \node[inner sep=0pt] (d)   at ($(pic.south)+( 0.35, 4.35)$)  {$|\Delta D_{xx}|~\si{[-]}$};
    \end{tikzpicture} 
  \end{subfigure}

  \vspace{1mm}

  \begin{subfigure}{.3\textwidth} 
    \centering 
    \includegraphics[width=\textwidth]{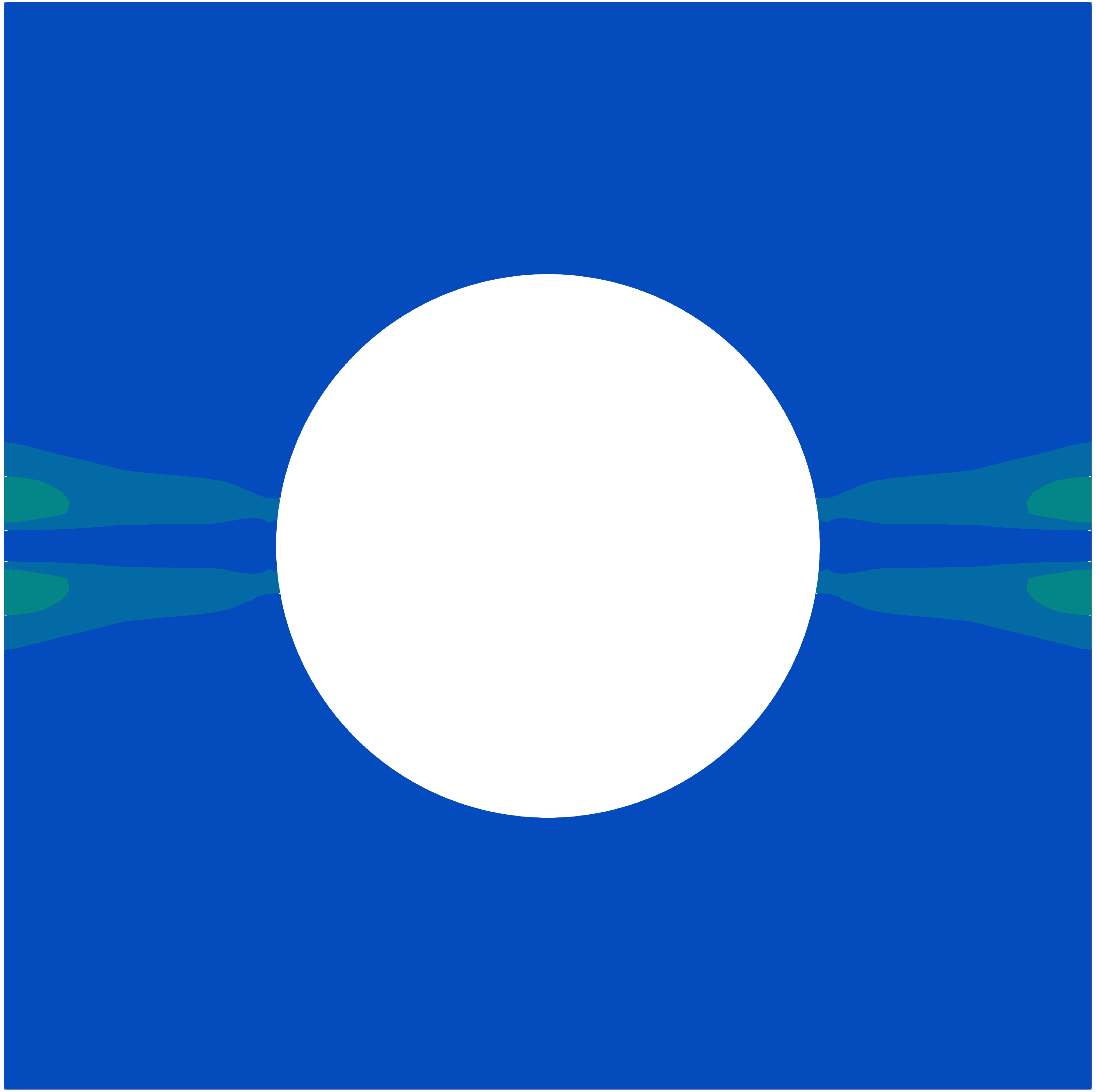}
  \end{subfigure}
  \begin{subfigure}{.3\textwidth} 
    \centering 
    \includegraphics[width=\textwidth]{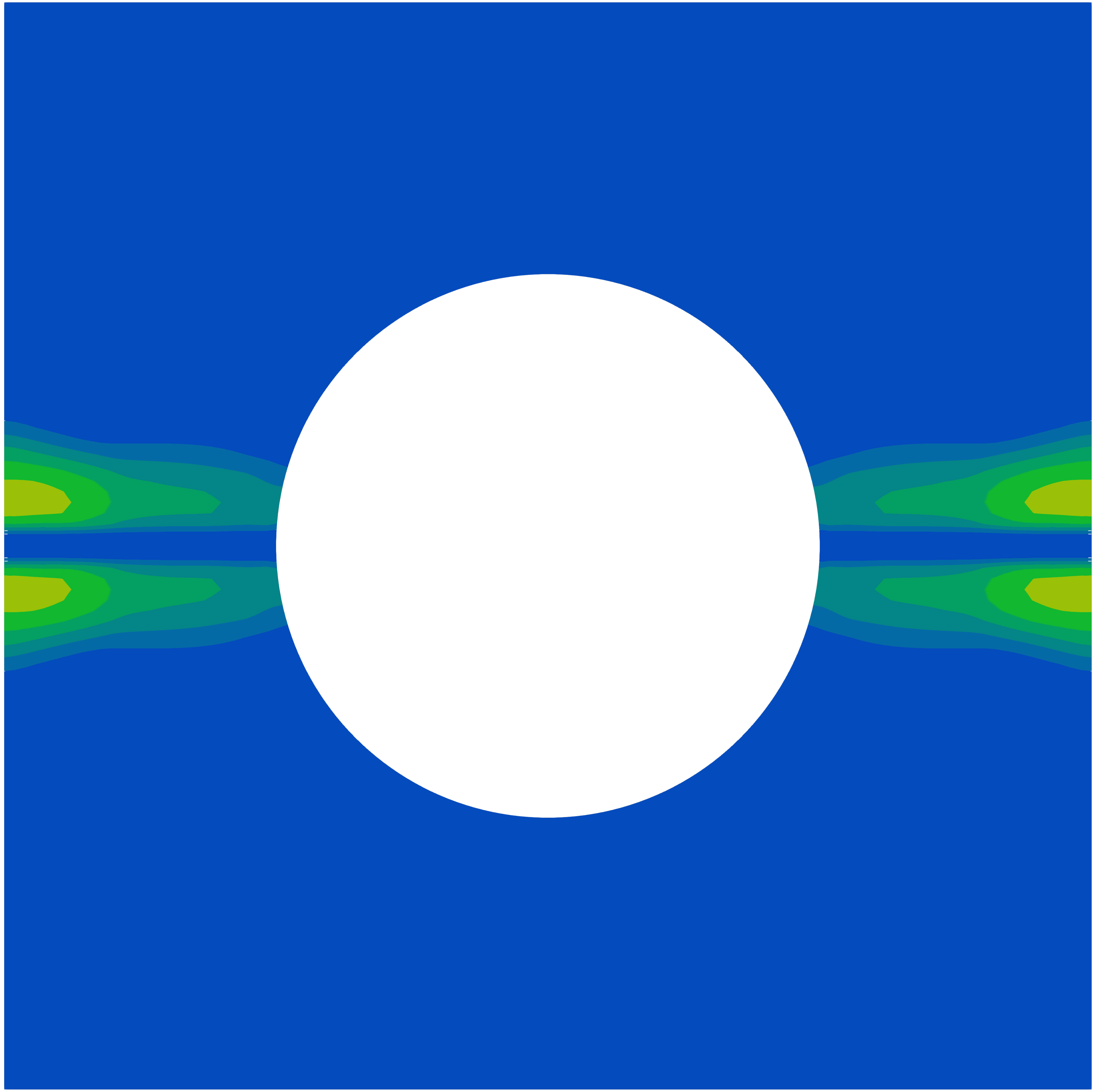}
  \end{subfigure}
  \begin{subfigure}{.3\textwidth} 
    \centering 
    \includegraphics[width=\textwidth]{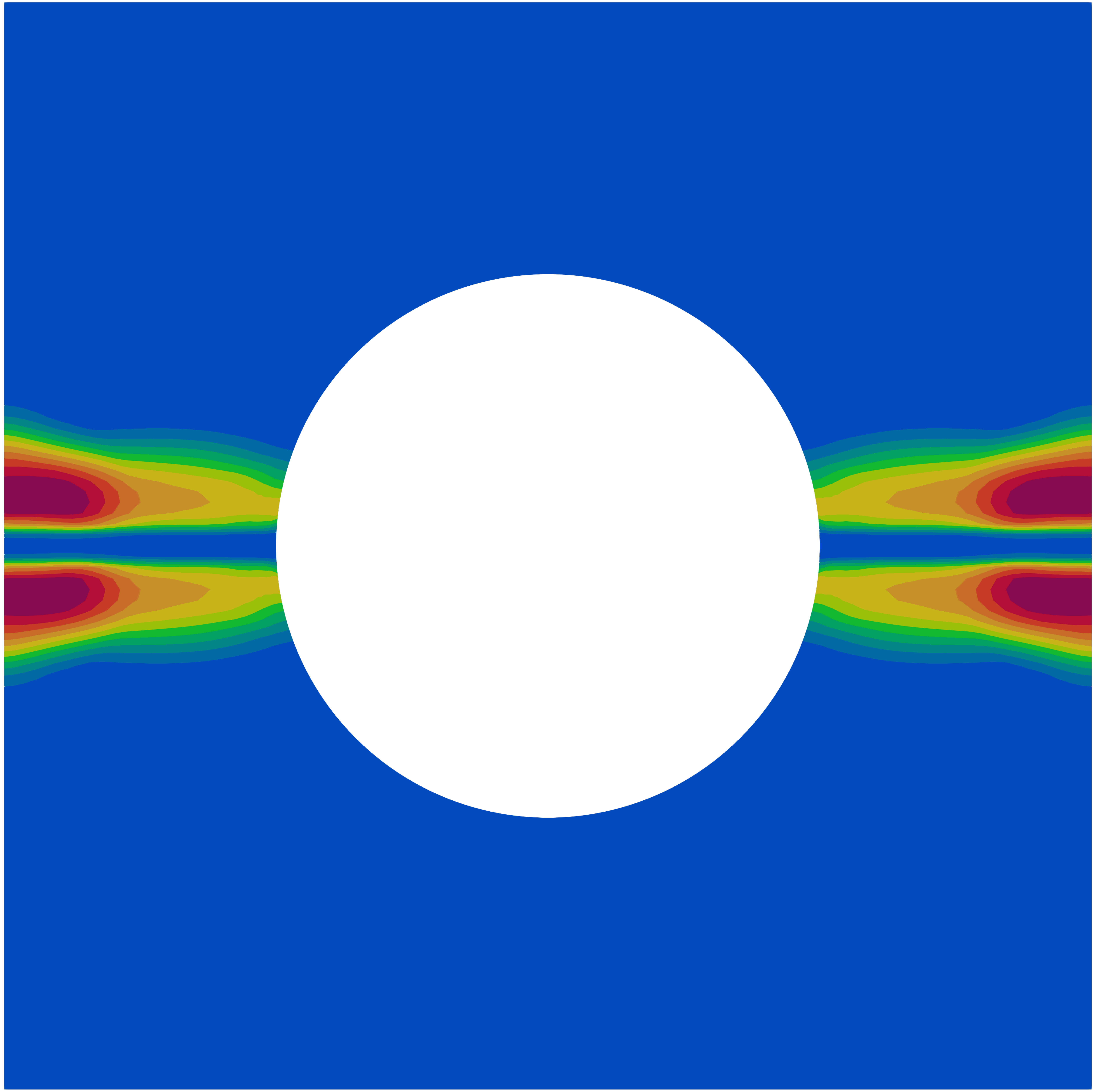}
  \end{subfigure}
  \begin{subfigure}{.08\textwidth} 
    \centering 
    \begin{tikzpicture}
      \node[inner sep=0pt] (pic) at (0,0) {\includegraphics[height=40mm, width=5mm]
      {02_Figures/03_Contour/00_Color_Maps/Damage_Step_Vertical.pdf}};
      \node[inner sep=0pt] (0)   at ($(pic.south)+( 0.50, 0.15)$)  {$0$};
      \node[inner sep=0pt] (1)   at ($(pic.south)+( 1.00, 3.80)$)  {$0.0395$};
      \node[inner sep=0pt] (d)   at ($(pic.south)+( 0.35, 4.35)$)  {$|\Delta D_{yy}|~\si{[-]}$};
    \end{tikzpicture} 
  \end{subfigure}

  \vspace{1mm}
  \begin{subfigure}{.3\textwidth} 
    \centering 
    \includegraphics[width=\textwidth]{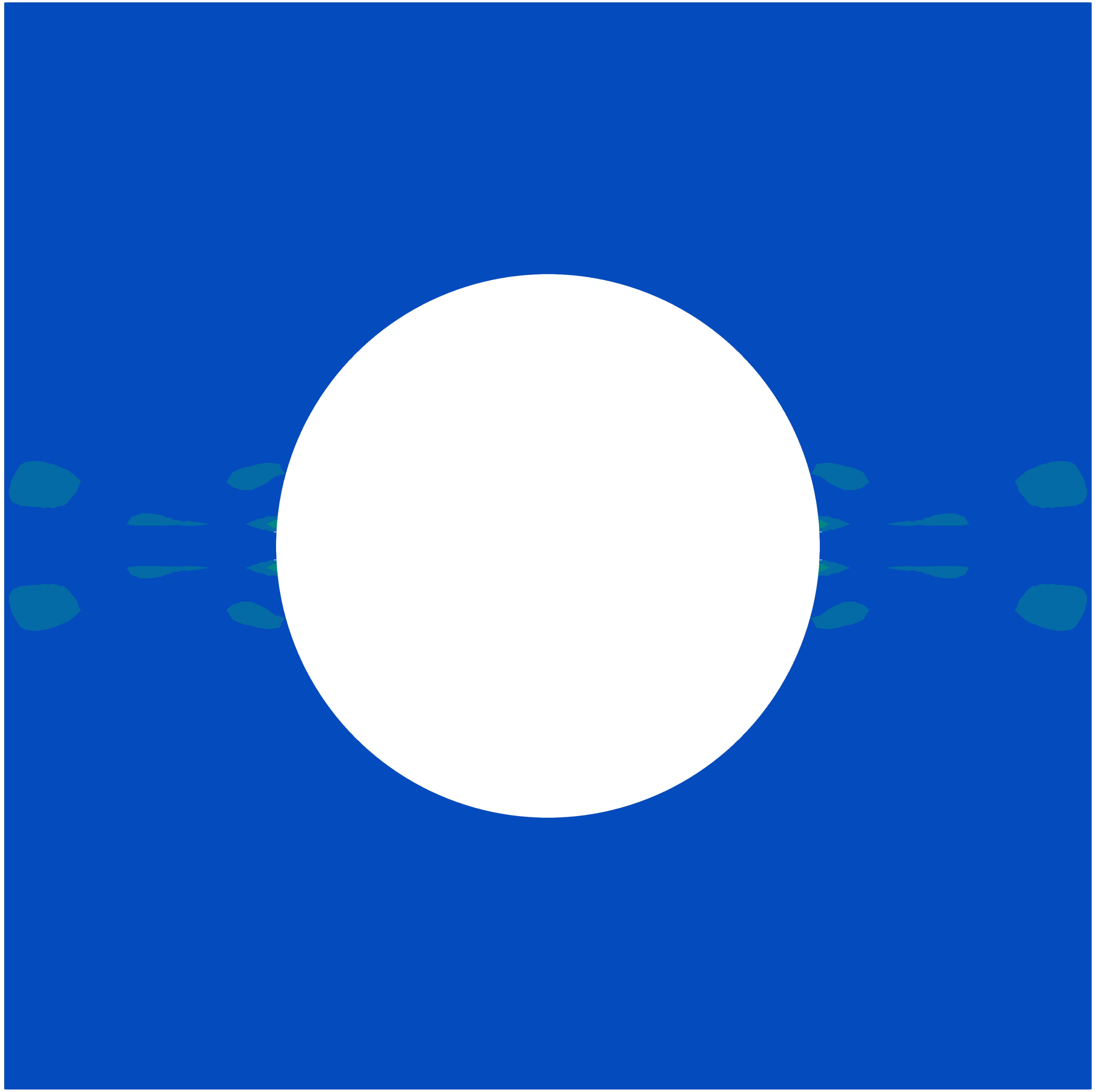}
    \caption{$\eta_v = 1~\text{vs.}~2~[\si{\N\s\per\square\mm}]$}
  \end{subfigure}
  \begin{subfigure}{.3\textwidth} 
    \centering 
    \includegraphics[width=\textwidth]{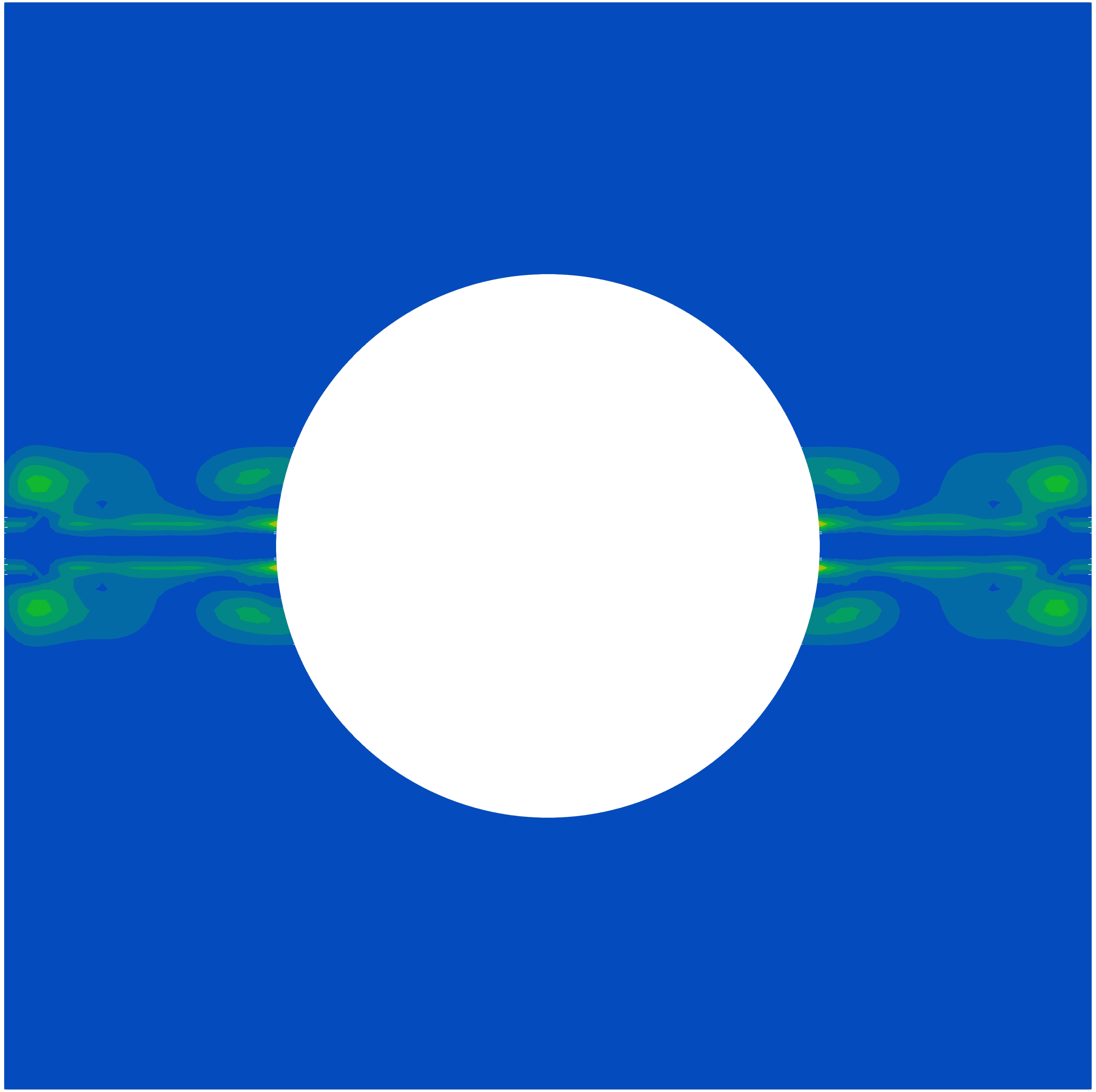}
    \caption{$\eta_v = 1~\text{vs.}~4~[\si{\N\s\per\square\mm}]$}
  \end{subfigure}
  \begin{subfigure}{.3\textwidth} 
    \centering 
    \includegraphics[width=\textwidth]{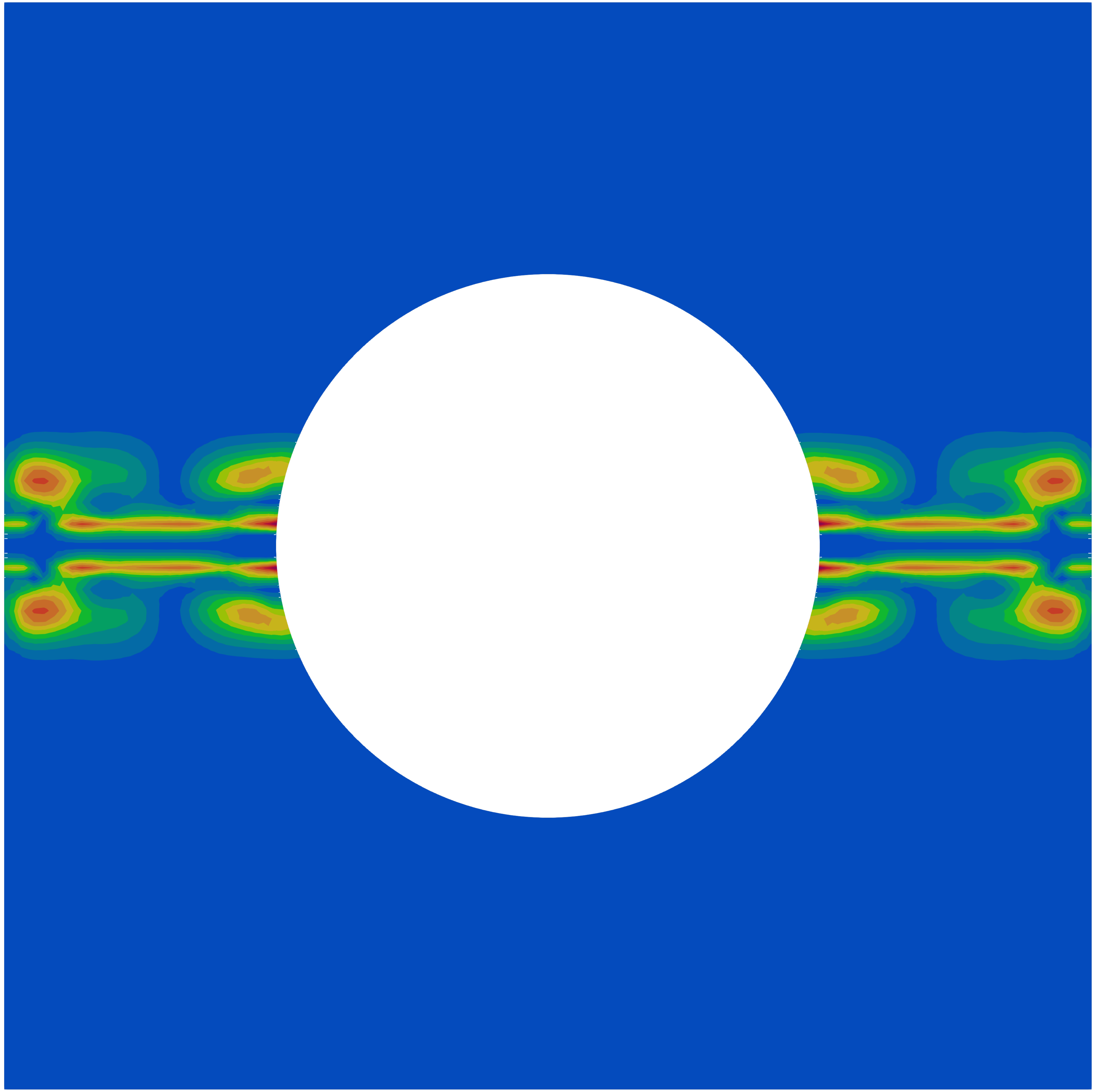}
    \caption{$\eta_v = 1~\text{vs.}~10~[\si{\N\s\per\square\mm}]$}
  \end{subfigure}
  \begin{subfigure}{.08\textwidth} 
    \centering 
    \begin{tikzpicture}
      \node[inner sep=0pt] (pic) at (0,0) {\includegraphics[height=40mm, width=5mm]
      {02_Figures/03_Contour/00_Color_Maps/Damage_Step_Vertical.pdf}};
      \node[inner sep=0pt] (0)   at ($(pic.south)+( 0.50, 0.15)$)  {$0$};
      \node[inner sep=0pt] (1)   at ($(pic.south)+( 1.00, 3.80)$)  {$0.0015$};
      \node[inner sep=0pt] (d)   at ($(pic.south)+( 0.35, 4.35)$)  {$|\Delta D_{xy}|~\si{[-]}$};
    \end{tikzpicture} 
    \hphantom{$\eta_v = 1~\text{vs.}~2$}
  \end{subfigure}
  
  \caption{Difference plot of the damage contours of the normal and shear components of the damage tensor for the plate with hole specimen using model~C for different values of the artificial viscosity $\eta_v$ at the end of the simulation of Fig.~\ref{fig:ExpwhFuEtav}.
  } 
  \label{fig:ExpwhDiffEtav}     
\end{figure}

Nevertheless, the model's response is obviously not completely independent of the choice of the artificial viscosity. Thus, we study the influence of the parameter $\eta_v$ in Figs.~\ref{fig:ExpwhFuEtav} and \ref{fig:ExpwhDiffEtav} using model~C. Fig.~\ref{fig:ExpwhFuEtav} shows the increasing the artificial viscosity leads to a less step drop in the force-displacement curve after reaching the maximum peak load and, also, to a higher residual force after the failure of the specimen. However, the maximum load bearing capacity of the structure is unaffected by a variation of $\eta_v$. Fig.~\ref{fig:ExpwhDiffEtav} shows the difference plots for the components of the damage tensor comparing the results of using $\eta_v = 1~[\si{\N\s\per\square\mm}]$ versus $\eta_v = 2/4/10~[\si{\N\s\per\square\mm}]$. Even for an increase of the artificial viscosity by a factor of ten, the maximum difference for the normal and shear components yields only values of $|\Delta D_{xx}|=0.0386~\si{[-]}$, $|\Delta D_{yy}|=0.0395~\si{[-]}$, and $|\Delta D_{xy}|=0.0015~\si{[-]}$.

These studies have proven the negligible influence of the artificial viscosity on the results of the simulation and justify its use in the present work to allow for a displacement-driven load control.

\subsection{Asymmetrically notched specimen}
\label{sec:Ex_an}

\begin{figure}[htbp] 
  \centering 
  \begin{subfigure}{.7\textwidth} 
      \centering 
      \begin{tikzpicture}
        \node (pic) at (0,0) {\includegraphics[width=\textwidth]{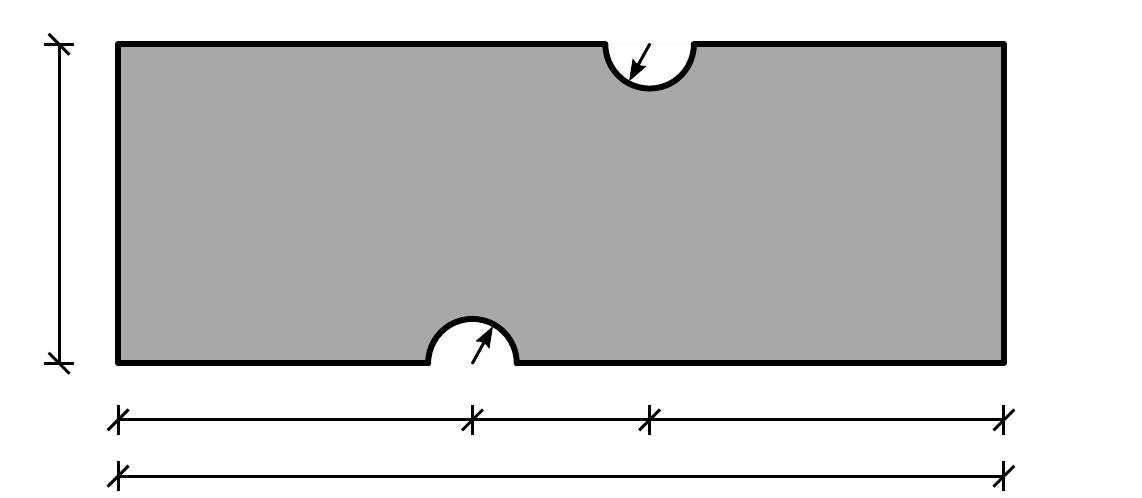}};
        \node (l1)  at ($(pic.south)+(-2.50, 0.63)$)  {$l_1$};
        \node (l1-2)at ($(pic.south)+( 2.50, 0.63)$)  {$l_1$};
        \node (l2)  at ($(pic.south)+( 0.00, 0.63)$)  {$l_2$};
        \node (l)   at ($(pic.south)+( 0.00, 0.10)$)  {$l$};
        \node (h)   at ($(pic.south)+(-5.30, 3.10)$)  {$h$};
        \node (r1)  at ($(pic.south)+( 0.92, 4.42)$)  {$r$};
        \node (r2)  at ($(pic.south)+(-1.10, 1.60)$)  {$r$};
      \end{tikzpicture} 
      \caption{Geometry}
      \label{fig:Ex_an_geom}
  \end{subfigure}

  \begin{subfigure}{.7\textwidth} 
      \centering 
      \begin{tikzpicture}
        \node[inner sep=0pt] (pic) at (0,0) {\includegraphics[width=\textwidth]{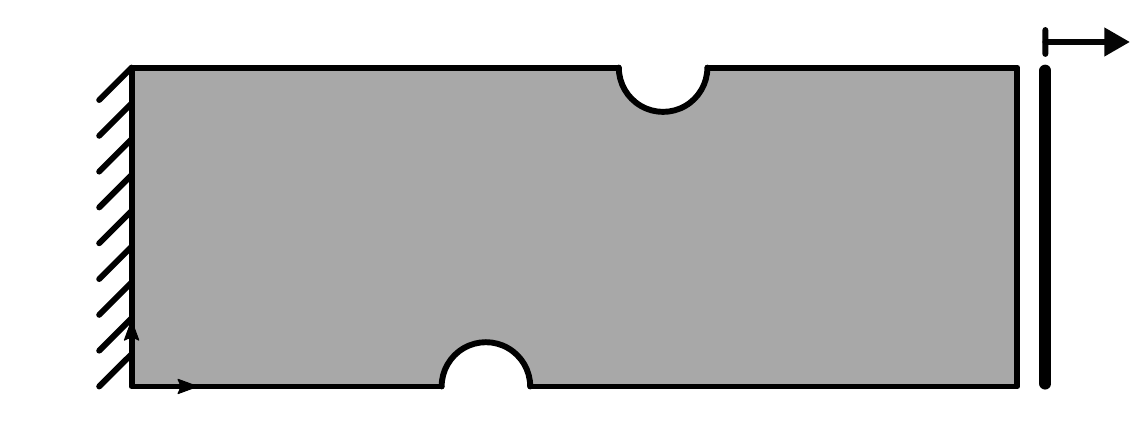}};
        \node[inner sep=0pt] (Fu) at ($(pic.south) +( 4.95, 4.20)$)  {$F,u$};
        \node[inner sep=0pt] (x)  at ($(pic.south) +(-3.65, 0.70)$)  {$x$};
        \node[inner sep=0pt] (y)  at ($(pic.south) +(-4.10, 1.15)$)  {$y$};
      \end{tikzpicture} 
      \caption{Boundary value problem}
      \label{fig:Ex_an_bvp}
  \end{subfigure}
  \caption{Geometry and boundary value problem of the asymmetrically notched specimen.} 
  \label{fig:Ex_an}
\end{figure}

The next example compares the three gradient-extensions for a combined tension and shear loading situation and considers an asymmetrically notched specimen. This example has also been investigated in e.g.~\cite{FriedleinMergheimEtAl2023}, \cite{FelderKopic-OsmanovicEtAl2022}, \citetalias{BarfuszBrepolsEtAl2021}[\citeyear{BarfuszBrepolsEtAl2021}], \cite{BrepolsWulfinghoffEtAl2020}, \cite{AmbatiKruseEtAl2016} and also in \citetalias{HolthusenBrepolsEtAl2022a}[\citeyear{HolthusenBrepolsEtAl2022a}], where the same boundary value problem with the same material parameters is solved for model~B using an arc-length controlled method yielding a double snap-back. These results serve in this section as an additional reference solution for model~B and confirm the displacement controlled simulation results using the artificial viscosity.

Fig.~\ref{fig:Ex_an} shows the geometry and the corresponding boundary value problem. The dimensions read $h = 36~[\si{\mm}]$, $l = 100~[\si{\mm}]$, $l_1 = 40~[\si{\mm}]$, $l_2 = 20~[\si{\mm}]$ and $r = 5~[\si{\mm}]$ with a thickness of $1~[\si{\mm}]$. The finite element meshes stem from \cite{FelderKopic-OsmanovicEtAl2022} and \citetalias{HolthusenBrepolsEtAl2022a}[\citeyear{HolthusenBrepolsEtAl2022a}]. The internal length scales of model B are chosen, analogously to \citetalias{HolthusenBrepolsEtAl2022a}[\citeyear{HolthusenBrepolsEtAl2022a}], as $A_i^\text{B} = 100~[\si{\MPa\mm\squared}]$ and the parameters of model~A and C are identified as $A_i^\text{A} = 330~[\si{\MPa\mm\squared}]$ and $A_i^\text{C} = 1100~[\si{\MPa\mm\squared}]$.

\begin{figure}[htbp]
  \centering
  \begin{subfigure}{.48\textwidth}
    \centering
    \includegraphics{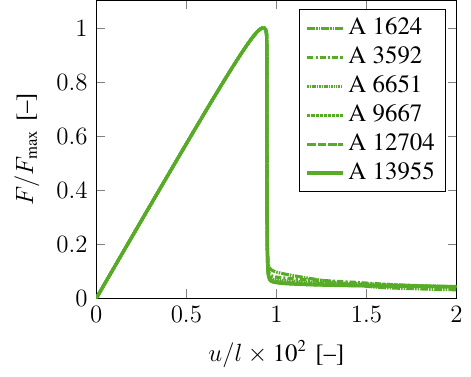}
    \vspace{-7mm}
    \caption{Model~A}
    \label{fig:ExanFuA}
  \end{subfigure}
  \quad
  \begin{subfigure}{.48\textwidth}
    \centering
    \includegraphics{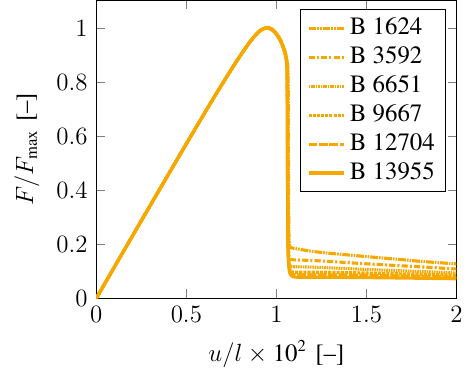}
    \vspace{-7mm}
    \caption{Model~B}
    \label{fig:ExanFuB}
  \end{subfigure}%
  \vspace{5mm} 
  \begin{subfigure}{.48\textwidth}
    \centering
    \includegraphics{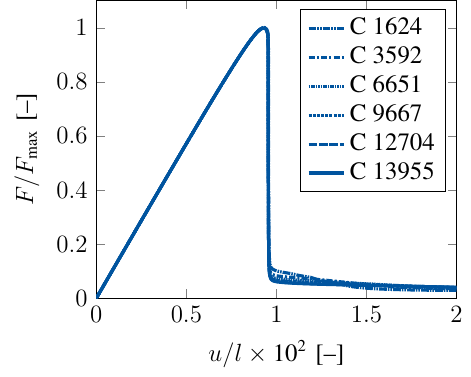}
    \vspace{-7mm}
    \caption{Model~C}
    \label{fig:ExanFuC}
  \end{subfigure}
  \quad
  \begin{subfigure}{.48\textwidth}
    \centering
    \includegraphics{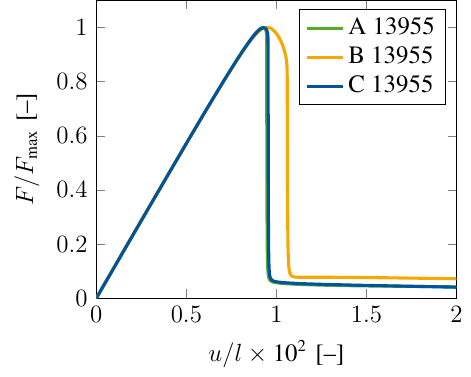}
    \vspace{-7mm}
    \caption{Model comparison}
    \label{fig:ExanFuComp}
  \end{subfigure}
  \caption{Mesh convergence studies for the asymmetrically notched specimen and model comparison. The forces are normalized with respect to the maximum force of the finest mesh (13955 elements) of model~B with $F_\text{max} = 3.7959 \times 10^4~[\si{\newton}]$.}
  \label{fig:ExanFu}
\end{figure}

\begin{figure}
  \centering 
  
  \begin{subfigure}{.3\textwidth} 
    \centering 
    \begin{tikzpicture}
      \node[inner sep=0pt] (pic) at (0,0) {\includegraphics[width=0.8\textwidth]{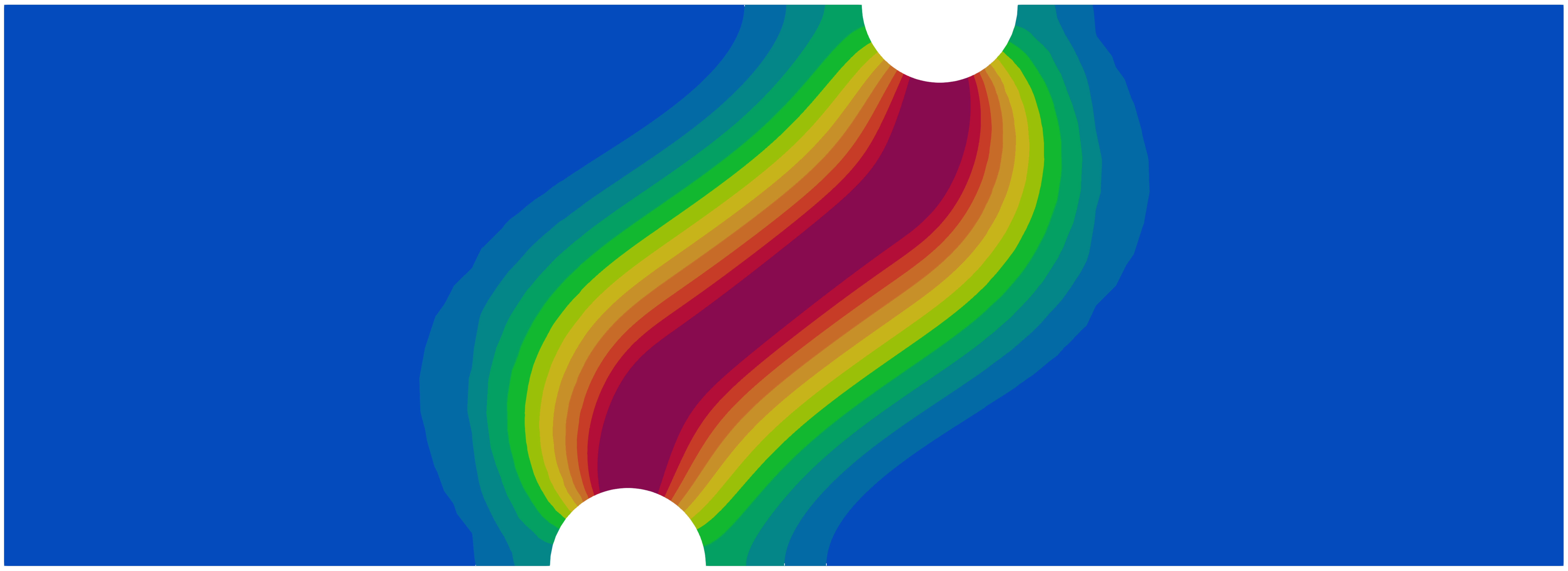}};
      \node[inner sep=0pt] (pic2) at (0,-2.65) {\includegraphics[width=\textwidth]{02_Figures/03_Contour/Ex2/mesh_convergence/Zoom_A-13955_Dxx.pdf}};
      \draw[draw=rwth8, line width=1.5pt] (-0.92,-0.70) rectangle ( 0.92, 0.70);
      \draw[draw=rwth8, line width=1.5pt] (-2.40,-4.40) rectangle ( 2.40,-0.90);
      \draw[draw=rwth8, line width=1.5pt] ( 0.00,-0.70) -- ( 0.00,-0.90);
    \end{tikzpicture} 
  \end{subfigure}
  \begin{subfigure}{.3\textwidth} 
    \centering 
    \begin{tikzpicture}
      \node[inner sep=0pt] (pic) at (0,0) {\includegraphics[width=0.8\textwidth]{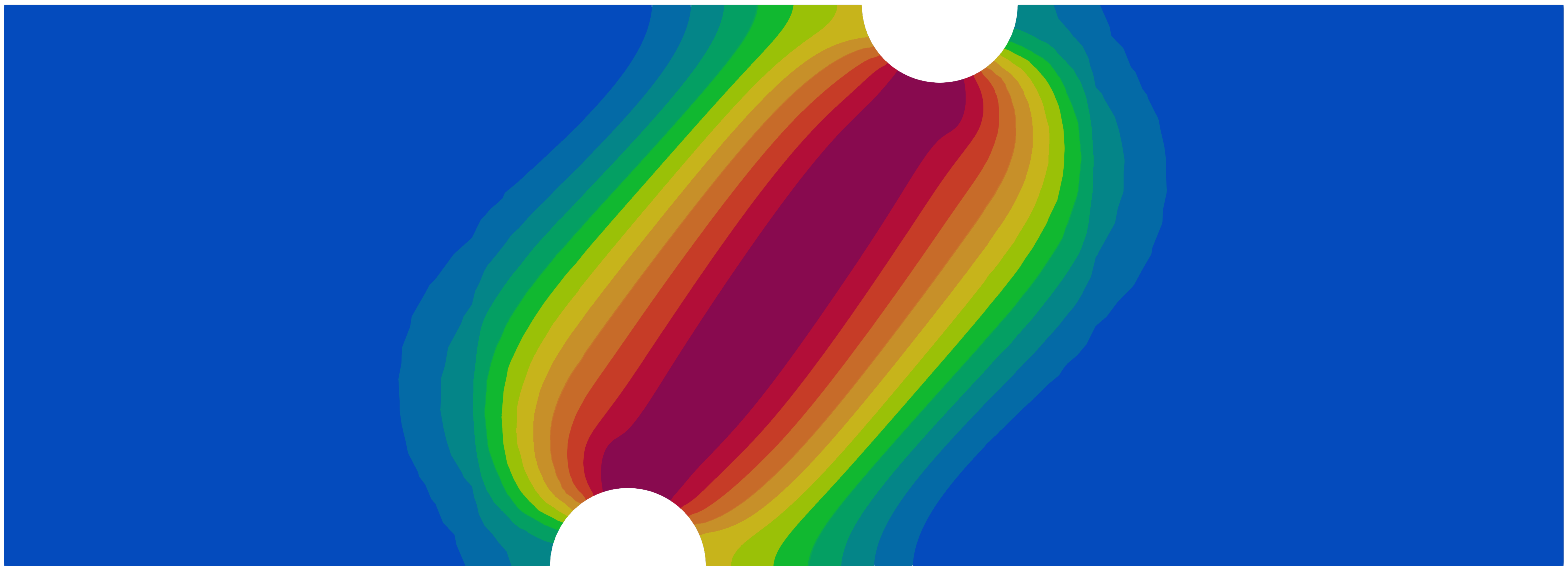}};
      \node[inner sep=0pt] (pic2) at (0,-2.65) {\includegraphics[width=\textwidth]{02_Figures/03_Contour/Ex2/mesh_convergence/Zoom_B-13955_Dxx.pdf}};
      \draw[draw=rwth8, line width=1.5pt] (-0.92,-0.70) rectangle ( 0.92, 0.70);
      \draw[draw=rwth8, line width=1.5pt] (-2.40,-4.40) rectangle ( 2.40,-0.90);
      \draw[draw=rwth8, line width=1.5pt] ( 0.00,-0.70) -- ( 0.00,-0.90);
    \end{tikzpicture} 
  \end{subfigure}
  \begin{subfigure}{.3\textwidth} 
    \centering 
    \begin{tikzpicture}
      \node[inner sep=0pt] (pic) at (0,0) {\includegraphics[width=0.8\textwidth]{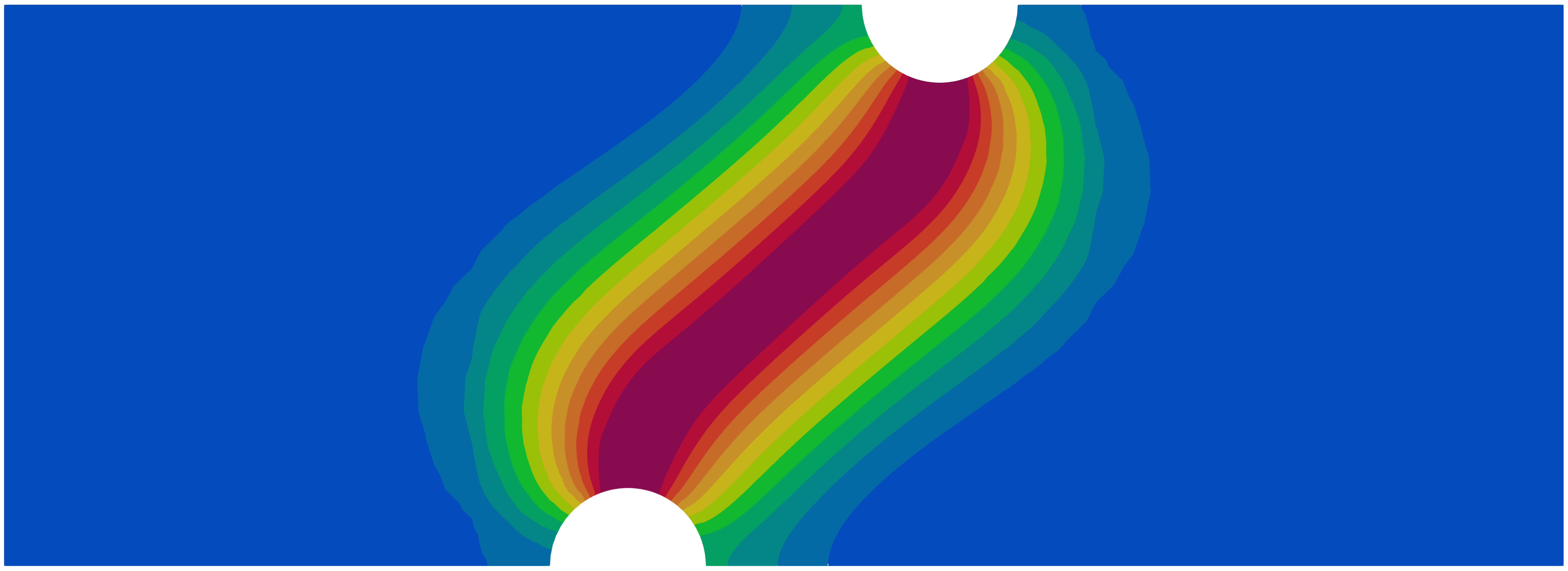}};
      \node[inner sep=0pt] (pic2) at (0,-2.65) {\includegraphics[width=\textwidth]{02_Figures/03_Contour/Ex2/mesh_convergence/Zoom_C-13955_Dxx.pdf}};
      \draw[draw=rwth8, line width=1.5pt] (-0.92,-0.70) rectangle ( 0.92, 0.70);
      \draw[draw=rwth8, line width=1.5pt] (-2.40,-4.40) rectangle ( 2.40,-0.90);
      \draw[draw=rwth8, line width=1.5pt] ( 0.00,-0.70) -- ( 0.00,-0.90);
    \end{tikzpicture} 
  \end{subfigure}
  \begin{subfigure}{.08\textwidth} 
    \centering 
    \begin{tikzpicture}
      \node[inner sep=0pt] (pic) at (0,0) {\includegraphics[height=40mm, width=5mm]
      {02_Figures/03_Contour/00_Color_Maps/Damage_Step_Vertical.pdf}};
      \node[inner sep=0pt] (0)   at ($(pic.south)+( 0.50, 0.15)$)  {$0$};
      \node[inner sep=0pt] (1)   at ($(pic.south)+( 0.50, 3.80)$)  {$1$};
      \node[inner sep=0pt] (d)   at ($(pic.south)+( 0.00, 4.35)$)  {$D_{xx}~\si{[-]}$};
    \end{tikzpicture} 
  \end{subfigure}

  \vspace{5mm}  

  \begin{subfigure}{.3\textwidth} 
    \centering 
    \begin{tikzpicture}
      \node[inner sep=0pt] (pic) at (0,0) {\includegraphics[width=0.8\textwidth]{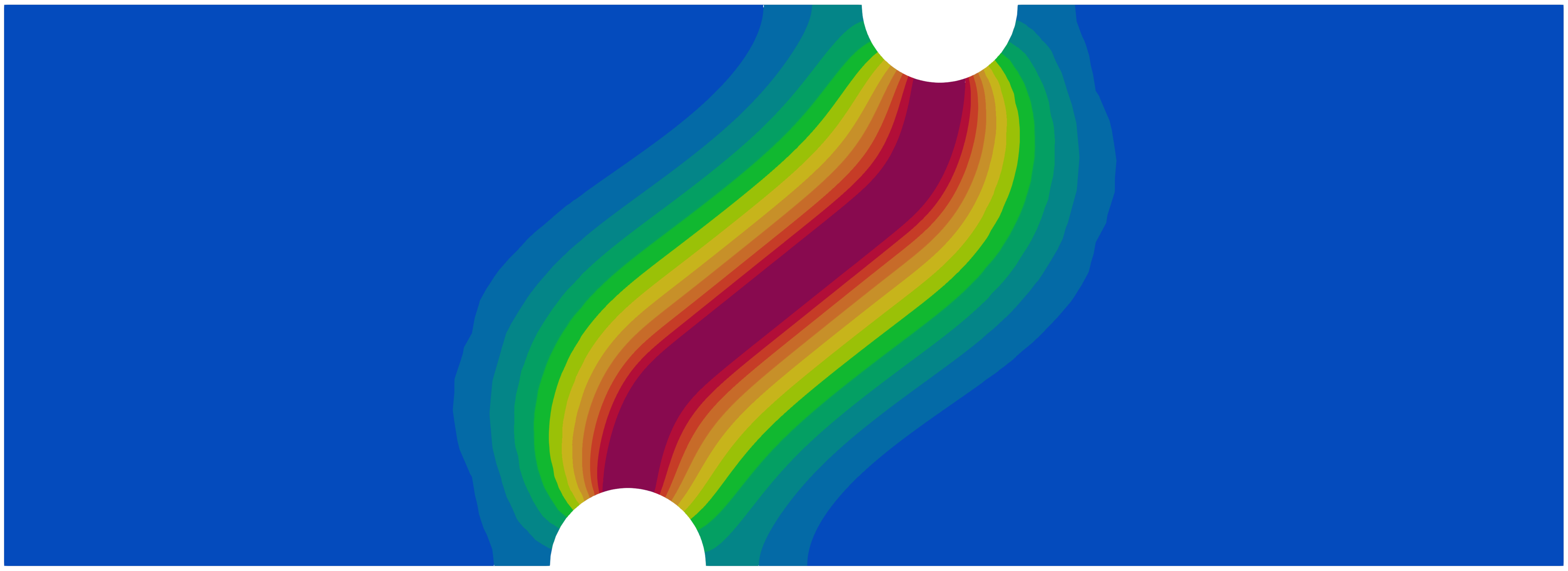}};
      \node[inner sep=0pt] (pic2) at (0,-2.65) {\includegraphics[width=\textwidth]{02_Figures/03_Contour/Ex2/mesh_convergence/Zoom_A-13955_Dyy.pdf}};
      \draw[draw=rwth8, line width=1.5pt] (-0.92,-0.70) rectangle ( 0.92, 0.70);
      \draw[draw=rwth8, line width=1.5pt] (-2.40,-4.40) rectangle ( 2.40,-0.90);
      \draw[draw=rwth8, line width=1.5pt] ( 0.00,-0.70) -- ( 0.00,-0.90);
    \end{tikzpicture} 
  \end{subfigure}
  \begin{subfigure}{.3\textwidth} 
    \centering 
    \begin{tikzpicture}
      \node[inner sep=0pt] (pic) at (0,0) {\includegraphics[width=0.8\textwidth]{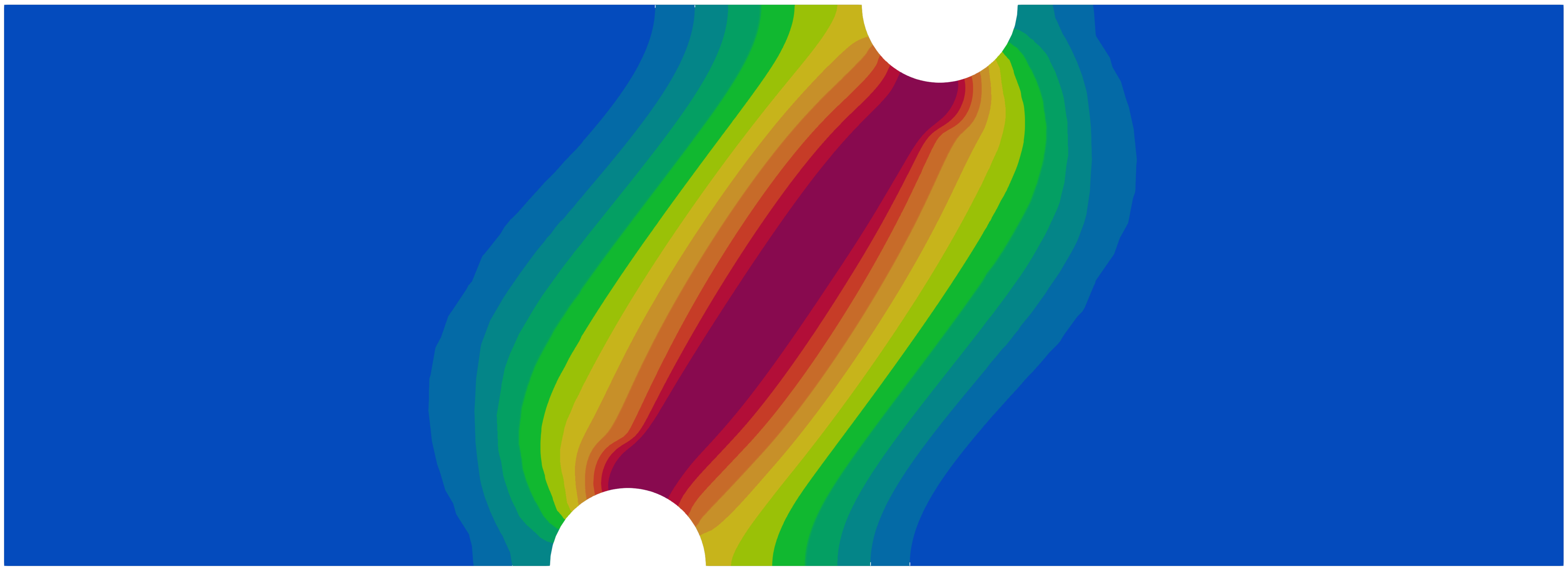}};
      \node[inner sep=0pt] (pic2) at (0,-2.65) {\includegraphics[width=\textwidth]{02_Figures/03_Contour/Ex2/mesh_convergence/Zoom_B-13955_Dyy.pdf}};
      \draw[draw=rwth8, line width=1.5pt] (-0.92,-0.70) rectangle ( 0.92, 0.70);
      \draw[draw=rwth8, line width=1.5pt] (-2.40,-4.40) rectangle ( 2.40,-0.90);
      \draw[draw=rwth8, line width=1.5pt] ( 0.00,-0.70) -- ( 0.00,-0.90);
    \end{tikzpicture} 
  \end{subfigure}
  \begin{subfigure}{.3\textwidth} 
    \centering 
    \begin{tikzpicture}
      \node[inner sep=0pt] (pic) at (0,0) {\includegraphics[width=0.8\textwidth]{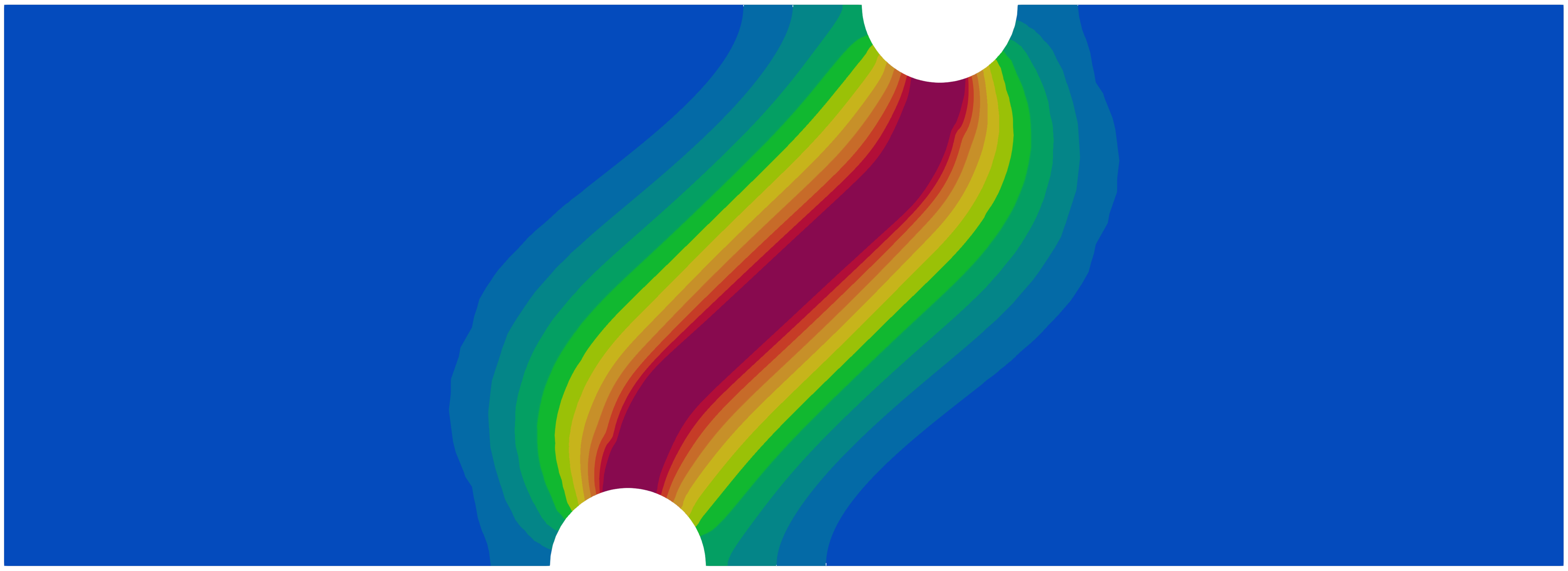}};
      \node[inner sep=0pt] (pic2) at (0,-2.65) {\includegraphics[width=\textwidth]{02_Figures/03_Contour/Ex2/mesh_convergence/Zoom_C-13955_Dyy.pdf}};
      \draw[draw=rwth8, line width=1.5pt] (-0.92,-0.70) rectangle ( 0.92, 0.70);
      \draw[draw=rwth8, line width=1.5pt] (-2.40,-4.40) rectangle ( 2.40,-0.90);
      \draw[draw=rwth8, line width=1.5pt] ( 0.00,-0.70) -- ( 0.00,-0.90);
    \end{tikzpicture} 
  \end{subfigure}
  \begin{subfigure}{.08\textwidth} 
    \centering 
    \begin{tikzpicture}
      \node[inner sep=0pt] (pic) at (0,0) {\includegraphics[height=40mm, width=5mm]
      {02_Figures/03_Contour/00_Color_Maps/Damage_Step_Vertical.pdf}};
      \node[inner sep=0pt] (0)   at ($(pic.south)+( 0.50, 0.15)$)  {$0$};
      \node[inner sep=0pt] (1)   at ($(pic.south)+( 0.50, 3.80)$)  {$1$};
      \node[inner sep=0pt] (d)   at ($(pic.south)+( 0.00, 4.35)$)  {$D_{yy}~\si{[-]}$};
    \end{tikzpicture} 
  \end{subfigure}

  \vspace{5mm}  
  
  \begin{subfigure}{.3\textwidth} 
    \centering 
    \begin{tikzpicture}
      \node[inner sep=0pt] (pic) at (0,0) {\includegraphics[width=0.8\textwidth]{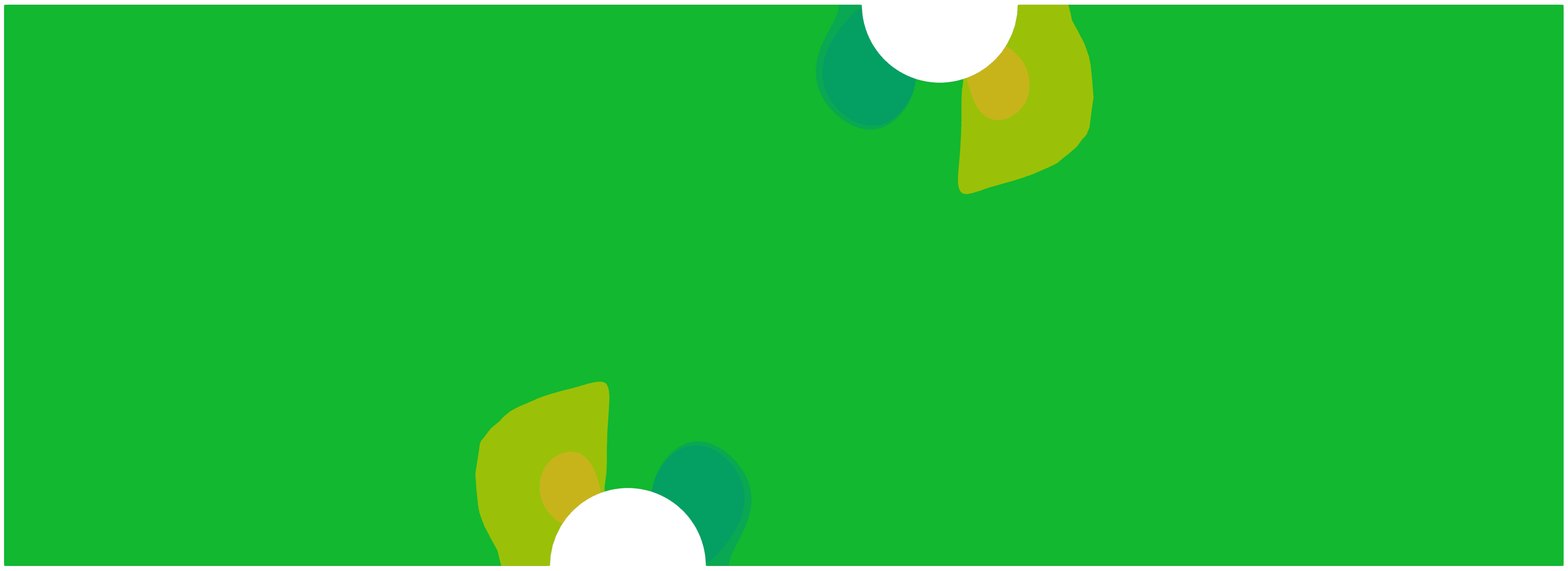}};
      \node[inner sep=0pt] (pic2) at (0,-2.65) {\includegraphics[width=\textwidth]{02_Figures/03_Contour/Ex2/mesh_convergence/Zoom_A-13955_Dxy.pdf}};
      \draw[draw=rwth8, line width=1.5pt] (-0.92,-0.70) rectangle ( 0.92, 0.70);
      \draw[draw=rwth8, line width=1.5pt] (-2.40,-4.40) rectangle ( 2.40,-0.90);
      \draw[draw=rwth8, line width=1.5pt] ( 0.00,-0.70) -- ( 0.00,-0.90);
    \end{tikzpicture} 
    \caption{Model~A}
    \label{fig:ExanDA}
  \end{subfigure}
  \begin{subfigure}{.3\textwidth} 
    \centering 
    \begin{tikzpicture}
      \node[inner sep=0pt] (pic) at (0,0) {\includegraphics[width=0.8\textwidth]{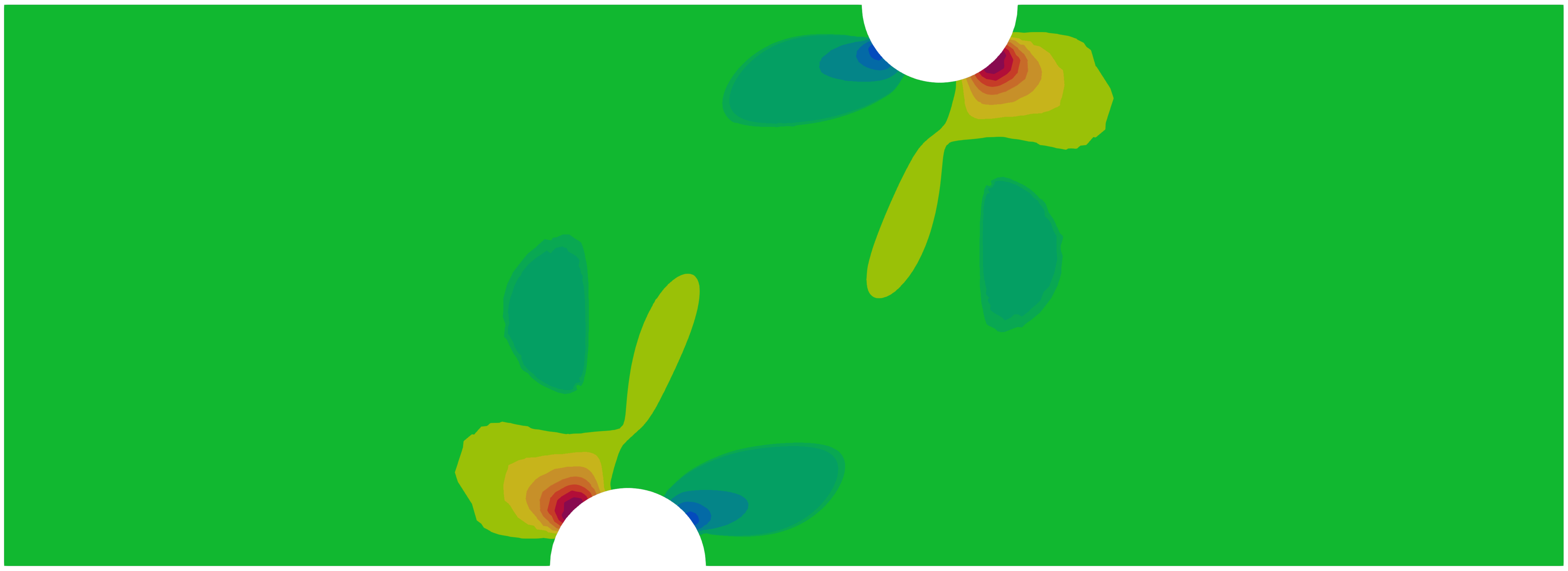}};
      \node[inner sep=0pt] (pic2) at (0,-2.65) {\includegraphics[width=\textwidth]{02_Figures/03_Contour/Ex2/mesh_convergence/Zoom_B-13955_Dxy.pdf}};
      \draw[draw=rwth8, line width=1.5pt] (-0.92,-0.70) rectangle ( 0.92, 0.70);
      \draw[draw=rwth8, line width=1.5pt] (-2.40,-4.40) rectangle ( 2.40,-0.90);
      \draw[draw=rwth8, line width=1.5pt] ( 0.00,-0.70) -- ( 0.00,-0.90);
    \end{tikzpicture} 
    \caption{Model~B}
    \label{fig:ExanDB}
  \end{subfigure}
  \begin{subfigure}{.3\textwidth} 
    \centering 
    \begin{tikzpicture}
      \node[inner sep=0pt] (pic) at (0,0) {\includegraphics[width=0.8\textwidth]{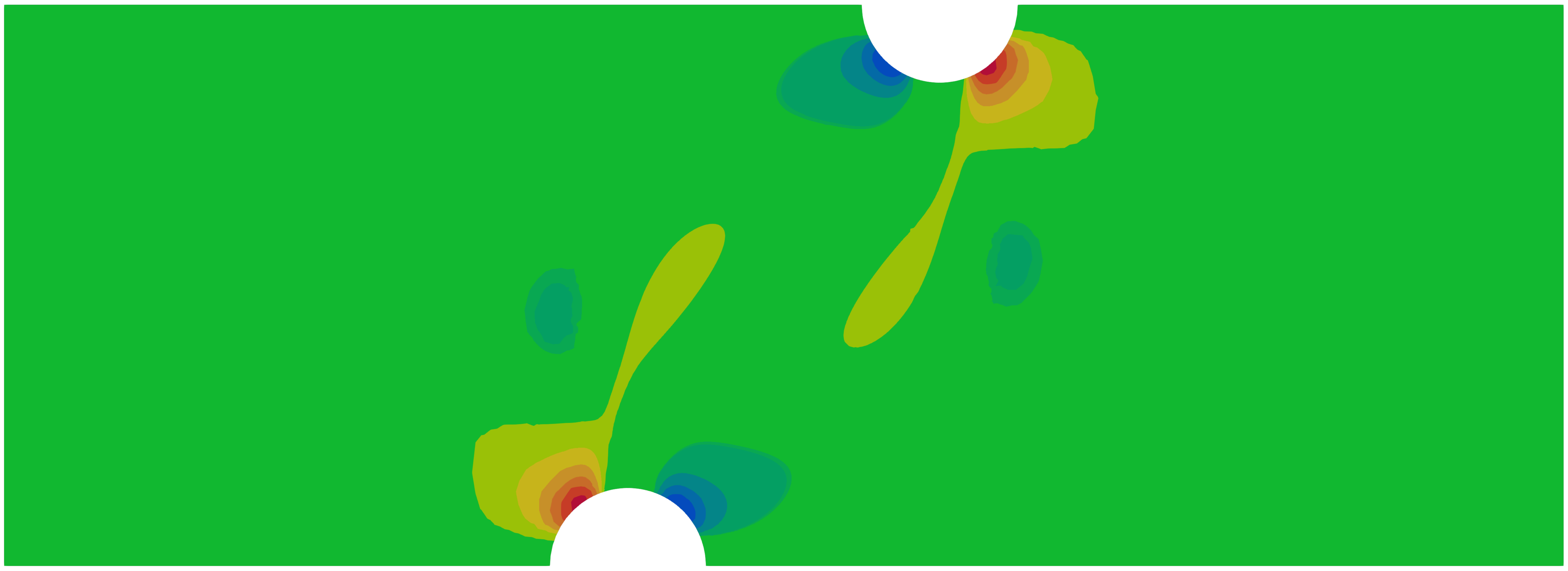}};
      \node[inner sep=0pt] (pic2) at (0,-2.65) {\includegraphics[width=\textwidth]{02_Figures/03_Contour/Ex2/mesh_convergence/Zoom_C-13955_Dxy.pdf}};
      \draw[draw=rwth8, line width=1.5pt] (-0.92,-0.70) rectangle ( 0.92, 0.70);
      \draw[draw=rwth8, line width=1.5pt] (-2.40,-4.40) rectangle ( 2.40,-0.90);
      \draw[draw=rwth8, line width=1.5pt] ( 0.00,-0.70) -- ( 0.00,-0.90);
    \end{tikzpicture} 
    \caption{Model~C}
    \label{fig:ExanDC}
  \end{subfigure}
  \begin{subfigure}{.08\textwidth} 
    \centering 
    \begin{tikzpicture}
      \node[inner sep=0pt] (pic) at (0,0) {\includegraphics[height=40mm, width=5mm]
      {02_Figures/03_Contour/00_Color_Maps/Damage_Step_Vertical.pdf}};
      \node[inner sep=0pt] (0)   at ($(pic.south)+( 1.00, 0.15)$)  {$-0.0627$};
      \node[inner sep=0pt] (1)   at ($(pic.south)+( 1.00, 3.80)$)  {$+0.1040$};
      \node[inner sep=0pt] (d)   at ($(pic.south)+( 0.00, 4.35)$)  {$D_{xy}~\si{[-]}$};
    \end{tikzpicture}
    \hphantom{Model~C}
  \end{subfigure}

  \caption{Contour plots of the normal and shear components of the damage tensor for the asymmetrically notched specimen at the end of the simulation.}
  \label{fig:ExanD}     
\end{figure}

Fig.~\ref{fig:ExanFu} shows the normalized force-displacement curves for the asymmetrically notched specimen and all models predict the maximum peak force also with the coarsest mesh (1624 elements) accurately. In the post-failure regime, models~A and C show with increasing mesh refinement less deviations from the final solution compared to model B (see Figs.~\ref{fig:ExanFuA}-\ref{fig:ExanFuC}). The model comparison in Fig.~\ref{fig:ExanFuComp} shows that, analogously to the tension dominated example in Section~\ref{sec:Ex_pwh}, the vertical drop of model~B is shifted to the right, i.e.~$u_{0.5 \, F_\text{max}}^\text{B} = 1.062~[\si{\mm}]$ compared to $u_{0.5 \, F_\text{max}}^\text{A} = 0.947~[\si{\mm}]$ and $u_{0.5 \, F_\text{max}}^\text{C} = 0.955~[\si{\mm}]$.

In Fig.~\ref{fig:ExanD}, the damage contour plots with a zoom to the center of the asymmetrically notched specimen are presented. All models demonstrate the formation of a shear crack between the notches as well as a more pronounced evolution of the damage component $D_{xx}$, since the $x$-direction corresponds to the loading direction. With regard to the normal components of the damage tensor, the results of models~A and C differ in shape and intensity compared to model~B. While models~A and C yield a sigmoidal crack pattern, model~B yields a straight shear crack. Moreover, the total width of the damage zone for model~B is greater than for models~A and C, which is in line with the findings of Section~\ref{sec:Ex_pwh}. When comparing the shear components of the damage tensor, model~A yields the evolution of $D_{xy}$ over a wider spread area compared to models~B and C, but exhibits no distinct peak values at the notches. The smoothed out distribution of $D_{xy}$ can result from the strict regularization properties of model~A that controls each component of the damage tensor individually.

\begin{figure}[htbp]
  \centering
  \includegraphics{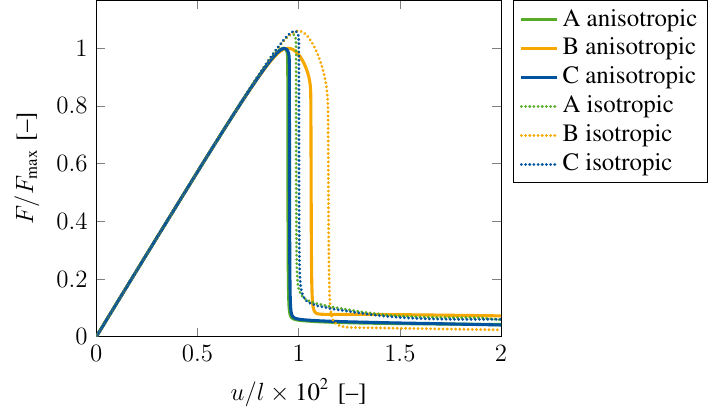}
  \caption{Comparison of the anisotropic and isotropic computation for the asymmetrically notched specimen (13955 elements). The forces are normalized with respect to the maximum force of the anisotropic computation of model~B with $F_\text{max} = 3.7959 \times 10^4~[\si{\newton}]$.}
  \label{fig:ExanFuIsotropic}
\end{figure}

The study comparing isotropic and anisotropic damage for the asymmetrically notched specimen is presented in Fig.~\ref{fig:ExanFuIsotropic}. The force-displacement curves yield also for this example a significant overestimation of the maximum peak force when considering only an isotropic damage formulation (A: $+4.86~[\si{\percent}]$, B: $+6.00~[\si{\percent}]$, C: $+6.03~[\si{\percent}]$) and corroborates that damage has to be modeled as an anisotropic phenomenon.

\begin{figure}[htbp]
  \centering
  \includegraphics{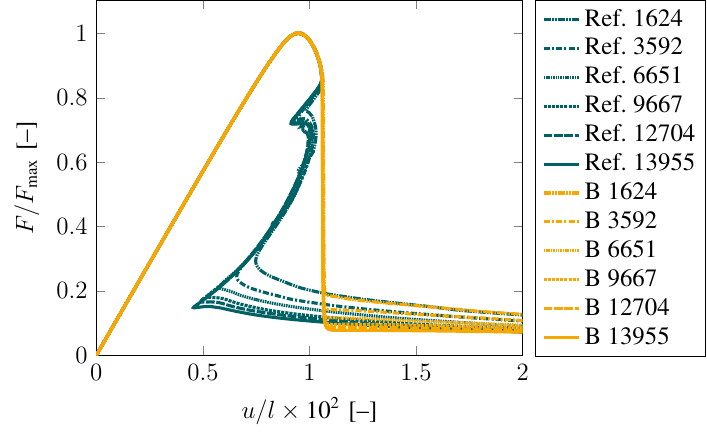}
  \caption{Comparison of the results of model~B for the asymmetrically notched specimen obtained by a displacement-controlled procedure using artificial viscosity  and by an arc-length controlled procedure (reference solution from \citetalias{HolthusenBrepolsEtAl2022a}[\citeyear{HolthusenBrepolsEtAl2022a}]). The forces are normalized with respect to the maximum force of the computation with the displacement-driven procedure (13955 elements) with $F_\text{max} = 3.7959 \times 10^4~[\si{\newton}]$.}
  \label{fig:ExanFuDispArcl}
\end{figure}

Finally, this example serves to compare the displacement driven load control using artificial viscosity to an arc-length driven load control without artificial viscosity for model~B. Fig.~\ref{fig:ExanFuDispArcl} shows the force-displacement curves for both load control procedures, where the arc-length controlled reference solution is obtained from \citetalias{HolthusenBrepolsEtAl2022a}[\citeyear{HolthusenBrepolsEtAl2022a}]. Both procedures yield the same maximum peak force also for coarse meshes. Then, the displacement driven procedure yields a vertical drop of the force-displacement curve while the arc-length controlled procedure yields a double snap-back during the force decrease. Thereafter, the curves again unite and are congruent with each other and, thus, proof that both control procedures, with and without artificial viscosity, are equally valid.

\subsection{Three-dimensional tensile specimen}
\label{sec:Ex_ts}

This example features the failure investigation of a three-dimensional I-shaped tensile specimen with models~A, B, and C. Previously, this example was investigated in \cite{FelderKopic-OsmanovicEtAl2022} in the context of thermo-mechanical coupling, in \cite{AmbatiKruseEtAl2016} numerically and experimentally, and in \citetalias{HolthusenBrepolsEtAl2022b}~\citeyear{HolthusenBrepolsEtAl2022b} with a ductile formulation of model~B.

\begin{figure}[htbp] 
  \centering 
  \begin{subfigure}{.7\textwidth} 
      \centering 
      \begin{tikzpicture}
        \node (pic) at (0,0) {\includegraphics[width=\textwidth]{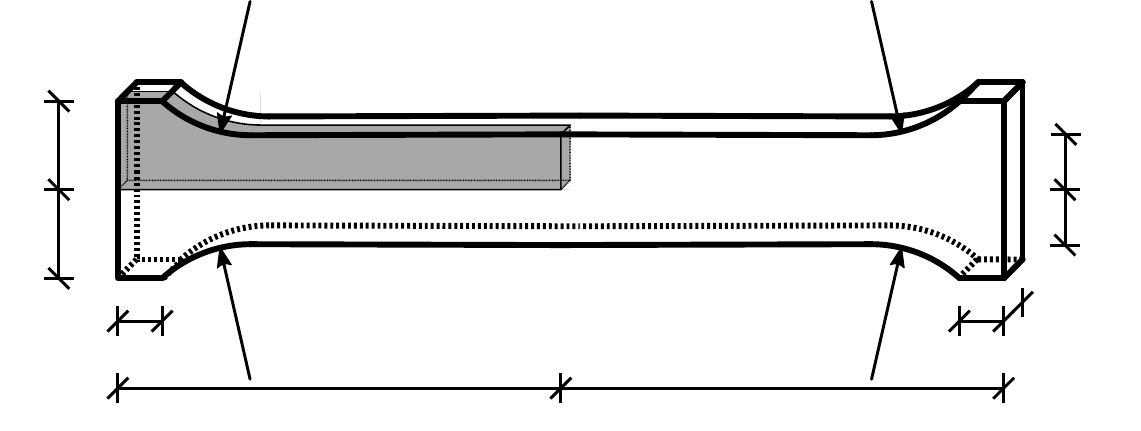}};
        \node (l1)  at ($(pic.south)+(-2.20, 0.20)$)  {$l$};
        \node (l2)  at ($(pic.south)+( 2.20, 0.20)$)  {$l$};
        \node (h1)  at ($(pic.south)+(-5.40, 2.05)$)  {$h_1$};
        \node (h1b) at ($(pic.south)+(-5.40, 2.90)$)  {$h_1$};        
        \node (h2)  at ($(pic.south)+( 5.30, 2.22)$)  {$h_2$};
        \node (h2b) at ($(pic.south)+( 5.30, 2.75)$)  {$h_2$};        
        \node (d1)  at ($(pic.south)+(-4.28, 0.92)$)  {$d$};
        \node (d2)  at ($(pic.south)+( 4.00, 0.92)$)  {$d$};
        \node (t2)  at ($(pic.south)+( 4.50, 1.00)$)  {$2t$};
        \node (r1)  at ($(pic.south)+( 2.80, 1.20)$)  {$r$};
        \node (r2)  at ($(pic.south)+( 2.80, 3.70)$)  {$r$};
        \node (r3)  at ($(pic.south)+(-2.95, 1.20)$)  {$r$};
        \node (r4)  at ($(pic.south)+(-2.95, 3.70)$)  {$r$};
      \end{tikzpicture} 
      \caption{Geometry}
      \label{fig:Ex_ts_geom}
  \end{subfigure}\\
  \vspace{10mm}
  \begin{subfigure}{.7\textwidth} 
      \centering 
      \hspace{-2.5mm}
      \begin{tikzpicture}
        \node[inner sep=0pt] (pic) at (0,0) {\includegraphics[width=\textwidth]{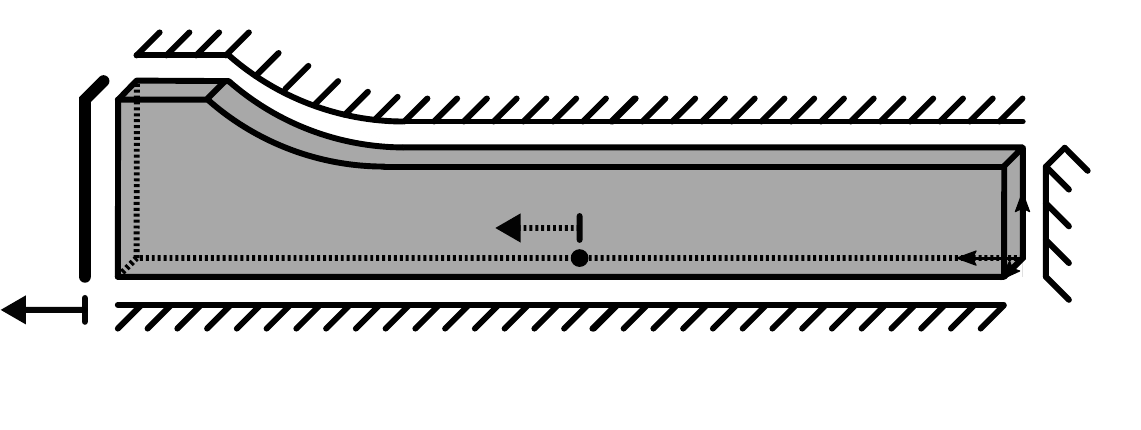}};
        \node[inner sep=0pt] (Fu) at ($(pic.south) +(-5.40, 1.65)$)  {$F,u_t$};
        \node[inner sep=0pt] (u)  at ($(pic.south) +(-0.25, 2.28)$)  {$u$};
        \node[inner sep=0pt] (x)  at ($(pic.south) +( 4.00, 2.25)$)  {$x$};
        \node[inner sep=0pt] (y)  at ($(pic.south) +( 3.70, 1.90)$)  {$y$};
        \node[inner sep=0pt] (z)  at ($(pic.south) +( 4.50, 1.35)$)  {$z$};
      \end{tikzpicture}%
      \vspace{-7mm}
      \caption{Boundary value problem}
      \label{fig:Ex_ts_bvp}
  \end{subfigure}
  \caption{Geometry and boundary value problem of the three-dimensional tensile specimen.} 
  \label{fig:Ex_ts}
\end{figure}
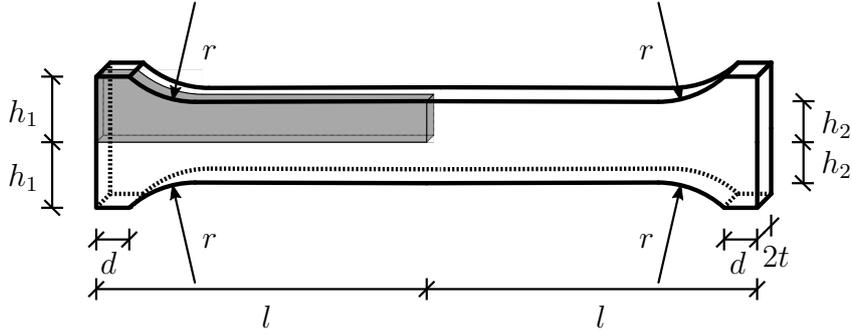
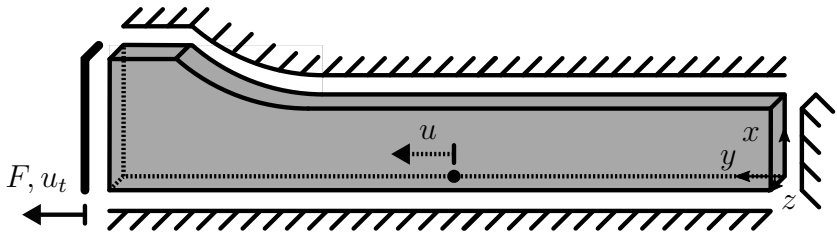

\begin{figure}[htbp]
  \centering
  \begin{subfigure}{.48\textwidth}
    \centering
    \includegraphics{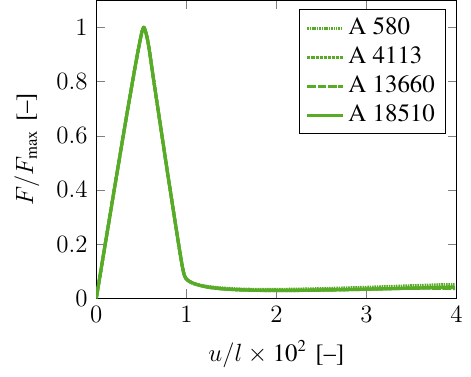}
    \vspace{-7mm}
    \caption{Model~A}
    \label{fig:ExtsFuA}
  \end{subfigure}
  \quad
  \begin{subfigure}{.48\textwidth}
    \centering
    \includegraphics{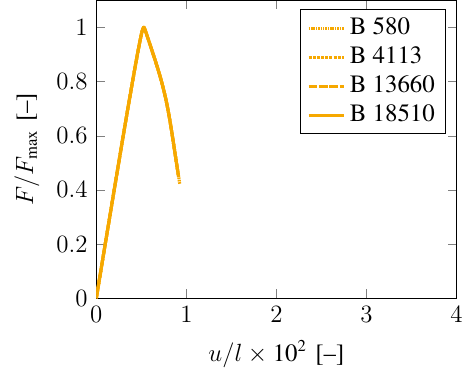}
    \vspace{-7mm}
    \caption{Model~B}
    \label{fig:ExtsFuB}
  \end{subfigure}%
  \vspace{5mm} 
  \begin{subfigure}{.48\textwidth}
    \centering
    \includegraphics{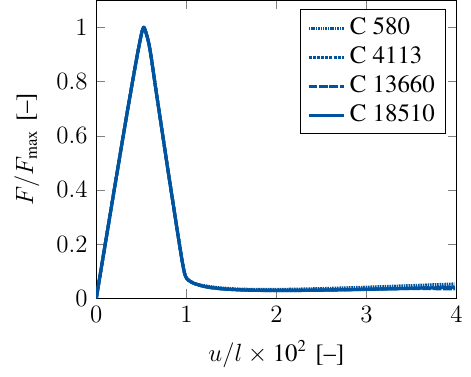}
    \vspace{-7mm}
    \caption{Model~C}
    \label{fig:ExtsFuC}
  \end{subfigure}
  \quad
  \begin{subfigure}{.48\textwidth}
    \centering
    \includegraphics{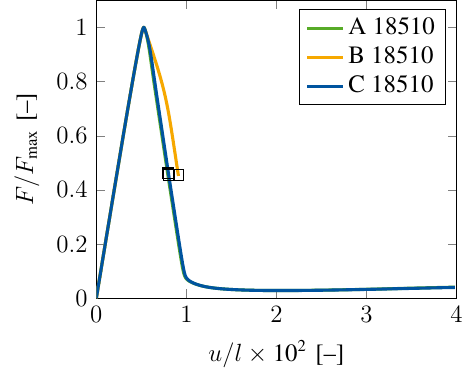}
    \vspace{-7mm}
    \caption{Model comparison}
    \label{fig:ExtsFuComp}
  \end{subfigure}
  \caption{Mesh convergence studies for the three-dimensional tensile specimen and model comparison. The forces are normalized with respect to the maximum force of the finest mesh (18510 elements) of model~B with $F_\text{max} = 1.1762 \times 10^4~[\si{\newton}]$. The black boxes in Fig.~\ref{fig:ExtsFuComp} indicate the points of comparison in Figs.~\ref{fig:ExtsDnormal} and \ref{fig:ExtsDshear}.}
  \label{fig:ExtsFu}
\end{figure}

Fig.~\ref{fig:Ex_ts} shows the geometry and the considered boundary value problem. Due to symmetry, only an eighth of the original specimen is considered in the simulation. The dimensions read $l = 50~[\si{\mm}]$, $h_1 = 10~[\si{\mm}]$, $h_2 = 6.25~[\si{\mm}]$, $d = 5~[\si{\mm}]$, $r = 15~[\si{\mm}]$ and $t = 1.5~[\si{\mm}]$. The finite element meshes stem from \citetalias{HolthusenBrepolsEtAl2022b}~\citeyear{HolthusenBrepolsEtAl2022b}. The internal length scales of model B are chosen as $A_i^\text{B} = 75~[\si{\MPa\mm\squared}]$ and the parameters of model~A and C are identified as $A_i^\text{A} = 180~[\si{\MPa\mm\squared}]$ and $A_i^\text{C} = 680~[\si{\MPa\mm\squared}]$. In the simulation we apply $u_t = 2~[\si{\mm}]$ at the end of the specimen and plot in Fig.~\ref{fig:ExtsFu} the reaction force $F$ over the displacement $u$ at position $x = 0~[\si{\mm}]$, $y = 25~[\si{\mm}]$, and $z = 0~[\si{\mm}]$ (cf. Fig.~\ref{fig:Ex_ts_bvp}).

In Fig.~\ref{fig:ExtsFu}, all models again yield in the force-displacement curves the same maximum peak force also for coarse mesh discretizations (580~elements). In this example, only models~A and C were able to compute converged solutions up to the final loading of $u_t = 2~[\si{\mm}]$. With model~B, no solution could be obtained due to local convergence problems beyond $u_t = 0.556~[\si{\mm}]$, which corresponds to $u = 0.456~[\si{\mm}]$ and $u / l \times 10^2 = 0.912~[\si{-}]$ (see Fig.~\ref{fig:ExtsFuB}).

\begin{figure}
  \centering
  \begin{subfigure}{.3\textwidth} 
    \centering 
    \includegraphics[width=\textwidth]{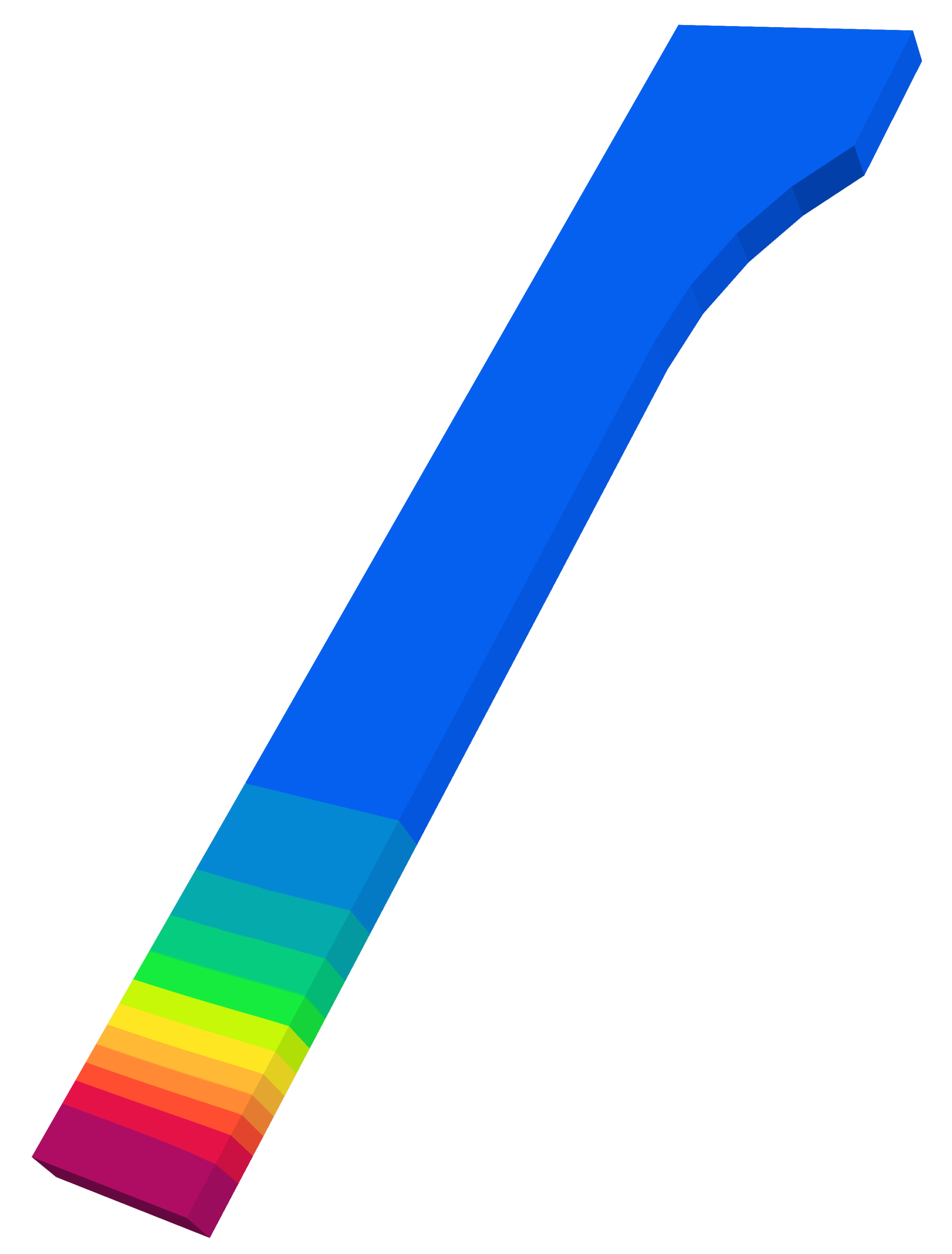}
  \end{subfigure}
  \begin{subfigure}{.3\textwidth} 
    \centering 
    \includegraphics[width=\textwidth]{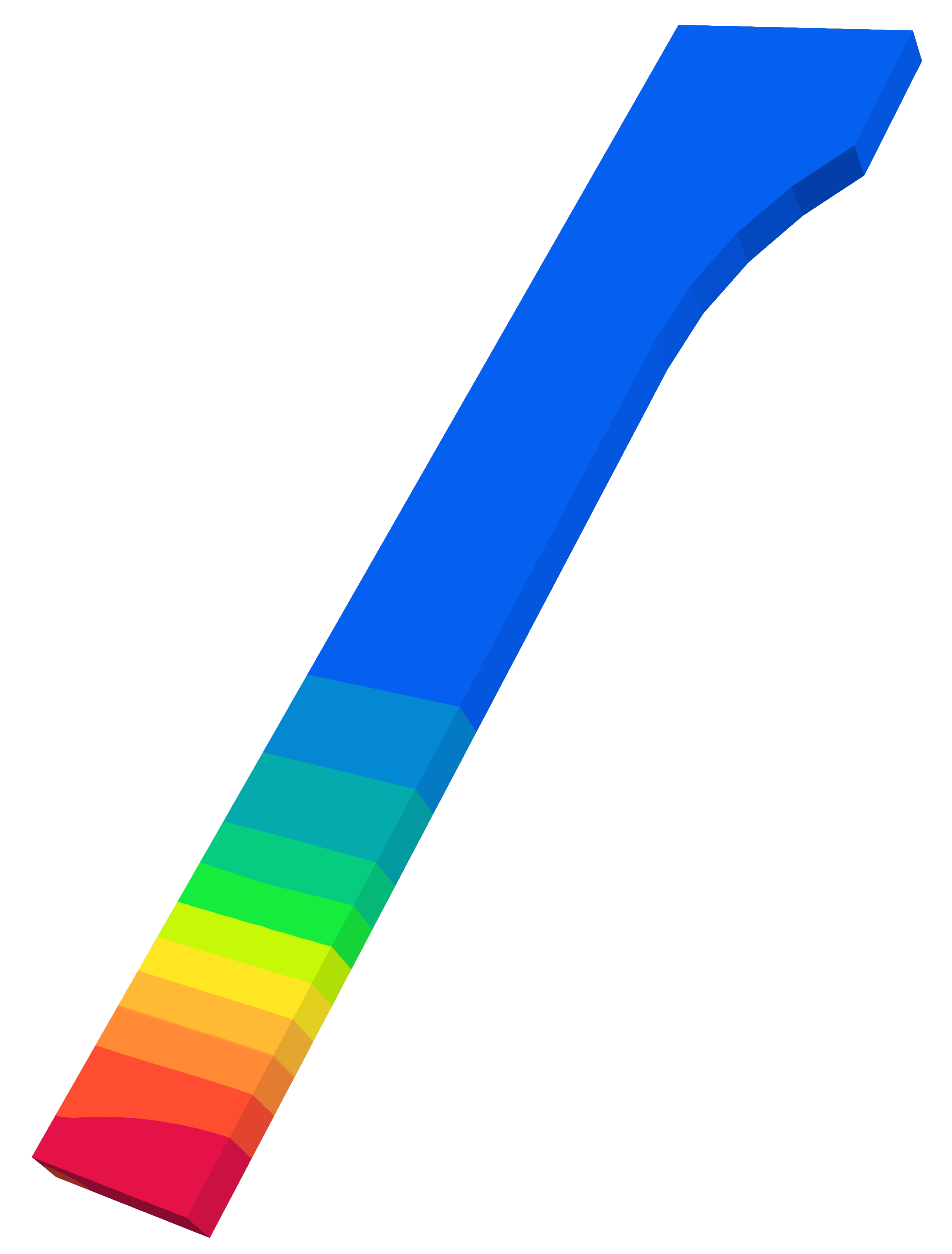}
  \end{subfigure}
  \begin{subfigure}{.3\textwidth} 
    \centering 
    \includegraphics[width=\textwidth]{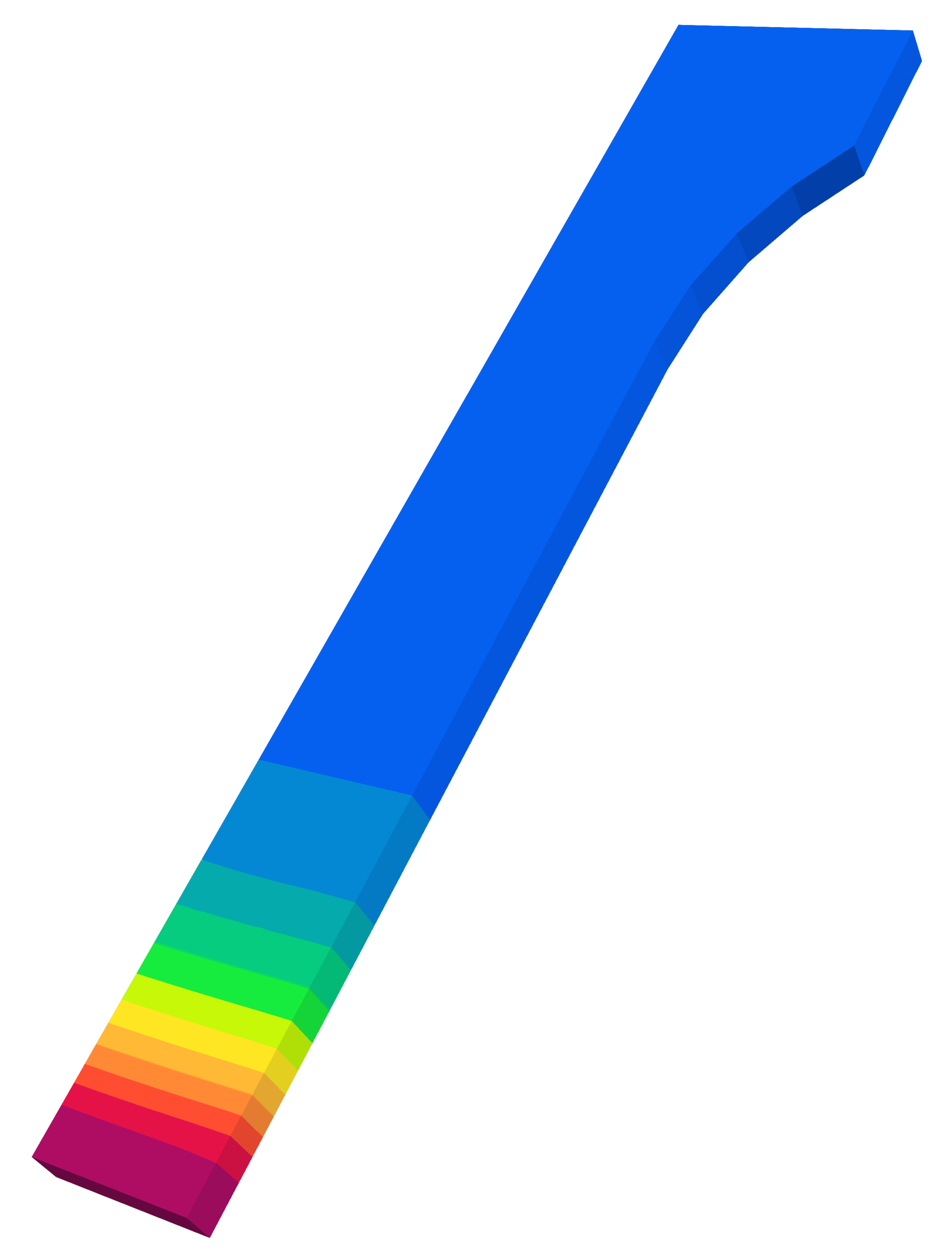}
  \end{subfigure}
  \begin{subfigure}{.08\textwidth} 
    \centering 
    \begin{tikzpicture}
      \node[inner sep=0pt] (pic) at (0,0) {\includegraphics[height=40mm, width=5mm]
      {02_Figures/03_Contour/00_Color_Maps/Damage_Step_Vertical.pdf}};
      \node[inner sep=0pt] (0)   at ($(pic.south)+( 0.50, 0.15)$)  {$0$};
      \node[inner sep=0pt] (1)   at ($(pic.south)+( 0.50, 3.80)$)  {$1$};
      \node[inner sep=0pt] (d)   at ($(pic.south)+( 0.00, 4.35)$)  {$D_{xx}~\si{[-]}$};
    \end{tikzpicture} 
  \end{subfigure}

  \vspace{4mm}

  \begin{subfigure}{.3\textwidth} 
    \centering 
    \includegraphics[width=\textwidth]{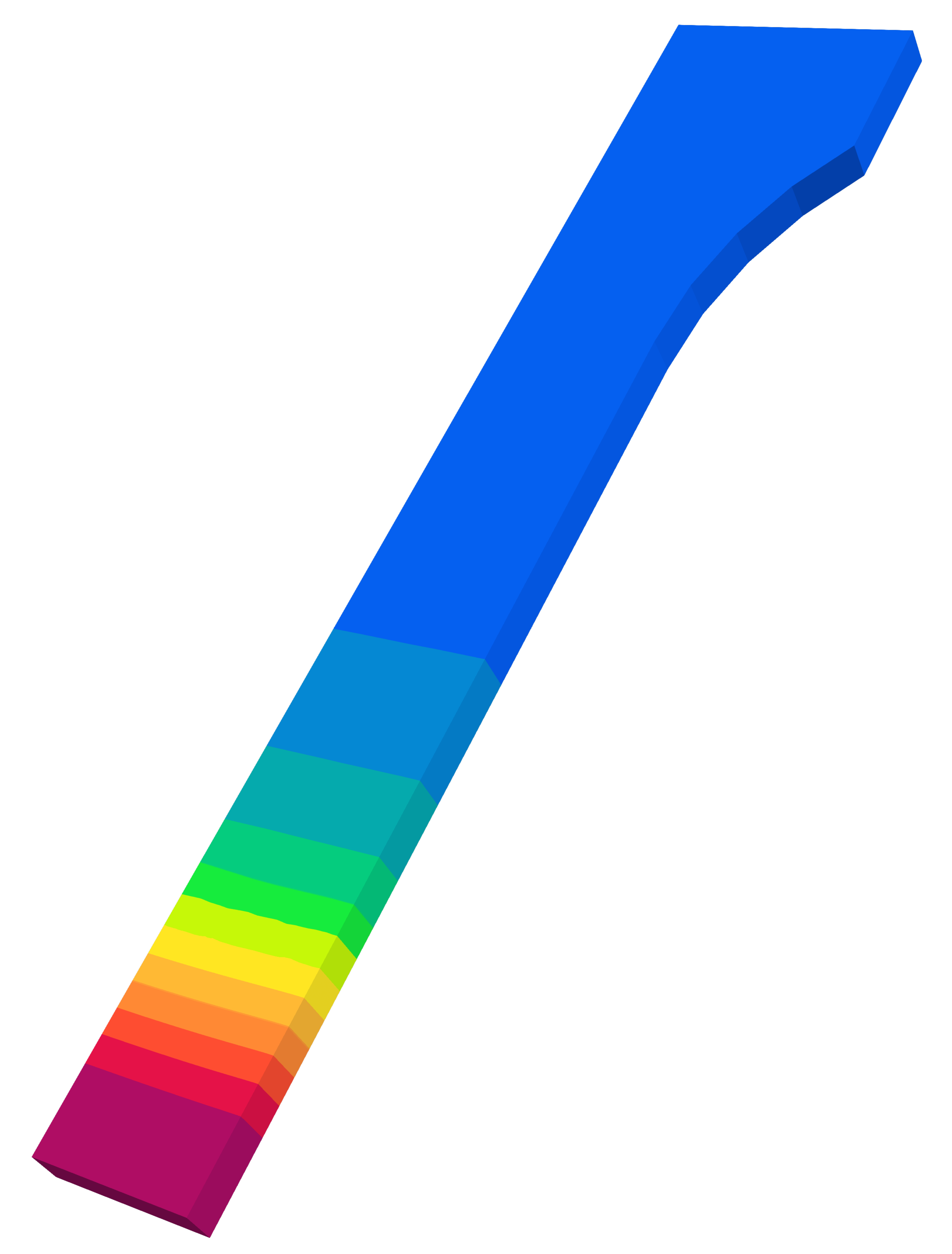}
  \end{subfigure}
  \begin{subfigure}{.3\textwidth} 
    \centering 
    \includegraphics[width=\textwidth]{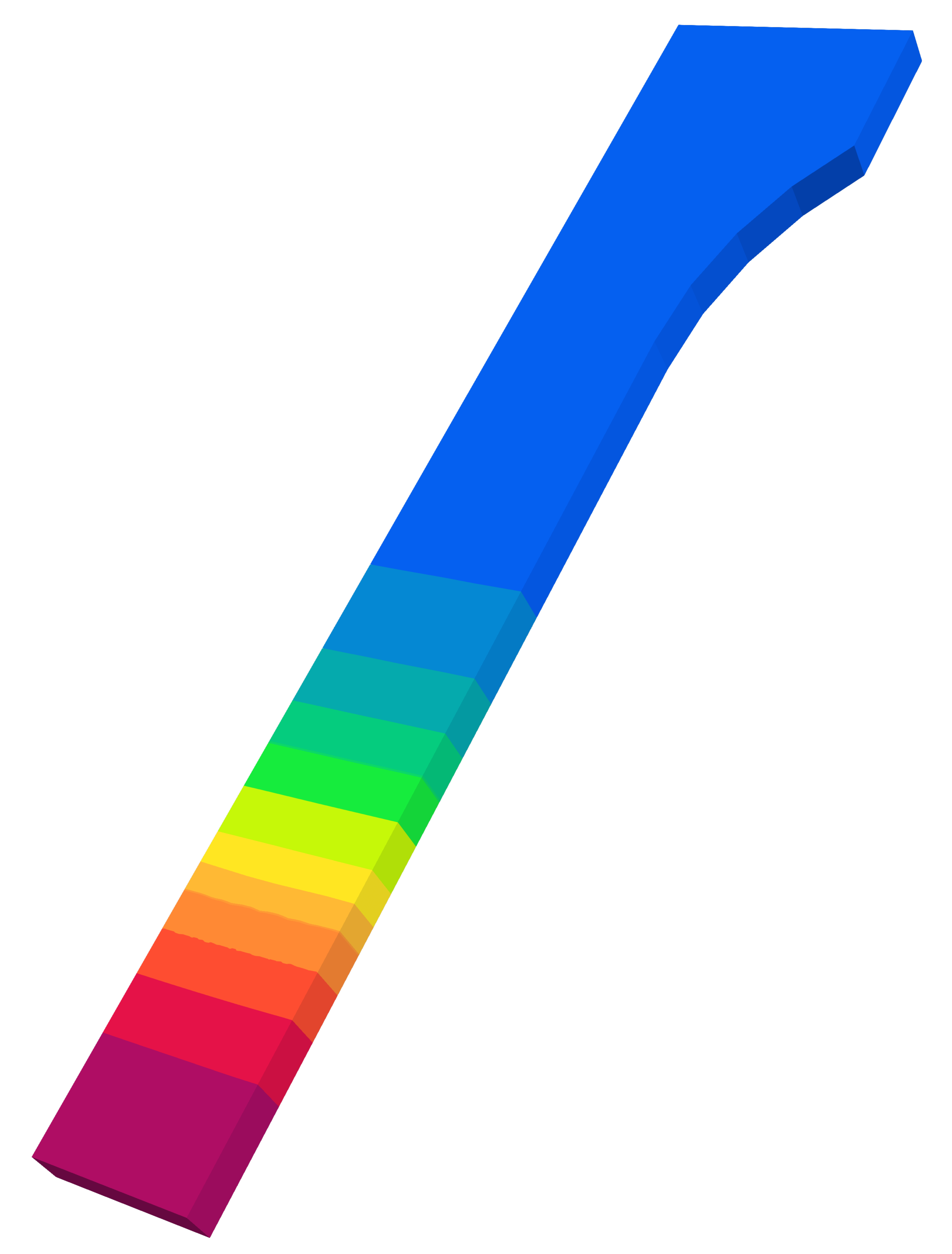}
  \end{subfigure}
  \begin{subfigure}{.3\textwidth} 
    \centering 
    \includegraphics[width=\textwidth]{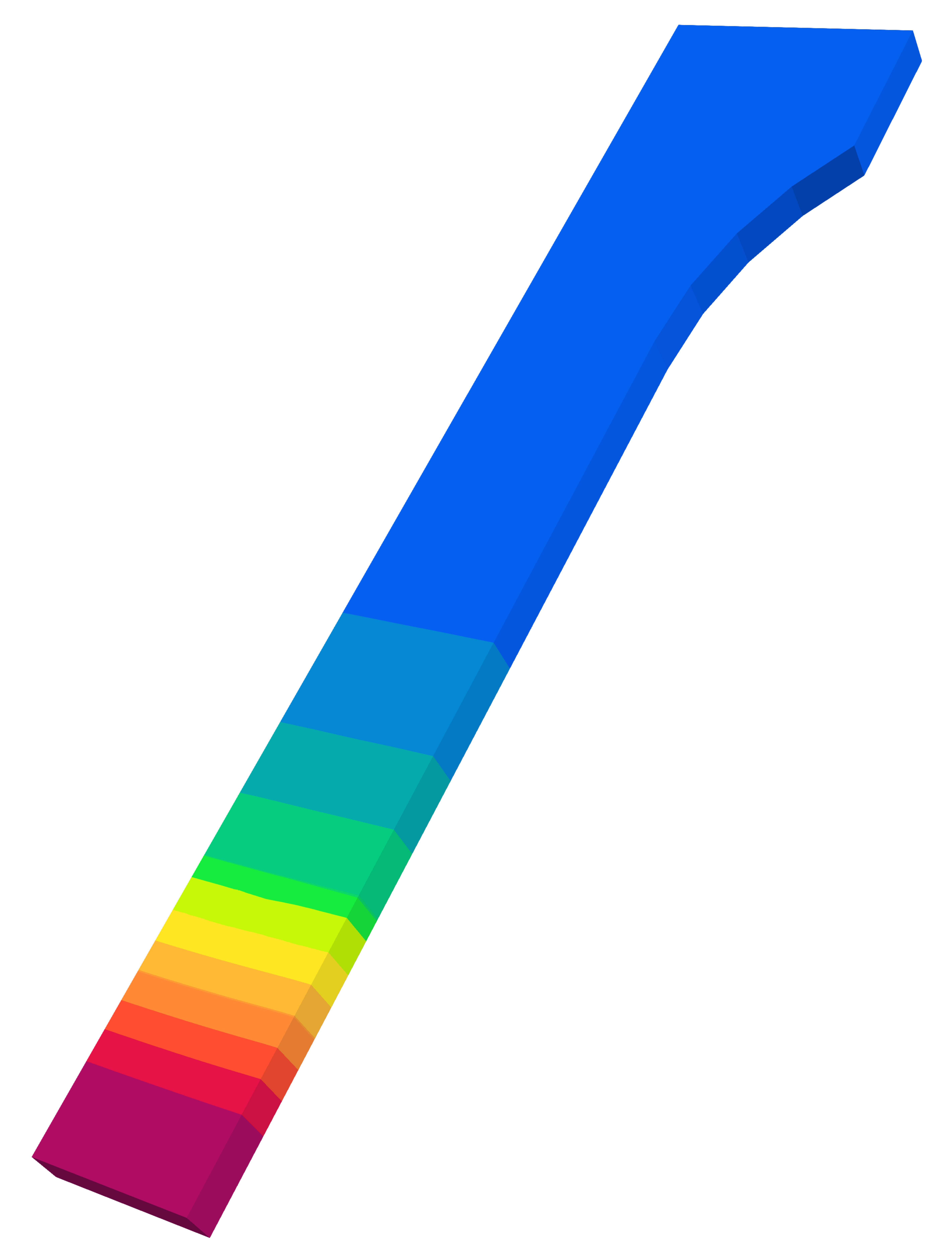}
  \end{subfigure}
  \begin{subfigure}{.08\textwidth} 
    \centering 
    \begin{tikzpicture}
      \node[inner sep=0pt] (pic) at (0,0) {\includegraphics[height=40mm, width=5mm]
      {02_Figures/03_Contour/00_Color_Maps/Damage_Step_Vertical.pdf}};
      \node[inner sep=0pt] (0)   at ($(pic.south)+( 0.50, 0.15)$)  {$0$};
      \node[inner sep=0pt] (1)   at ($(pic.south)+( 0.50, 3.80)$)  {$1$};
      \node[inner sep=0pt] (d)   at ($(pic.south)+( 0.00, 4.35)$)  {$D_{yy}~\si{[-]}$};
    \end{tikzpicture} 
  \end{subfigure}

  \vspace{4mm}

  \begin{subfigure}{.3\textwidth} 
    \centering 
    \includegraphics[width=\textwidth]{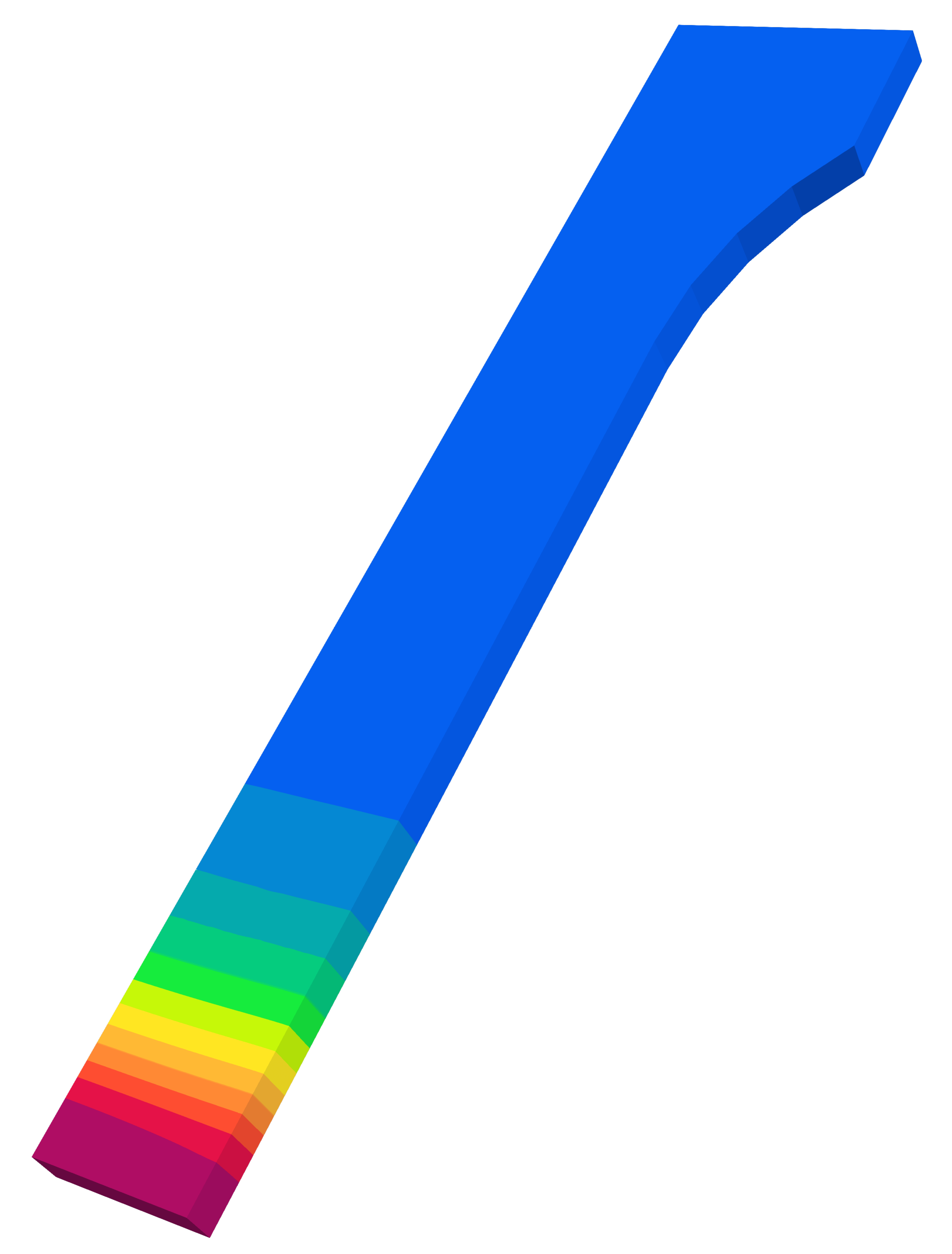}
    \caption{Model~A}
    \label{fig:ExtsDnormalA}
  \end{subfigure}
  \begin{subfigure}{.3\textwidth} 
    \centering 
    \includegraphics[width=\textwidth]{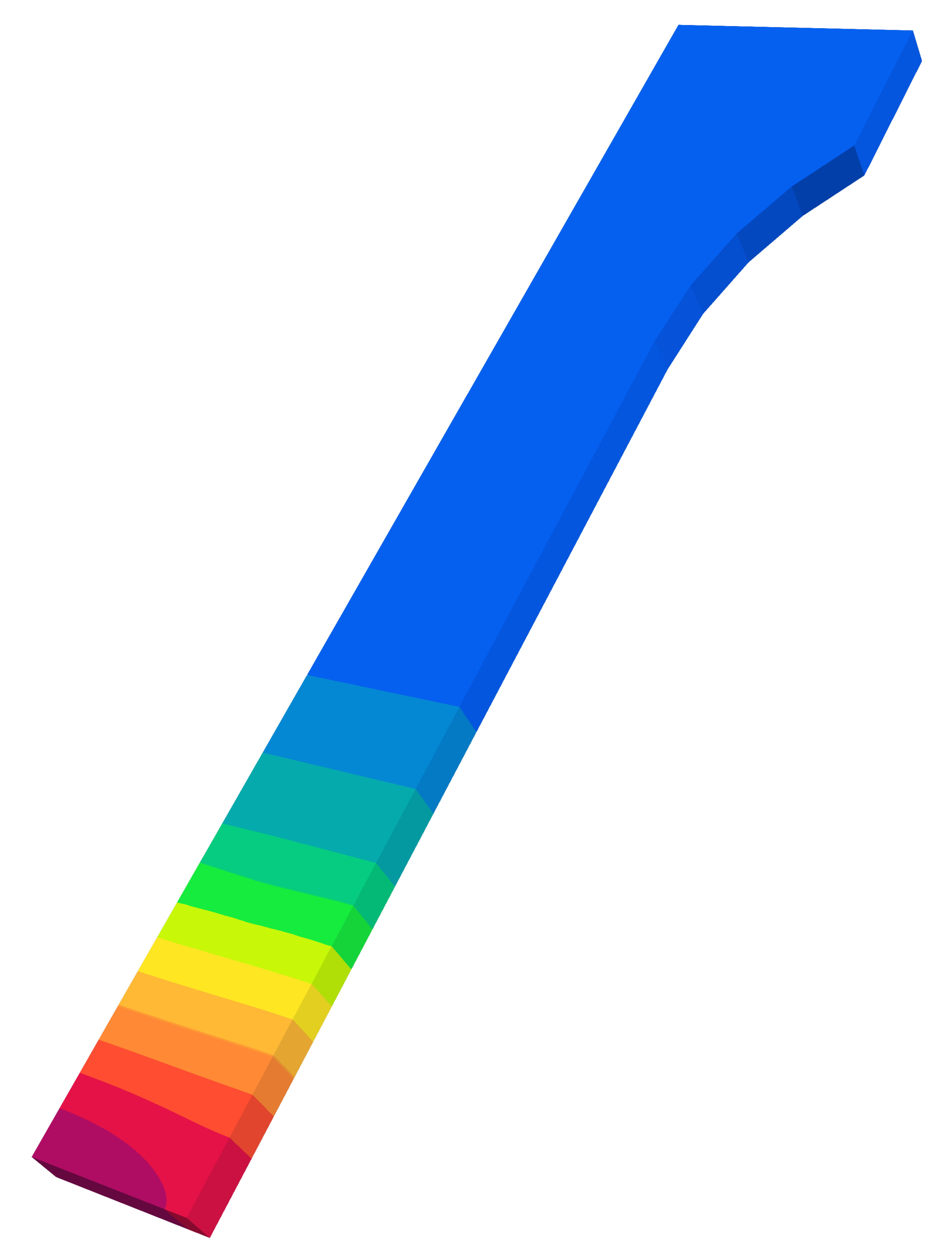}
    \caption{Model~B}
    \label{fig:ExtsDnormalB}
  \end{subfigure}
  \begin{subfigure}{.3\textwidth} 
    \centering 
    \includegraphics[width=\textwidth]{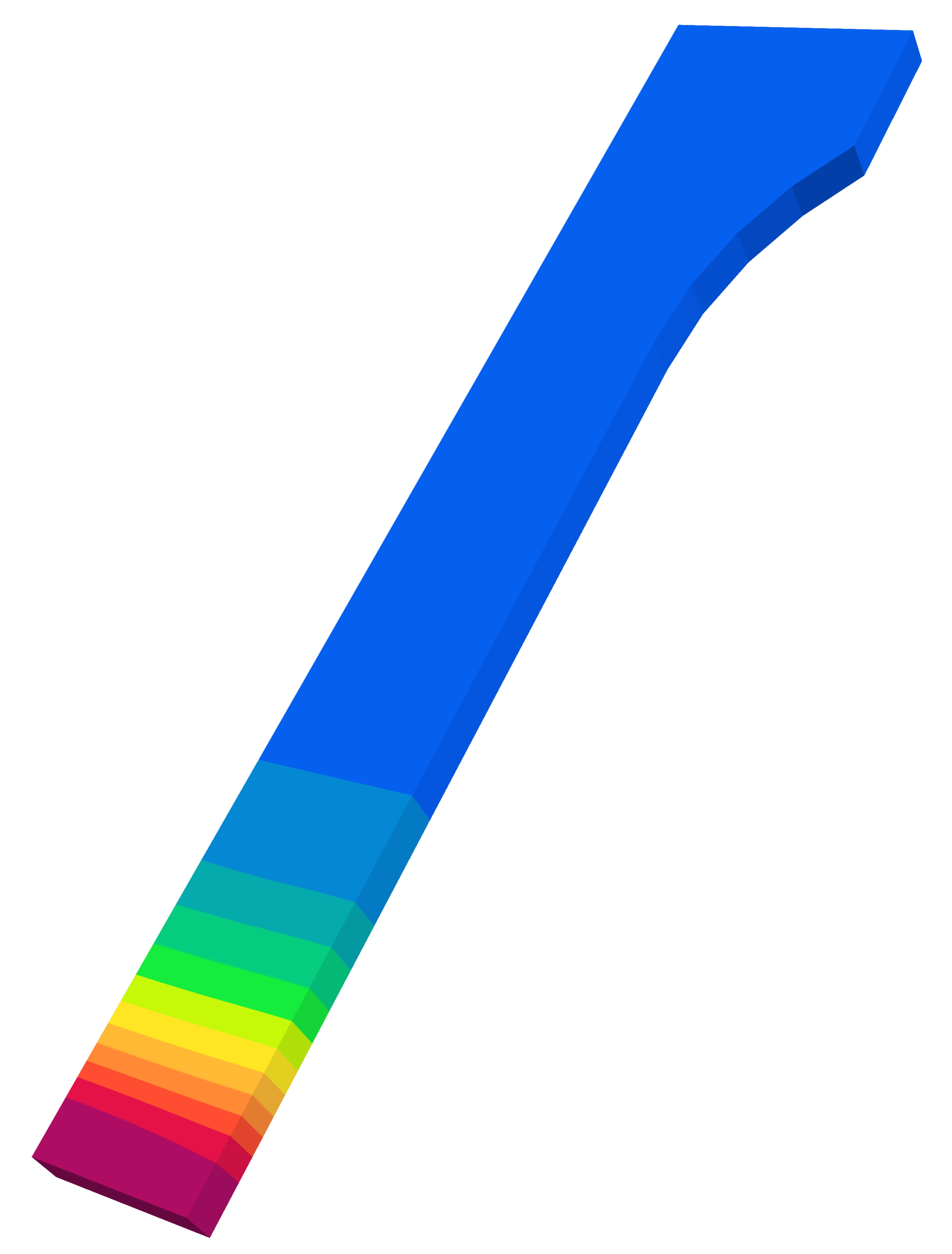}
    \caption{Model~C}
    \label{fig:ExtsDnormalC}
  \end{subfigure}
  \begin{subfigure}{.08\textwidth} 
    \centering 
    \begin{tikzpicture}
      \node[inner sep=0pt] (pic) at (0,0) {\includegraphics[height=40mm, width=5mm]
      {02_Figures/03_Contour/00_Color_Maps/Damage_Step_Vertical.pdf}};
      \node[inner sep=0pt] (0)   at ($(pic.south)+( 0.50, 0.15)$)  {$0$};
      \node[inner sep=0pt] (1)   at ($(pic.south)+( 0.50, 3.80)$)  {$1$};
      \node[inner sep=0pt] (d)   at ($(pic.south)+( 0.00, 4.35)$)  {$D_{zz}~\si{[-]}$};
    \end{tikzpicture} 
    \hphantom{Model~C}
  \end{subfigure}
  
  \caption{Contour plots of the normal components of the damage tensor for the three-dimensional tensile specimen at the point of comparison indicated in Fig.~\ref{fig:ExtsFuComp}.} 
  \label{fig:ExtsDnormal}     
\end{figure}

\begin{figure}
  \centering
  \begin{subfigure}{.3\textwidth} 
    \centering 
    \includegraphics[width=\textwidth]{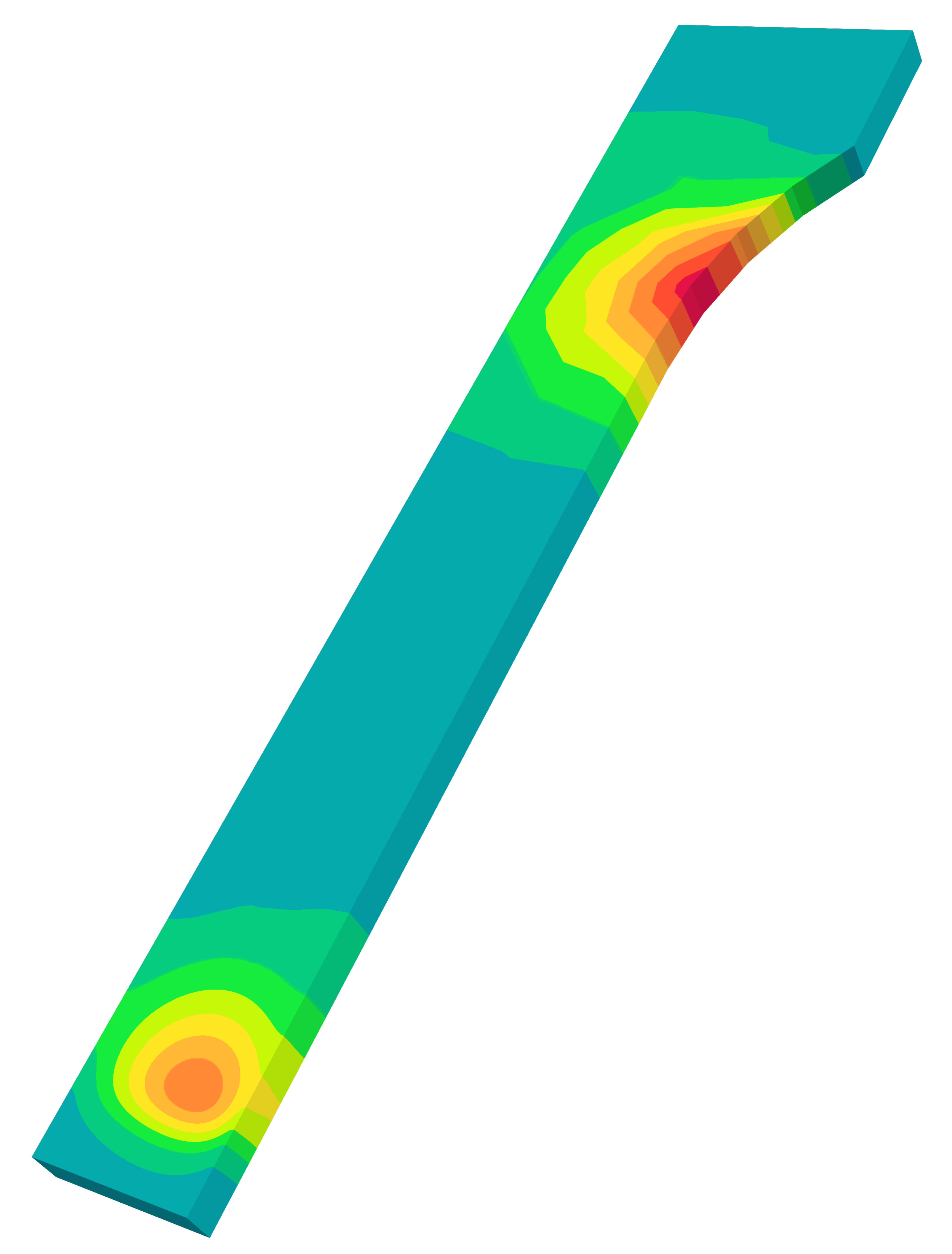}
  \end{subfigure}
  \begin{subfigure}{.3\textwidth} 
    \centering 
    \includegraphics[width=\textwidth]{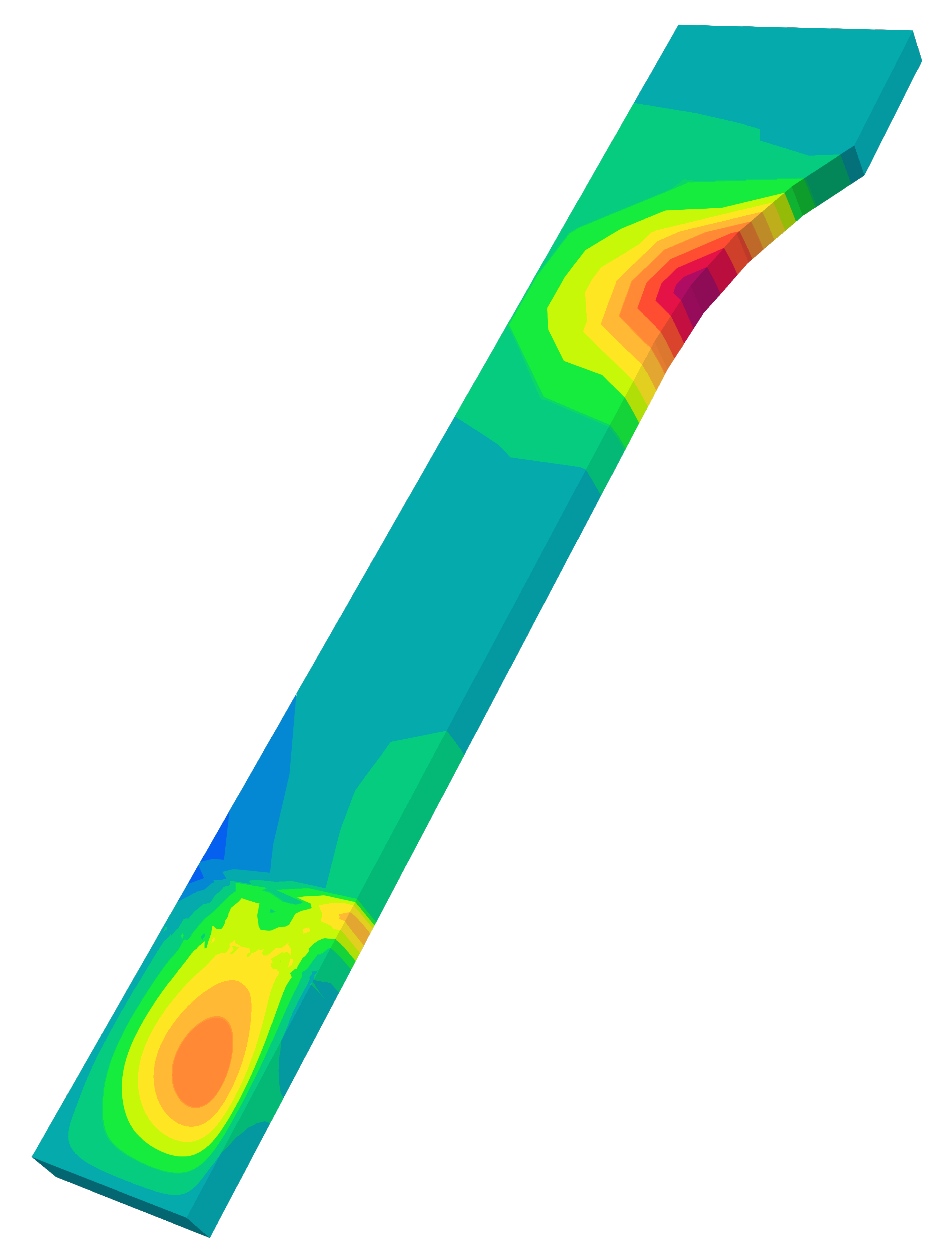}
  \end{subfigure}
  \begin{subfigure}{.3\textwidth} 
    \centering 
    \includegraphics[width=\textwidth]{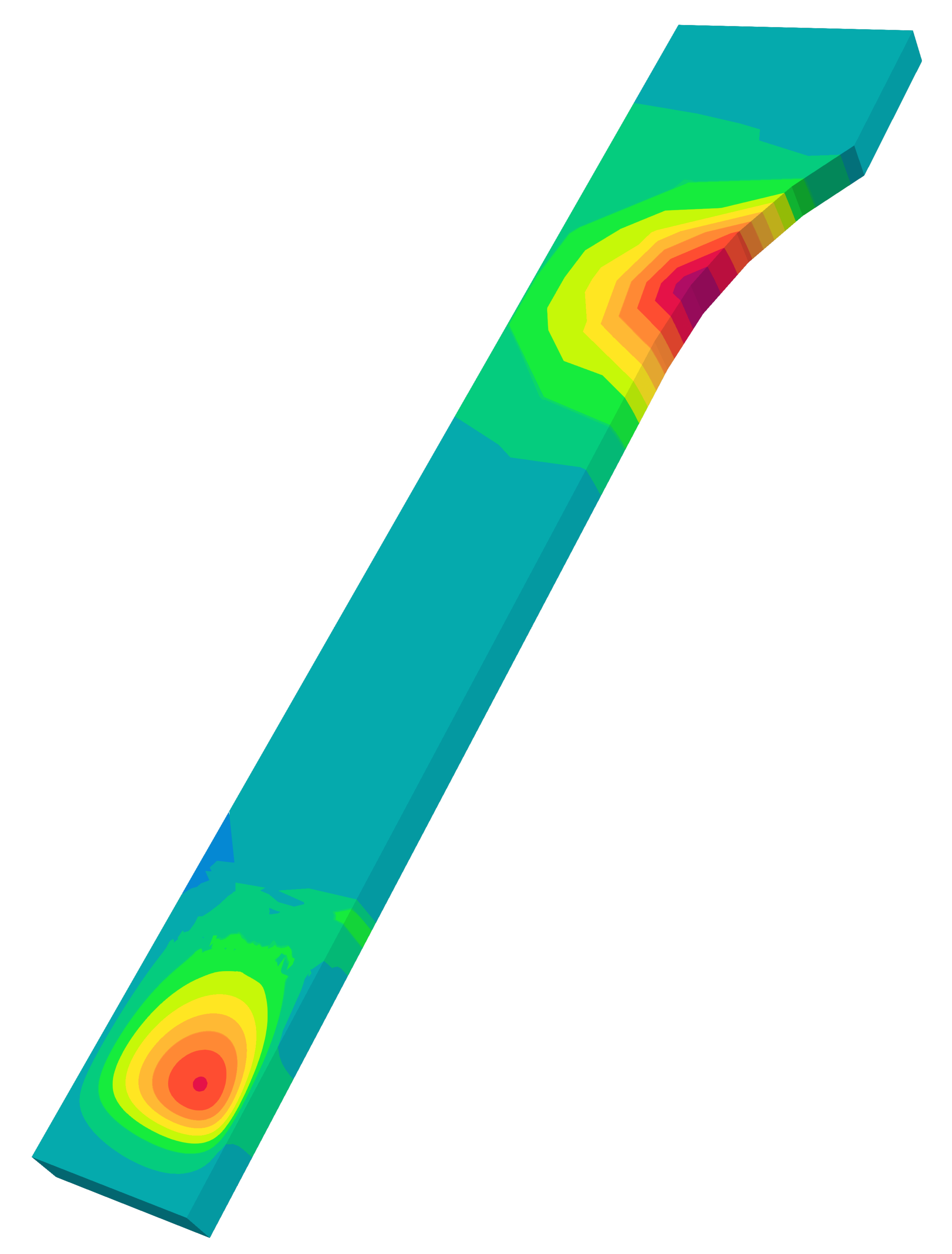}
  \end{subfigure}
  \begin{subfigure}{.08\textwidth} 
    \centering 
    \begin{tikzpicture}
      \node[inner sep=0pt] (pic) at (0,0) {\includegraphics[height=40mm, width=5mm]
      {02_Figures/03_Contour/00_Color_Maps/Damage_Step_Vertical.pdf}};
      \node[inner sep=0pt] (0)   at ($(pic.south)+( 1.10, 0.15)$)  {$-0.00088$};
      \node[inner sep=0pt] (1)   at ($(pic.south)+( 1.10, 3.80)$)  {$+0.00339$};
      \node[inner sep=0pt] (d)   at ($(pic.south)+( 1.10, 4.35)$)  {$D_{xy}~\si{[-]}$};
    \end{tikzpicture} 
  \end{subfigure}

  \vspace{4mm}

  \begin{subfigure}{.3\textwidth} 
    \centering 
    \includegraphics[width=\textwidth]{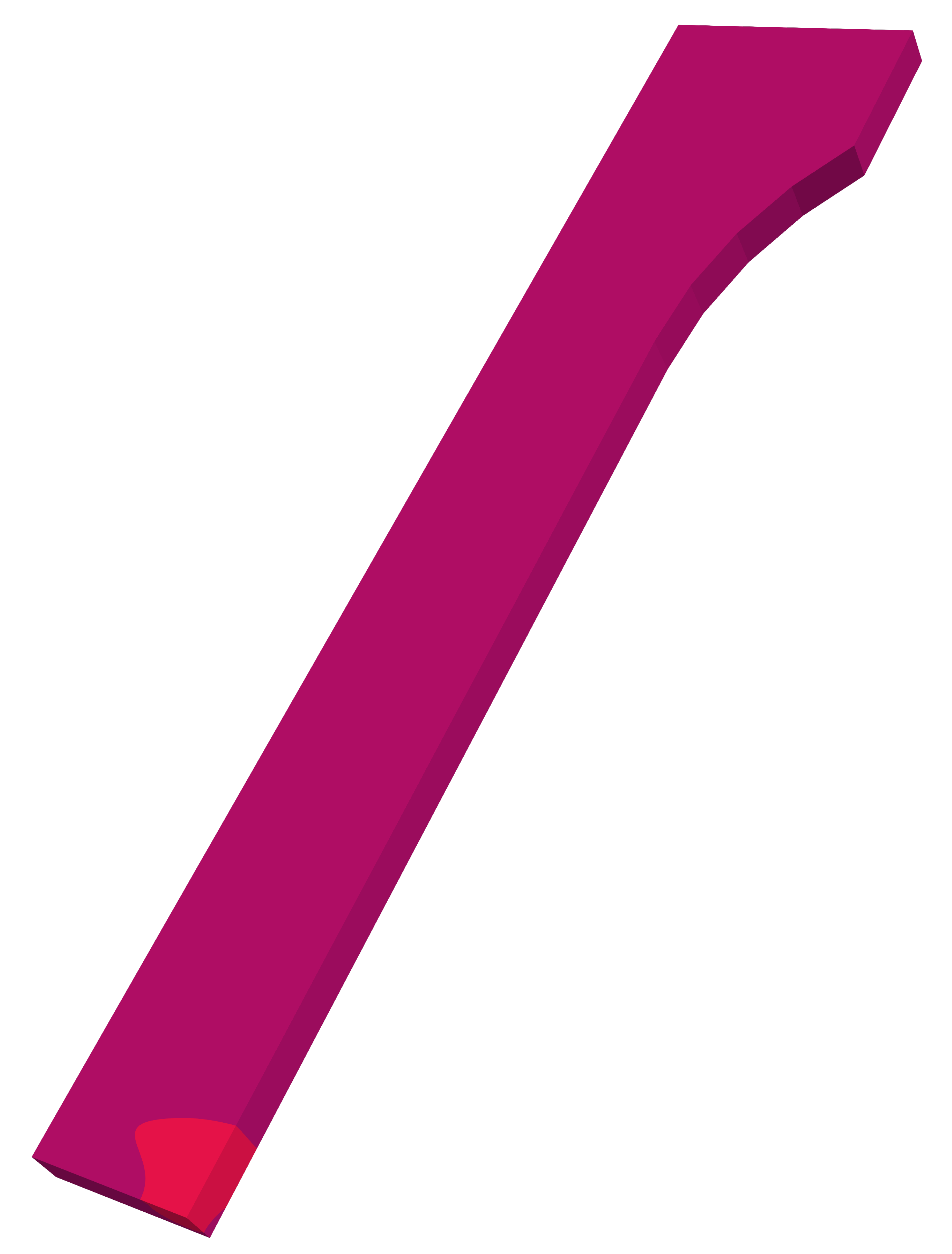}
  \end{subfigure}
  \begin{subfigure}{.3\textwidth} 
    \centering 
    \includegraphics[width=\textwidth]{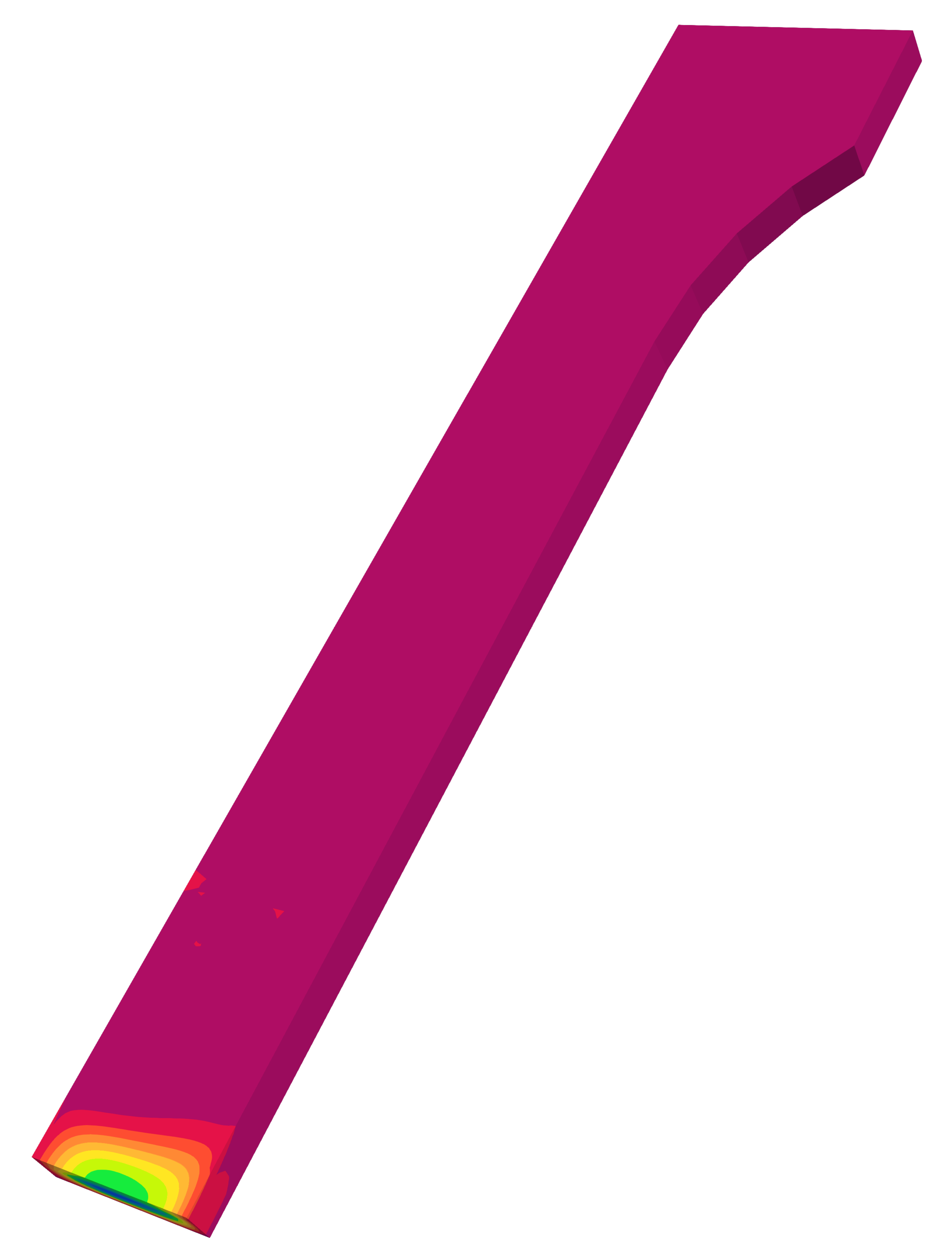}
  \end{subfigure}
  \begin{subfigure}{.3\textwidth} 
    \centering 
    \includegraphics[width=\textwidth]{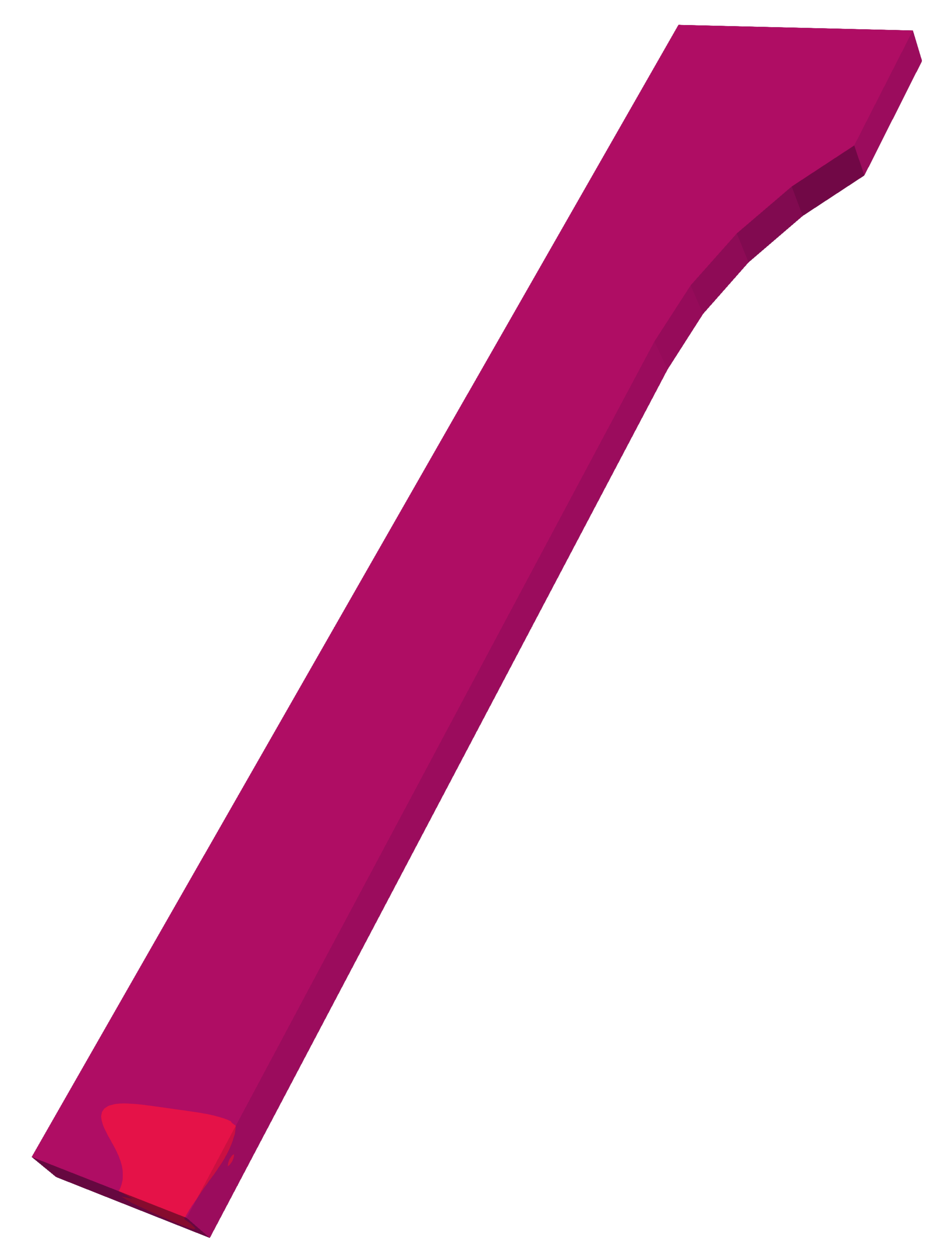}
  \end{subfigure}
  \begin{subfigure}{.08\textwidth} 
    \centering 
    \begin{tikzpicture}
      \node[inner sep=0pt] (pic) at (0,0) {\includegraphics[height=40mm, width=5mm]
      {02_Figures/03_Contour/00_Color_Maps/Damage_Step_Vertical.pdf}};
      \node[inner sep=0pt] (0)   at ($(pic.south)+( 1.10, 0.15)$)  {$-0.00403$};
      \node[inner sep=0pt] (1)   at ($(pic.south)+( 1.10, 3.80)$)  {$+0.00029$};
      \node[inner sep=0pt] (d)   at ($(pic.south)+( 1.10, 4.35)$)  {$D_{xz}~\si{[-]}$};
    \end{tikzpicture} 
  \end{subfigure}

  \vspace{4mm}

  \begin{subfigure}{.3\textwidth} 
    \centering 
    \includegraphics[width=\textwidth]{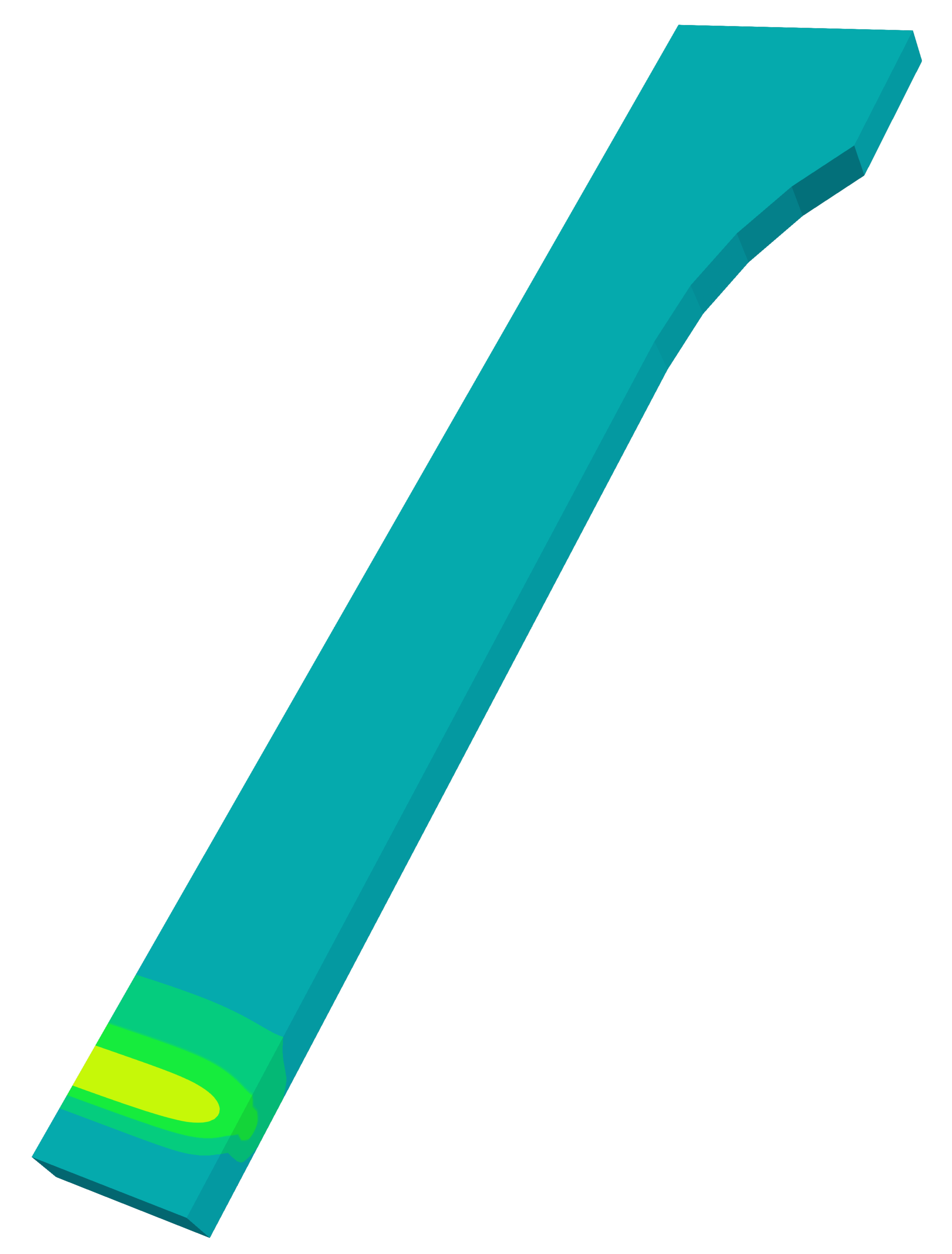}
    \caption{Model~A}
    \label{fig:ExtsDshearA}
  \end{subfigure}
  \begin{subfigure}{.3\textwidth} 
    \centering 
    \includegraphics[width=\textwidth]{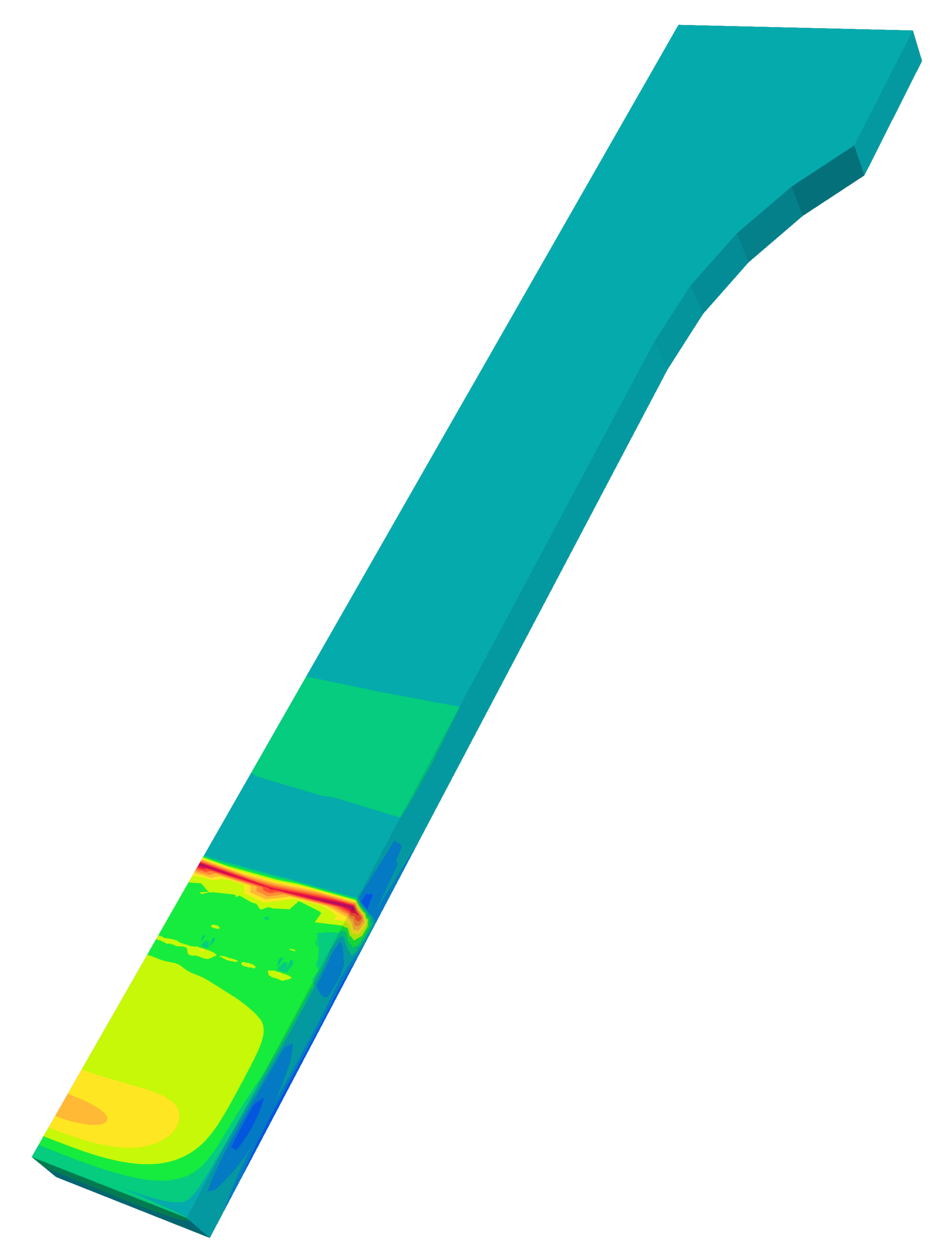}
    \caption{Model~B}
    \label{fig:ExtsDshearB}
  \end{subfigure}
  \begin{subfigure}{.3\textwidth} 
    \centering 
    \includegraphics[width=\textwidth]{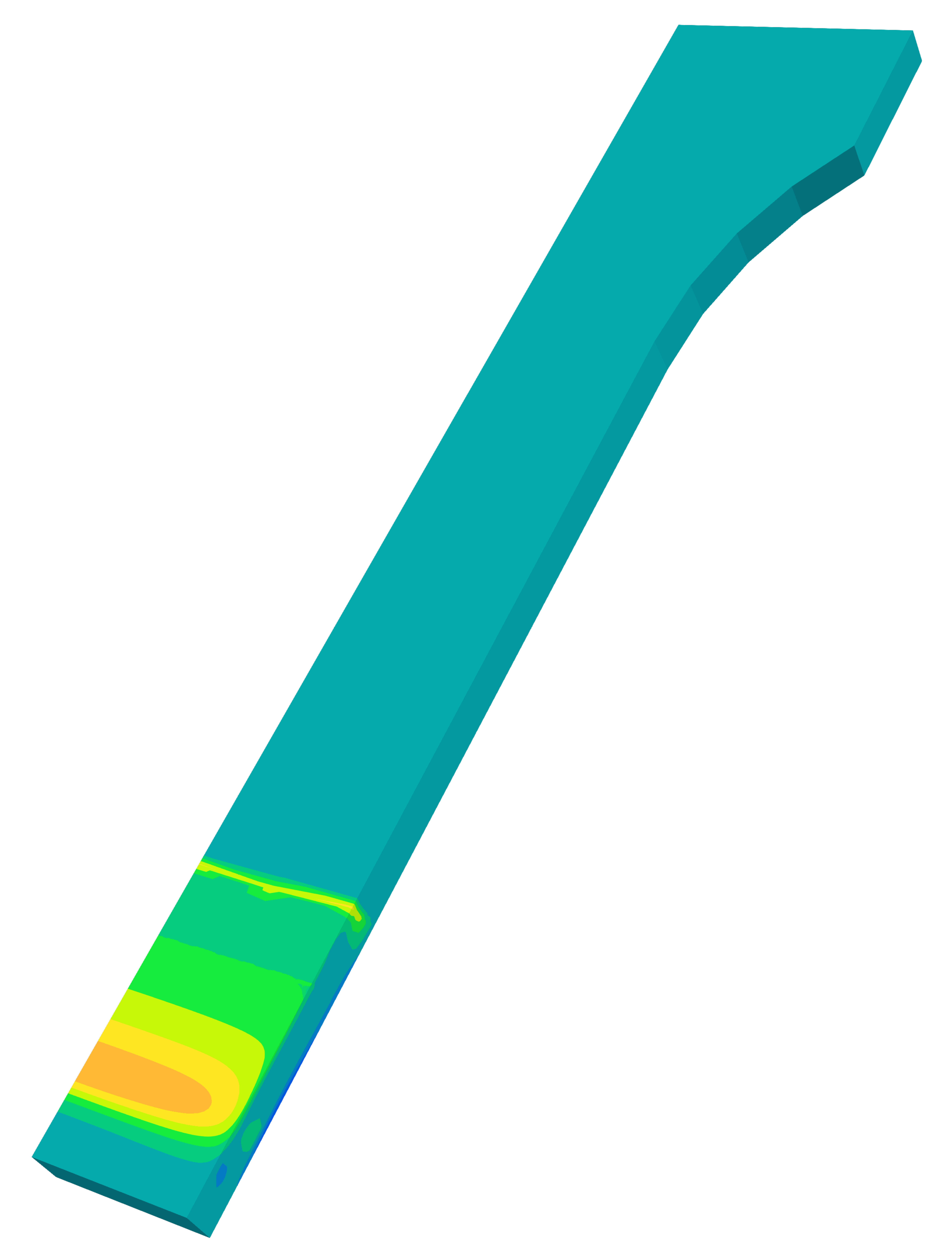}
    \caption{Model~C}
    \label{fig:ExtsDshearC}
  \end{subfigure}
  \begin{subfigure}{.08\textwidth} 
    \centering 
    \begin{tikzpicture}
      \node[inner sep=0pt] (pic) at (0,0) {\includegraphics[height=40mm, width=5mm]
      {02_Figures/03_Contour/00_Color_Maps/Damage_Step_Vertical.pdf}};
      \node[inner sep=0pt] (0)   at ($(pic.south)+( 1.10, 0.15)$)  {$-0.00018$};
      \node[inner sep=0pt] (1)   at ($(pic.south)+( 1.10, 3.80)$)  {$+0.00062$};
      \node[inner sep=0pt] (d)   at ($(pic.south)+( 1.10, 4.35)$)  {$D_{yz}~\si{[-]}$};
    \end{tikzpicture} 
    \hphantom{Model~C}
  \end{subfigure}
  
  \caption{Contour plots of the shear components of the damage tensor for the three-dimensional tensile specimen at the point of comparison indicated in Fig.~\ref{fig:ExtsFuComp}.}
  \label{fig:ExtsDshear}     
\end{figure}

The model comparison in Fig.~\ref{fig:ExtsFuComp} shows again an excellent agreement between models~A and C, while model~B analogously to the previous Sections~\ref{sec:Ex_pwh} and \ref{sec:Ex_an}, yields a higher energy dissipation. The points of comparison for the damage contour plots in Figs.~\ref{fig:ExtsDnormal} and \ref{fig:ExtsDshear} are indicated by the black boxes in Fig.~\ref{fig:ExtsFuComp}.

As already reported in \citetalias{HolthusenBrepolsEtAl2022b}~\citeyear{HolthusenBrepolsEtAl2022b}, the damage tensor component $D_{yy}$, i.e.~the degradation of the plane perpendicular to the loading direction evolves most pronounced for all models (see Fig.~\ref{fig:ExtsDnormal}). And again, the damage zone of model~B spreads furthest and, thus, dissipates the largest amount of energy. Moreover, the contour plots for the normal components agree well for models~A and C.

In Fig.~\ref{fig:ExtsDshear}, the study of the shear components $D_{xy}$, a plane parallel to the loading direction, reveals a concentration at the shoulder of the specimen for all models. The study of the shear components $D_{xz}$, i.e.~the plane perpendicular to the loading direction, yields a uniform distribution, except for the center of the specimen with model~B. The study of the shear components $D_{yz}$, i.e.~the second plane perpendicular to the loading direction, reveals a localization for model B at the transition from the fine to the coarse mesh, which cannot be observed for the full regularization with model~A.

\subsection{Smiley specimen}
\label{sec:Ex_ss}

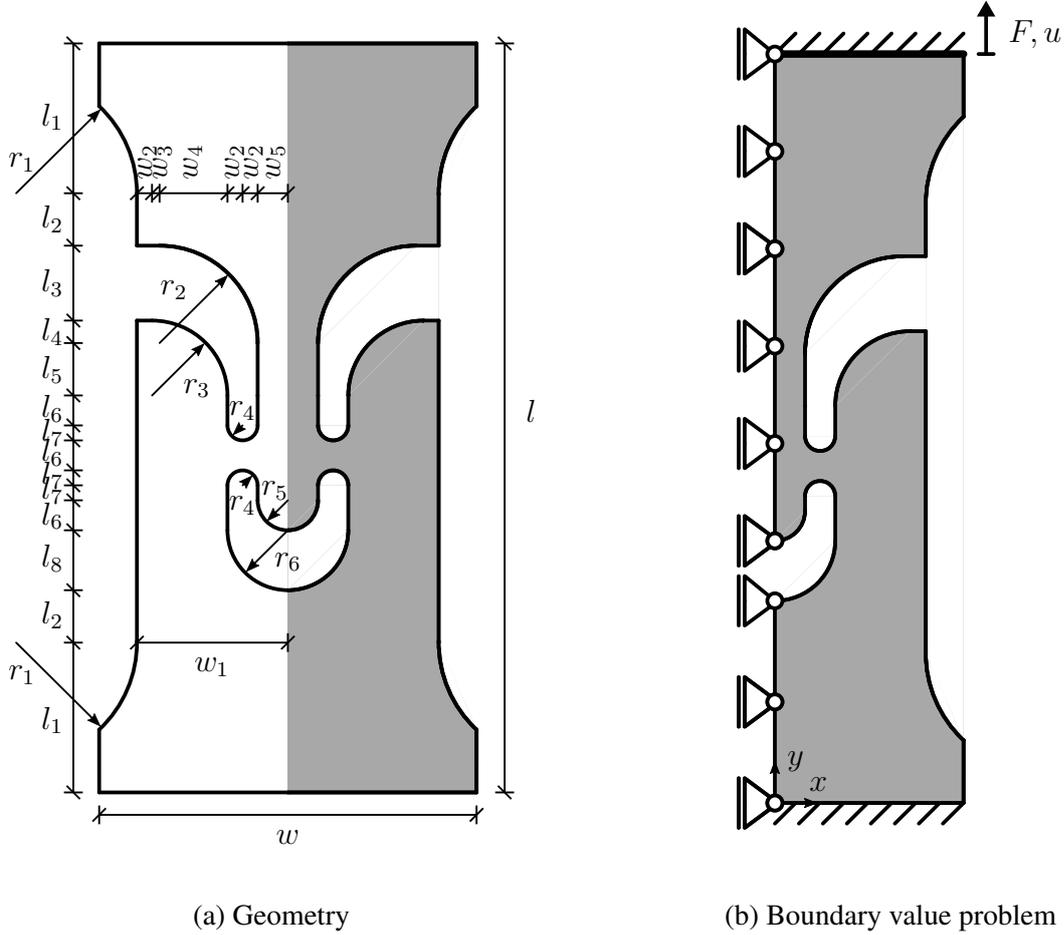
\begin{figure}[htbp] 
  \centering 
  \begin{subfigure}{.45\textwidth} 
      \centering 
      \begin{tikzpicture}
        \node (pic) at (0,0) {\includegraphics[width=\textwidth]{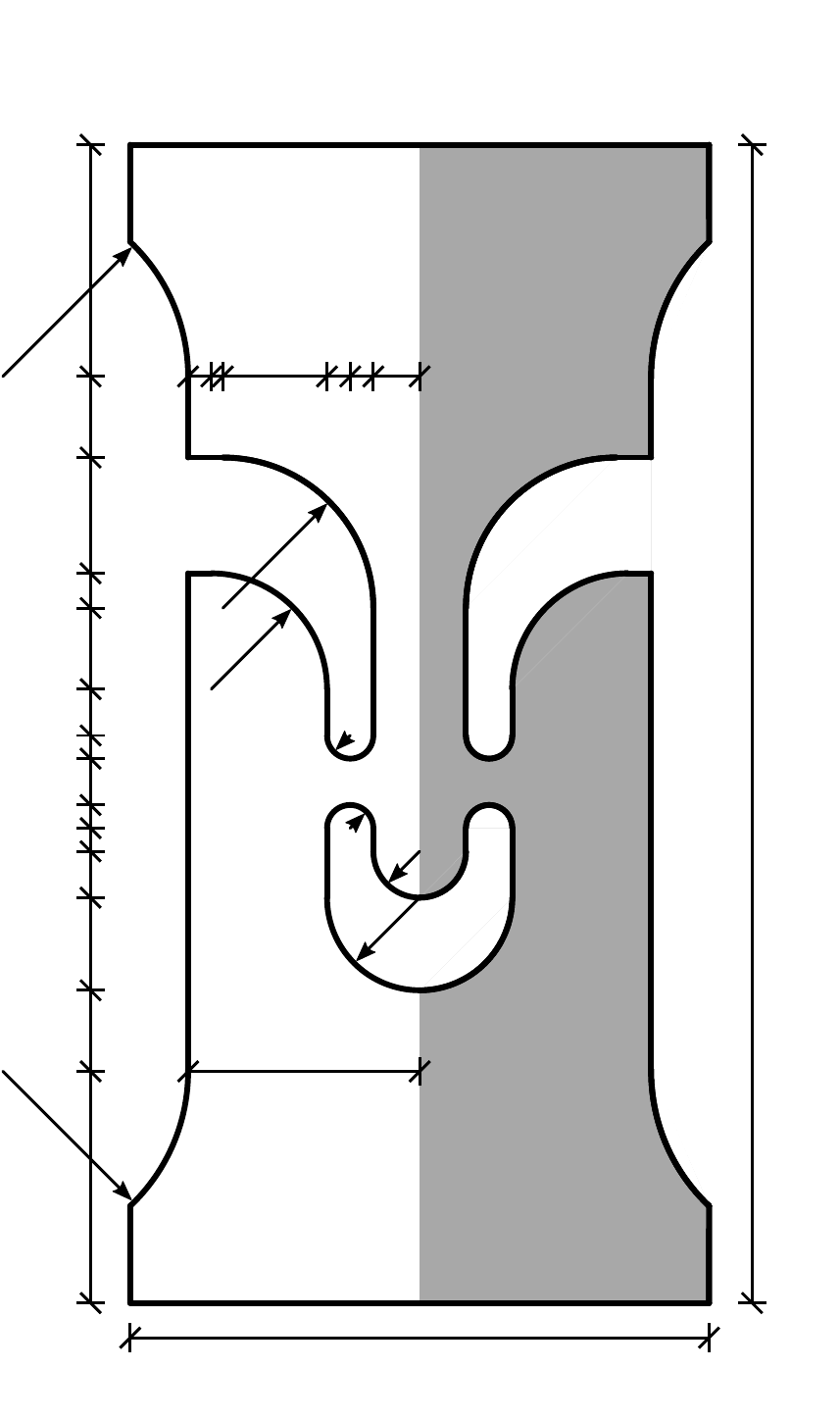}};
        \node at ($(pic.south) +( 3.20, 6.20)$)  {$l$};
        \node at ($(pic.south) +(-3.10,10.10)$)  {$l_1$};
        \node at ($(pic.south) +(-3.10, 2.10)$)  {$l_1$};
        \node at ($(pic.south) +(-3.10, 8.75)$)  {$l_2$};
        \node at ($(pic.south) +(-3.10, 3.50)$)  {$l_2$};
        \node at ($(pic.south) +(-3.10, 7.90)$)  {$l_3$};
        \node at ($(pic.south) +(-3.10, 7.25)$)  {$l_4$};
        \node at ($(pic.south) +(-3.10, 6.75)$)  {$l_5$};
        \node at ($(pic.south) +(-3.10, 6.23)$)  {$l_6$};
        \node at ($(pic.south) +(-3.10, 5.92)$)  {$l_7$};
        \node at ($(pic.south) +(-3.10, 5.65)$)  {$l_6$};
        \node at ($(pic.south) +(-3.10, 5.35)$)  {$l_7$};
        \node at ($(pic.south) +(-3.10, 5.15)$)  {$l_7$};
        \node at ($(pic.south) +(-3.10, 4.81)$)  {$l_6$};
        \node at ($(pic.south) +(-3.10, 4.25)$)  {$l_8$};
        \node at ($(pic.south) +( 0.00, 0.55)$)  {$w$};
        \node at ($(pic.south) +(-1.00, 2.80)$)  {$w_1$};
        \node[rotate=90] at ($(pic.south) +(-1.85, 9.50)$)  {$w_2$};
        \node[rotate=90] at ($(pic.south) +(-1.65, 9.50)$)  {$w_3$};
        \node[rotate=90] at ($(pic.south) +(-1.25, 9.50)$)  {$w_4$};
        \node[rotate=90] at ($(pic.south) +(-0.70, 9.50)$)  {$w_2$};
        \node[rotate=90] at ($(pic.south) +(-0.45, 9.50)$)  {$w_2$};
        \node[rotate=90] at ($(pic.south) +(-0.16, 9.50)$)  {$w_5$};
        \node at ($(pic.south) +(-3.50, 2.70)$)  {$r_1$};
        \node at ($(pic.south) +(-3.50, 9.50)$)  {$r_1$};
        \node at ($(pic.south) +(-1.50, 7.80)$)  {$r_2$};
        \node at ($(pic.south) +(-1.20, 6.50)$)  {$r_3$};
        \node at ($(pic.south) +(-0.60, 6.20)$)  {$r_4$};
        \node at ($(pic.south) +(-0.60, 4.95)$)  {$r_4$};
        \node at ($(pic.south) +(-0.17, 5.15)$)  {$r_5$};
        \node at ($(pic.south) +(-0.00, 4.20)$)  {$r_6$};
      \end{tikzpicture} 
      \caption{Geometry}
      \label{fig:Ex_ss_geom}
  \end{subfigure}
  \qquad
  \begin{subfigure}{.45\textwidth} 
    \centering 
    \begin{tikzpicture}
      \node[inner sep=0pt] (pic) at (0,0) {\includegraphics[width=\textwidth]{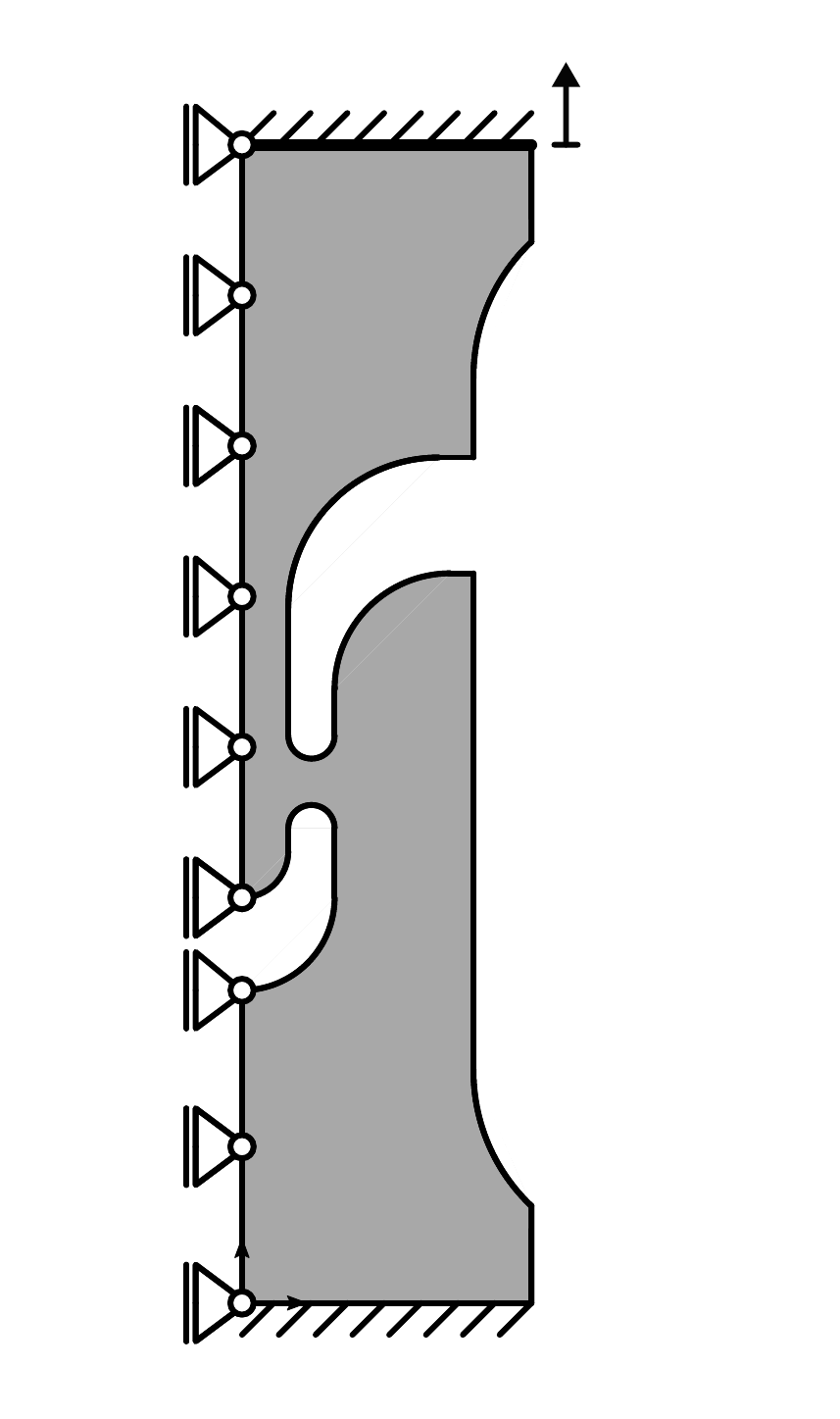}};
      \node[inner sep=0pt] (Fu) at ($(pic.south) +( 1.90,11.20)$)  {$F,u$};
      \node[inner sep=0pt] (x)  at ($(pic.south) +(-0.95, 1.25)$)  {$x$};
      \node[inner sep=0pt] (y)  at ($(pic.south) +(-1.25, 1.55)$)  {$y$};
    \end{tikzpicture} 
    \caption{Boundary value problem}
    \label{fig:Ex_ss_bvp}
  \end{subfigure}
  \caption{Geometry and boundary value problem of the smiley specimen.} 
  \label{fig:Ex_ss}
\end{figure}

The final example serves for the investigation of a complex combination of normal and shear stress states. Inspired by \cite{GerkeZistlEtAl2020}, \cite{RothMohr2016}, \cite{Tancogne-DejeanRothEtAl2016}, \cite{TillHackl2013}, and \cite{Miyauchi1984}, we designed a smiley specimen where the normal and shear load carrying cross sections are equal. The design, further, features smooth transitions from arcs to straight lines to avoid stress singularities at these points. Furthermore, this example illustrates the necessity to investigate the eigenvalues of the damage tensor in order to accurately study the degradation of the specimen and, again, compares the differing results of the isotropic and the anisotropic damage model.

Fig.~\ref{fig:Ex_ss} shows the geometry and the considered boundary value problem. The dimensions read
$l = 50~[\si{\mm}]$, $l_1 = 10~[\si{\mm}]$, $l_2 = 3.5~[\si{\mm}]$, $l_3 = 5~[\si{\mm}]$, $l_4 = 1.5~[\si{\mm}]$, $l_5 = 4.5~[\si{\mm}]$, $l_6 = 2~[\si{\mm}]$, $l_7 = 1~[\si{\mm}]$, $l_8 = 4~[\si{\mm}]$, 
$w = 25~[\si{\mm}]$, $w_1 = 10~[\si{\mm}]$, $w_2 = 1~[\si{\mm}]$, $w_3 = 0.5~[\si{\mm}]$, $w_4 = 4.5~[\si{\mm}]$, $w_5 = 2~[\si{\mm}]$, 
$r_1 = 8~[\si{\mm}]$, $r_2 = 6.5~[\si{\mm}]$, $r_3 = 5~[\si{\mm}]$, $r_4 = 1~[\si{\mm}]$, $r_5 = 2~[\si{\mm}]$,  and $r_6 = 4~[\si{\mm}]$ 
with a thickness of $1~[\si{\mm}]$. Due to symmetry, only one half of the specimen with clamped ends is modeled in the simulation. The internal length scales of model B are chosen as $A_i^\text{B} = 75~[\si{\MPa\mm\squared}]$ and the parameters of model~A and C are identified as $A_i^\text{A} = 220~[\si{\MPa\mm\squared}]$ and $A_i^\text{C} = 790~[\si{\MPa\mm\squared}]$.

\begin{figure}[htbp]
  \centering
  \begin{subfigure}{.48\textwidth}
    \centering
    \includegraphics{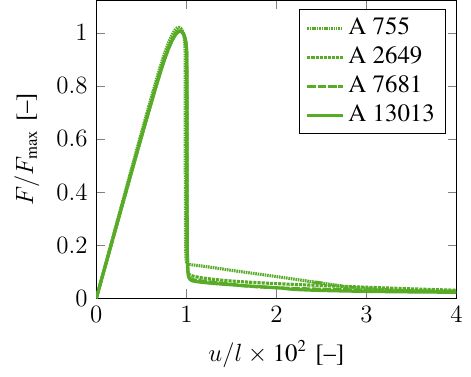}
    \vspace{-7mm}
    \caption{Model~A}
    \label{fig:ExssFuA}
  \end{subfigure}
  \quad
  \begin{subfigure}{.48\textwidth}
    \centering
    \includegraphics{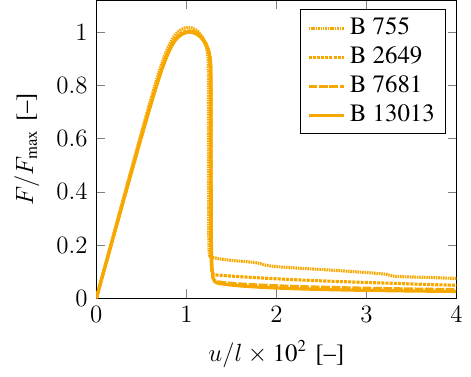}
    \vspace{-7mm}
    \caption{Model~B}
    \label{fig:ExssFuB}
  \end{subfigure}%
  \vspace{5mm} 
  \begin{subfigure}{.48\textwidth}
    \centering
    \includegraphics{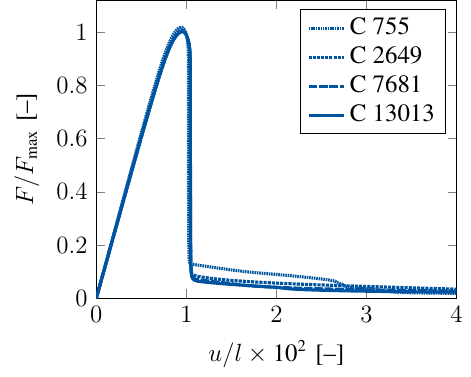}
    \vspace{-7mm}
    \caption{Model~C}
    \label{fig:ExssFuC}
  \end{subfigure}
  \quad
  \begin{subfigure}{.48\textwidth}
    \centering
    \includegraphics{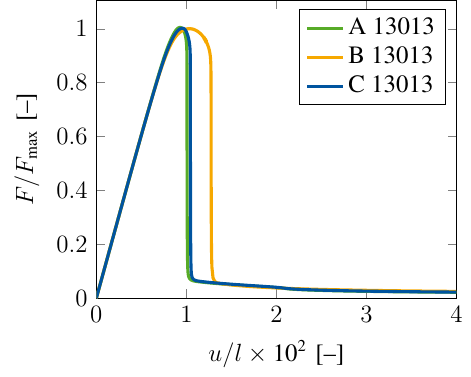}
    \vspace{-7mm}
    \caption{Model comparison}
    \label{fig:ExssFuComp}
  \end{subfigure}
  \caption{Mesh convergence studies for the smiley specimen and model comparison. The forces are normalized with respect to the maximum force of the finest mesh (13013 elements) of model~B with $F_\text{max} = 2.9590 \times 10^3~[\si{\newton}]$.}
  \label{fig:ExssFu}
\end{figure}

Fig.~\ref{fig:ExssFu} shows the normalized force-displacement curves. In this example, no model obtains convergence with respect to the maximum peak force using the coarsest mesh (755~elements), only upon mesh refinement this is achieved. The model comparison in Fig.~\ref{fig:ExssFuComp} yields a distinct horizontal offset to the right of the vertical drop for model~B at $u_{0.5 F_\text{max}}^\text{B}/l \times \text{10}^\text{2} = 1.275~\si{[-]}$. For this example, a difference in the force drop can also be observed for models~A and C with $u_{0.5 F_\text{max}}^\text{A}/l \times \text{10}^\text{2} = 1.008~\si{[-]}$ compared to $u_{0.5 F_\text{max}}^\text{C}/l \times \text{10}^\text{2} = 1.045~\si{[-]}$.

\begin{figure}
  \centering 

  \begin{subfigure}{.22\textwidth} 
    \centering 
    \includegraphics[width=\textwidth]{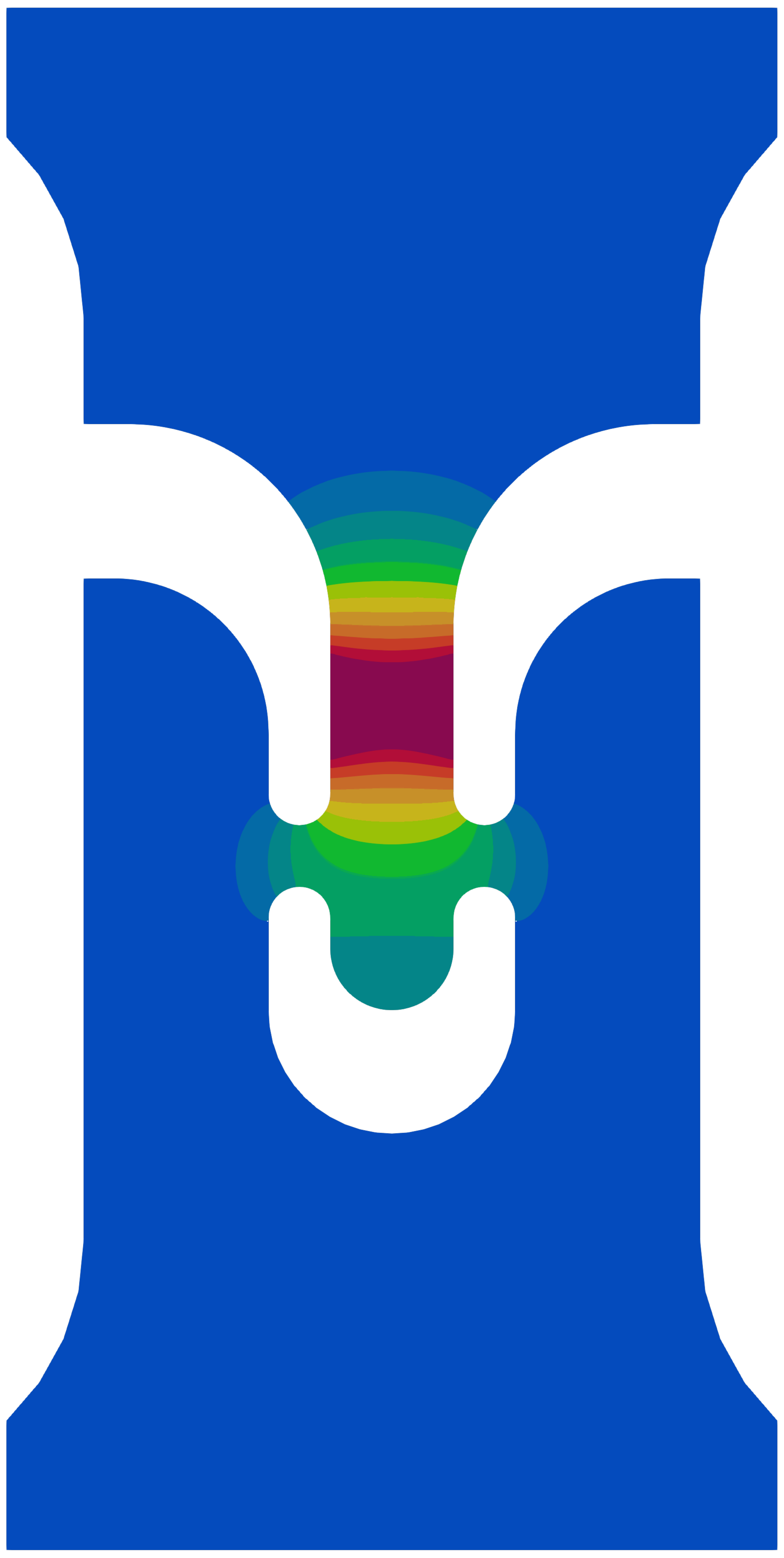}
  \end{subfigure}
  \begin{subfigure}{.22\textwidth} 
    \centering 
    \includegraphics[width=\textwidth]{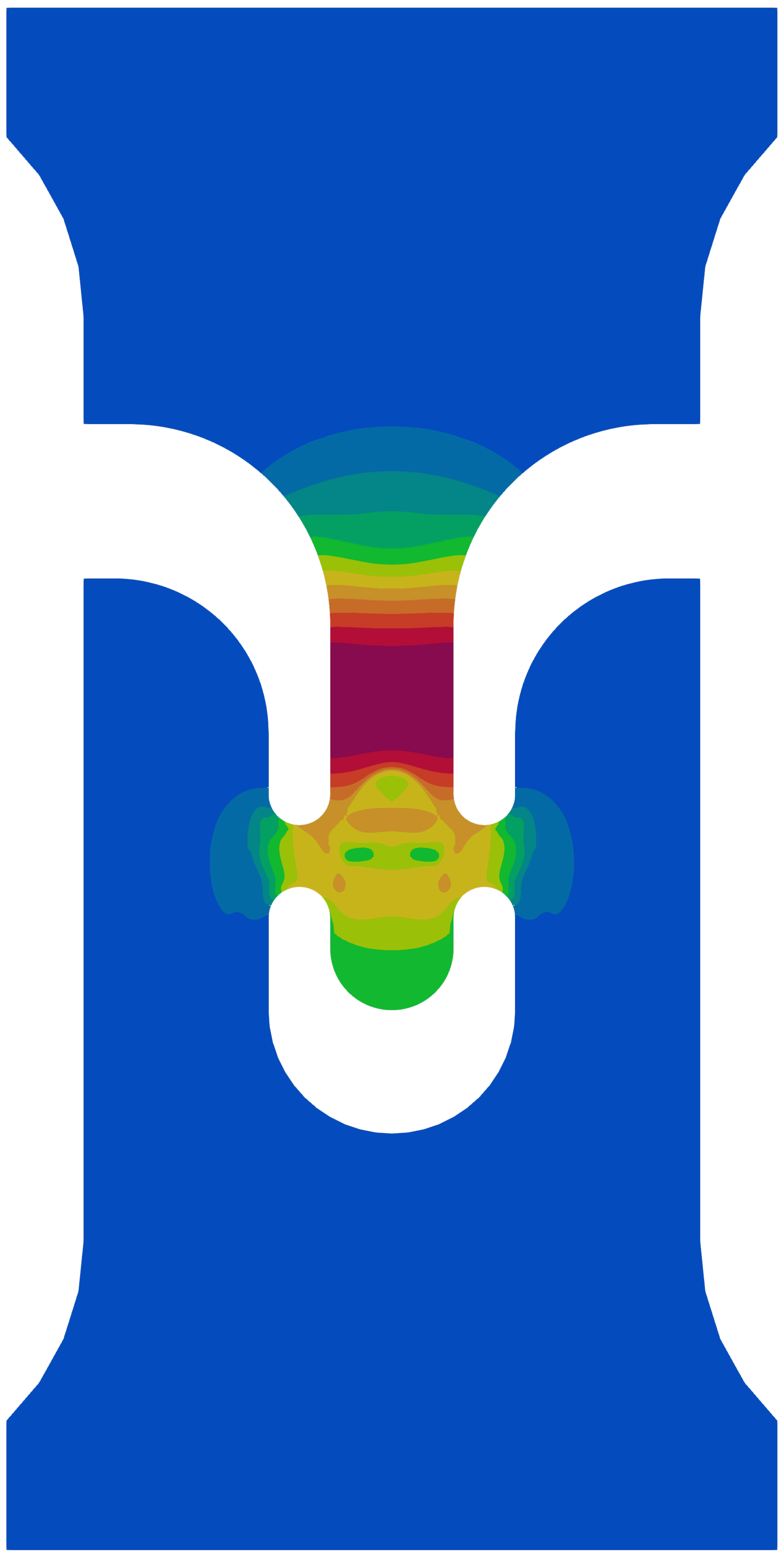}
  \end{subfigure}
  \begin{subfigure}{.22\textwidth} 
    \centering 
    \includegraphics[width=\textwidth]{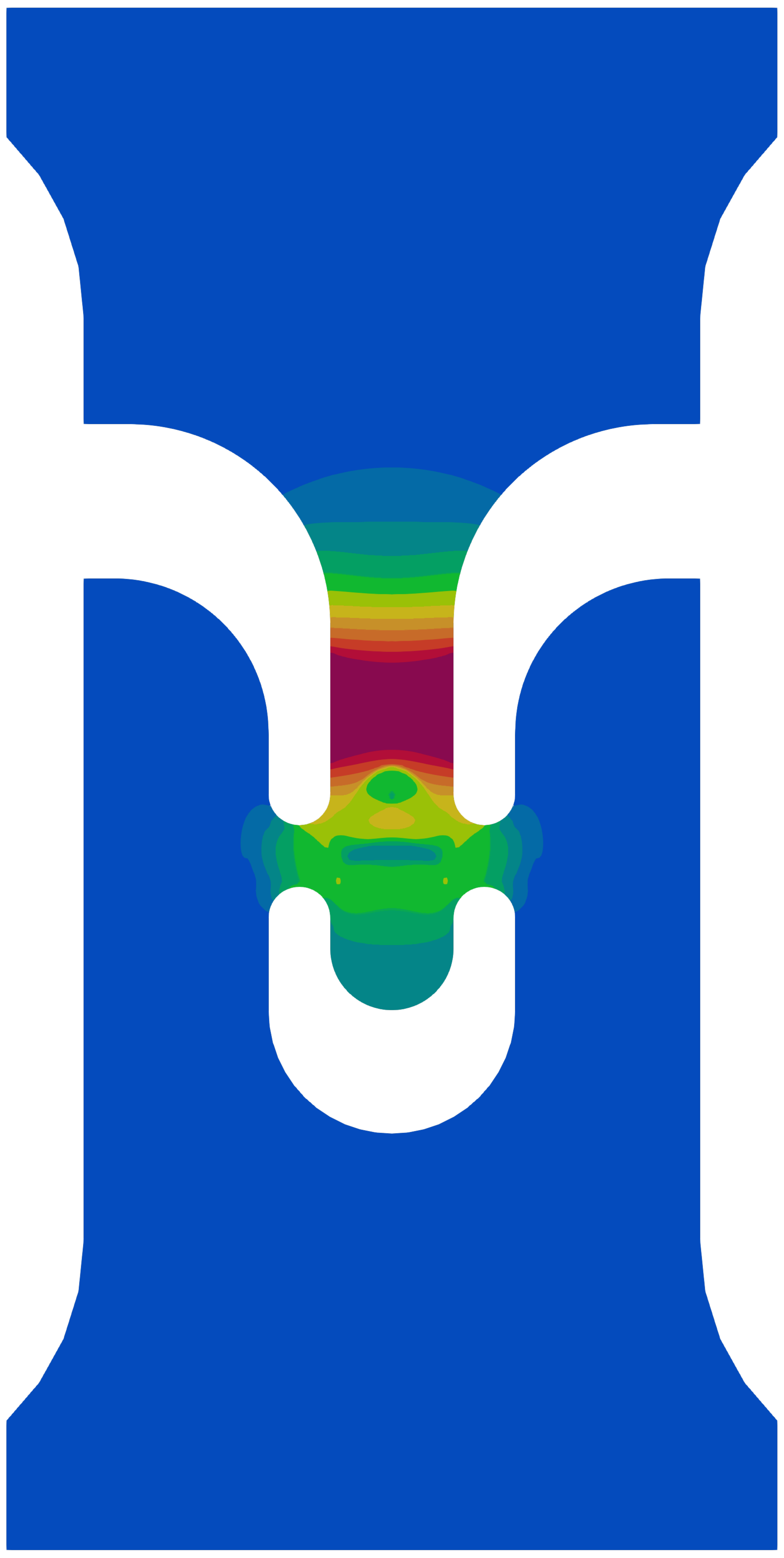}
  \end{subfigure}
  \begin{subfigure}{.08\textwidth} 
    \centering 
    \begin{tikzpicture}
      \node[inner sep=0pt] (pic) at (0,0) {\includegraphics[height=40mm, width=5mm]
      {02_Figures/03_Contour/00_Color_Maps/Damage_Step_Vertical.pdf}};
      \node[inner sep=0pt] (0)   at ($(pic.south)+( 0.50, 0.15)$)  {$0$};
      \node[inner sep=0pt] (1)   at ($(pic.south)+( 0.50, 3.80)$)  {$1$};
      \node[inner sep=0pt] (d)   at ($(pic.south)+( 0.00, 4.35)$)  {$D_{xx}~\si{[-]}$};
    \end{tikzpicture} 
  \end{subfigure}

  \vspace{1mm}

  \begin{subfigure}{.22\textwidth} 
    \centering 
    \includegraphics[width=\textwidth]{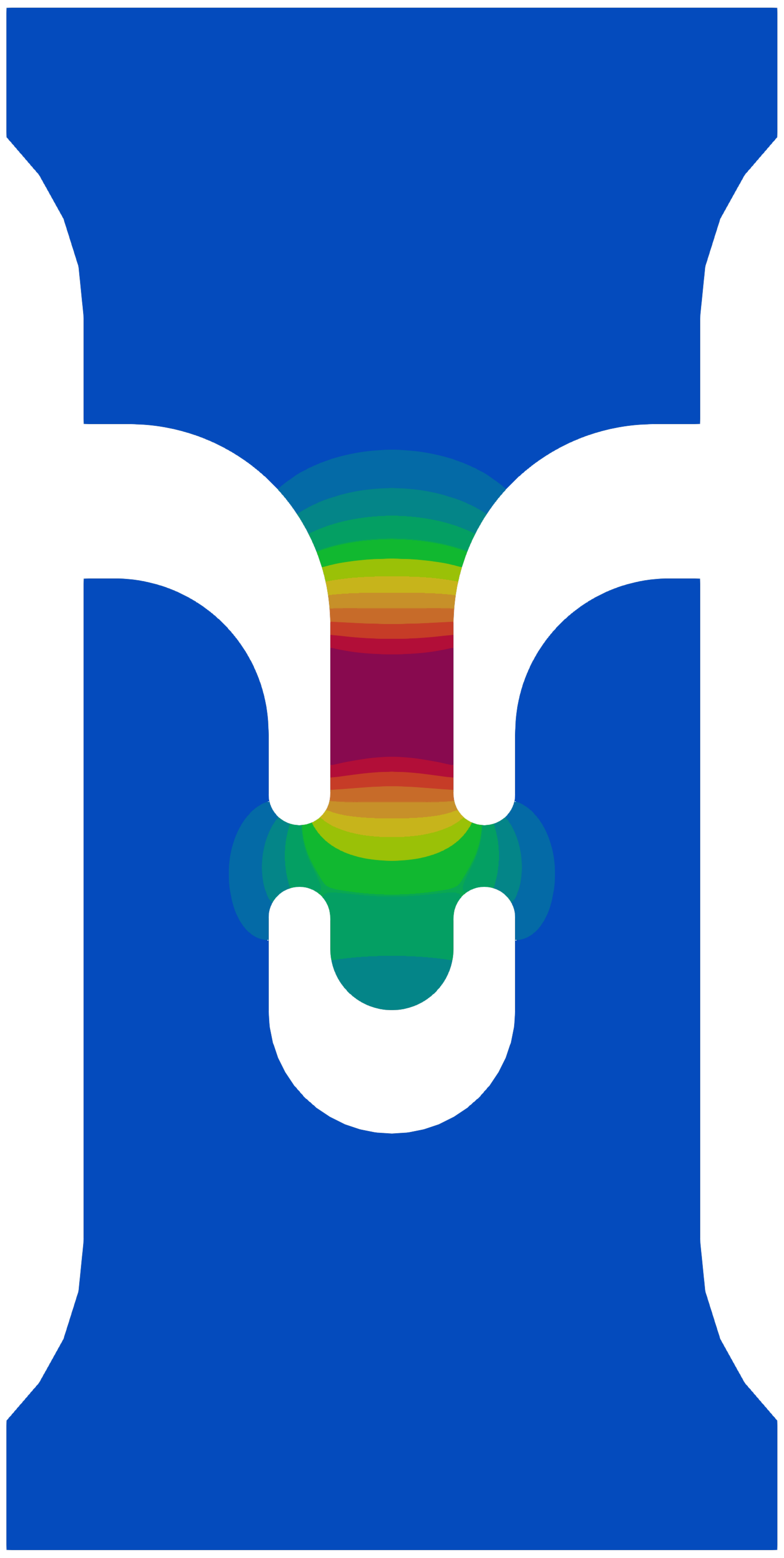}
  \end{subfigure}
  \begin{subfigure}{.22\textwidth} 
    \centering 
    \includegraphics[width=\textwidth]{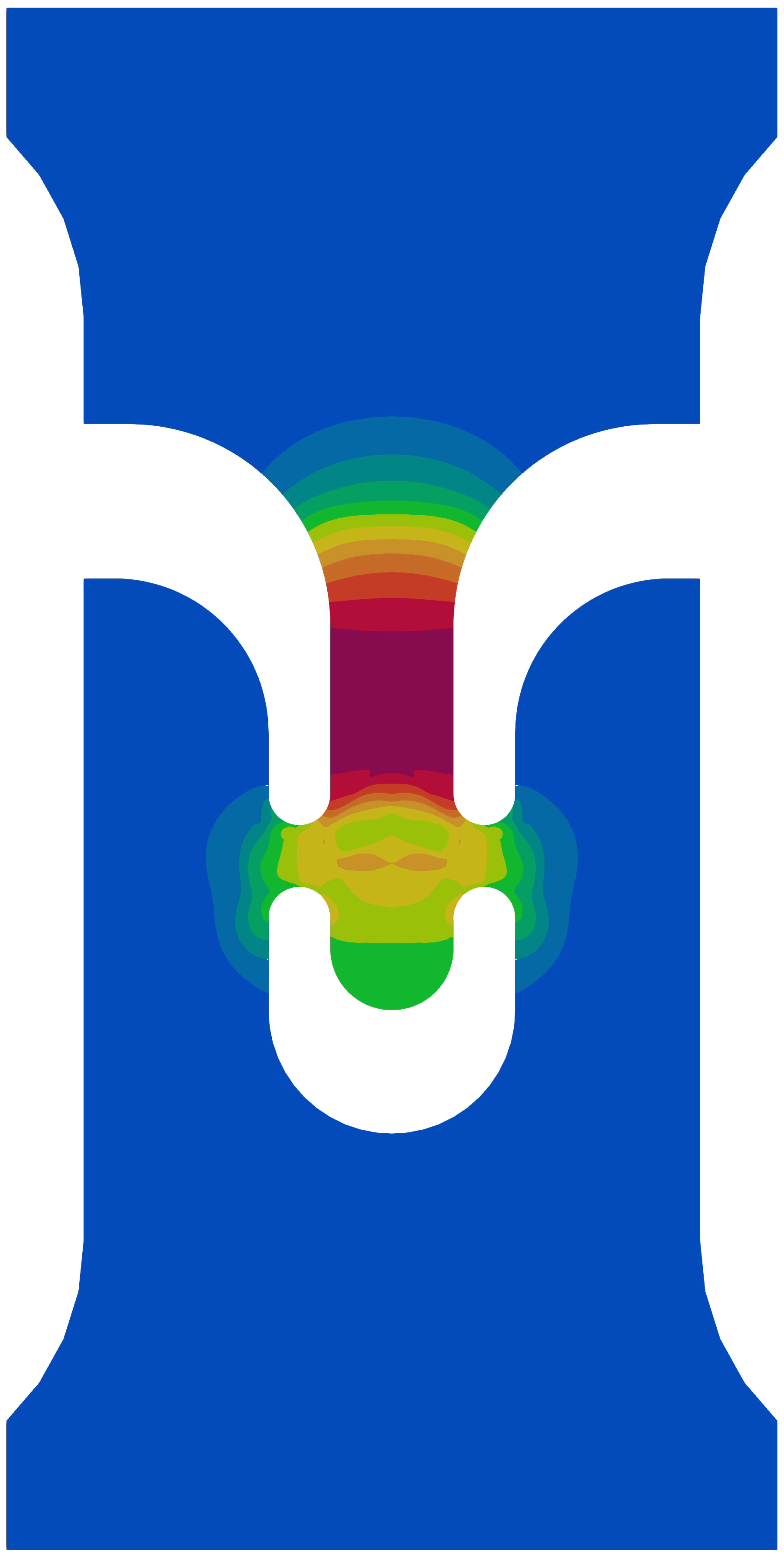}
  \end{subfigure}
  \begin{subfigure}{.22\textwidth} 
    \centering 
    \includegraphics[width=\textwidth]{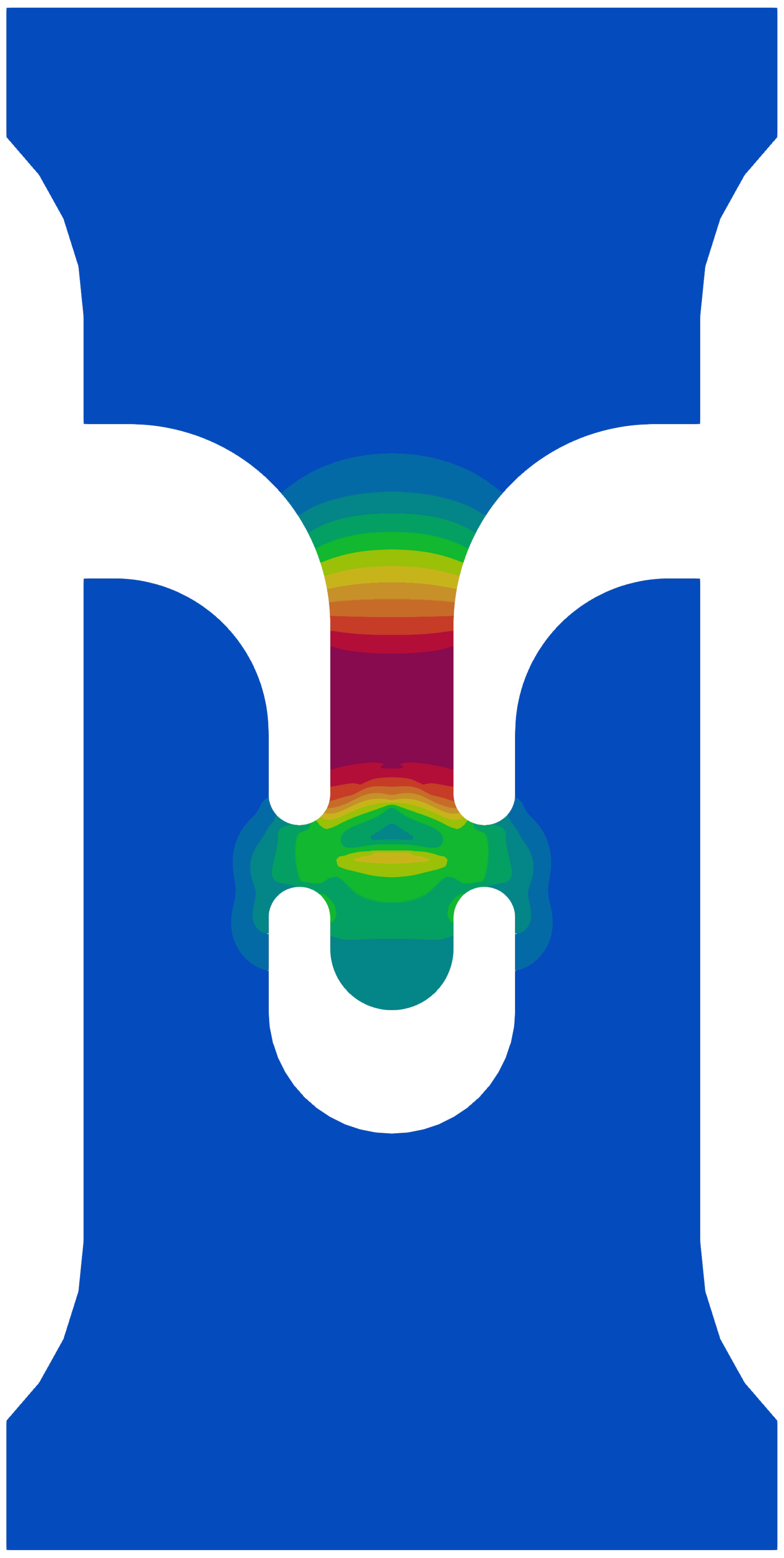}
  \end{subfigure}
  \begin{subfigure}{.08\textwidth} 
    \centering 
    \begin{tikzpicture}
      \node[inner sep=0pt] (pic) at (0,0) {\includegraphics[height=40mm, width=5mm]
      {02_Figures/03_Contour/00_Color_Maps/Damage_Step_Vertical.pdf}};
      \node[inner sep=0pt] (0)   at ($(pic.south)+( 0.50, 0.15)$)  {$0$};
      \node[inner sep=0pt] (1)   at ($(pic.south)+( 0.50, 3.80)$)  {$1$};
      \node[inner sep=0pt] (d)   at ($(pic.south)+( 0.00, 4.35)$)  {$D_{yy}~\si{[-]}$};
    \end{tikzpicture} 
  \end{subfigure}

  \vspace{1mm}

  \begin{subfigure}{.22\textwidth} 
    \centering 
    \includegraphics[width=\textwidth]{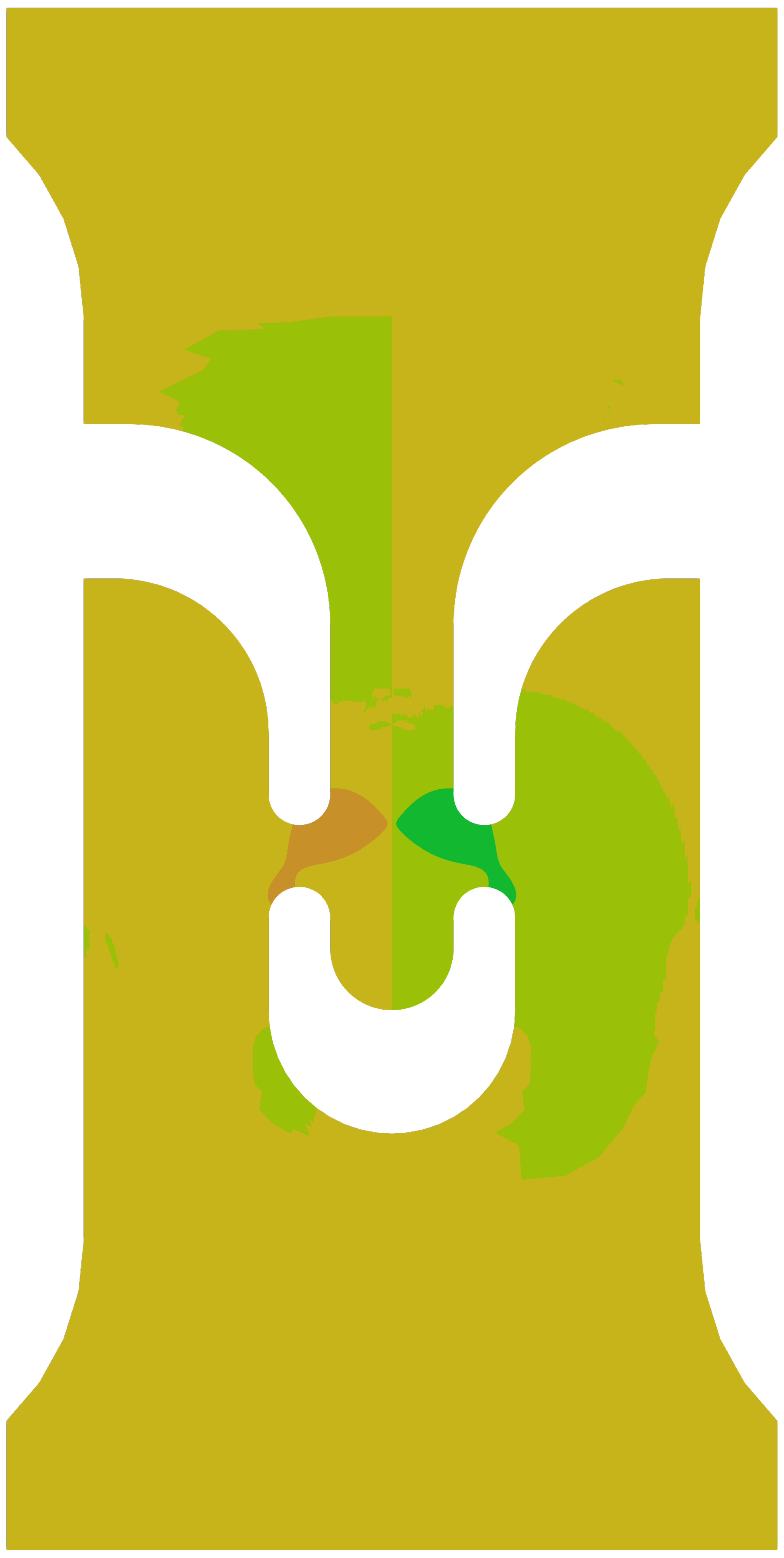}
    \caption{Model~A}
    \label{fig:ExssDA}
  \end{subfigure}
  \begin{subfigure}{.22\textwidth} 
    \centering 
    \includegraphics[width=\textwidth]{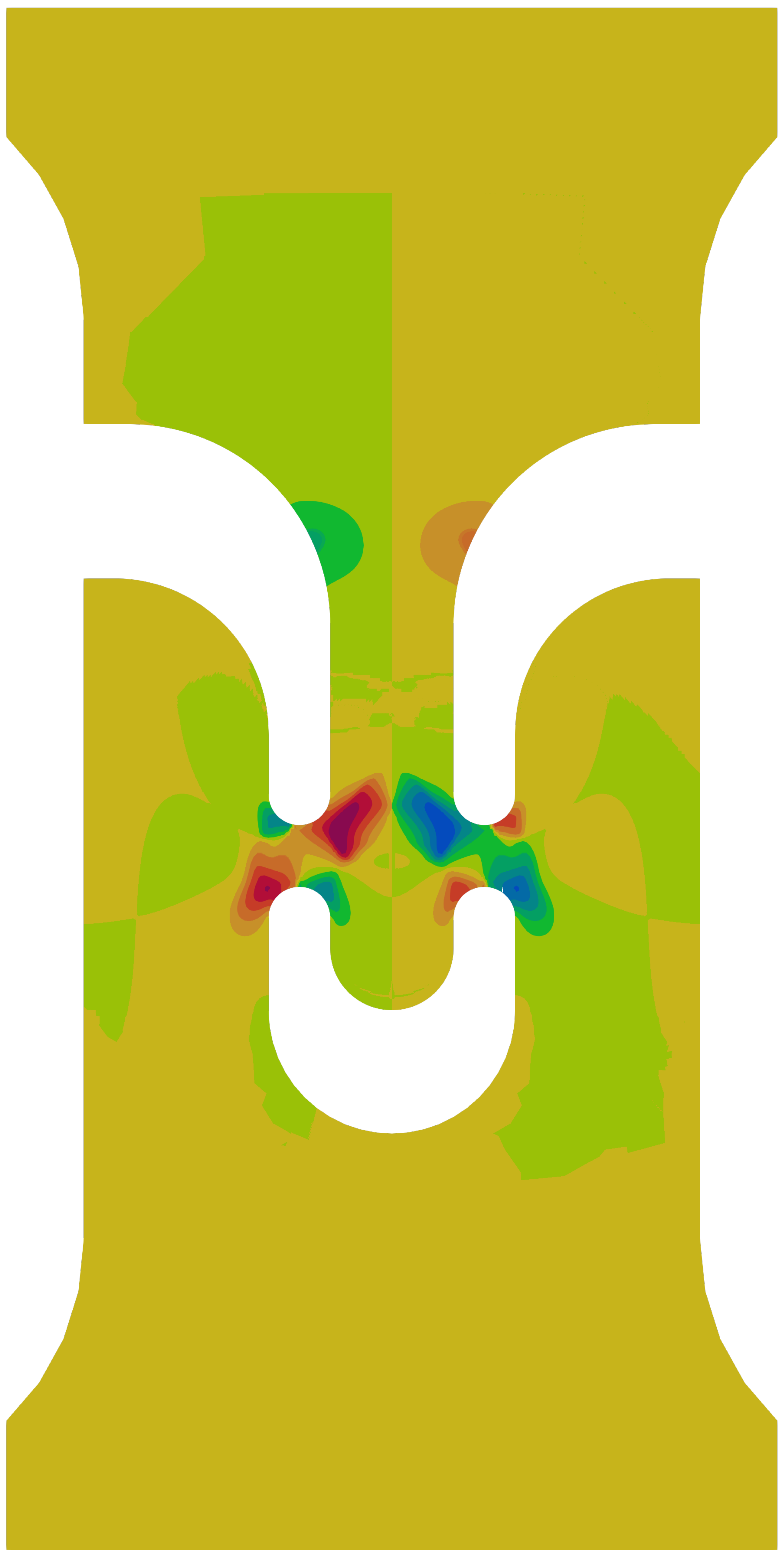}
    \caption{Model~B}
    \label{fig:ExssDB}
  \end{subfigure}
  \begin{subfigure}{.22\textwidth} 
    \centering 
    \includegraphics[width=\textwidth]{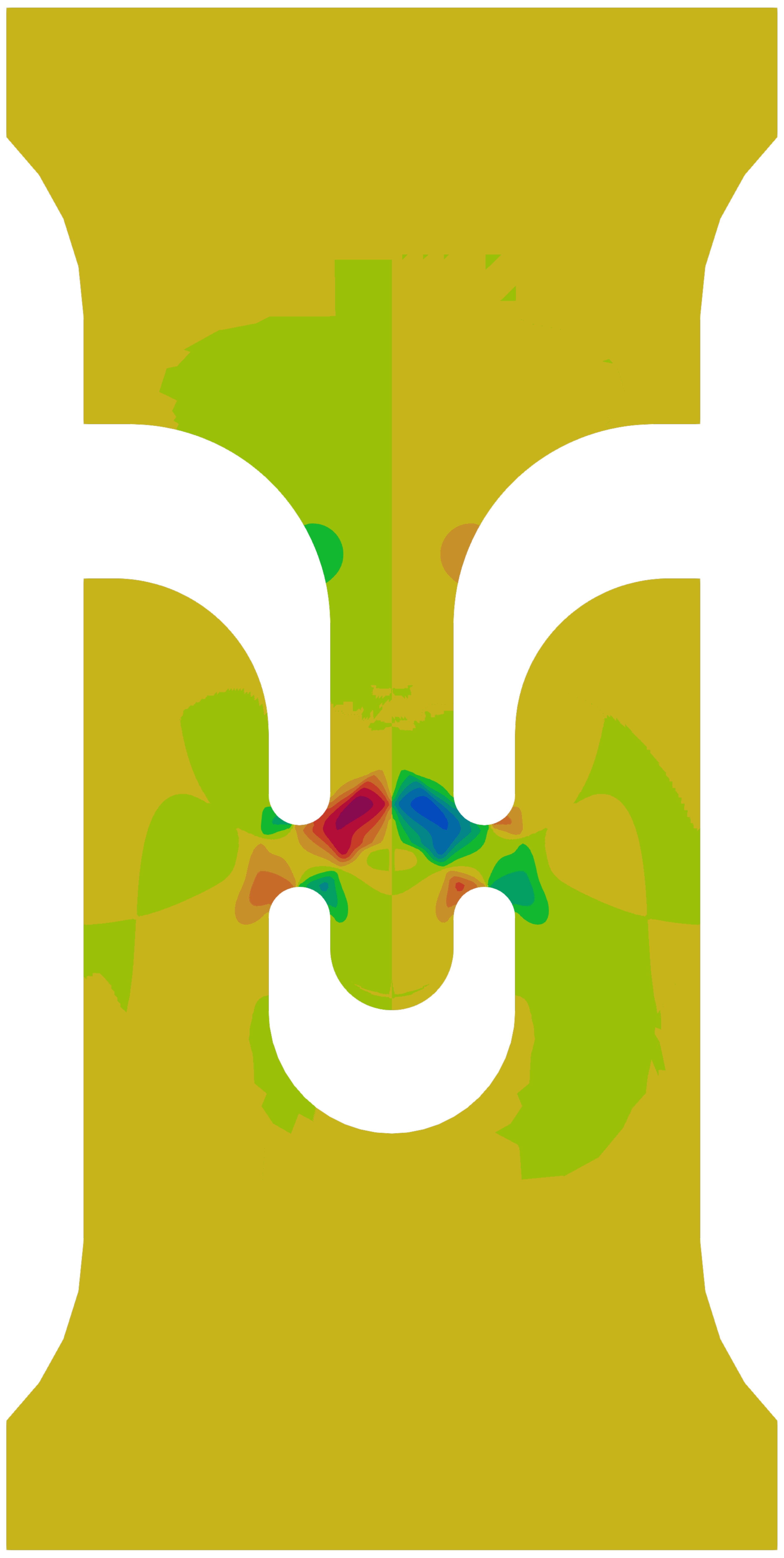}
    \caption{Model~C}
    \label{fig:ExssDC}
  \end{subfigure}
  \begin{subfigure}{.08\textwidth} 
    \centering 
    \begin{tikzpicture}
      \node[inner sep=0pt] (pic) at (0,0) {\includegraphics[height=40mm, width=5mm]
      {02_Figures/03_Contour/00_Color_Maps/Damage_Step_Vertical.pdf}};
      \node[inner sep=0pt] (0)   at ($(pic.south)+( 1.00, 0.15)$)  {$-0.2174$};
      \node[inner sep=0pt] (1)   at ($(pic.south)+( 1.00, 3.80)$)  {$+0.2174$};
      \node[inner sep=0pt] (d)   at ($(pic.south)+( 0.00, 4.35)$)  {$D_{xy}~\si{[-]}$};
    \end{tikzpicture} 
    \hphantom{Model~C}
  \end{subfigure}
  
  \caption{Contour plots of the normal and shear components of the damage tensor for the smiley specimen at the end of the simulation.}

  \label{fig:ExssD}     
\end{figure}

The damage contour plots in Fig.~\ref{fig:ExssD} reveal a tension dominated failure with all models, where model~B shows the largest damage zone. More models~B and C exhibit concentrated peak values for the shear component of the damage tensor $D_{xy}$ while model~A yields a smooth distribution.

\begin{figure}
  \centering 

  \begin{subfigure}{.22\textwidth} 
    \centering 
    \includegraphics[width=\textwidth]{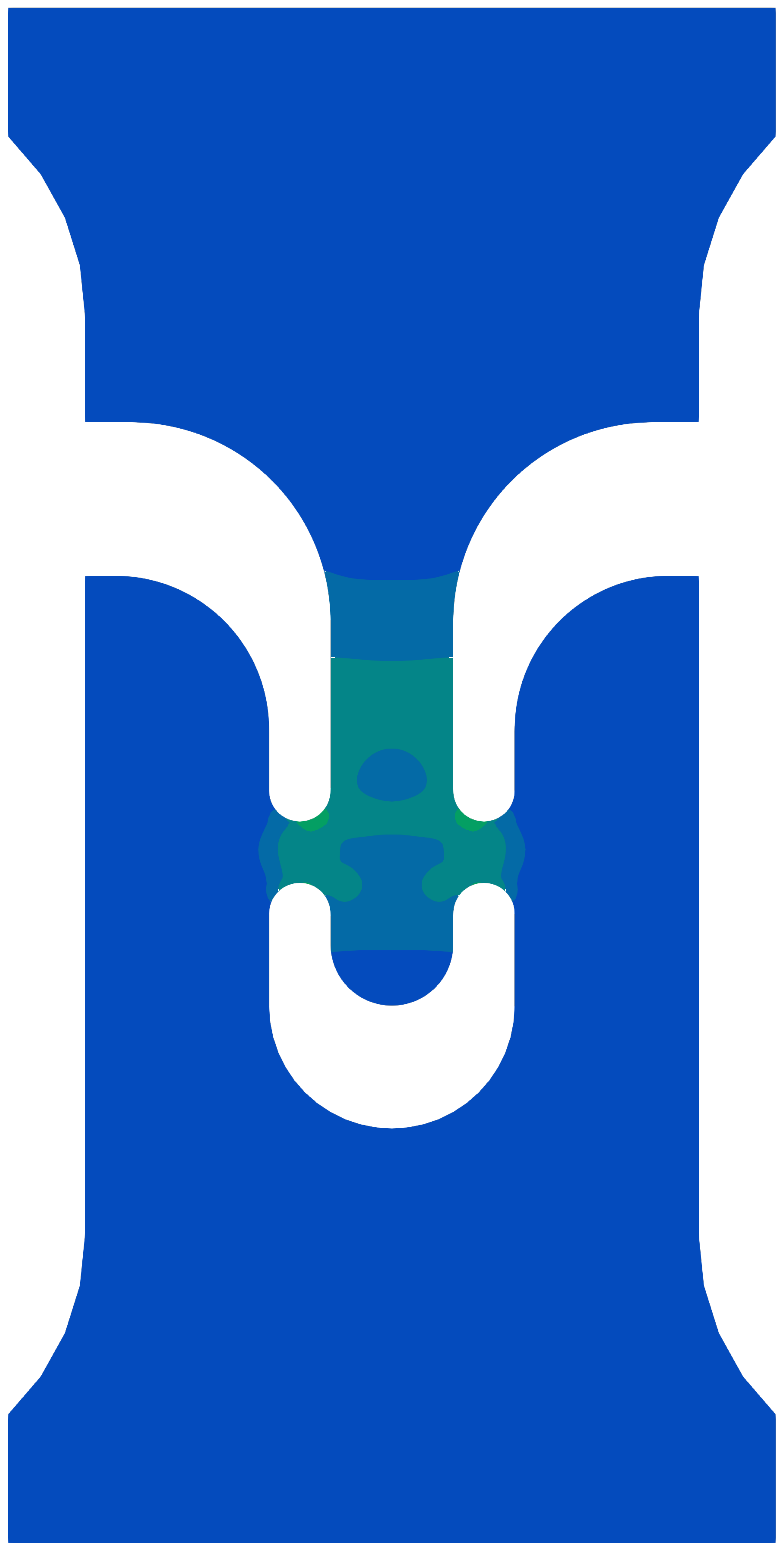}
  \end{subfigure}
  \begin{subfigure}{.22\textwidth} 
    \centering 
    \includegraphics[width=\textwidth]{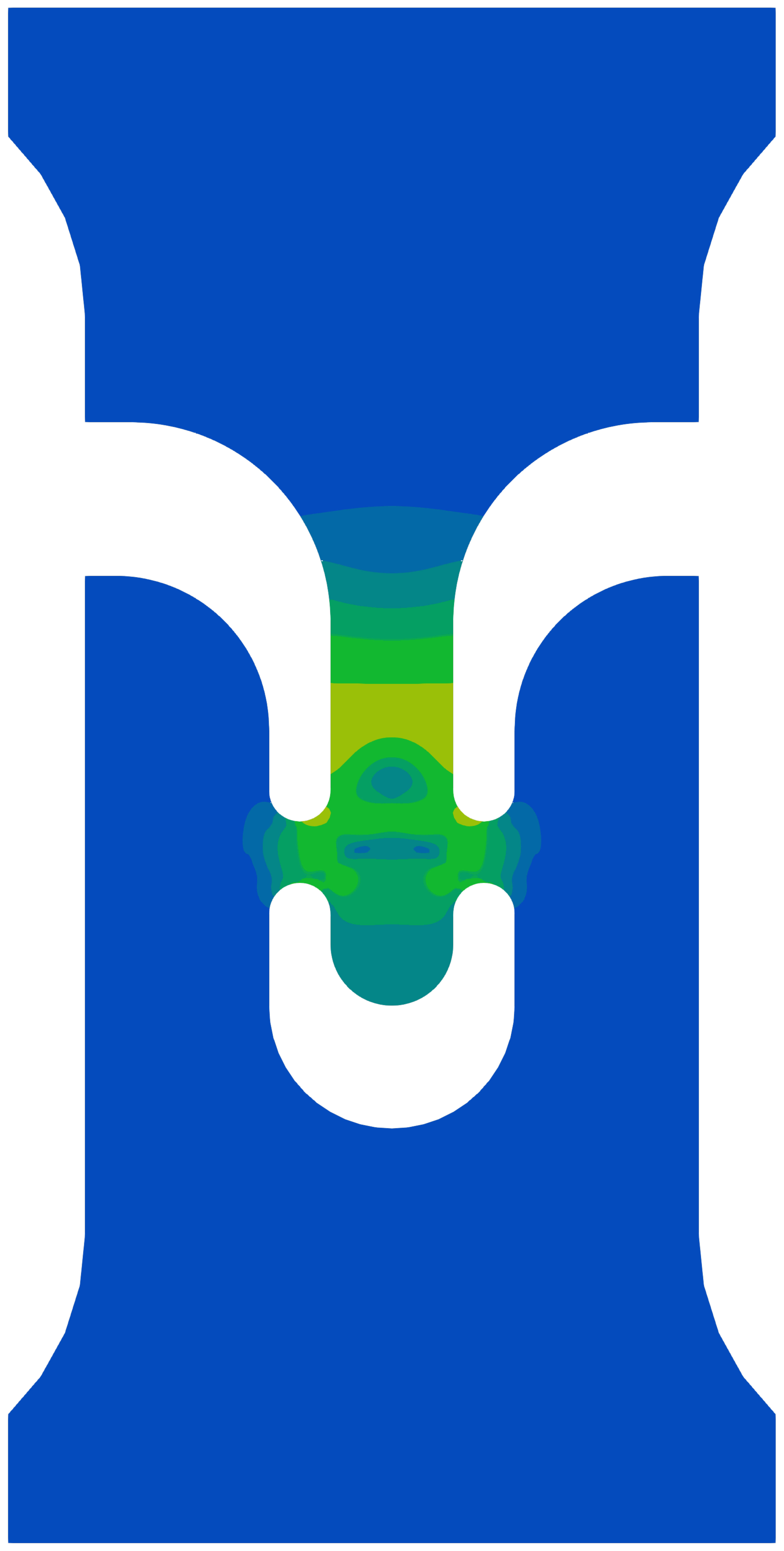}
  \end{subfigure}
  \begin{subfigure}{.22\textwidth} 
    \centering 
    \includegraphics[width=\textwidth]{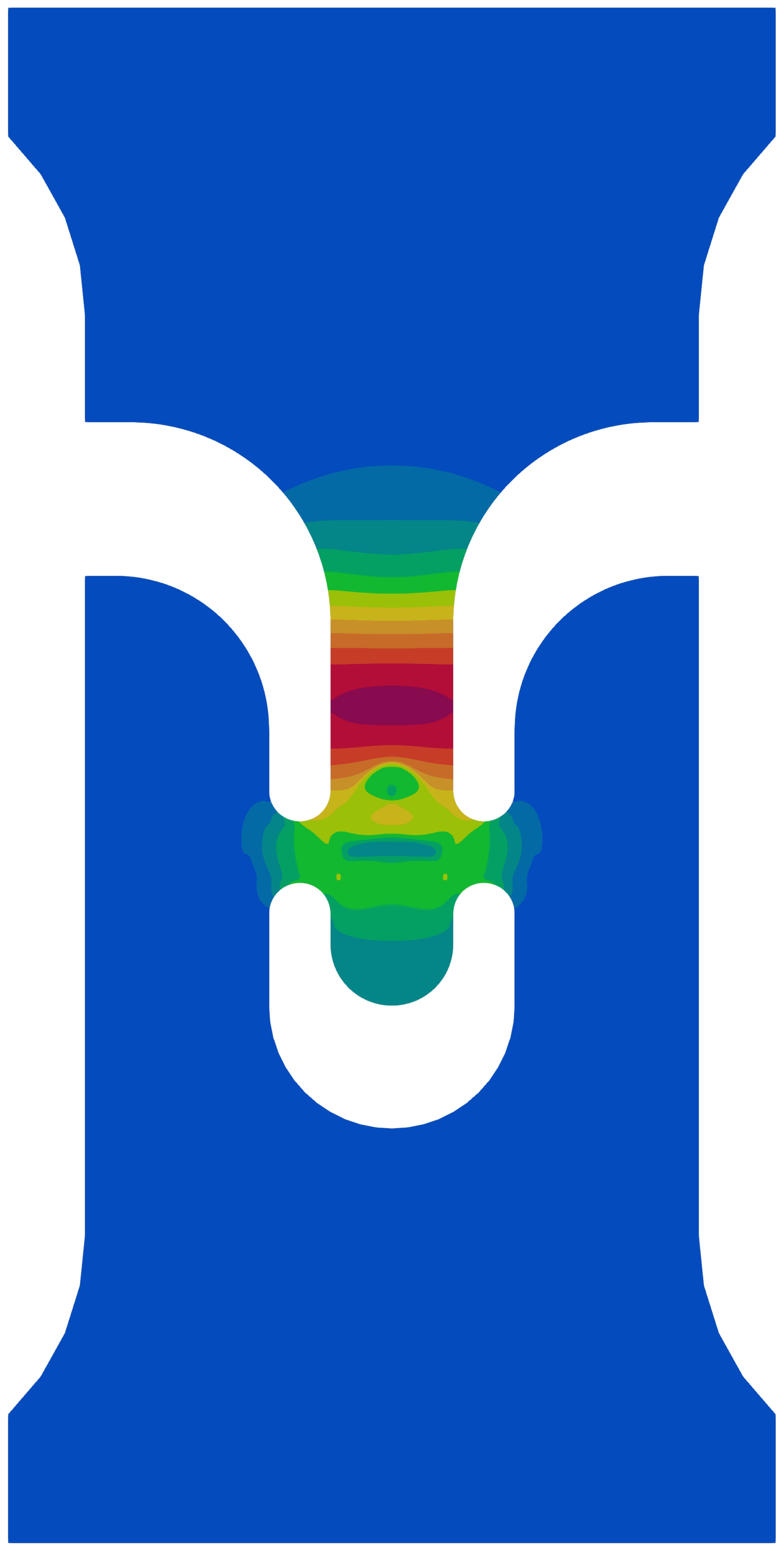}
  \end{subfigure}
  \begin{subfigure}{.22\textwidth} 
    \centering 
    \includegraphics[width=\textwidth]{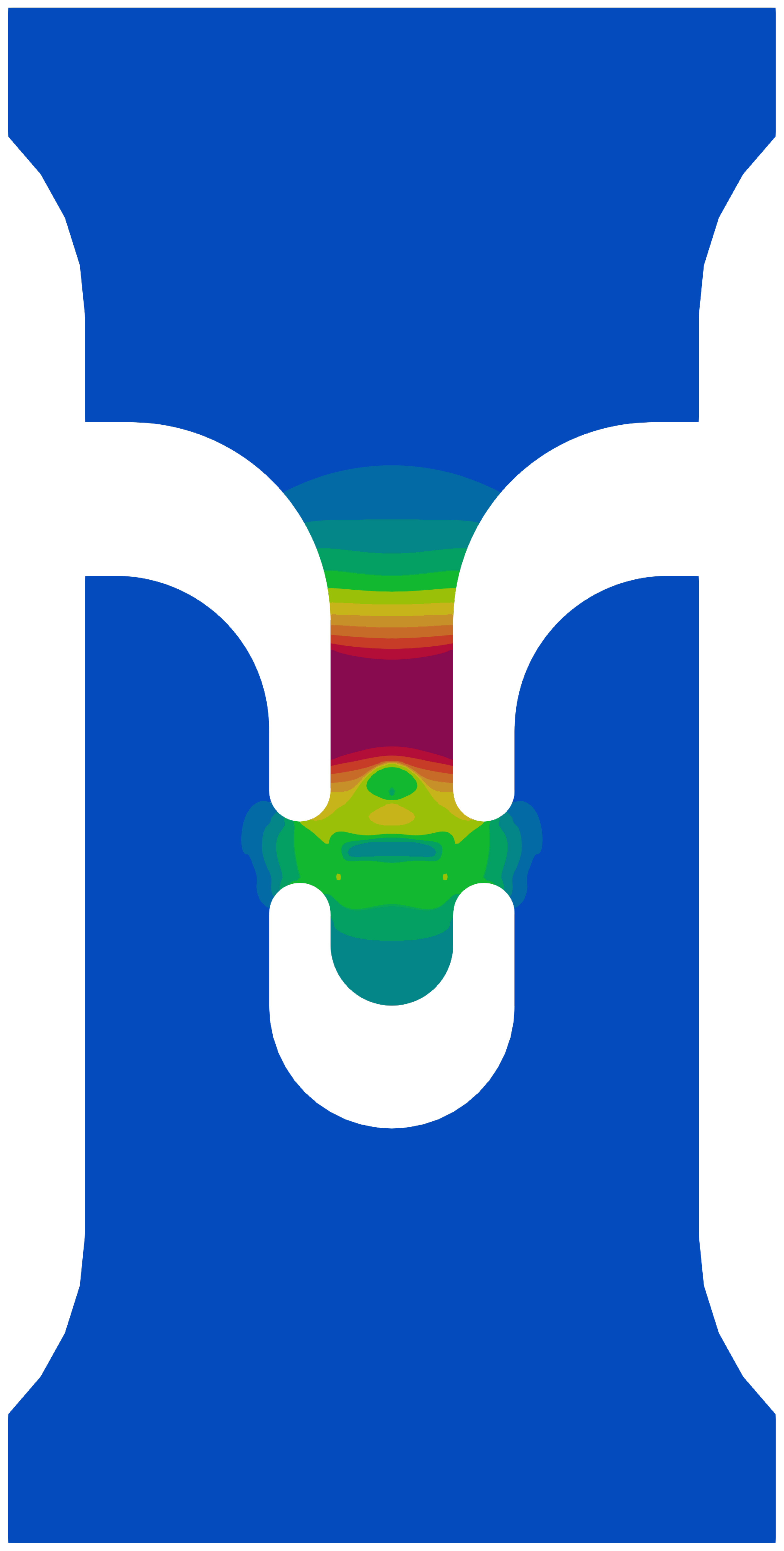}
  \end{subfigure}
  \begin{subfigure}{.08\textwidth} 
    \centering 
    \begin{tikzpicture}
      \node[inner sep=0pt] (pic) at (0,0) {\includegraphics[height=40mm, width=5mm]
      {02_Figures/03_Contour/00_Color_Maps/Damage_Step_Vertical.pdf}};
      \node[inner sep=0pt] (0)   at ($(pic.south)+( 0.50, 0.15)$)  {$0$};
      \node[inner sep=0pt] (1)   at ($(pic.south)+( 0.50, 3.80)$)  {$1$};
      \node[inner sep=0pt] (d)   at ($(pic.south)+( 0.00, 4.35)$)  {$D_{xx}~\si{[-]}$};
    \end{tikzpicture} 
  \end{subfigure}

  \vspace{1mm}

  \begin{subfigure}{.22\textwidth} 
    \centering 
    \includegraphics[width=\textwidth]{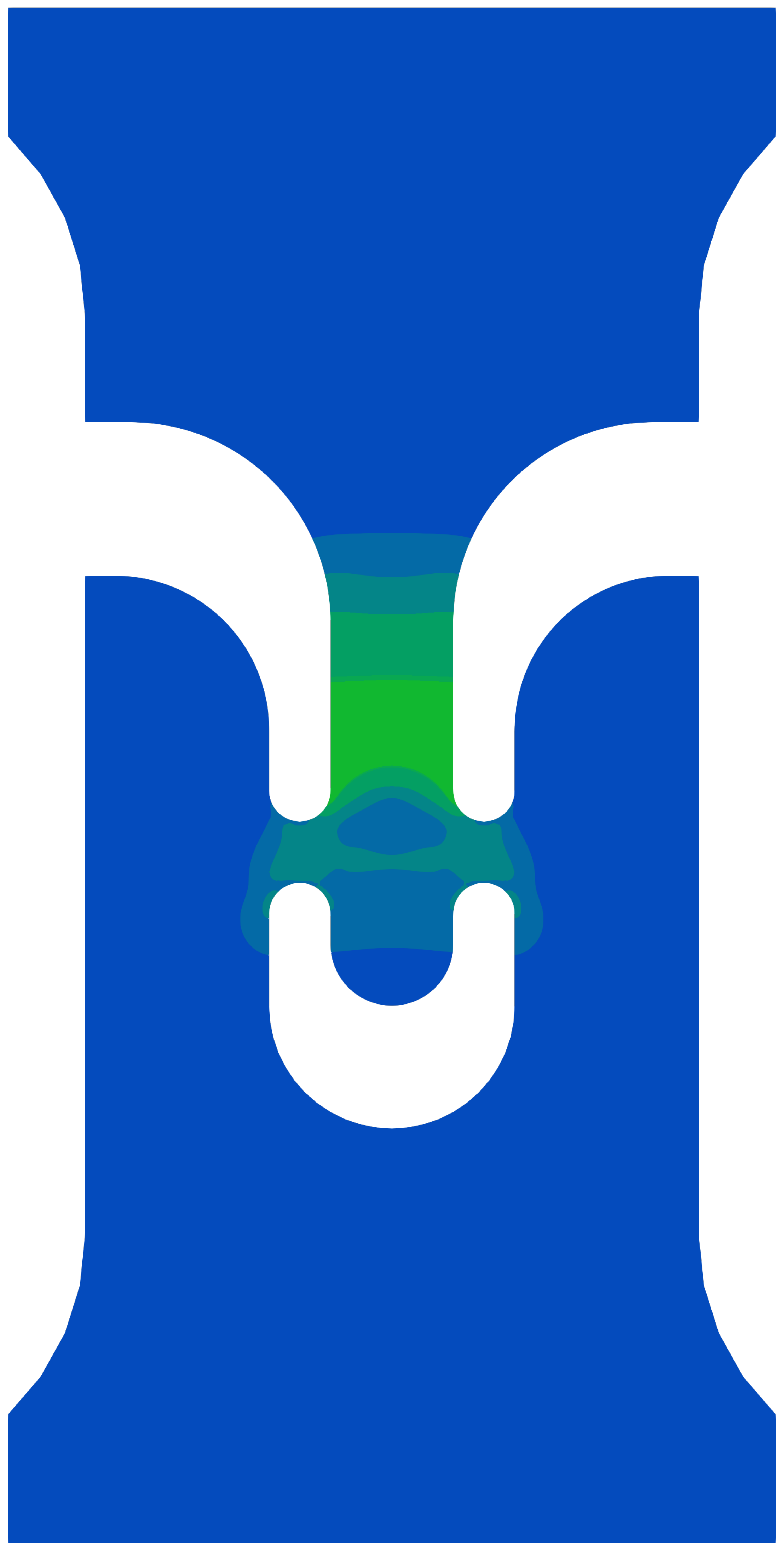}
  \end{subfigure}
  \begin{subfigure}{.22\textwidth} 
    \centering 
    \includegraphics[width=\textwidth]{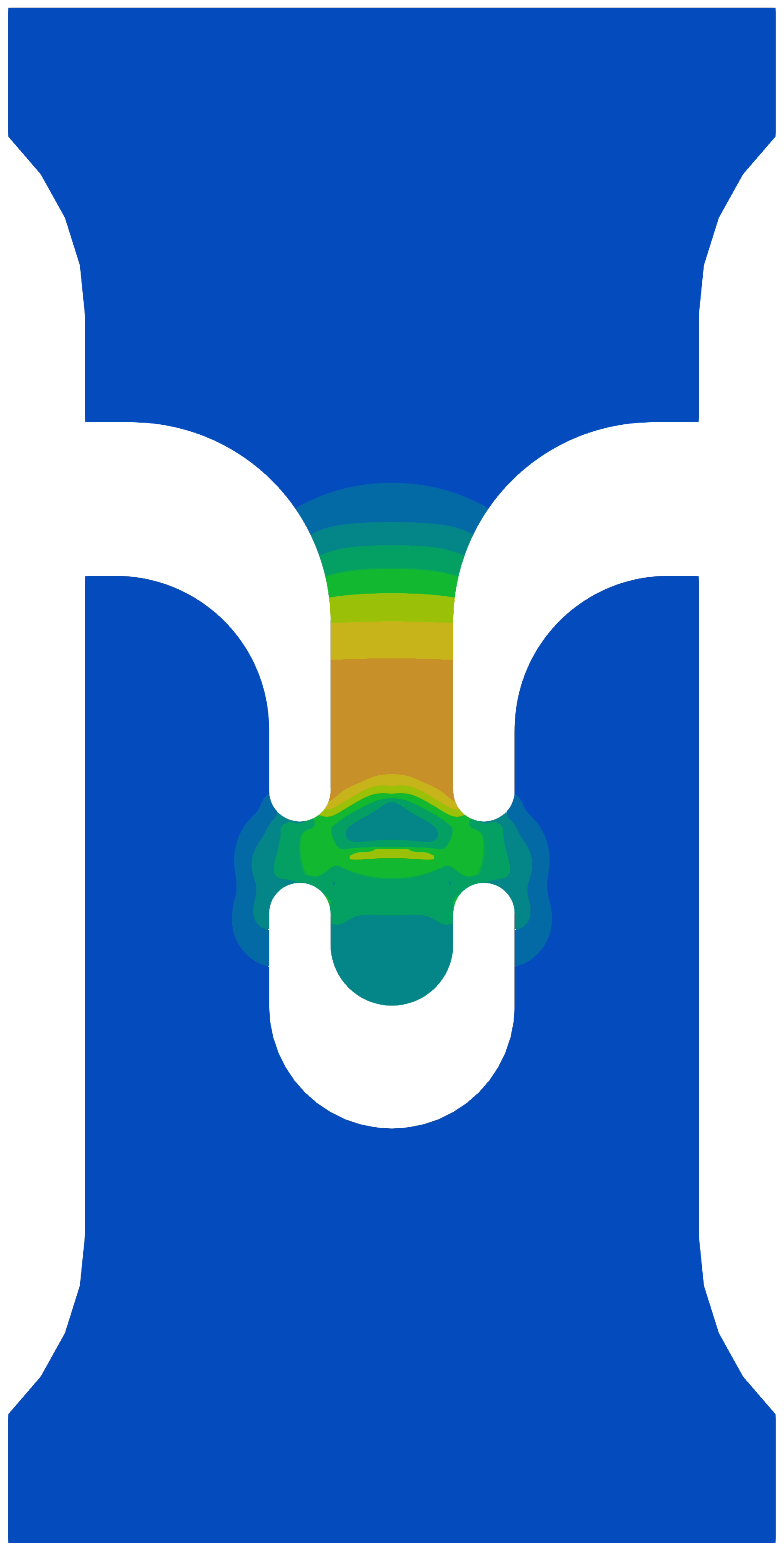}
  \end{subfigure}
  \begin{subfigure}{.22\textwidth} 
    \centering 
    \includegraphics[width=\textwidth]{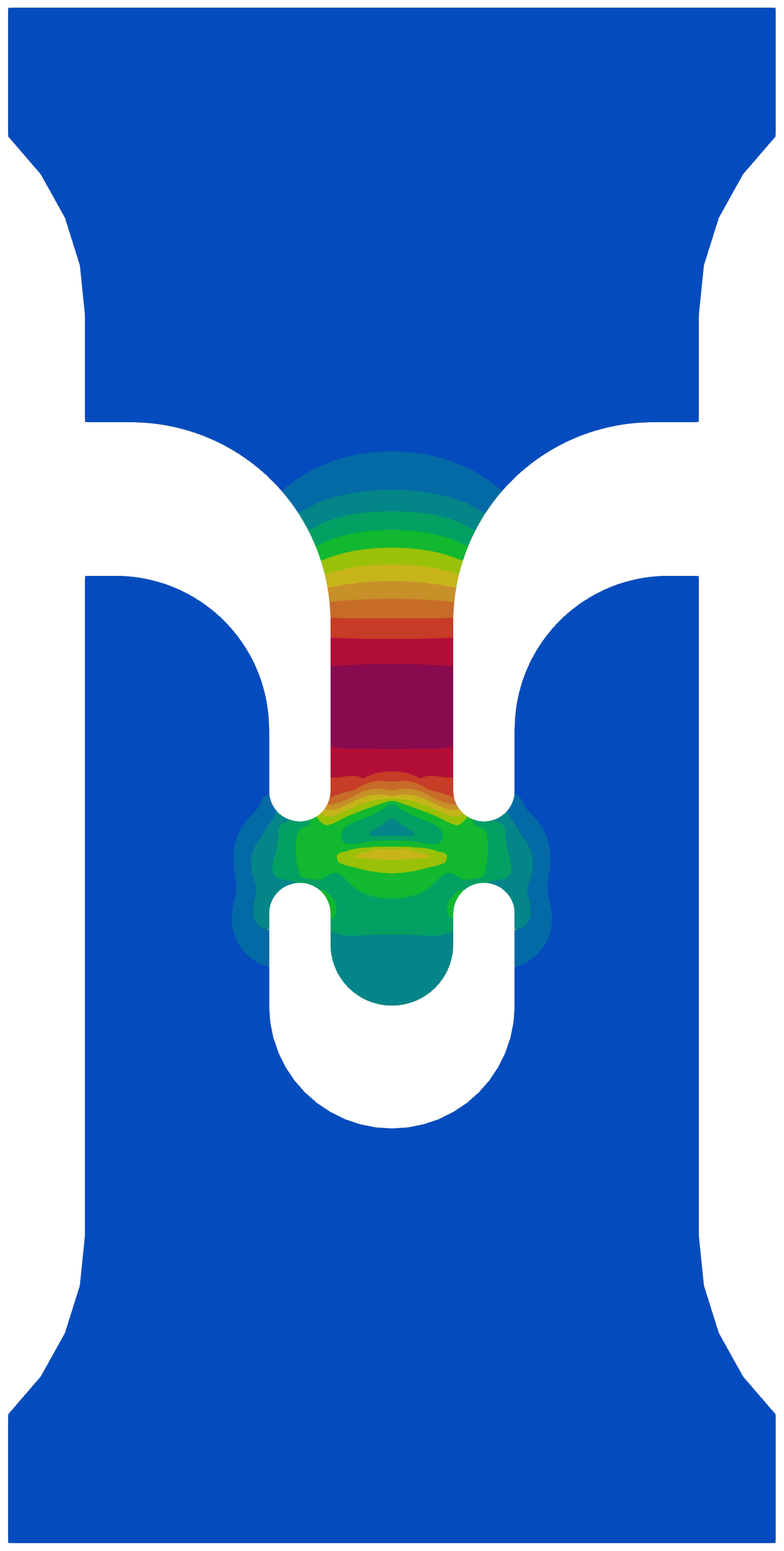}
  \end{subfigure}
  \begin{subfigure}{.22\textwidth} 
    \centering 
    \includegraphics[width=\textwidth]{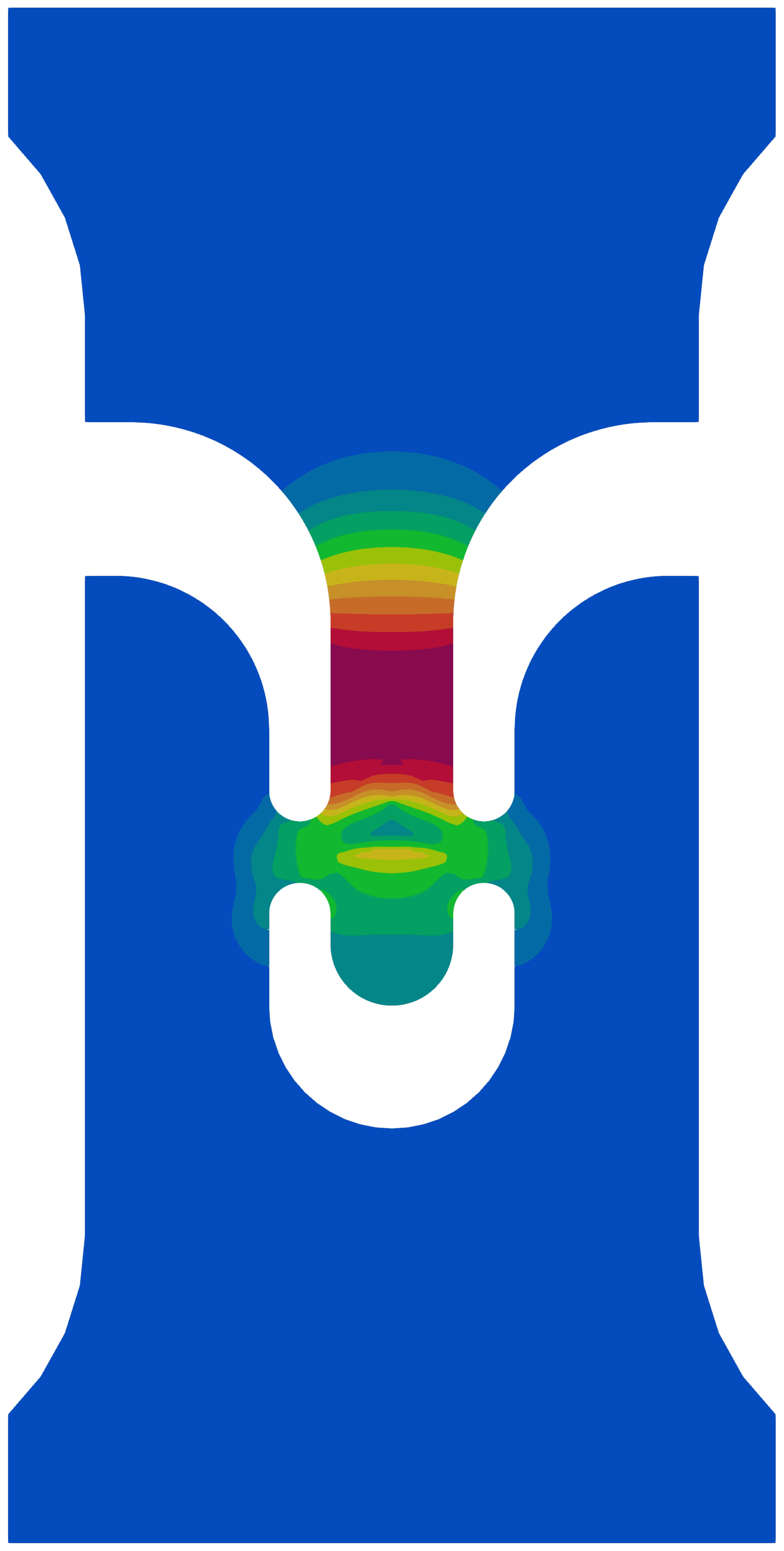}
  \end{subfigure}
  \begin{subfigure}{.08\textwidth} 
    \centering 
    \begin{tikzpicture}
      \node[inner sep=0pt] (pic) at (0,0) {\includegraphics[height=40mm, width=5mm]
      {02_Figures/03_Contour/00_Color_Maps/Damage_Step_Vertical.pdf}};
      \node[inner sep=0pt] (0)   at ($(pic.south)+( 0.50, 0.15)$)  {$0$};
      \node[inner sep=0pt] (1)   at ($(pic.south)+( 0.50, 3.80)$)  {$1$};
      \node[inner sep=0pt] (d)   at ($(pic.south)+( 0.00, 4.35)$)  {$D_{yy}~\si{[-]}$};
    \end{tikzpicture} 
  \end{subfigure}

  \vspace{1mm}

  \begin{subfigure}{.22\textwidth} 
    \centering 
    \includegraphics[width=\textwidth]{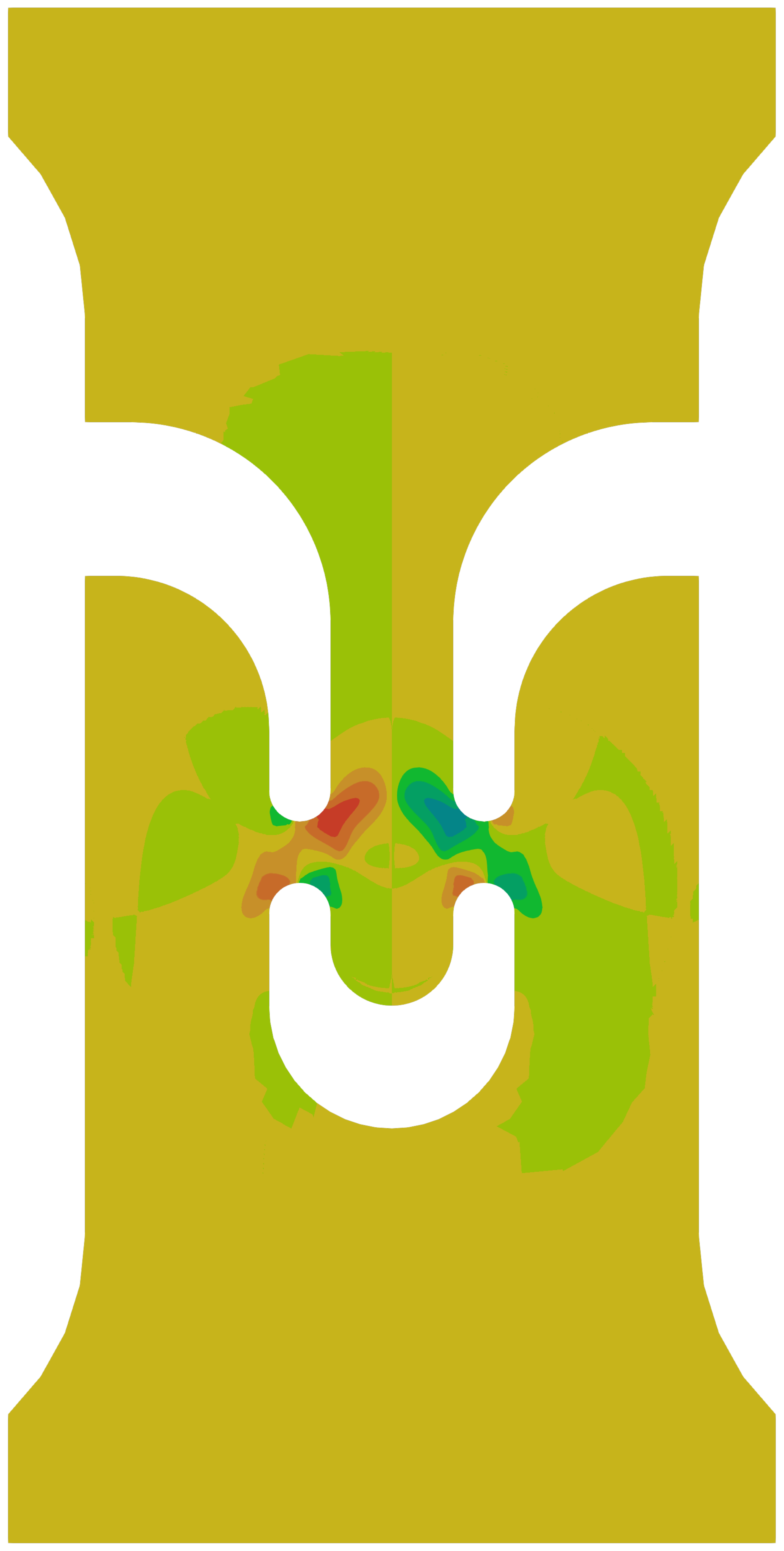}
    \caption{}
  \end{subfigure}
  \begin{subfigure}{.22\textwidth} 
    \centering 
    \includegraphics[width=\textwidth]{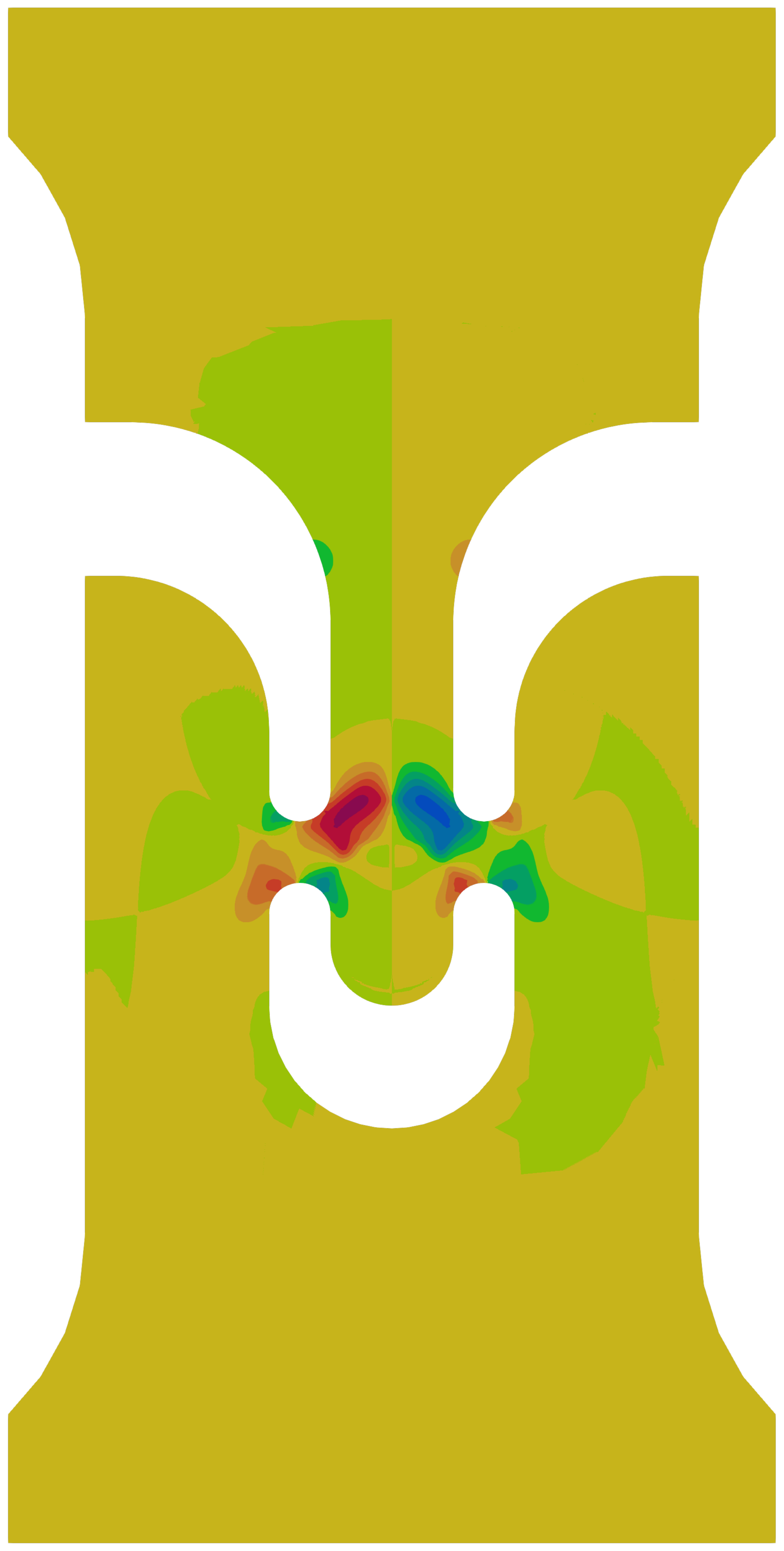}
    \caption{}
  \end{subfigure}
  \begin{subfigure}{.22\textwidth} 
    \centering 
    \includegraphics[width=\textwidth]{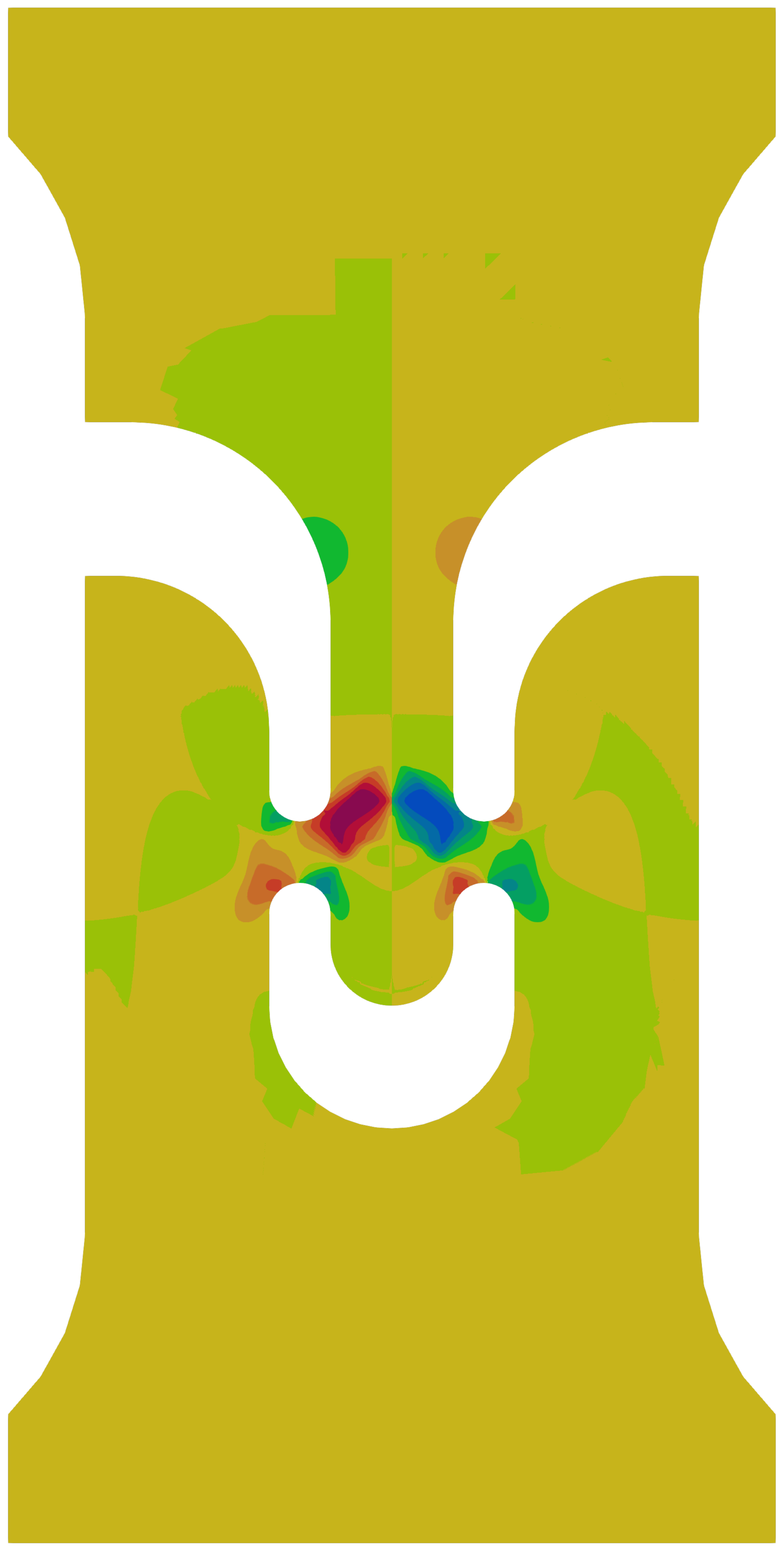}
    \caption{}
  \end{subfigure}
  \begin{subfigure}{.22\textwidth} 
    \centering 
    \includegraphics[width=\textwidth]{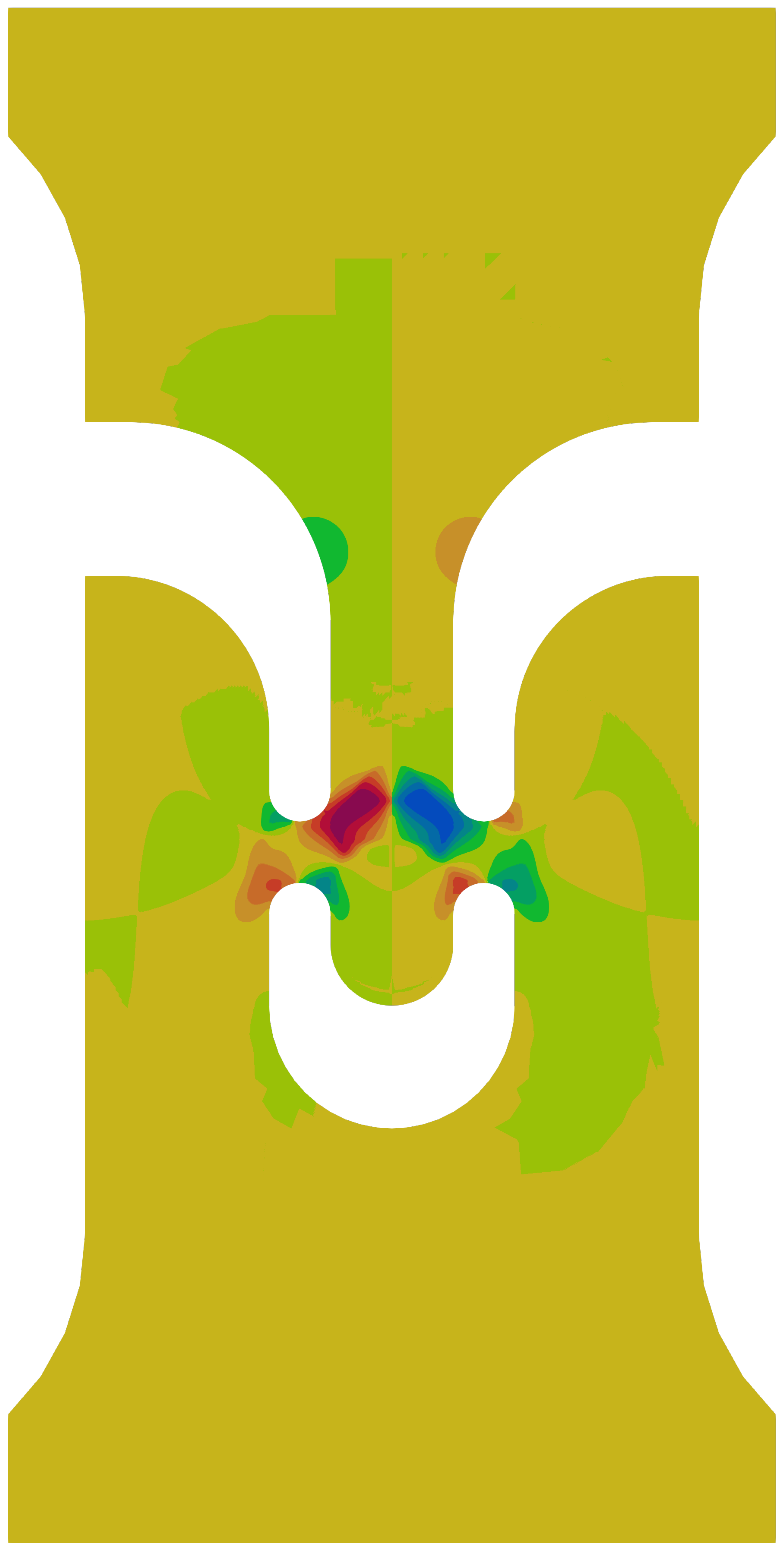}
    \caption{}
  \end{subfigure}
  \begin{subfigure}{.08\textwidth} 
    \centering 
    \begin{tikzpicture}
      \node[inner sep=0pt] (pic) at (0,0) {\includegraphics[height=40mm, width=5mm]
      {02_Figures/03_Contour/00_Color_Maps/Damage_Step_Vertical.pdf}};
      \node[inner sep=0pt] (0)   at ($(pic.south)+( 1.00, 0.15)$)  {$-0.2002$};
      \node[inner sep=0pt] (1)   at ($(pic.south)+( 1.00, 3.80)$)  {$+0.2002$};
      \node[inner sep=0pt] (d)   at ($(pic.south)+( 0.00, 4.35)$)  {$D_{xy}~\si{[-]}$};
    \end{tikzpicture} 
    \hphantom{(d)}
  \end{subfigure}

  \caption{Contour plots of the evolution of the normal and shear components of the damage tensor for the smiley specimen (model~C).}
  \label{fig:ExssevolutionDten}     
\end{figure}

For the smiley specimen, we also study the evolution of the components of the damage tensor in Fig.~\ref{fig:ExssevolutionDten}, where we restrict ourselves to the presentation of model~C. In the initial damage state, the normal component $D_{xx}$ evolves equally at the tension and shear load carrying cross sections. In the intermediate damage states, the evolution of $D_{xx}$ concentrates in the normal load carrying cross section up to total failure. The evolution of the normal component $D_{yy}$ occurs predominantly in the normal load carrying cross section during the entire loading. Finally, the evolution of the shear component $D_{xy}$ primarily happens at the inner side of the shear load carrying cross section.

\begin{figure}
  \centering 

  \begin{subfigure}{.22\textwidth} 
    \centering 
    \includegraphics[width=\textwidth]{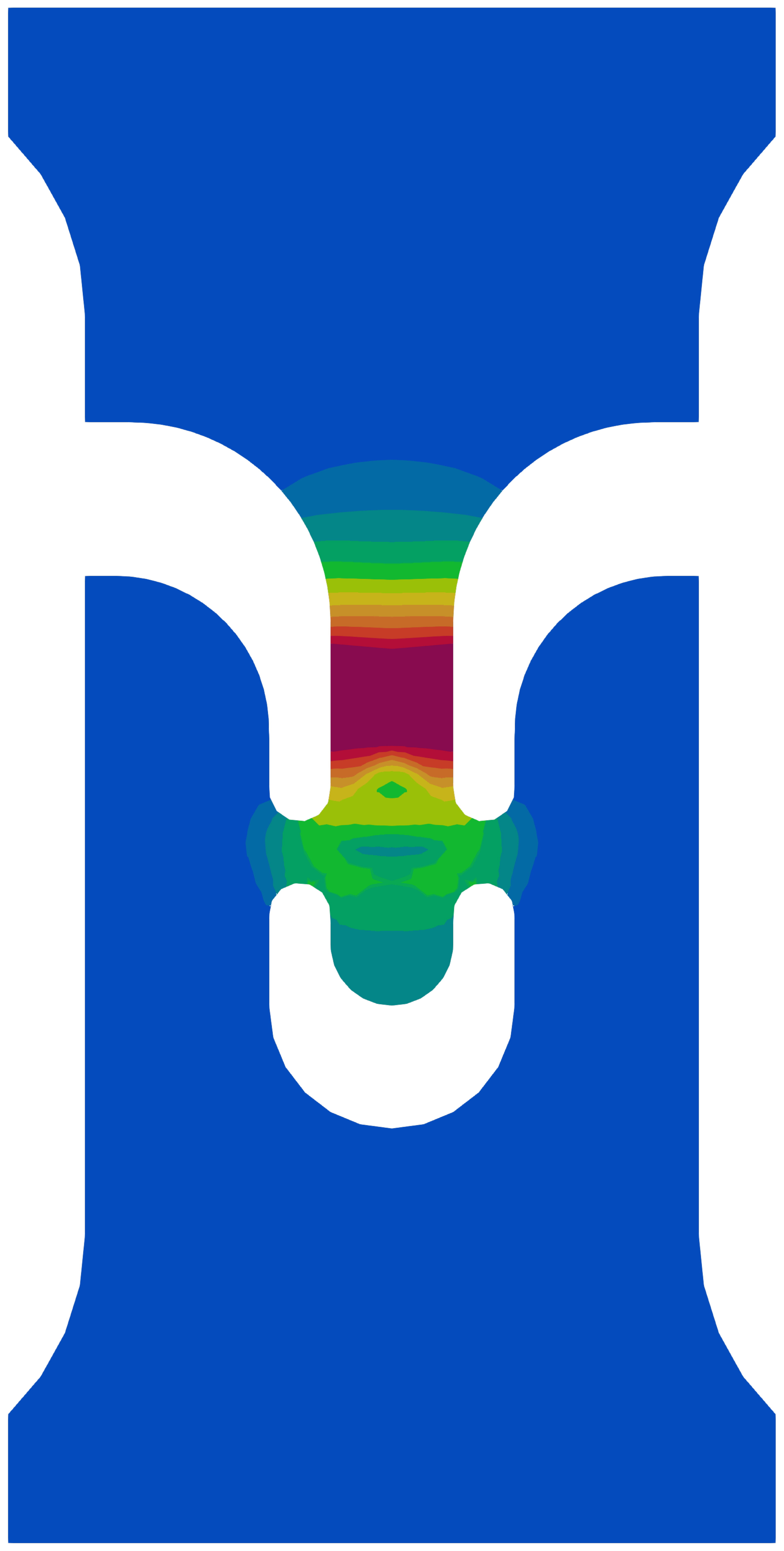}
  \end{subfigure}
  \begin{subfigure}{.22\textwidth} 
    \centering 
    \includegraphics[width=\textwidth]{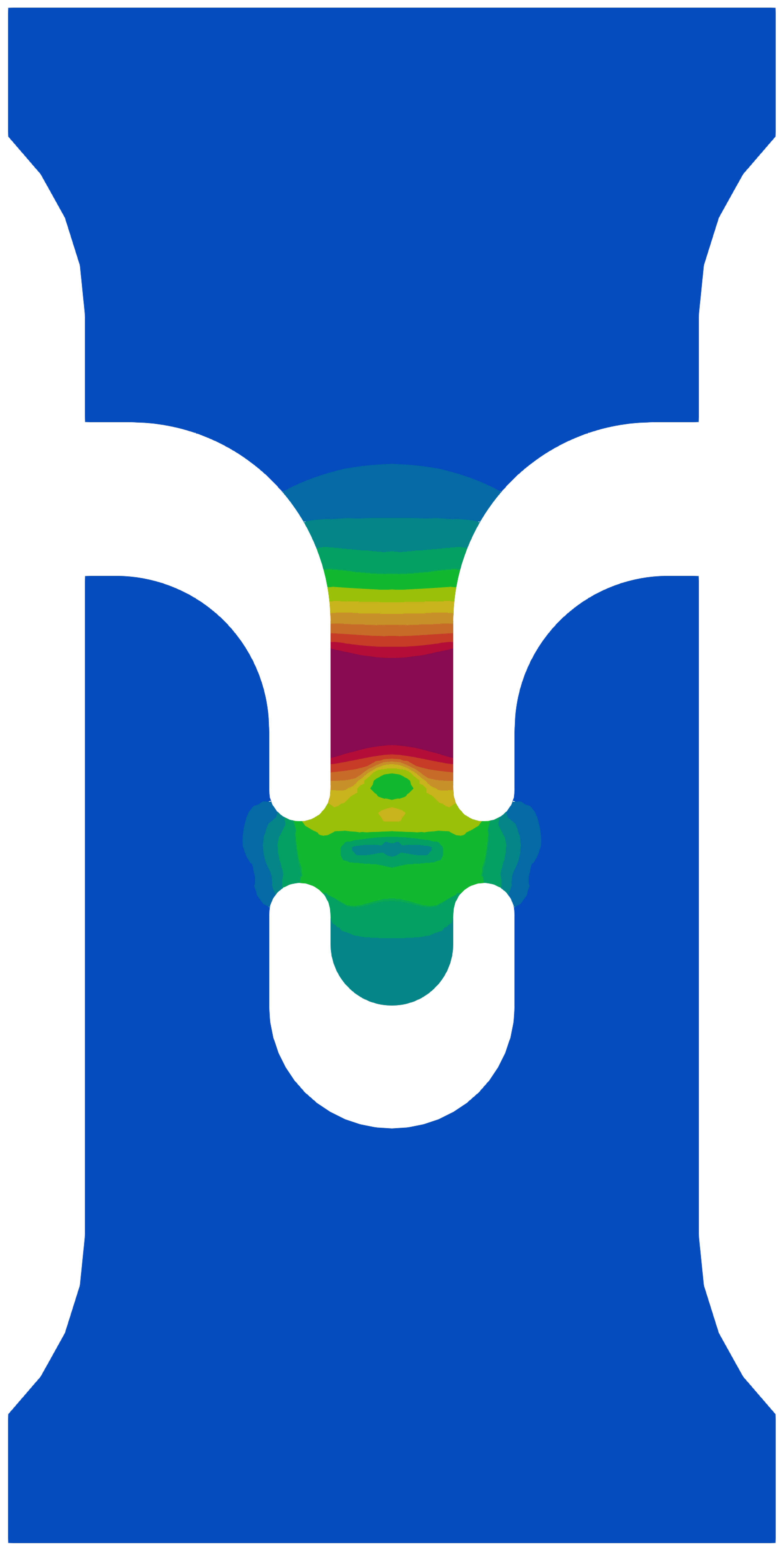}
  \end{subfigure}
  \begin{subfigure}{.22\textwidth} 
    \centering 
    \includegraphics[width=\textwidth]{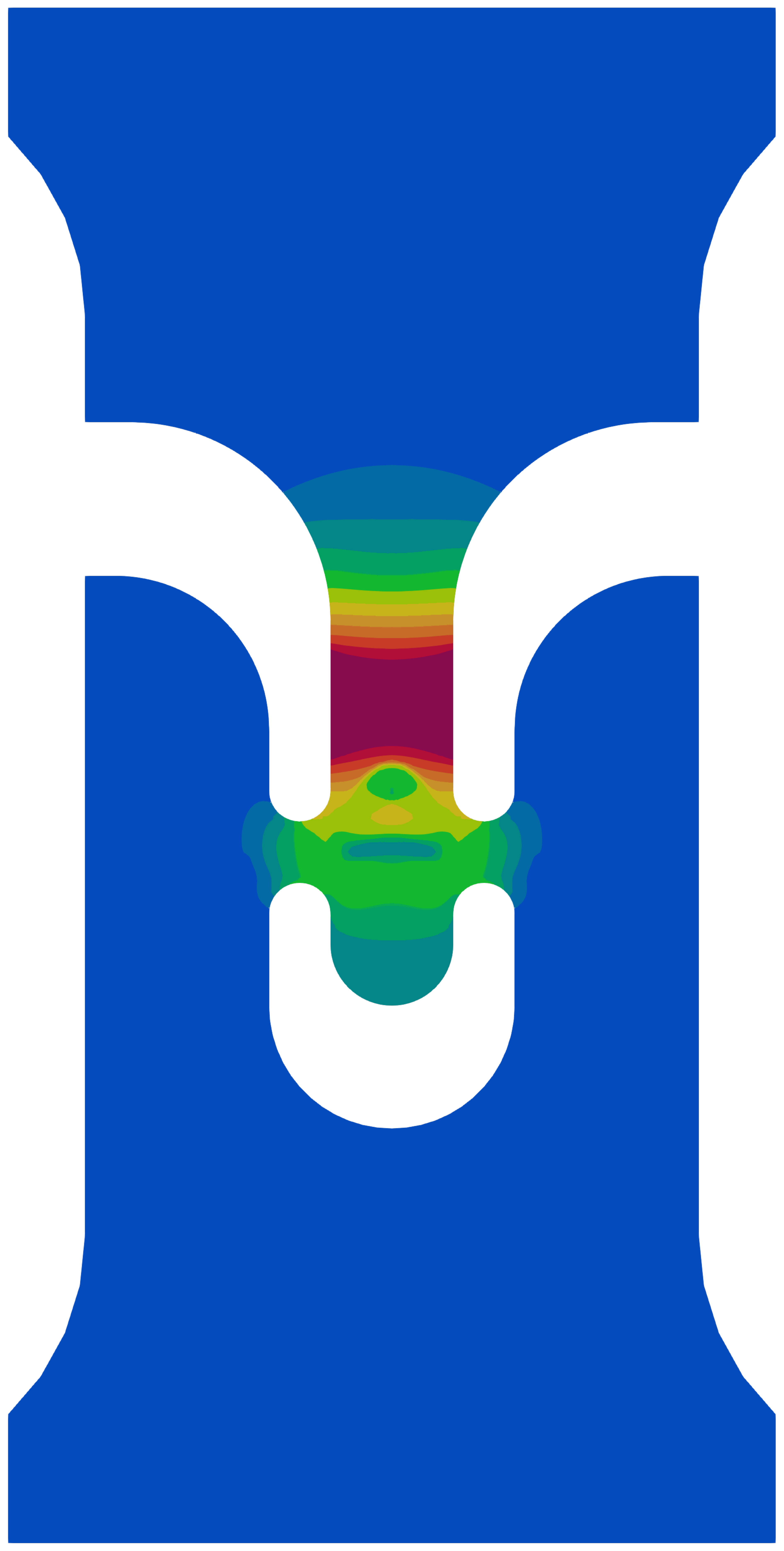}
  \end{subfigure}
  \begin{subfigure}{.22\textwidth} 
    \centering 
    \includegraphics[width=\textwidth]{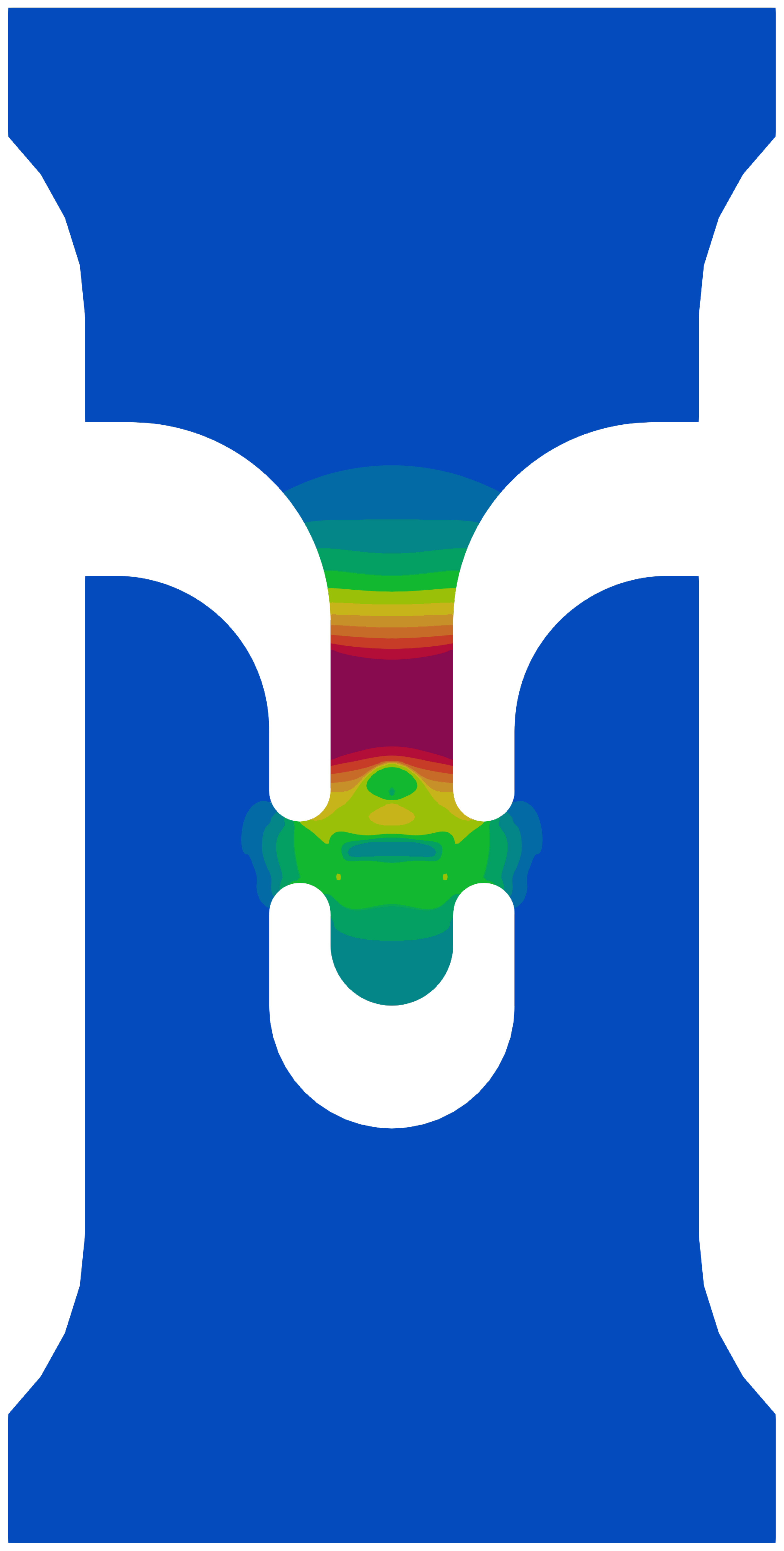}
  \end{subfigure}
  \begin{subfigure}{.08\textwidth} 
    \centering 
    \begin{tikzpicture}
      \node[inner sep=0pt] (pic) at (0,0) {\includegraphics[height=40mm, width=5mm]
      {02_Figures/03_Contour/00_Color_Maps/Damage_Step_Vertical.pdf}};
      \node[inner sep=0pt] (0)   at ($(pic.south)+( 0.50, 0.15)$)  {$0$};
      \node[inner sep=0pt] (1)   at ($(pic.south)+( 0.50, 3.80)$)  {$1$};
      \node[inner sep=0pt] (d)   at ($(pic.south)+( 0.00, 4.35)$)  {$D_{xx}~\si{[-]}$};
    \end{tikzpicture} 
  \end{subfigure}

  \vspace{1mm}

  \begin{subfigure}{.22\textwidth} 
    \centering 
    \includegraphics[width=\textwidth]{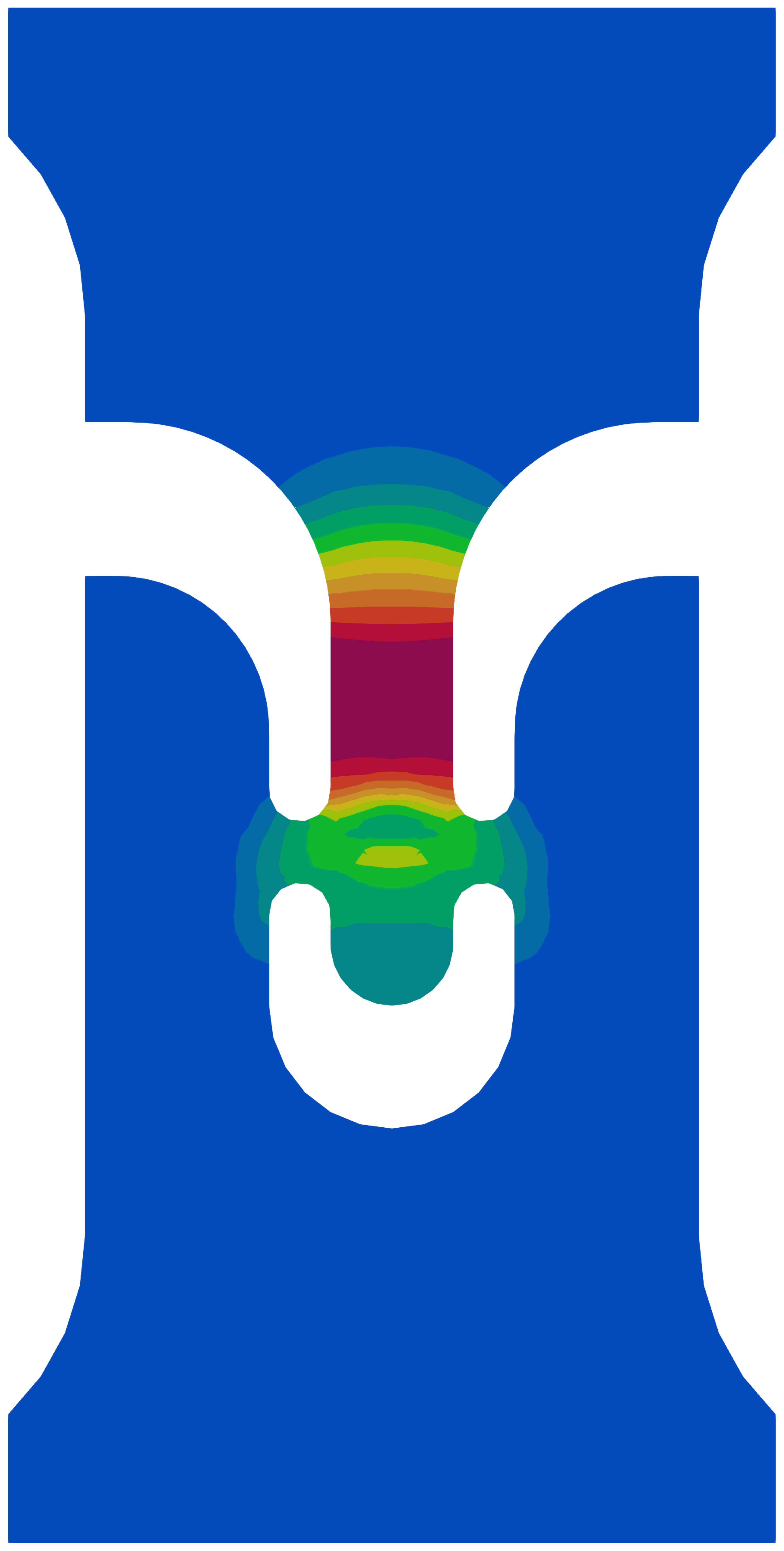}
  \end{subfigure}
  \begin{subfigure}{.22\textwidth} 
    \centering 
    \includegraphics[width=\textwidth]{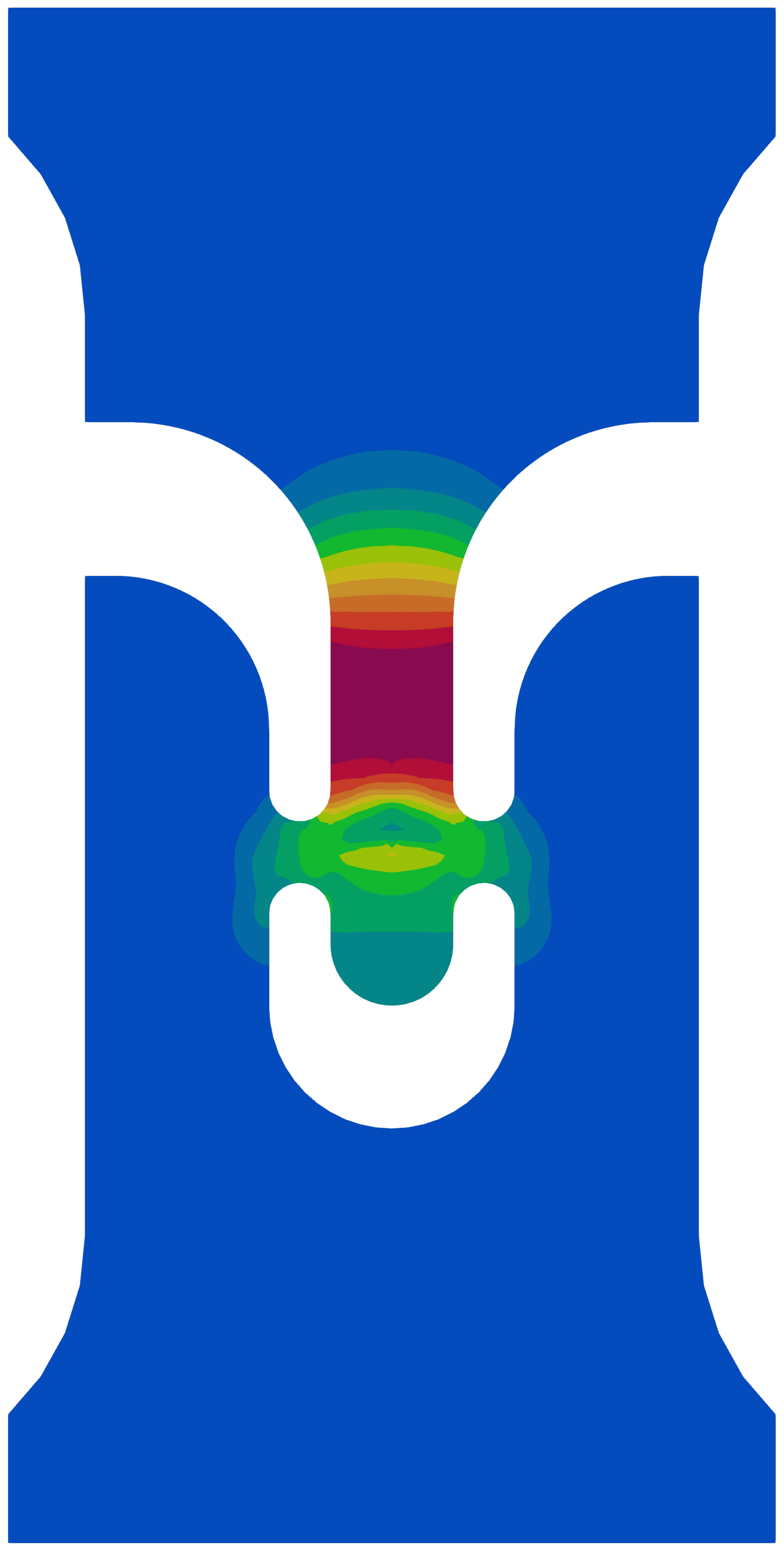}
  \end{subfigure}
  \begin{subfigure}{.22\textwidth} 
    \centering 
    \includegraphics[width=\textwidth]{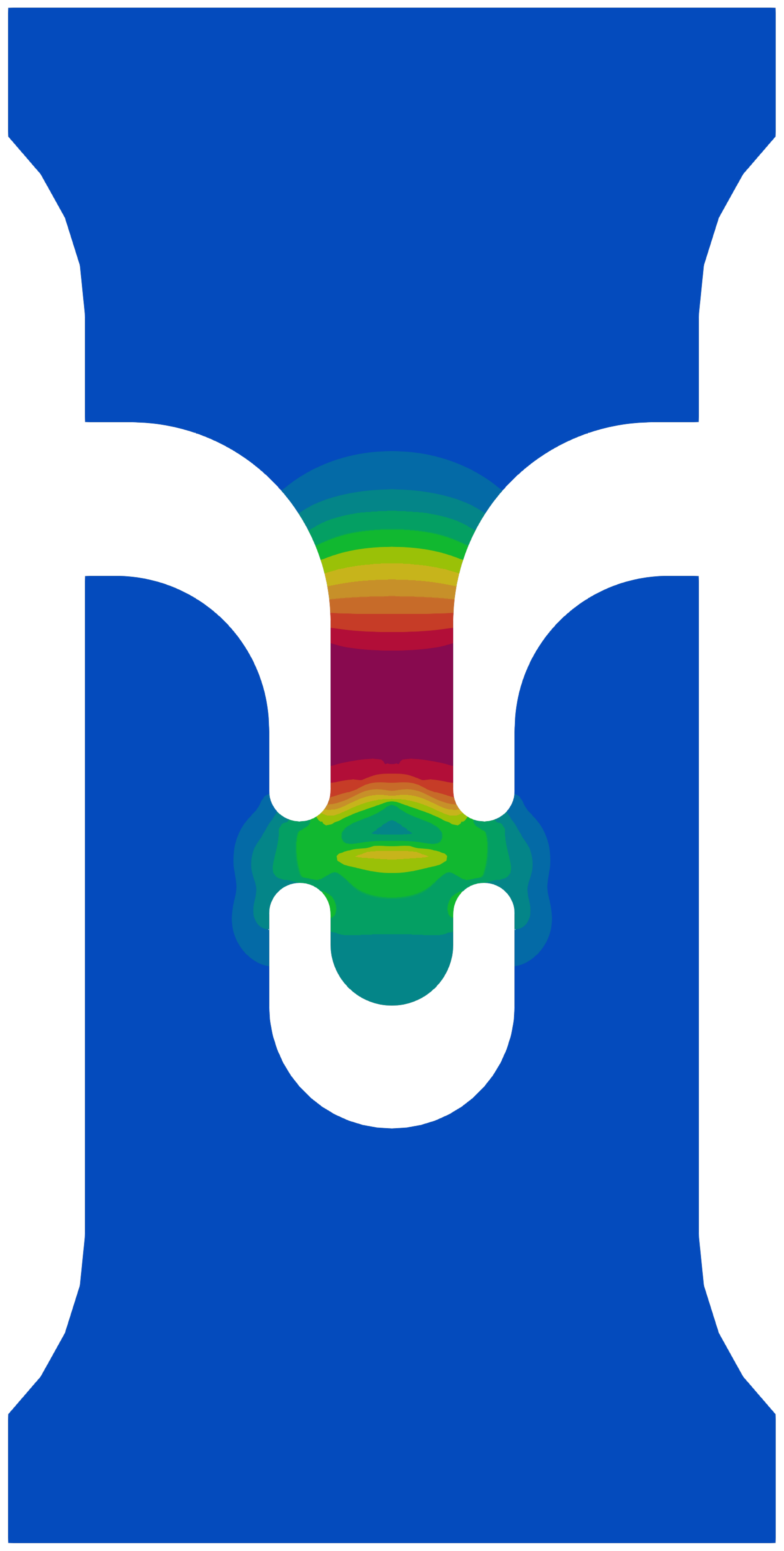}
  \end{subfigure}
  \begin{subfigure}{.22\textwidth} 
    \centering 
    \includegraphics[width=\textwidth]{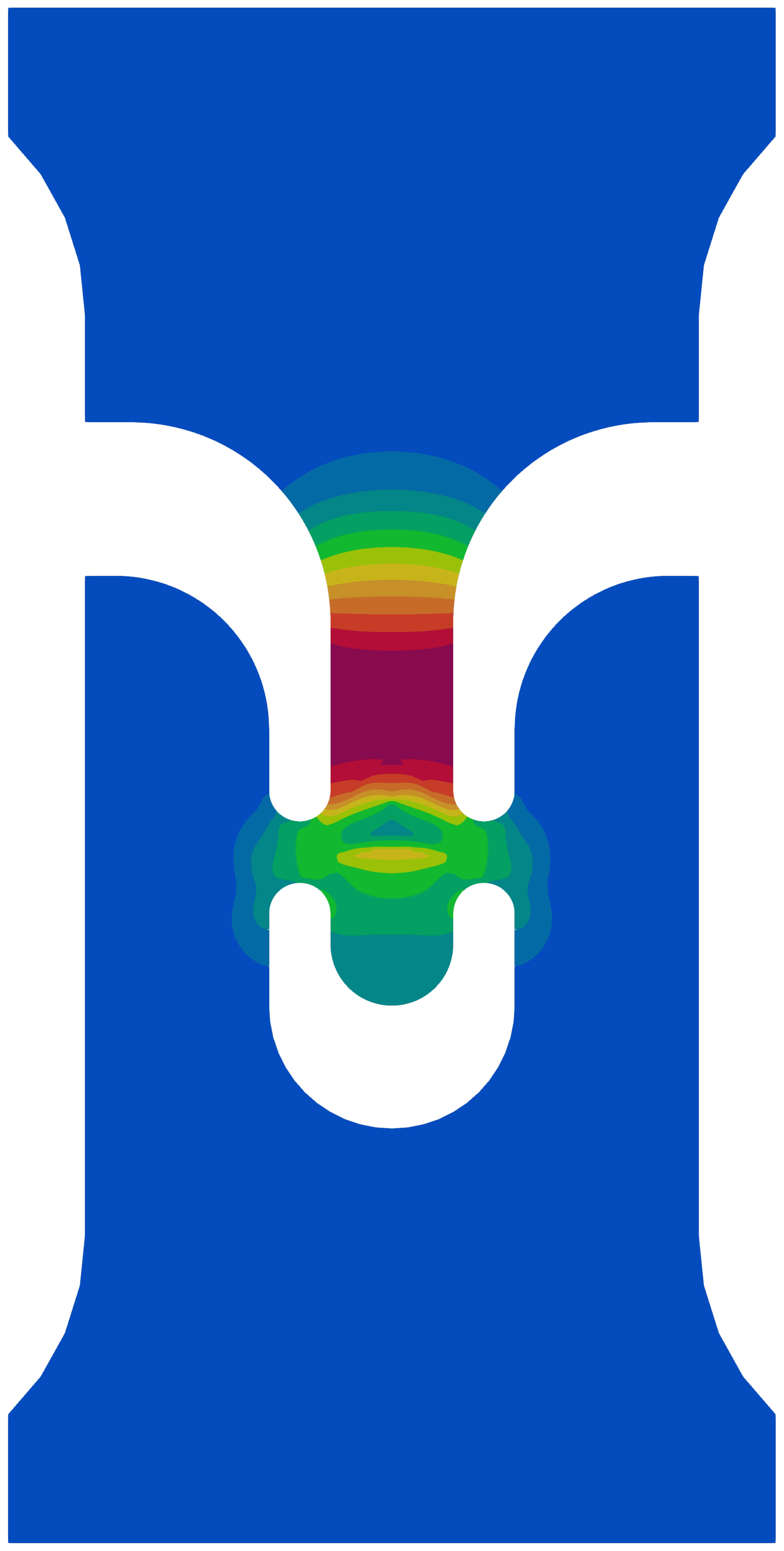}
  \end{subfigure}
  \begin{subfigure}{.08\textwidth} 
    \centering 
    \begin{tikzpicture}
      \node[inner sep=0pt] (pic) at (0,0) {\includegraphics[height=40mm, width=5mm]
      {02_Figures/03_Contour/00_Color_Maps/Damage_Step_Vertical.pdf}};
      \node[inner sep=0pt] (0)   at ($(pic.south)+( 0.50, 0.15)$)  {$0$};
      \node[inner sep=0pt] (1)   at ($(pic.south)+( 0.50, 3.80)$)  {$1$};
      \node[inner sep=0pt] (d)   at ($(pic.south)+( 0.00, 4.35)$)  {$D_{yy}~\si{[-]}$};
    \end{tikzpicture} 
  \end{subfigure}

  \vspace{1mm}

  \begin{subfigure}{.22\textwidth} 
    \centering 
    \includegraphics[width=\textwidth]{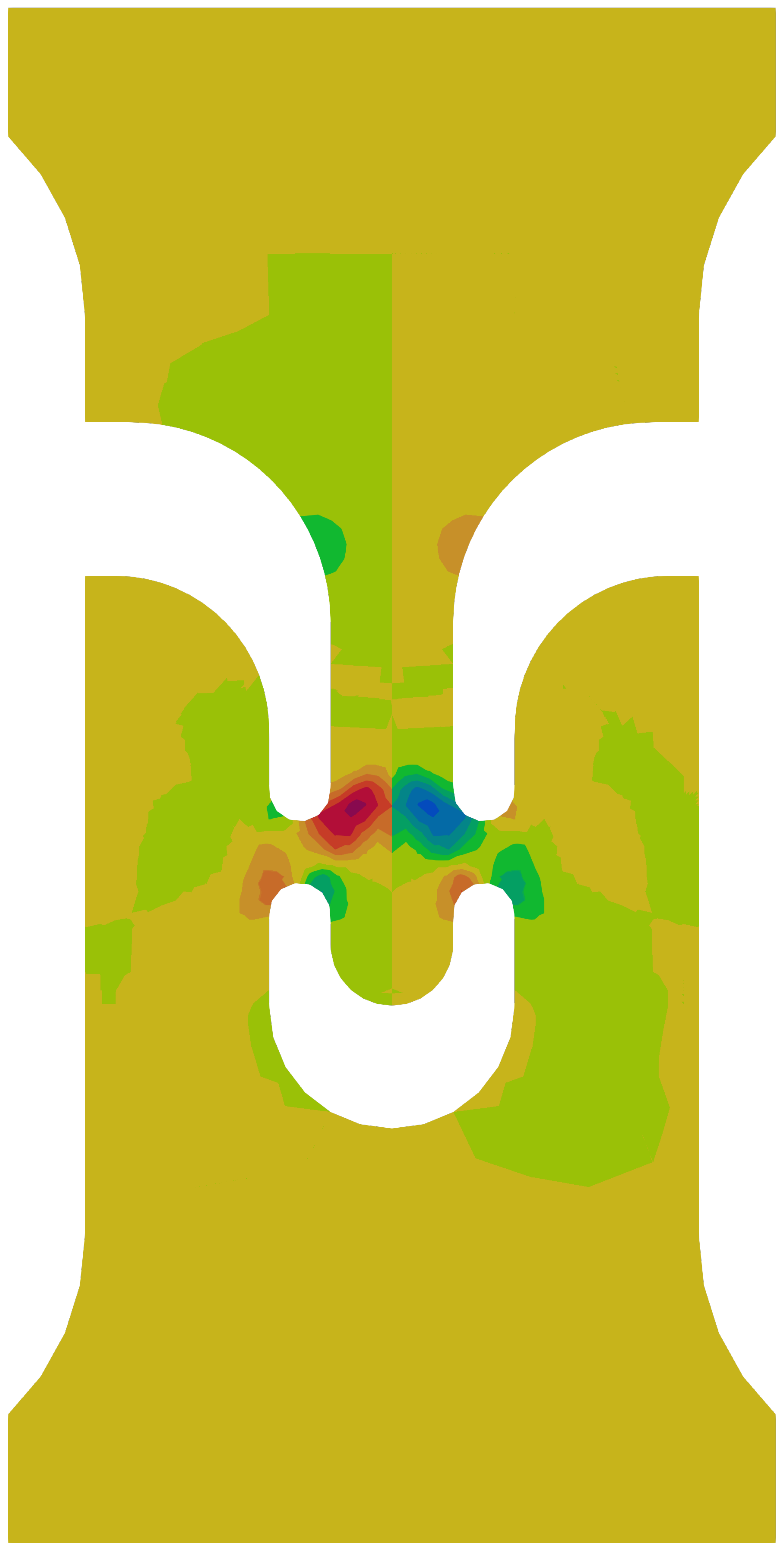}
    \caption{755}
    \label{fig:ExssmeshconvergenceDten755}     
  \end{subfigure}
  \begin{subfigure}{.22\textwidth} 
    \centering 
    \includegraphics[width=\textwidth]{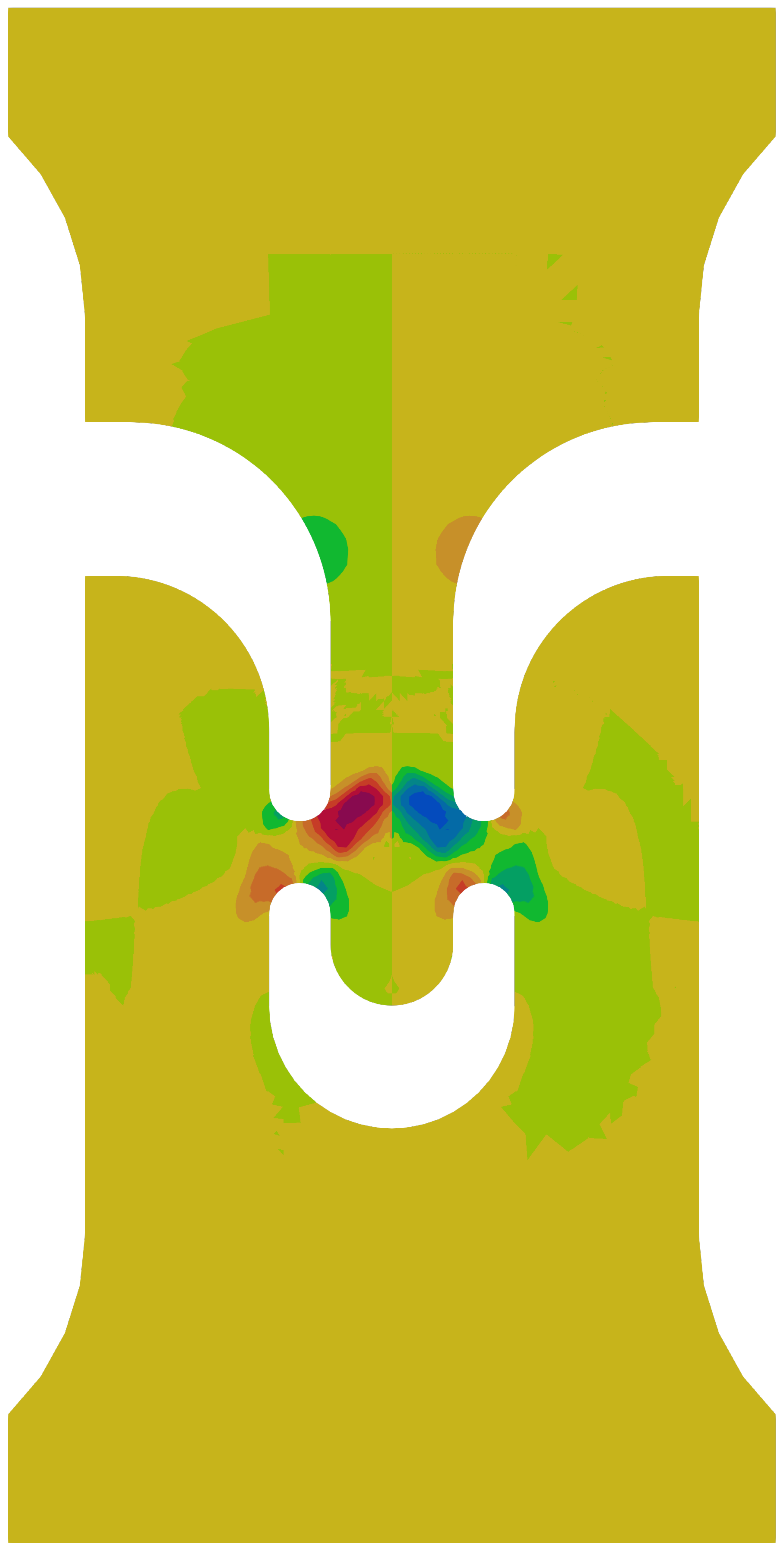}
    \caption{2649}
    \label{fig:ExssmeshconvergenceDten2649}     
  \end{subfigure}
  \begin{subfigure}{.22\textwidth} 
    \centering 
    \includegraphics[width=\textwidth]{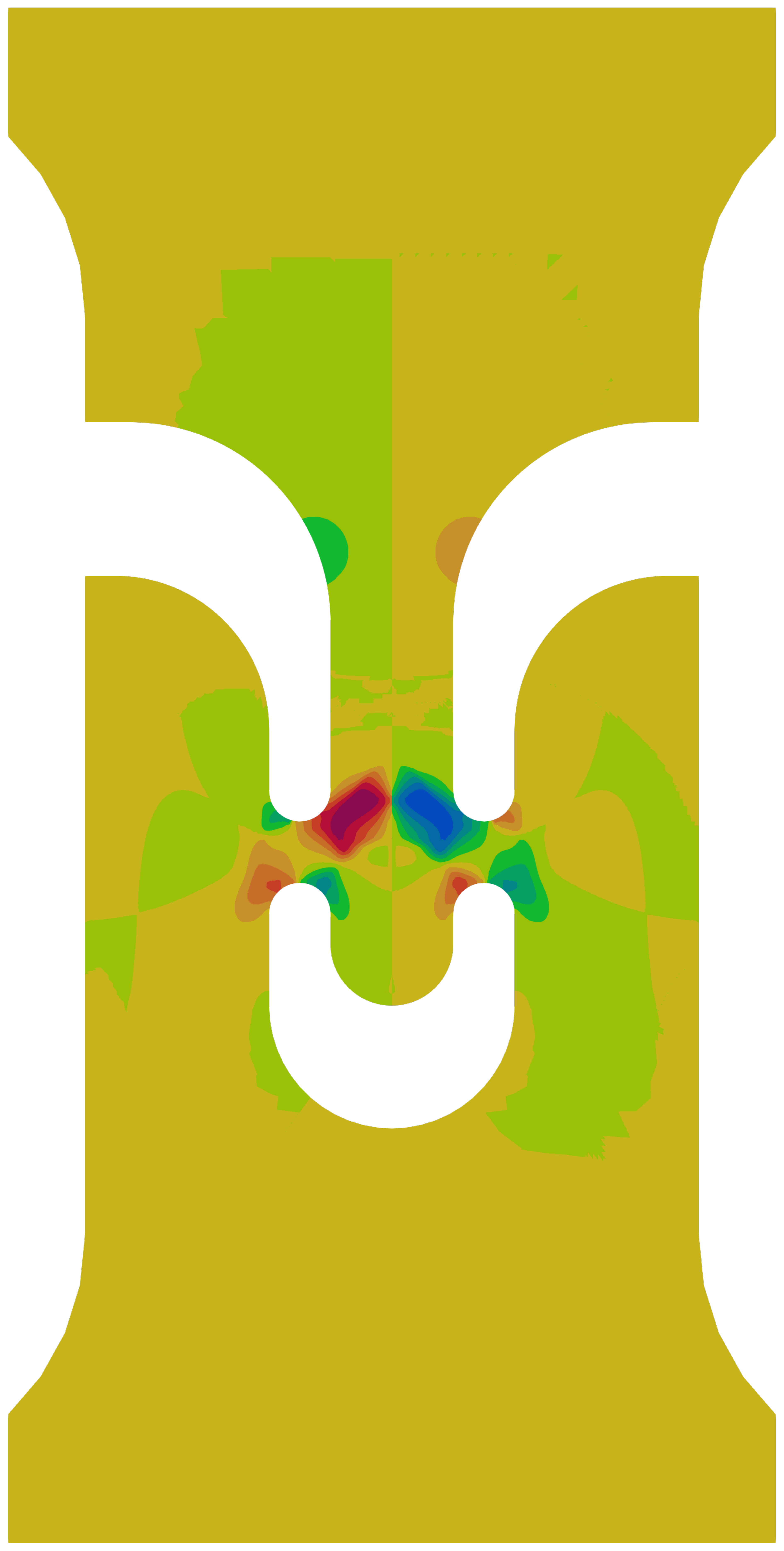}
    \caption{7681}
    \label{fig:ExssmeshconvergenceDten7681}
  \end{subfigure}
  \begin{subfigure}{.22\textwidth} 
    \centering 
    \includegraphics[width=\textwidth]{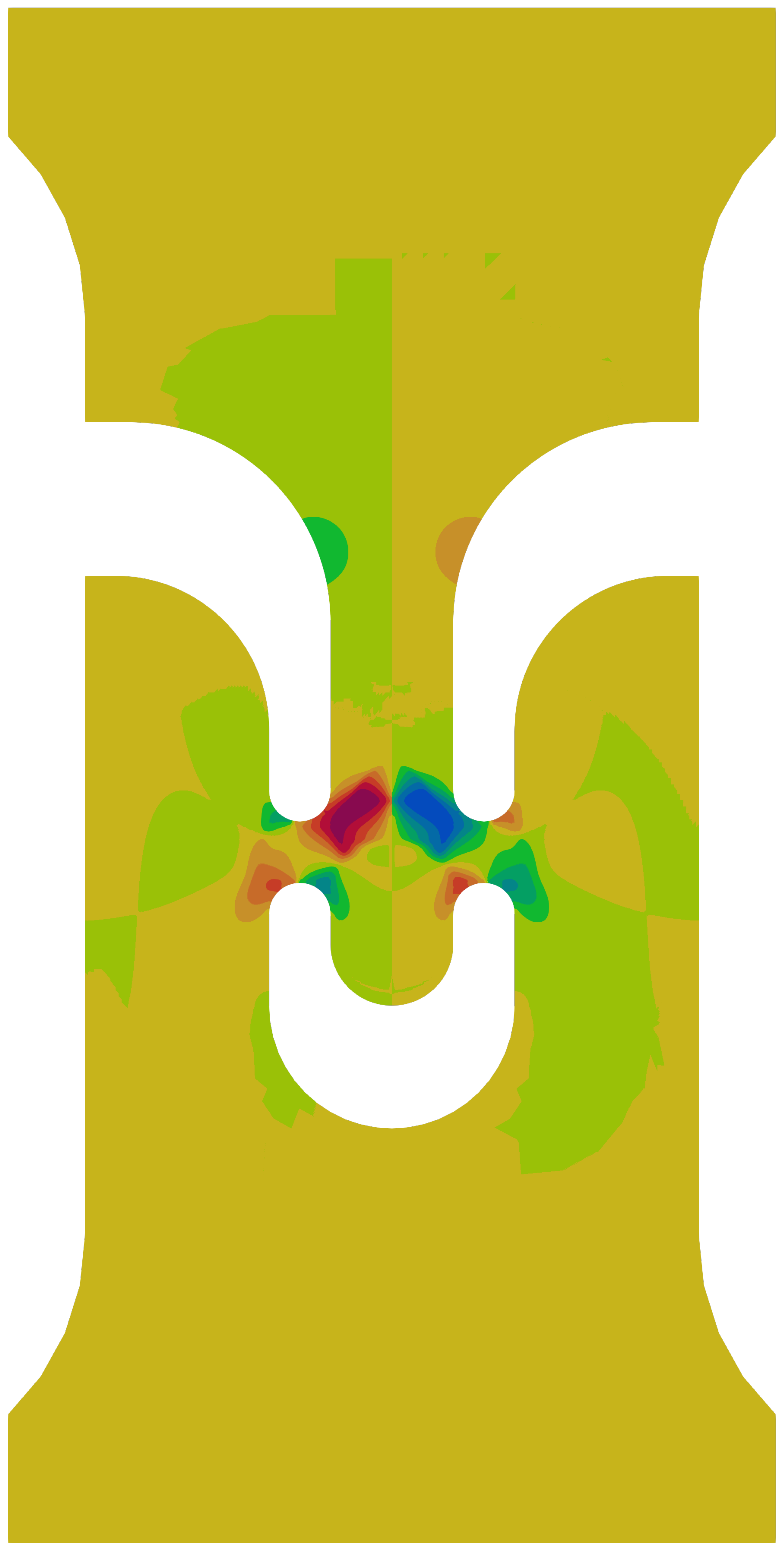}
    \caption{13013}
    \label{fig:ExssmeshconvergenceDten13013}
  \end{subfigure}
  \begin{subfigure}{.08\textwidth} 
    \centering 
    \begin{tikzpicture}
      \node[inner sep=0pt] (pic) at (0,0) {\includegraphics[height=40mm, width=5mm]
      {02_Figures/03_Contour/00_Color_Maps/Damage_Step_Vertical.pdf}};
      \node[inner sep=0pt] (0)   at ($(pic.south)+( 1.00, 0.15)$)  {$-0.2002$};
      \node[inner sep=0pt] (1)   at ($(pic.south)+( 1.00, 3.80)$)  {$+0.2002$};
      \node[inner sep=0pt] (d)   at ($(pic.south)+( 0.00, 4.35)$)  {$D_{xy}~\si{[-]}$};
    \end{tikzpicture} 
    \hphantom{13013}
  \end{subfigure}

  \caption{Mesh convergence study of the damage contour plots of the normal and shear components of the damage tensor for the smiley specimen at the end of the simulation (model~C).}
  \label{fig:ExssmeshconvergenceDten}     
\end{figure}

Next, Fig.\ref{fig:ExssmeshconvergenceDten} shows the mesh convergence of the components of the damage tensor, where we again restrict ourselves to the presentation of model~C. As indicated by the force-displacement curves in Fig.~\ref{fig:ExssFuC}, differences can be observed in the damage contour plots obtained with the coarsest mesh (Fig.~\ref{fig:ExssmeshconvergenceDten755}) compared to the results obtained with the refined meshes (Figs.~\ref{fig:ExssmeshconvergenceDten2649}-\ref{fig:ExssmeshconvergenceDten13013}). However, the results obtained with the refined meshes hardly deviate and are, thus, considered converged.

\begin{figure}
  \centering 

  \begin{subfigure}{.8\textwidth} 
    \centering 
    \begin{tikzpicture}
      \node[inner sep=0pt] (pic)  at (0,0) {\includegraphics[height=70mm]{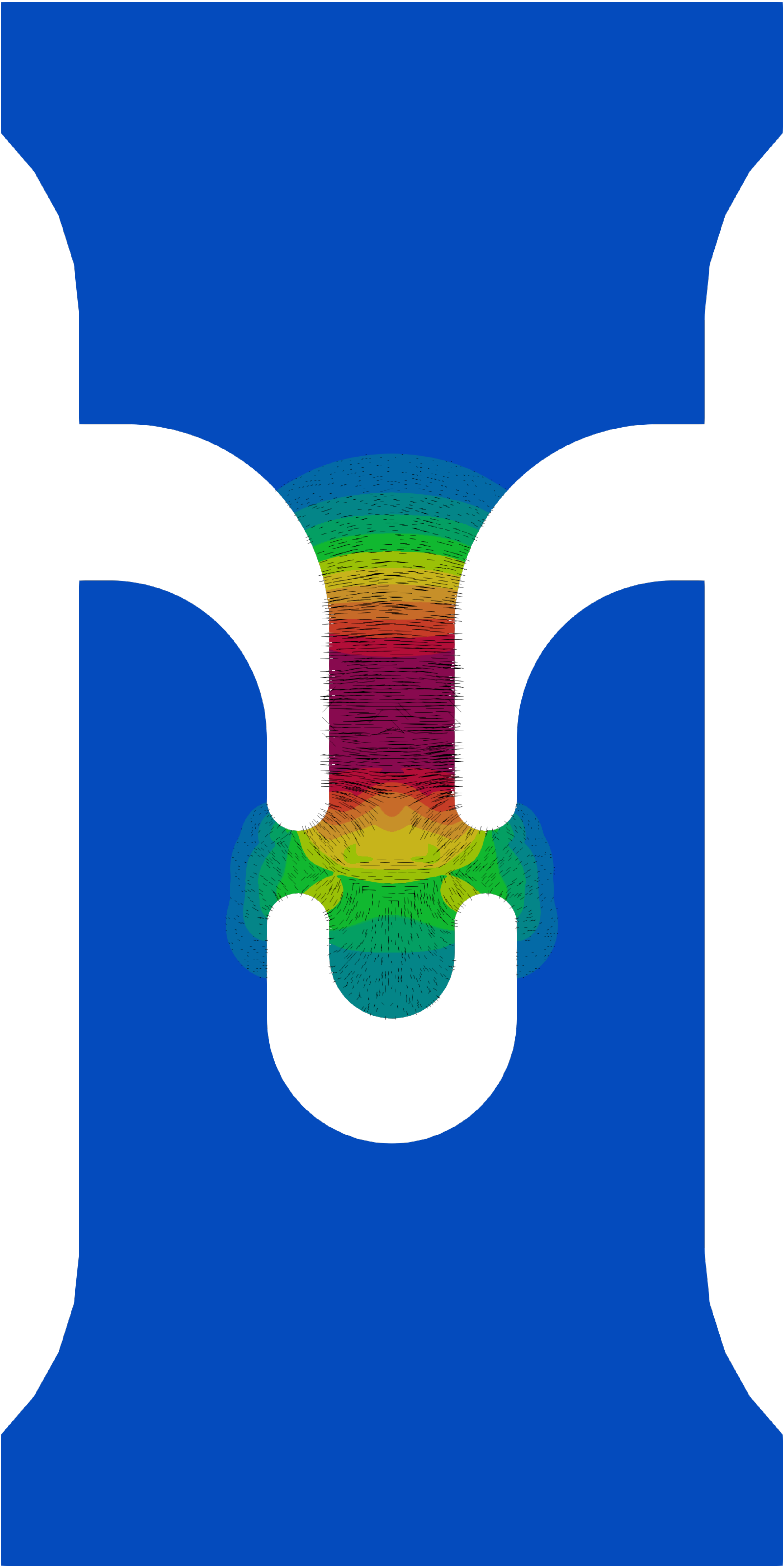}};
      \node[inner sep=0pt] (pic2) at (6,0) {\includegraphics[height=70mm]{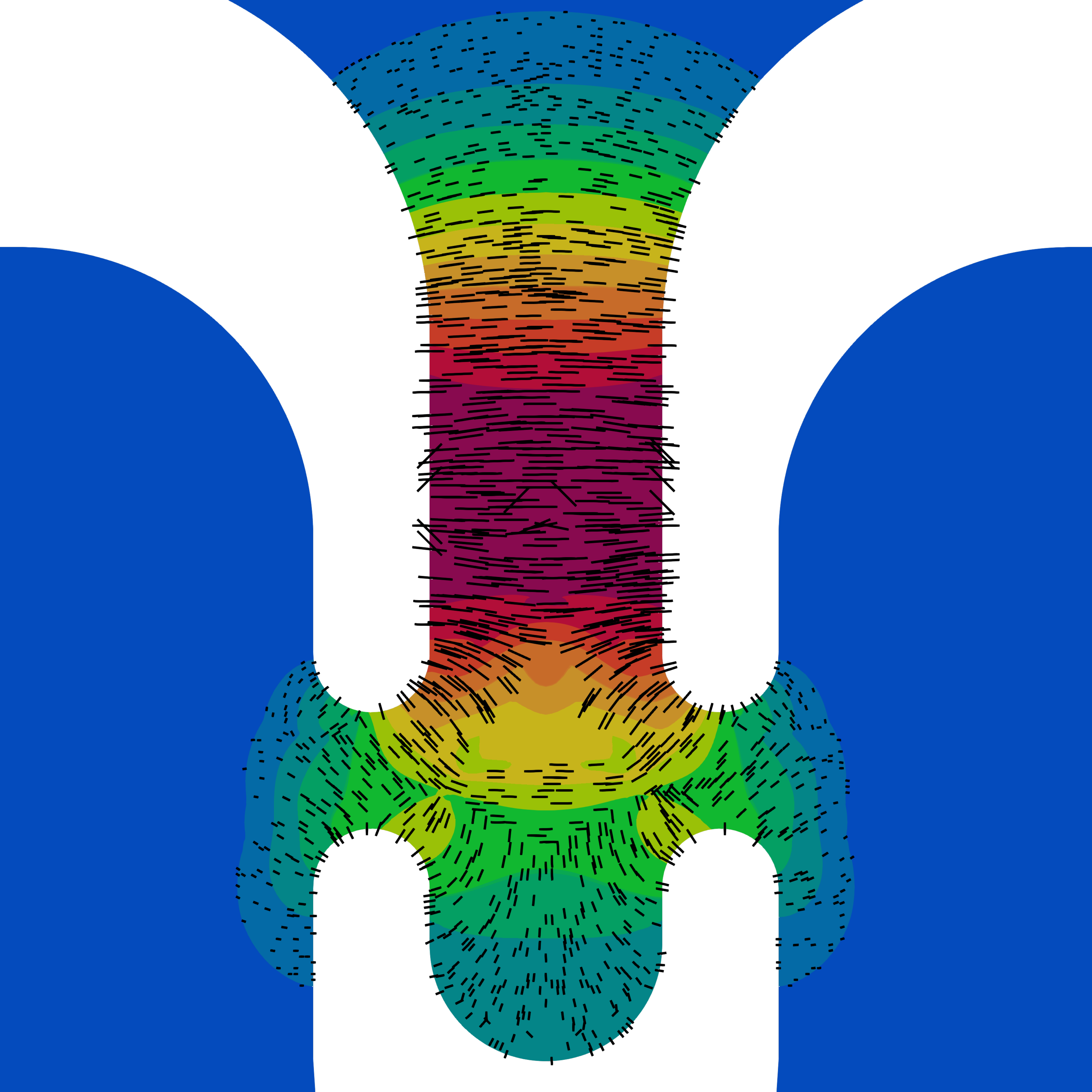}};
      \draw[draw=rwth8, line width=1.5pt] (-1.30,-1.10) rectangle ( 1.30, 1.50);
      \draw[draw=rwth8, line width=1.5pt] ( 2.50,-3.50) rectangle ( 9.50, 3.50);
      \draw[draw=rwth8, line width=1.5pt] ( 1.30, 0.00) -- ( 2.50, 0.00);
    \end{tikzpicture} 
  \end{subfigure}
  \begin{subfigure}{.08\textwidth} 
    \centering 
    \begin{tikzpicture}
      \node[inner sep=0pt] (pic) at (0,0) {\includegraphics[height=40mm, width=5mm]
      {02_Figures/03_Contour/00_Color_Maps/Damage_Step_Vertical.pdf}};
      \node[inner sep=0pt] (0)   at ($(pic.south)+( 0.50, 0.15)$)  {$0$};
      \node[inner sep=0pt] (1)   at ($(pic.south)+( 0.50, 3.80)$)  {$1$};
      \node[inner sep=0pt] (d)   at ($(pic.south)+( 0.10, 4.35)$)  {$D_{1}~\si{[-]}$};
    \end{tikzpicture} 
  \end{subfigure}

  \vspace{1mm}

  \begin{subfigure}{.8\textwidth} 
    \centering 
    \begin{tikzpicture}
      \node[inner sep=0pt] (pic)  at (0,0) {\includegraphics[height=70mm]{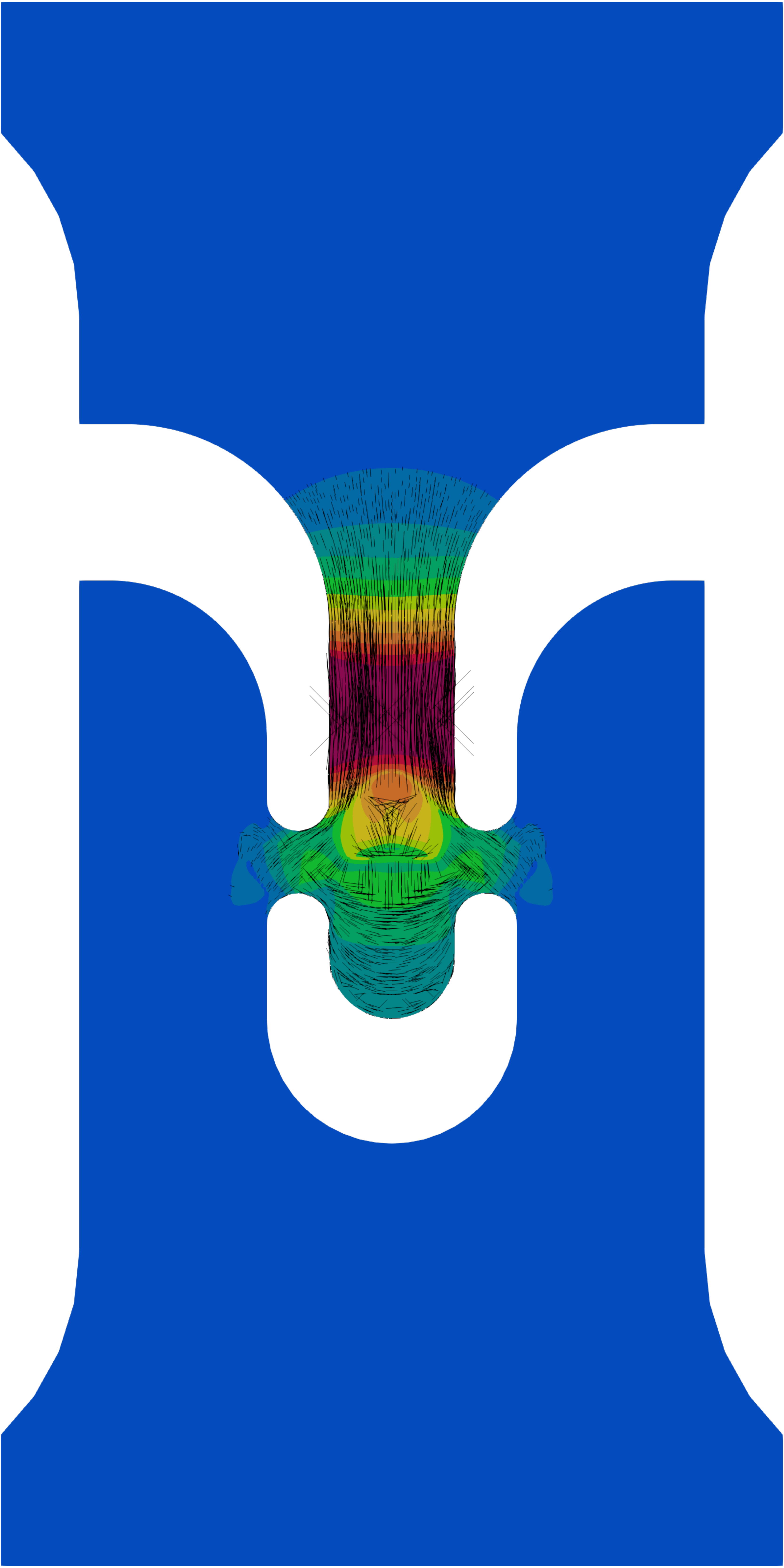}};
      \node[inner sep=0pt] (pic2) at (6,0) {\includegraphics[height=70mm]{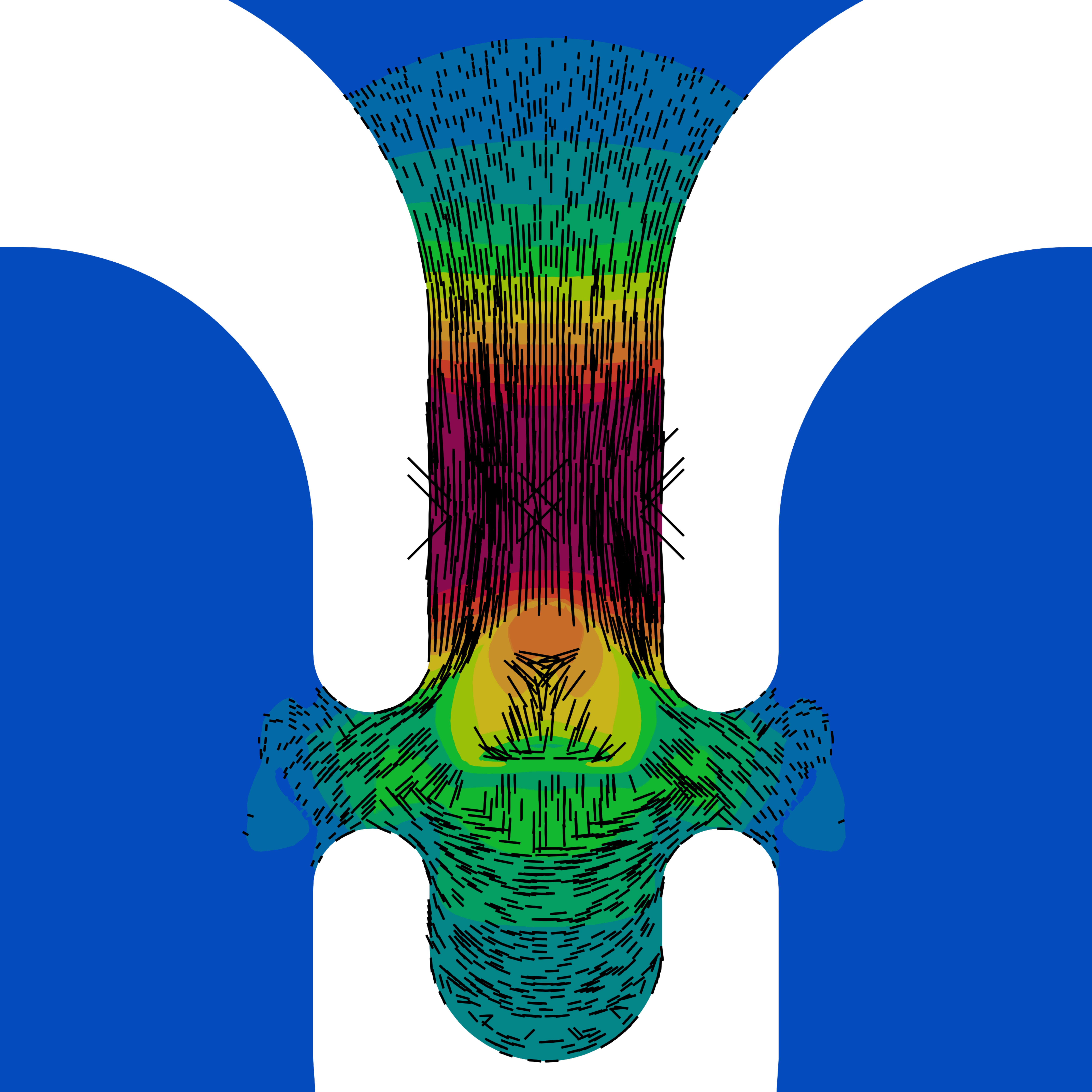}};
      \draw[draw=rwth8, line width=1.5pt] (-1.30,-1.10) rectangle ( 1.30, 1.50);
      \draw[draw=rwth8, line width=1.5pt] ( 2.50,-3.50) rectangle ( 9.50, 3.50);
      \draw[draw=rwth8, line width=1.5pt] ( 1.30, 0.00) -- ( 2.50, 0.00);
    \end{tikzpicture} 
  \end{subfigure}
  \begin{subfigure}{.08\textwidth} 
    \centering 
    \begin{tikzpicture}
      \node[inner sep=0pt] (pic) at (0,0) {\includegraphics[height=40mm, width=5mm]
      {02_Figures/03_Contour/00_Color_Maps/Damage_Step_Vertical.pdf}};
      \node[inner sep=0pt] (0)   at ($(pic.south)+( 0.50, 0.15)$)  {$0$};
      \node[inner sep=0pt] (1)   at ($(pic.south)+( 0.50, 3.80)$)  {$1$};
      \node[inner sep=0pt] (d)   at ($(pic.south)+( 0.10, 4.35)$)  {$D_{2}~\si{[-]}$};
    \end{tikzpicture} 
  \end{subfigure}

  \vspace{1mm}

  \begin{subfigure}{.8\textwidth} 
    \centering 
    \begin{tikzpicture}
      \node[inner sep=0pt] (pic)  at (0,0) {\includegraphics[height=70mm]{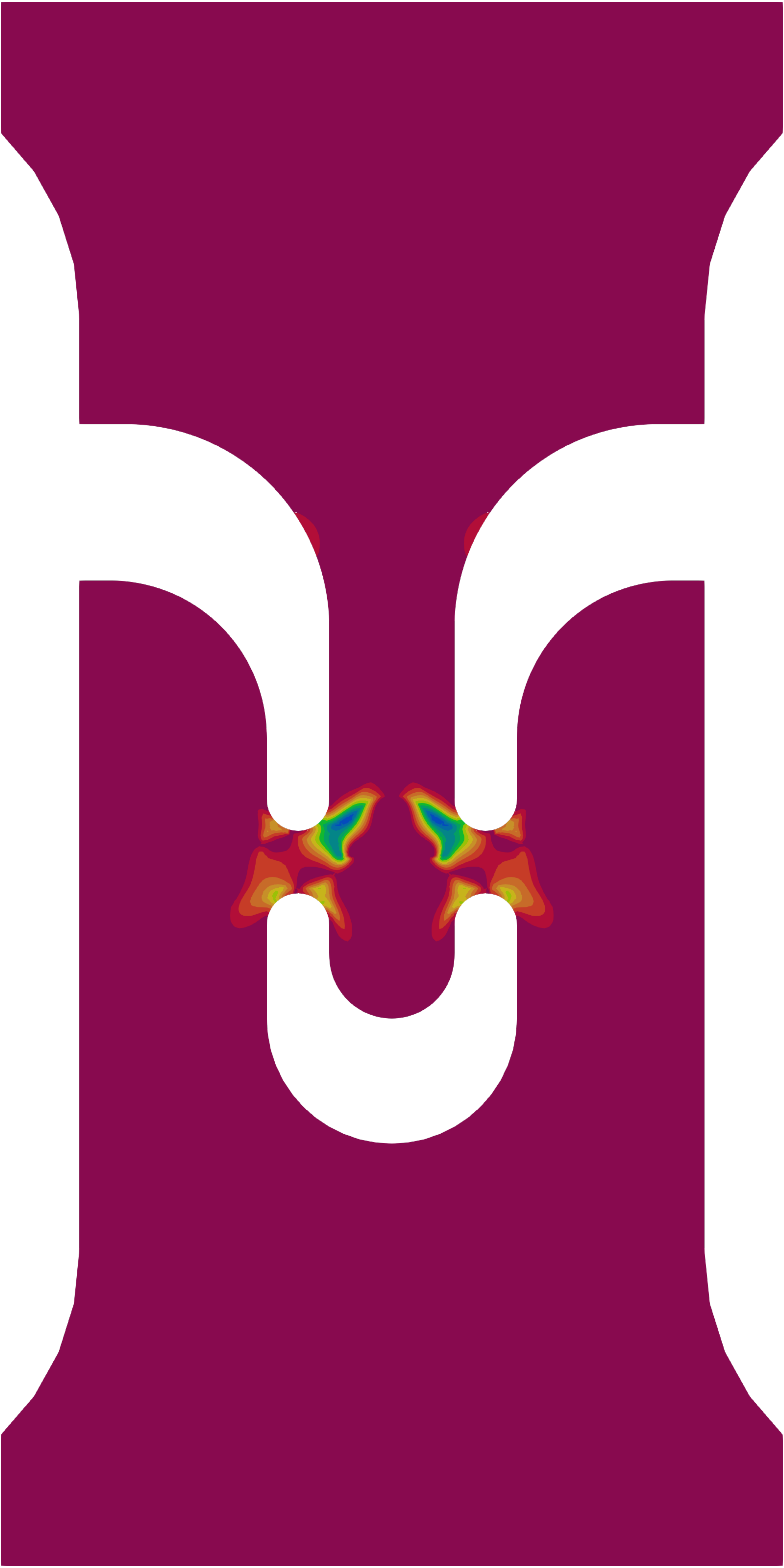}};
      \node[inner sep=0pt] (pic2) at (6,0) {\includegraphics[height=70mm]{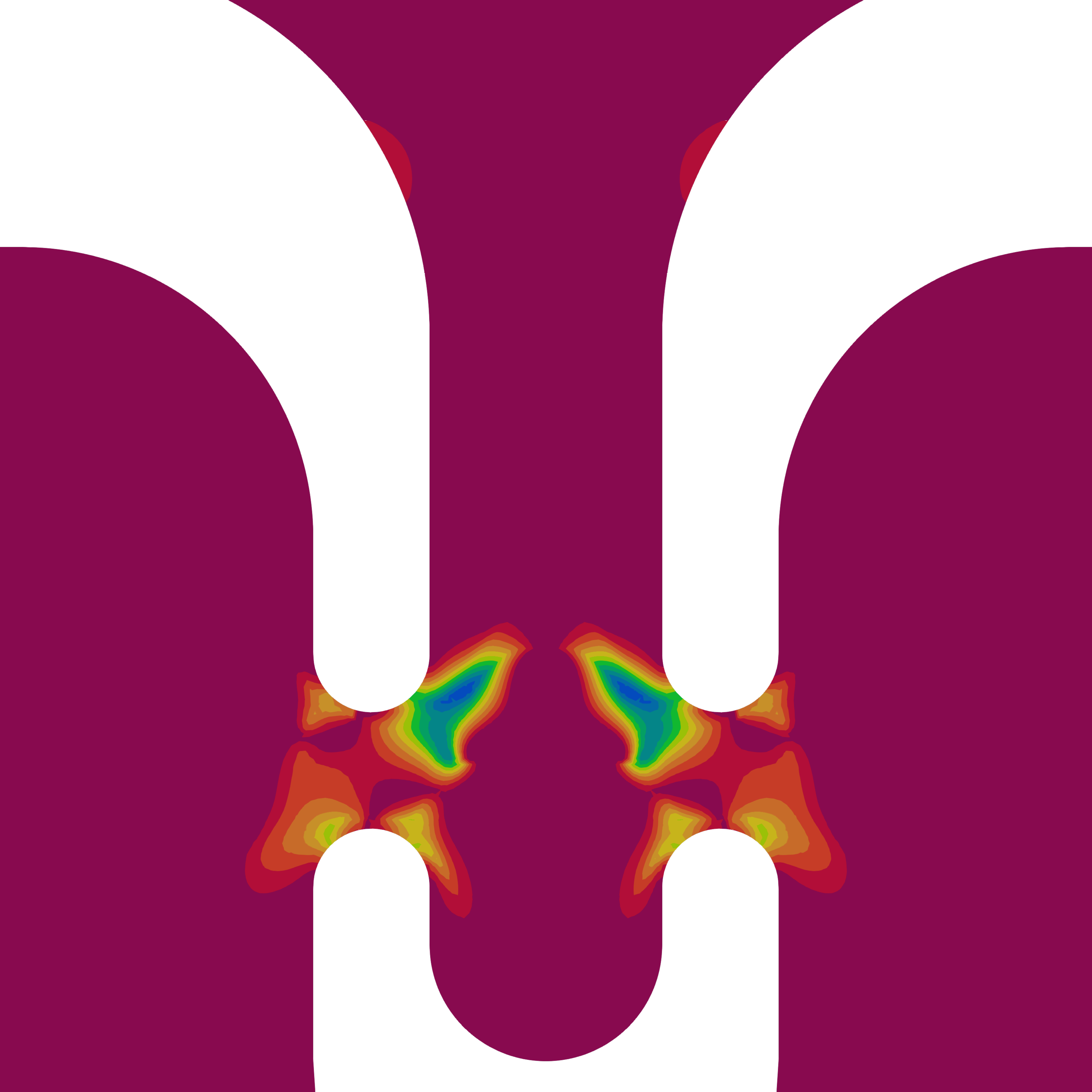}};
      \draw[draw=rwth8, line width=1.5pt] (-1.30,-1.10) rectangle ( 1.30, 1.50);
      \draw[draw=rwth8, line width=1.5pt] ( 2.50,-3.50) rectangle ( 9.50, 3.50);
      \draw[draw=rwth8, line width=1.5pt] ( 1.30, 0.00) -- ( 2.50, 0.00);
    \end{tikzpicture} 
  \end{subfigure}
  \begin{subfigure}{.08\textwidth} 
    \centering 
    \begin{tikzpicture}
      \node[inner sep=0pt] (pic) at (0,0) {\includegraphics[height=40mm, width=5mm]
      {02_Figures/03_Contour/00_Color_Maps/Damage_Step_Vertical.pdf}};
      \node[inner sep=0pt] at ($(pic.south)+( 1.00, 0.15)$)  {$-0.1926$};
      \node[inner sep=0pt] at ($(pic.south)+( 0.50, 3.80)$)  {$0$};
      \node[inner sep=0pt] at ($(pic.south)+( 1.00, 4.52)$)  {$ \underset{i \in (x,y,z)}{\text{max}(D_{ii})} -D_{1}~\si{[-]}$};
    \end{tikzpicture} 
  \end{subfigure}
  
  \caption{Contour plots of the first and second eigenvalue of the damage tensor and of the absolute difference between the maximum normal component and the first eigenvalue for the smiley specimen at the end of the simulation (model~C). The black lines indicate the scaled normals to the first (top) and second eigenvector (middle) of the damage tensor.}
  \label{fig:ExsseigD}     
\end{figure}

Now, we study the eigenvalues of the damage tensor for model~C. Fig.~\ref{fig:ExsseigD} shows the first eigenvalue $D_1$ (top) and second eigenvalue $D_2$ (middle) as well as the scaled normals to the corresponding eigenvectors in the $x$-$y$-plane. These normals are supposed to indicate the orientation and the density of the anisotropic micro cracks. Hence, the micro cracks associated with the largest eigenvalue $D_1$ are perpendicular to the loading direction and exhibit the highest density in the completely damaged zone. Due to the orthogonality of eigenvectors and an in-plane loading, the micro cracks associated with the second eigenvalue $D_2$ are perpendicular to the micro cracks associated with the first eigenvalue $D_1$.

Finally, Fig.~\ref{fig:ExsseigD} (bottom) shows the difference between the maximum of the normal components $D_{xx}$, $D_{yy}$, and $D_{zz}$ and the largest eigenvalue $D_1$. Evidently, a significant underestimation of the material degradation up to a value of $-0.1926~\si{[-]}$ occurs in the shear load dominated regions, when only considering the normal components of the Cartesian coordinate system.

\begin{figure}[htbp]
  \centering
    \includegraphics{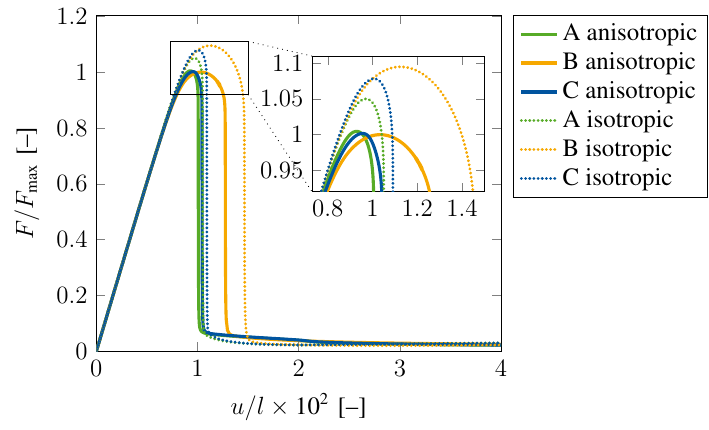}
  \caption{Comparison of the anisotropic and isotropic computation for the smiley specimen (13013 elements). The forces are normalized with respect to the maximum force of the anisotropic computation of model~B with $F_\text{max} = 2.9590 \times 10^3~[\si{\newton}]$.}
  \label{fig:ExssFuIsotropic}
\end{figure}

The last study is concerned with the comparison of isotropic and anisotropic damage for the smiley specimen. Fig.~\ref{fig:ExssFuIsotropic} shows the normalized force-displacement curves for the isotropic and anisotropic models and for all models the isotropic formulation overestimates the maximum peak force (A: $+4.52~[\si{\percent}]$, B: $+9.49~[\si{\percent}]$, C: $+7.65~[\si{\percent}]$).

\begin{figure}
  \centering 

  \begin{subfigure}{.22\textwidth} 
    \centering 
    \includegraphics[width=\textwidth]{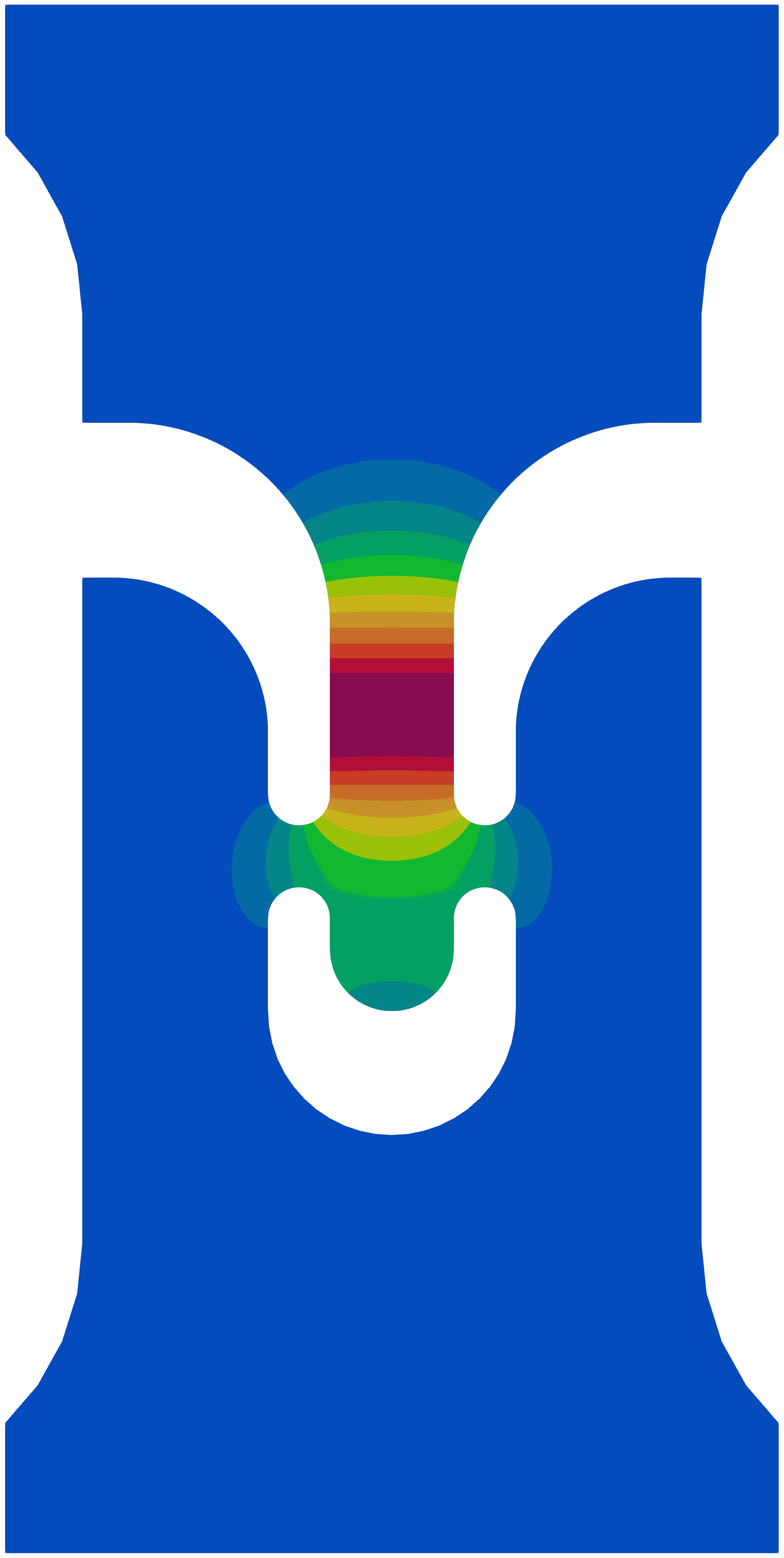}
  \end{subfigure}
  \begin{subfigure}{.22\textwidth} 
    \centering 
    \includegraphics[width=\textwidth]{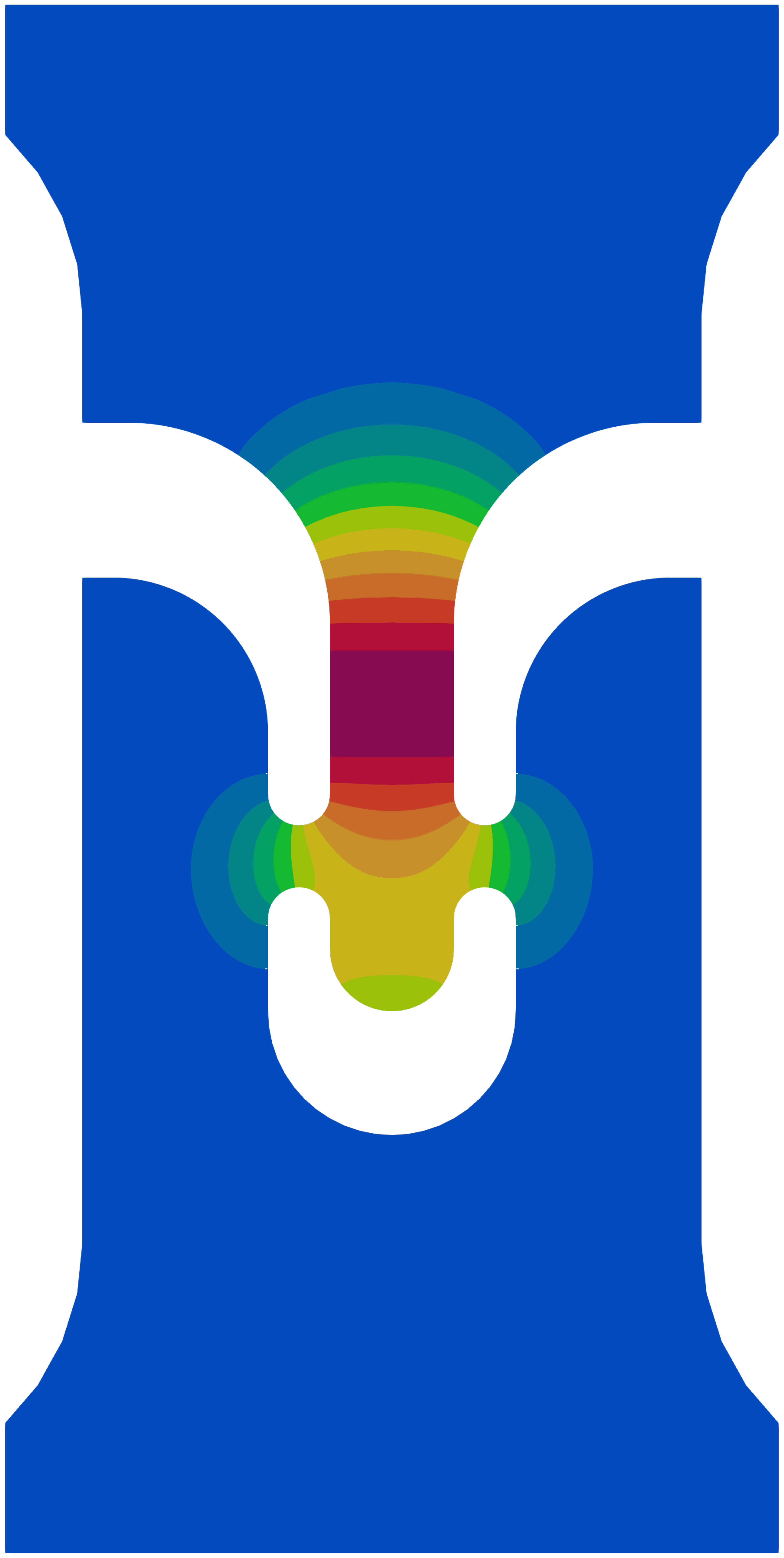}
  \end{subfigure}
  \begin{subfigure}{.22\textwidth} 
    \centering 
    \includegraphics[width=\textwidth]{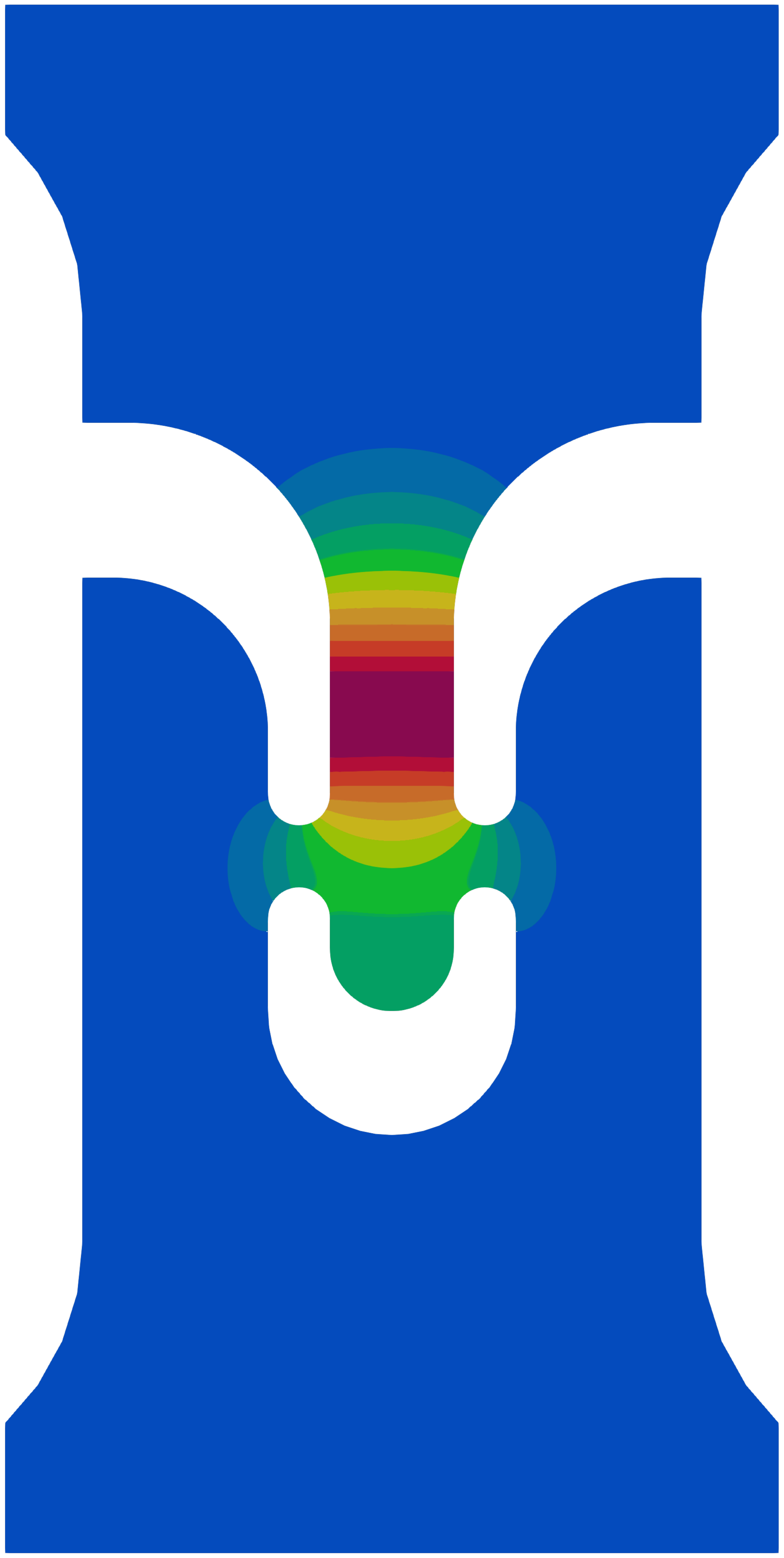}
  \end{subfigure}
  \begin{subfigure}{.08\textwidth} 
    \centering 
    \begin{tikzpicture}
      \node[inner sep=0pt] (pic) at (0,0) {\includegraphics[height=40mm, width=5mm]
      {02_Figures/03_Contour/00_Color_Maps/Damage_Step_Vertical.pdf}};
      \node[inner sep=0pt] (0)   at ($(pic.south)+( 0.50, 0.15)$)  {$0$};
      \node[inner sep=0pt] (1)   at ($(pic.south)+( 0.50, 3.80)$)  {$1$};
      \node[inner sep=0pt] (d)   at ($(pic.south)+(-0.15, 4.35)$)  {$D~\si{[-]}$};
    \end{tikzpicture} 
  \end{subfigure}

  \vspace{1mm}

  \begin{subfigure}{.22\textwidth} 
    \centering 
    \includegraphics[width=\textwidth]{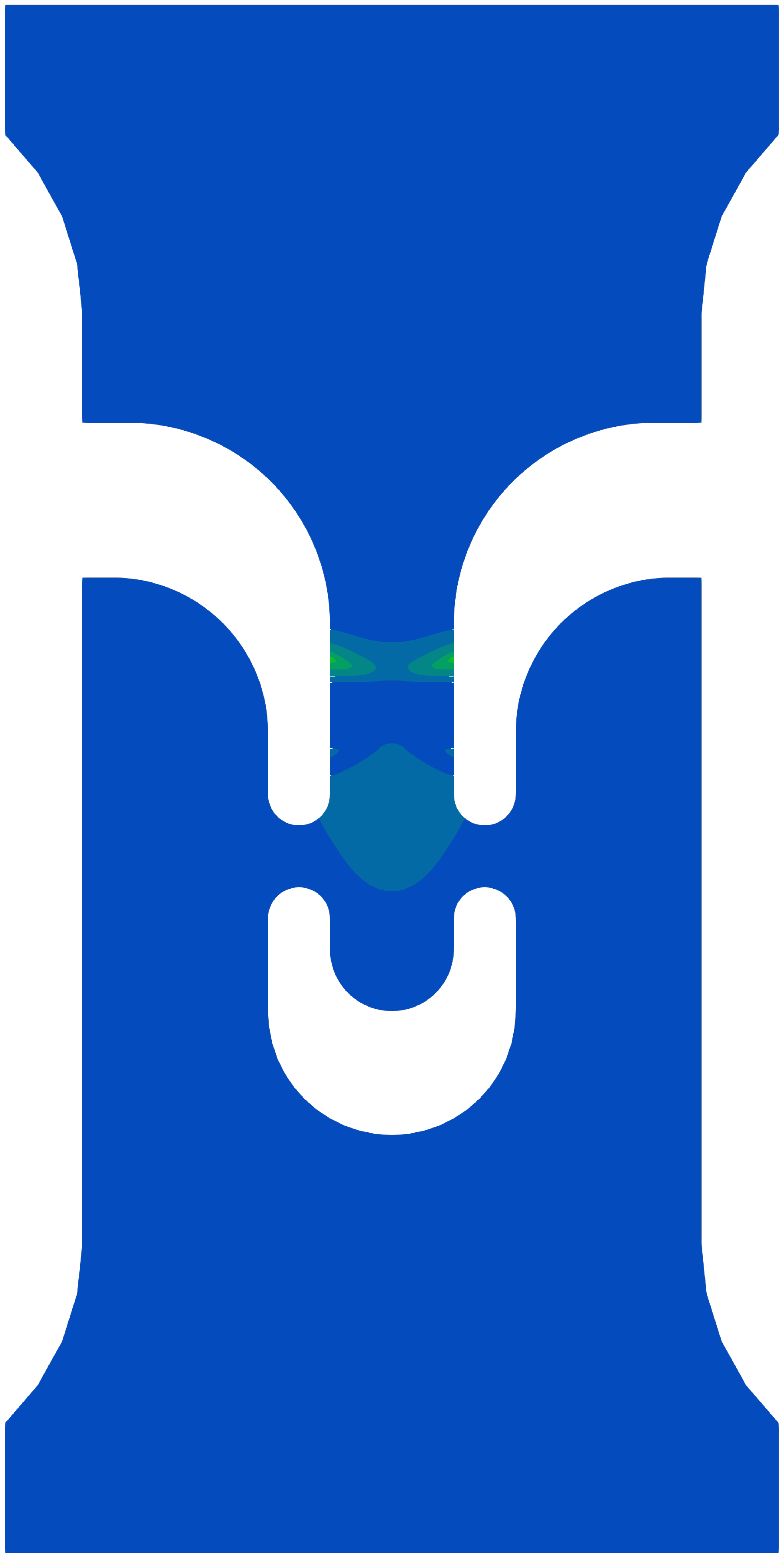}
  \end{subfigure}
  \begin{subfigure}{.22\textwidth} 
    \centering 
    \includegraphics[width=\textwidth]{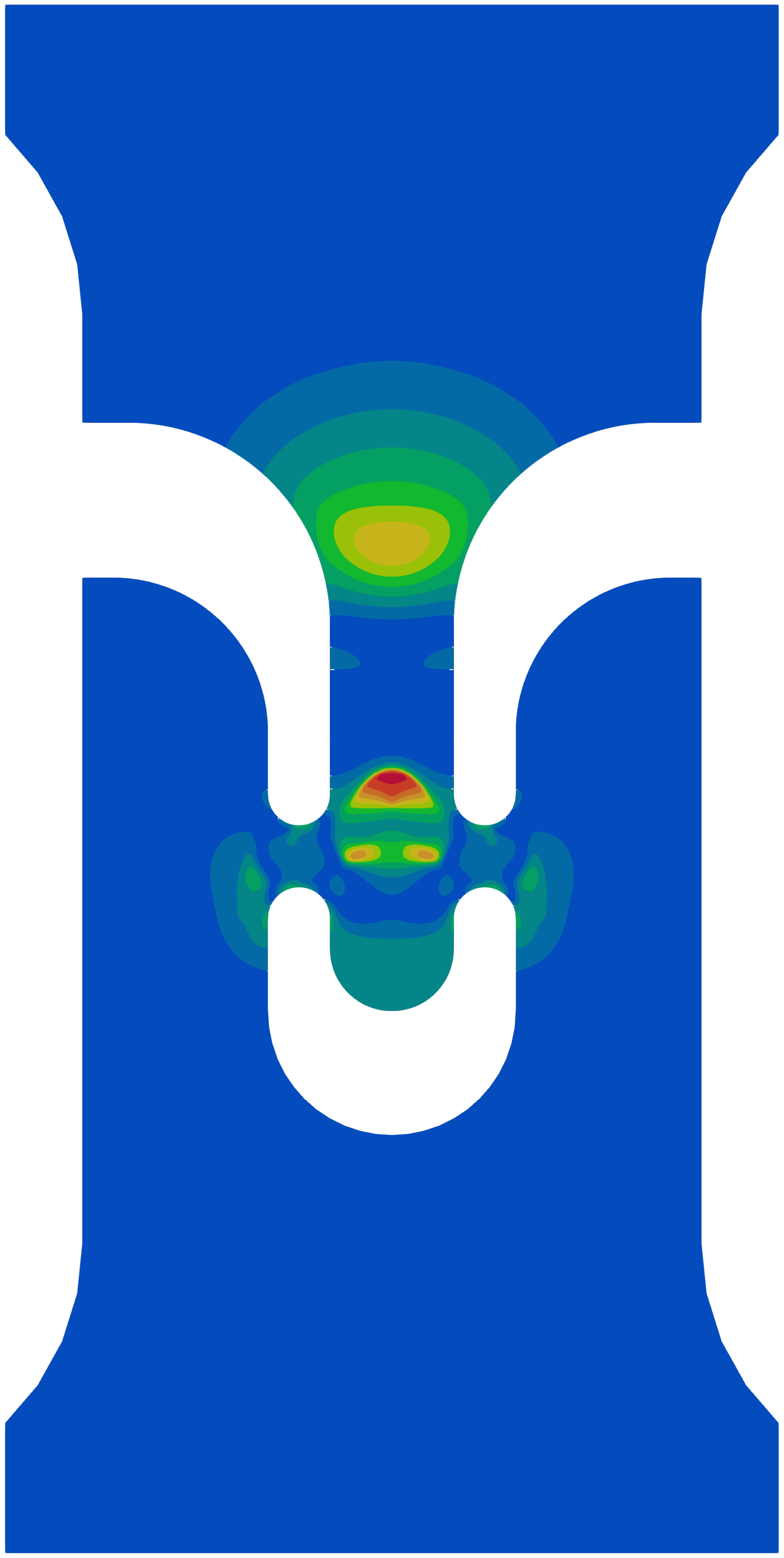}
  \end{subfigure}
  \begin{subfigure}{.22\textwidth} 
    \centering 
    \includegraphics[width=\textwidth]{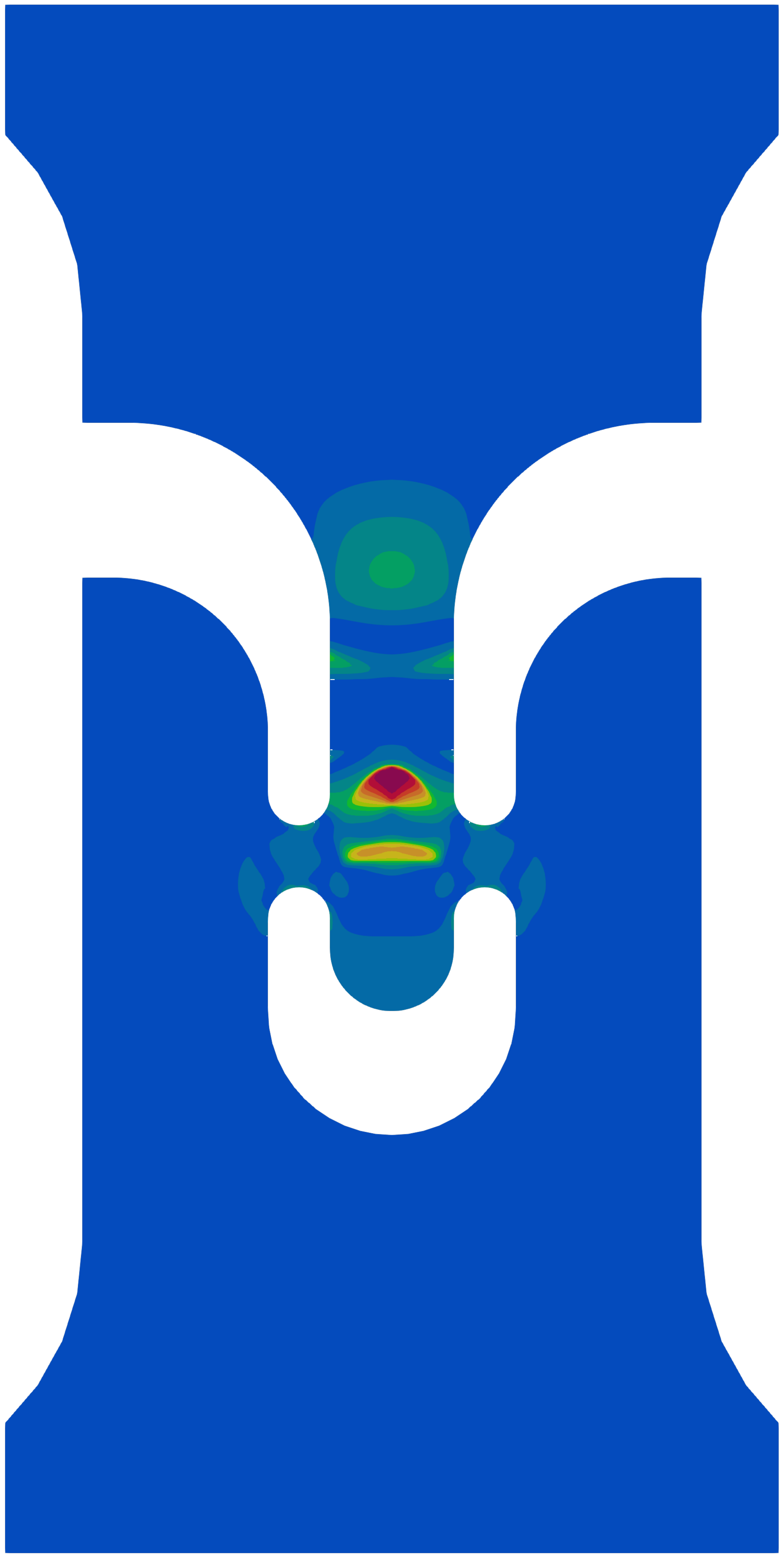}
  \end{subfigure}
  \begin{subfigure}{.08\textwidth} 
    \centering 
    \begin{tikzpicture}
      \node[inner sep=0pt] (pic) at (0,0) {\includegraphics[height=40mm, width=5mm]
      {02_Figures/03_Contour/00_Color_Maps/Damage_Step_Vertical.pdf}};
      \node[inner sep=0pt] (0)   at ($(pic.south)+( 0.53, 0.15)$)  {$0$};
      \node[inner sep=0pt] (1)   at ($(pic.south)+( 1.00, 3.80)$)  {$0.4158$};
      \node[inner sep=0pt] (d)   at ($(pic.south)+( 0.56, 4.35)$)  {$|D-D_{xx}|~\si{[-]}$};
    \end{tikzpicture} 
  \end{subfigure}

  \vspace{1mm}

  \begin{subfigure}{.22\textwidth} 
    \centering 
    \includegraphics[width=\textwidth]{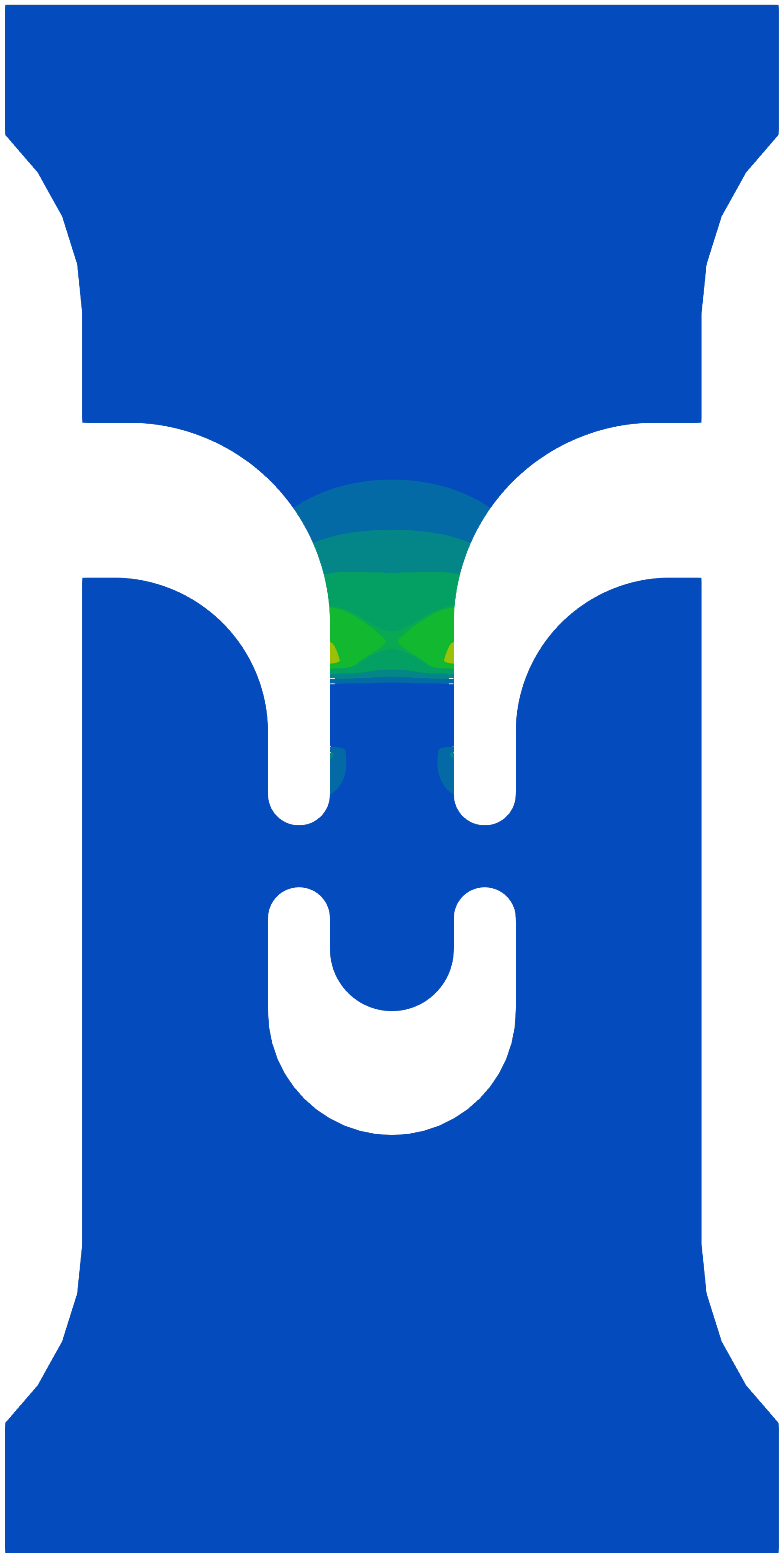}
    \caption{Model~A}
  \end{subfigure}
  \begin{subfigure}{.22\textwidth} 
    \centering 
    \includegraphics[width=\textwidth]{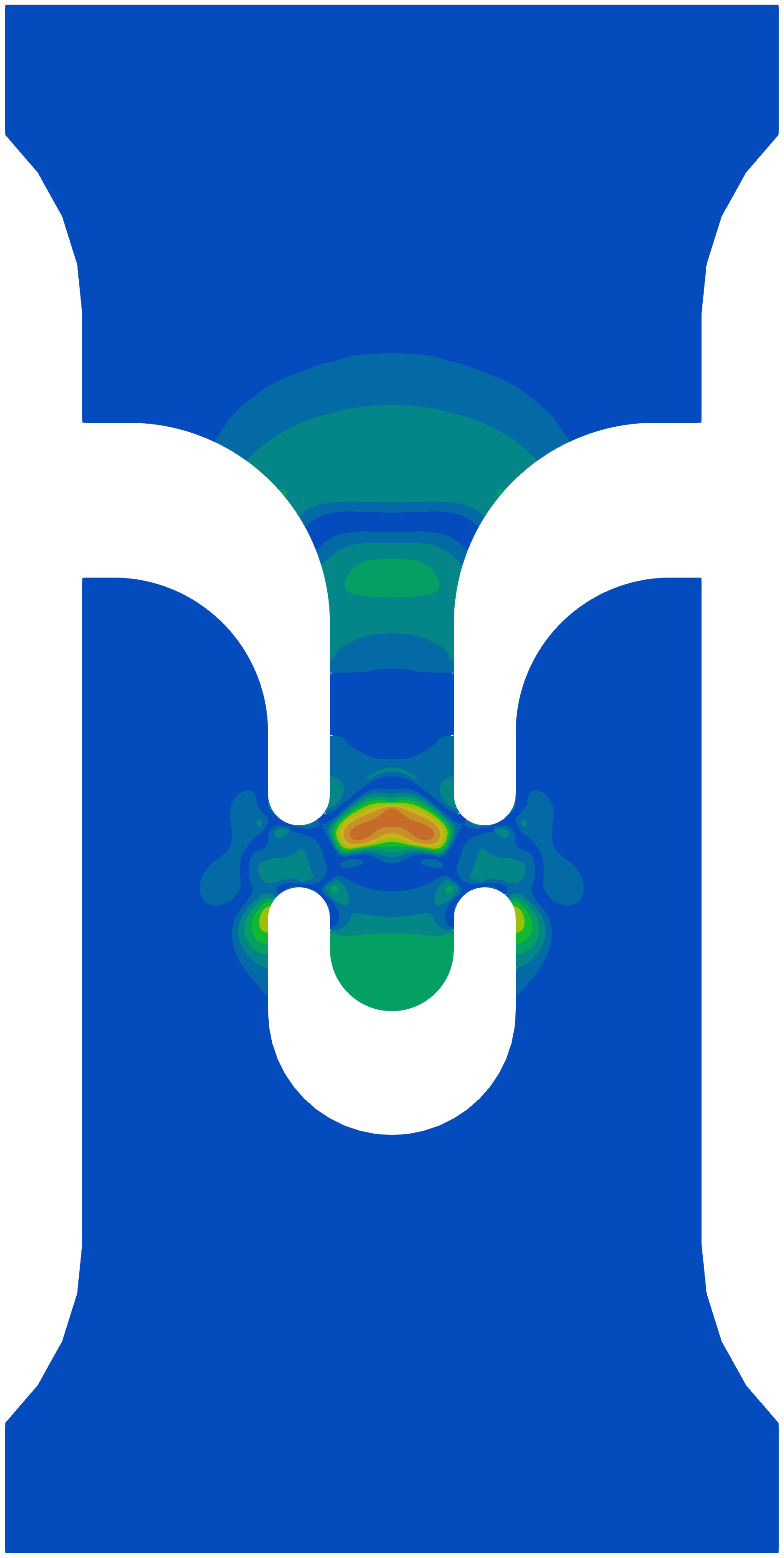}
    \caption{Model~B}
  \end{subfigure}
  \begin{subfigure}{.22\textwidth} 
    \centering 
    \includegraphics[width=\textwidth]{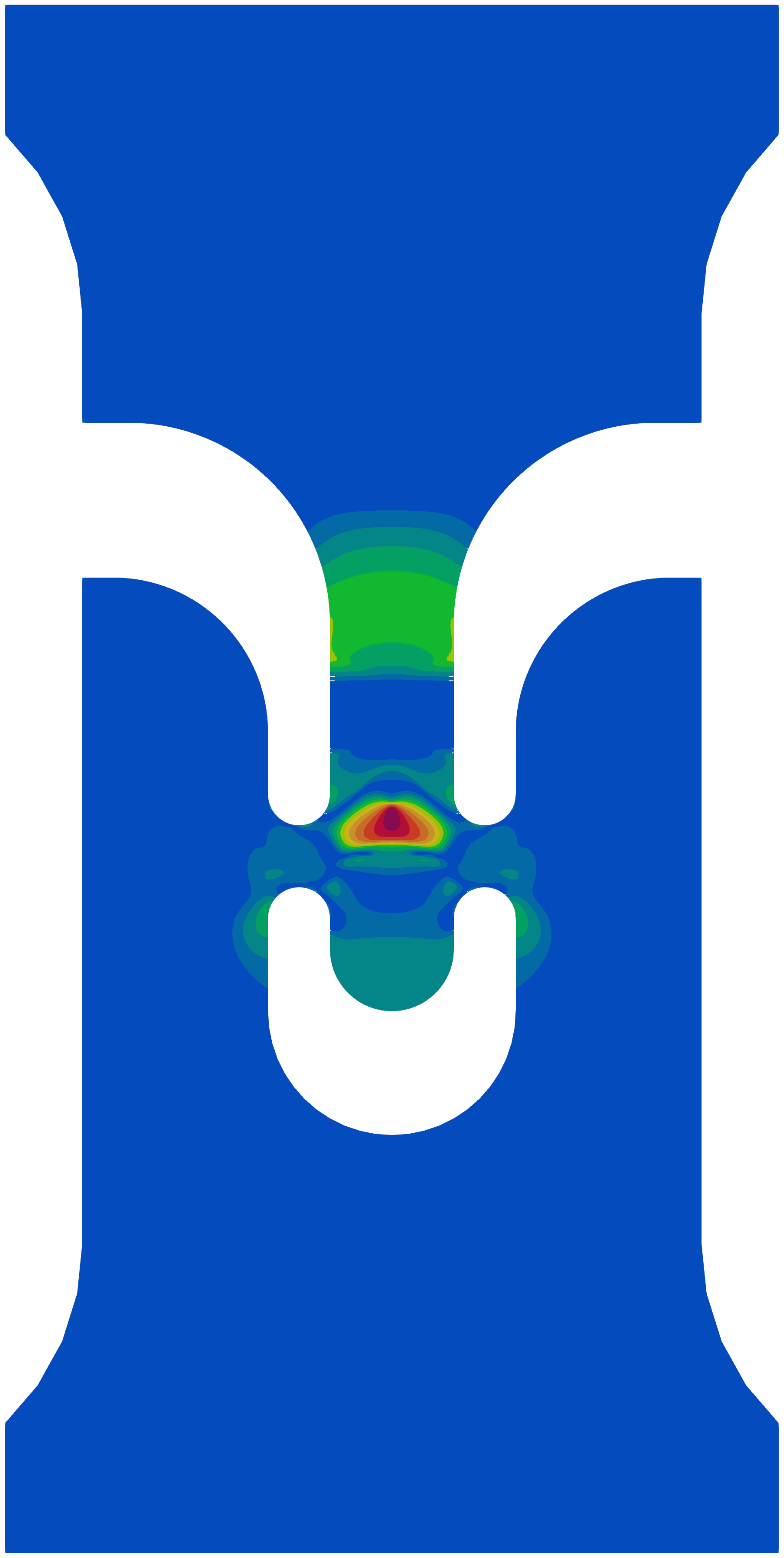}
    \caption{Model~C}
  \end{subfigure}
  \begin{subfigure}{.08\textwidth} 
    \centering 
    \begin{tikzpicture}
      \node[inner sep=0pt] (pic) at (0,0) {\includegraphics[height=40mm, width=5mm]
      {02_Figures/03_Contour/00_Color_Maps/Damage_Step_Vertical.pdf}};
      \node[inner sep=0pt] (0)   at ($(pic.south)+( 0.53, 0.15)$)  {$0$};
      \node[inner sep=0pt] (1)   at ($(pic.south)+( 1.00, 3.80)$)  {$0.3293$};
      \node[inner sep=0pt] (d)   at ($(pic.south)+( 0.56, 4.35)$)  {$|D-D_{yy}|~\si{[-]}$};
    \end{tikzpicture} 
    \hphantom{Model~C}
  \end{subfigure}

  \caption{Contour plots of the isotropic damage value and its absolute difference to the normal components of the damage tensor for the smiley specimen at the end of the simulation.}
  \label{fig:ExssDiso}     
\end{figure}

The corresponding isotropic damage contour plots are presented in Fig.~\ref{fig:ExssDiso} (top row) and, also for the isotropic models, total failure occurs in the tension load carrying cross section. However, the absolute difference of the isotropic damage value to the normals components of the damage tensor for the anisotropic computation (see Fig.~\ref{fig:ExssDiso} (middle and bottom row)) amounts up to $0.4158~\si{[-]}$ for $|D-D_{xx}|$ and to $0.3293~\si{[-]}$ for $|D-D_{yy}|$, which is in line with the observations in Fig.~\ref{fig:ExssFuIsotropic}.

\begin{figure}
  \centering 
  \begin{subfigure}{.8\textwidth} 
    \centering 
    \begin{tikzpicture}
      \node[inner sep=0pt] (pic)  at (0,0) {\includegraphics[height=70mm]{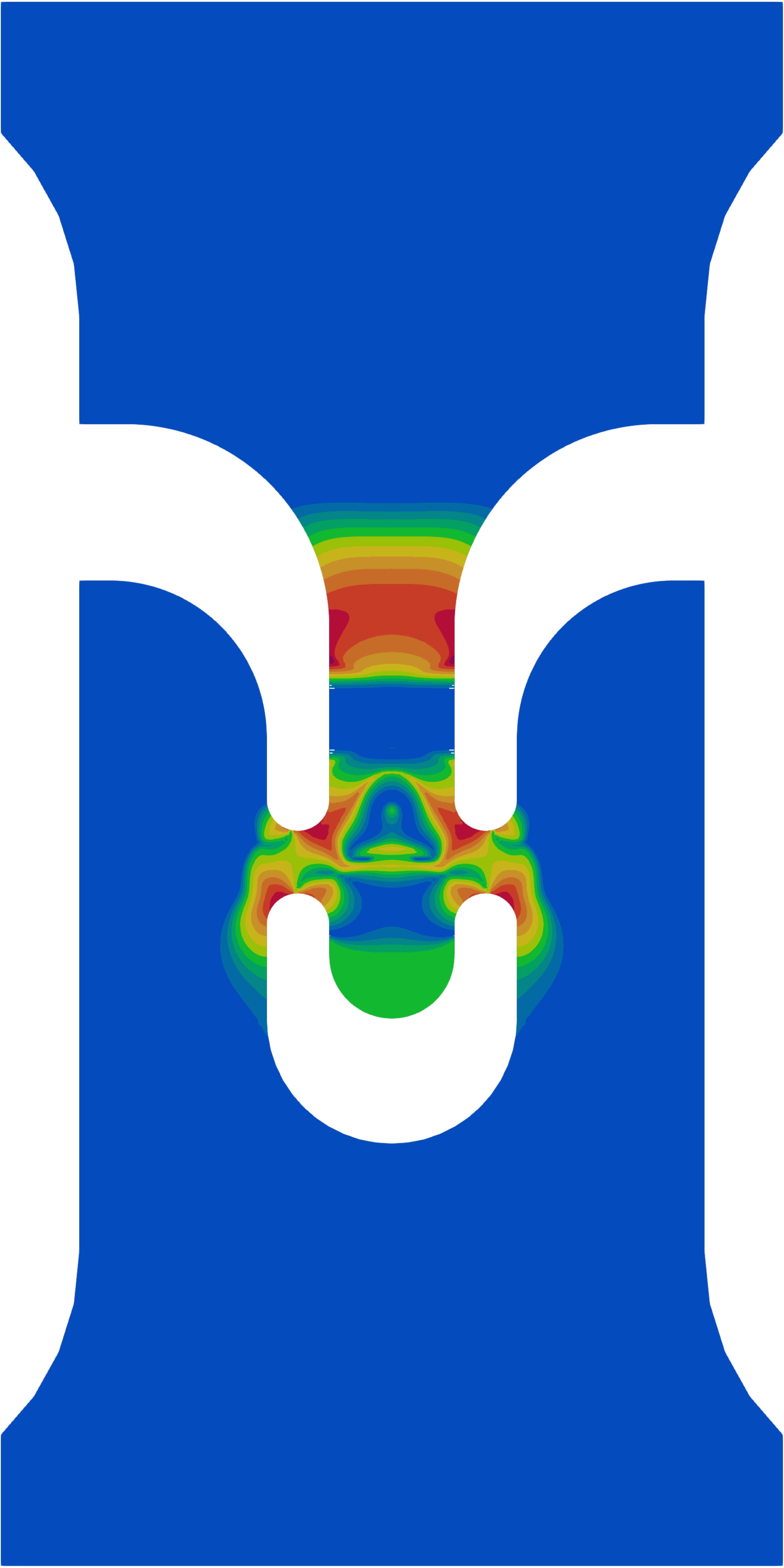}};
      \node[inner sep=0pt] (pic2) at (6,0) {\includegraphics[height=70mm]{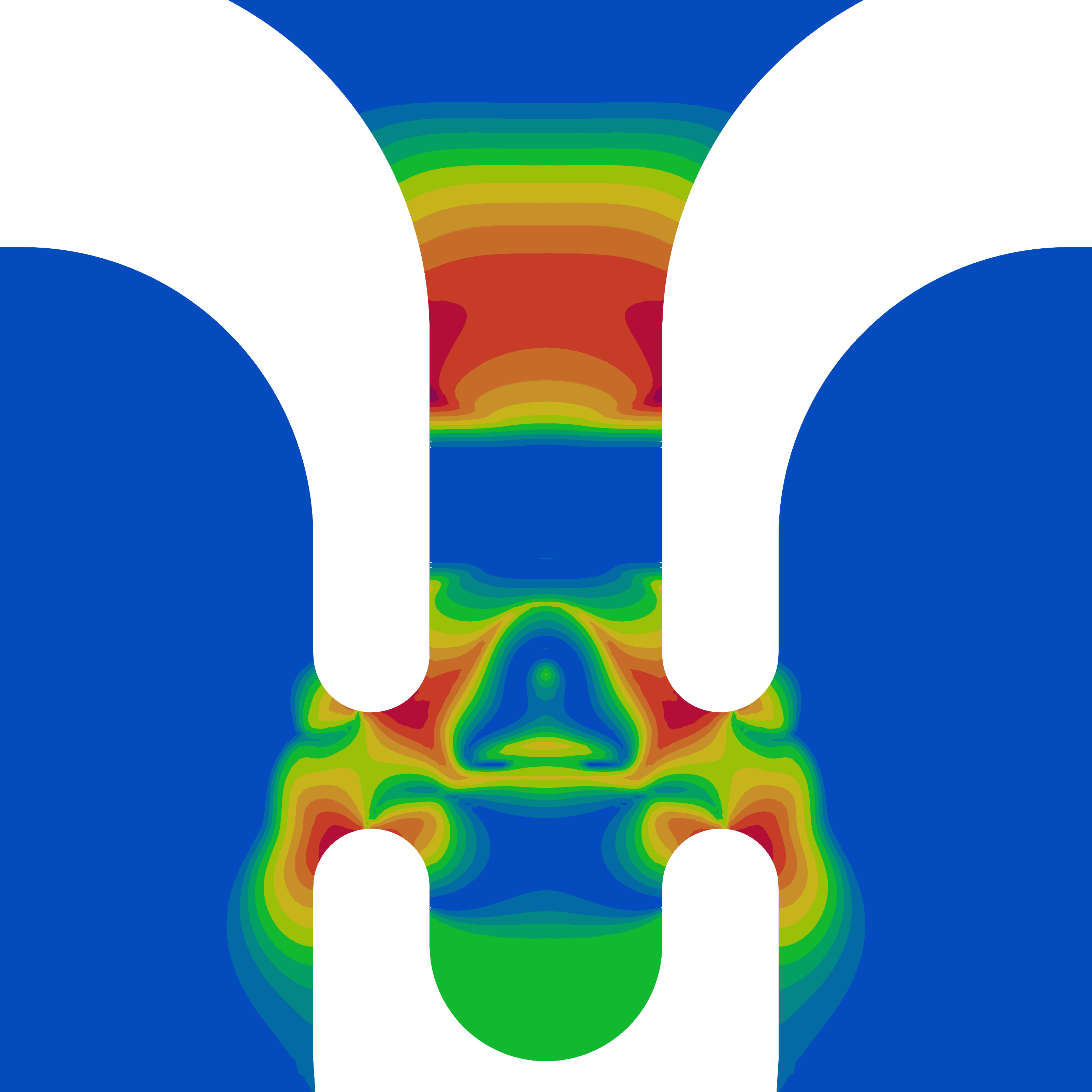}};
      \draw[draw=rwth8, line width=1.5pt] (-1.30,-1.10) rectangle ( 1.30, 1.50);
      \draw[draw=rwth8, line width=1.5pt] ( 2.50,-3.50) rectangle ( 9.50, 3.50);
      \draw[draw=rwth8, line width=1.5pt] ( 1.30, 0.00) -- ( 2.50, 0.00);
    \end{tikzpicture} 
  \end{subfigure}
  \begin{subfigure}{.08\textwidth} 
    \centering 
    \begin{tikzpicture}
      \node[inner sep=0pt] (pic) at (0,0) {\includegraphics[height=40mm, width=5mm]
      {02_Figures/03_Contour/00_Color_Maps/Damage_Step_Vertical.pdf}};
      \node[inner sep=0pt] (0)   at ($(pic.south)+( 0.50, 0.15)$)  {$0$};
      \node[inner sep=0pt] (1)   at ($(pic.south)+( 0.97, 3.80)$)  {$0.1581$};
      \node[inner sep=0pt] (d)   at ($(pic.south)+( 0.60, 4.35)$)  {$|D-D_{1}|~\si{[-]}$};
    \end{tikzpicture} 
  \end{subfigure}
  \caption{Contour plot of the absolute difference between the isotropic damage value and the first eigenvalue of the damage tensor for the smiley specimen at the end of the simulation (model~C).} 
  \label{fig:ExssDiffeigD1Diso}     
\end{figure}

Last, the absolute difference of the isotropic damage value and the largest eigenvalue of the damage tensor for the anisotropic calculation for model~C is shown in Fig.~\ref{fig:ExssDiffeigD1Diso}. The value of $|D-D_{1}|$ reaches up to $0.1581~\si{[-]}$ and, thus, underlines the significant difference between isotropic and anisotropic damage.

\subsection*{Summary of the numerical results}

The following most important results were obtained for model~A (full regularization, six micromorphic degrees of freedom), model~B (reduced regularization, three micromorphic degrees of freedom), and model~C (reduced regularization, two micromorphic degrees of freedom) in the numerical examples:
\begin{itemize}
  \item Models~A, B and C effectively prevent localization in the structural force-displacement response (Figs.~\ref{fig:ExpwhFu}, \ref{fig:ExanFu}, \ref{fig:ExtsFu}, and \ref{fig:ExssFu}).
  \item Models~A and C coincide in the structural response, while model~B yields a higher energy dissipation and a horizontal offset of the vertical force drop to the right for the same maximum peak load (Figs.~\ref{fig:ExpwhFuComp}, \ref{fig:ExanFuComp}, \ref{fig:ExtsFuComp}, and \ref{fig:ExssFuComp}).
  \item Models~A and C prevent localization of the normal and shear components of the damage tensor (Figs.~\ref{fig:ExpwhDA}, \ref{fig:ExpwhDC}, \ref{fig:ExanDA}, \ref{fig:ExanDC}, \ref{fig:ExtsDnormalA}, \ref{fig:ExtsDnormalC}, \ref{fig:ExtsDshearA}, \ref{fig:ExtsDshearC}, \ref{fig:ExssDA}, and \ref{fig:ExssDC}). Model~B also prevents localization of the normal components of the damage tensor (Figs.~\ref{fig:ExpwhDB}, \ref{fig:ExanDB}, \ref{fig:ExtsDnormalB}, and \ref{fig:ExssDB}), but a localization of one shear component occurred in a single example (Fig.~\ref{fig:ExtsDshearB}).
  \item The damage zones obtained with model~B are thicker and, thus, dissipate more energy than the damage zones obtained with models~A and C (Figs.~\ref{fig:ExpwhD}, \ref{fig:ExanD}, \ref{fig:ExtsDnormal}, and \ref{fig:ExssD}).
  \item The consideration of isotropic damage continuously yields an overestimation of the structure's load bearing capacity (Figs.~\ref{fig:ExpwhFuIsotropic}, \ref{fig:ExanFuIsotropic}, and \ref{fig:ExssFuIsotropic}).
  \item The influence of the artificial viscosity on the regularization, the structural response, and the damage distribution is ruled out (Figs.~\ref{fig:ExpwhFuLocal}, \ref{fig:ExpwhDLocal}, \ref{fig:ExpwhFuEtav}, \ref{fig:ExpwhDiffEtav}, and \ref{fig:ExanFuDispArcl}).
\end{itemize}

\section{Conclusion}
\label{sec:conclusion}

This work investigated different gradient-extensions for tensor-valued internal variable based inelastic material models. Here, we specifically focused on the regularization of anisotropic damage at finite strains through a micromorphic gradient-extension of the damage driving force. Three different gradient-extensions with full (six micromorphic degrees of freedom) and reduced regularization (three and two micromorphic degrees of freedom) of the damage tensor were compared theoretically and numerically in the present study.

A high level of agreement was obtained between the results of the model with full regularization of all six independent components of the damage tensor and the model with a reduced regularization of the volumetric and deviatoric part of the damage tensor, which only utilizes two micromorphic degrees of freedom. Thereby, an efficient, yet effective, regularization for anisotropic damage at finite strains was identified.

The utilized anisotropic damage model features a flexible formulation that incorporates isotropic, kinematic, and distortional damage hardening and fulfills the damage growth criterion for finite strains. Therefore, it can be considered as a general inelastic local material model of a tensor-valued internal variable based formulation.

Further investigations should verify the numerical results by experimental validations and could apply the gradient-extensions to the regularization of other inelastic localizing phenomena.

\subsection*{Acknowledgements}

Funding granted by the German Research Foundation (DFG) for projects number 453715964 (RE 1057/51-1), 417002380 (CRC~280 - A01) and 453596084 (CRC~339 - B05) is gratefully acknowledged.

\bibliographystyle{agsm}
\bibliography{literature}

\end{document}